\documentclass[lineno]{jfm}

\usepackage{graphicx}
\usepackage{newtxtext}
\usepackage{newtxmath}
\usepackage{natbib}
\usepackage{hyperref}
\hypersetup{
    colorlinks = true,
    urlcolor   = blue,
    citecolor  = black,
}

\newcommand{\RomanNumeralCaps}[1]
\linenumbers

\usepackage[caption=false]{subfig}
\usepackage{multirow}
\usepackage{cases}
\usepackage{booktabs}
\usepackage{multicol}
\usepackage{booktabs}
\usepackage{tabularx}
\usepackage{anyfontsize}

\usepackage{algorithm2e,epsfig,natbib,listings,color,cleveref}
\usepackage{subfig,float}
\usepackage{adjustbox}
\usepackage{xcolor}

\usepackage{color}
\definecolor{orange}{rgb}{1,0.5,0}
\definecolor{amethyst}{rgb}{0.6,0.4,0.8}
\definecolor{aureolin}{rgb}{0.99,0.93,0.0}
\definecolor{awesome}{rgb}{1.0,0.13,0.32}
\definecolor{ao-green}{rgb}{0.0, 0.5, 0.0}

\title{Linear stability analysis of wall-bounded high-pressure transcritical fluids}

\author{Marc Bernades\aff{1}
  \corresp{\email{marc.bernades@upc.edu}},
  Francesco Capuano\aff{1}
 \and Llu\'is Jofre\aff{1}}

\affiliation{\aff{1}Department of Fluid Mechanics, Universitat Polit\`ecnica de Catalunya $\cdot$ BarcelonaTech (UPC), Barcelona 08019, Spain}

\begin{document}
\maketitle

\begin{abstract}
%All papers should feature a single-paragraph abstract of no more than 250 words which must not spill onto the seond page of the manuscript.
Mixing and heat transfer rates are typically enhanced when operating at high-pressure transcritical turbulent flow regimes. The rapid variation of thermophysical properties in the vicinity of the pseudo-boiling region can be leveraged to significantly increase the Reynolds numbers and destabilize the flow.
The underlying physical mechanism responsible for this destabilization is the presence of a baroclinic torque mainly driven by large localized density gradients across the pseudo-boiling line.
As a result, the enstrophy levels are enhanced compared to equivalent low-pressure cases, and the flow physics behavior deviates from standard wall turbulence characteristics.
In this work, the nature of this instability is carefully analyzed and characterized by means of linear stability theory.
It is found that, at isothermal wall-bounded transcritical conditions, the non-linear thermodynamics exhibited near the pseudo-boiling region propitiates the laminar-to-turbulent transition with respect to sub- and super-critical thermodynamic states.
This transition is further exacerbated for non-isothermal flows even at low Brinkman numbers.
Particularly, neutral curve sensitivity to Brinkman numbers and perturbation profiles of dynamic and thermodynamic unstable modes based on modal and non-modal analysis, which trigger the early flow destabilization, confirm this phenomenon.
Nonetheless, a non-isothermal setup is a necessary condition for transition when operating at low-Mach/Reynolds-number regimes.
In detail, on equal Brinkman number, turbulence transition is accelerated and algebraic growth enhanced in comparison to isothermal cases.
Consequently, high-pressure transcritical setups result in larger kinetic energy budgets due to larger production rates and lower viscous dissipation.
\end{abstract}


%{\bf MSC Codes }  {\it(Optional)} Please enter your MSC Codes here

\section{Introduction}  \label{sec:introduction}	

% Paragraph 1: Introduction of transcritical fluids and liase with particular microconfined DNS flow physics results (barcoclinic instability)
High-pressure transcritical fluids operate within thermodynamic spaces in which supercritical gas- and liquid-like states can be differentiated across the pseudo-boiling line~\citep{Jofre2020-A,Jofre2021-A}.
The variation of thermophysical properties across this region can be leveraged to significantly increase the Reynolds number with respect to atmospheric pressure conditions~\citep{Bernades2022-A}.
The use of high-pressure supercritical fluids is a mature field in thermo-fluid engineering as they are utilized in a wide range of applications, such as liquid rocket engines, gas turbines, and supercritical water-cooled reactors~\citep{Yoo2013-A,Jofre2021-A}.
Moreover, utilizing direct numerical simulation (DNS) approaches, their inherent capacity to achieve microconfined turbulence~\citep{Bernades2023-A} has been recently demonstrated, which enables to significantly increase mixing and heat transfer rates in microfluidic applications.
The resulting flow physics differs significantly from the typical behavior of turbulent wall-bounded flows due to the presence of localized baroclinic torques responsible for remarkably increasing flow rotation.
As a result, the flow becomes unstable and rotation is transformed into a wide range of scales (i.e., turbulent flow motions) through vortex stretching mechanisms.
However, the phenomena responsible for destabilizing the flow are still not fully characterized yet.
To this end, this work aims to conduct linear stability and transient growth analyses of relatively low-Reynolds-number wall-bounded flows at high-pressure transcritical fluid conditions to carefully identify and quantify the underlying flow mechanisms.

% Paragraph 2: LST framework, literature survey of cases of LST on this case and mention about the current linear stability theory fundamentals with the citations
Historically, the study of hydrodynamic stability in wall-bounded configurations was first established for incompressible parallel shear flows.
In this context, linear stability theory (LST) gave rise to the well-known Orr-Sommerfeld equation~\citep{Orr1907-B,Sommerfeld1908-A} and related classical modal results, such as the critical Reynolds number $Re_c = 5772.22$ for plane Poiseuille flow~\citep{Thomas1953-A,Orzag1971-A}. Alternatively, energy stability-based methods yielded $Re_c = 49.2$~\citep{Busse1969-A,Joseph1969-A}, below which no energy growth was observed~\citep{Reddy1993-A}.
Research on the instability of ideal-gas compressible flows started later.
For instance,~\cite{Malik2006-A} carried out LST selecting density and temperature as state variables along with the velocity vector and corresponding Jacobian matrices for compressible flows.
They characterized the Y-shaped spectrum [the so-called \textit{branches}~\citep{Mack1976-A}] and related even and odd modes.
Over the past decades, variable-viscosity studies consisting of stratified or Poiseuille flows with temperature dependency have been performed based on a modified set of Orr-Sommerfeld equations~\citep{Govindarajan2014-A,Potter1972-A} ignoring, however, perturbations emerging from temperature and viscosity variations.
Related to viscosity effects,~\cite{Wall1996-A} investigated the stability limits at different conditions, whereas~\cite{Malik2008-A} demonstrated that viscosity stratification improves stability in compressible Couette flow, which was later confirmed by~\cite{Saikia2017-A}.
Nonetheless, these studies were limited to either incompressible flow or ideal-gas thermodynamics with temperature-dependant transport coefficients.

% Paragraph 3: LST on compressible flows and/or real-gas frameworks
In this regard,~\cite{Ren2019b-A} have recently introduced a LST framework to study Poiseuille flows of non-ideal fluids.
This study, in particular, studied the effects of varying: (i) the dominant dimensionless numbers that characterize the flow; and (ii) the temperature of walls and bulk pressure.
The isothermal limit (low Eckert and Prandtl numbers) results presented showed a good collapse with incompressible flow reference data.
However, at larger Prandtl and Eckert numbers, the effects of large variations of the thermodynamic and transport properties driven by viscous heating of the flow became important.
In particular, the sub-/trans-/supercritical conditions cases were compared against ideal-gas scenarios. The results indicated that the base flow was modally more unstable in the subcritical regime, inviscid unstable in the transcritical regime, and significantly more stable in the supercritical regime.
Similar conclusions were extracted from algebraic growth analysis based on non-modal optimal energy growth with three-dimensional (3D) perturbations.
To that end,~\cite{Ren2019-A} extended these analyses to compressible boundary layers over adiabatic walls with fluids at supercritical pressure.
A second co-existing mode was found, causing flow destabilization when crossing the pseudo-boiling line for two-dimensional (2D) perturbations.
In particular, the mode disappeared at supercritical conditions far from the pseudo-boiling region and at subcritical pressures.
However, the effect of this mode on the value of the transition Reynolds number has not been characterized yet.

% Paragraph 4: Describe the objective problem of interest of subsonic (low Mach) at high-pressure by means of LST and DNS at different cases
Therefore, the overall objective of this work is to perform linear stability and energy amplification analyses to characterize the mechanisms driving the baroclinic instability observed in non-isothermal wall-bounded high-pressure transcritical turbulence by~\citet{Bernades2023-A}.
In particular, notably higher levels of enstrophy were identified near the hot/top wall ($50\times$ larger) in comparison to the cold/bottom wall.
In addition, the enstrophy levels near the hot/top wall were $100\times$ and $10\times$ larger than equivalent laminar (iso-volumetric input power) and turbulent (iso-friction Reynolds number at cold/bottom wall) low-pressure systems, respectively.
Likewise,~\cite{Sahu2010-A} and~\cite{Srivastava2017-A} have explored the stability of low-pressure non-isothermal channel flow with viscous heating and found that increasing temperature across walls destabilized the flow. Nevertheless, these investigations were limited to incompressible flow conditions utilizing the Boussinesq approximation.
Furthermore, to identify optimum perturbation profiles, which yield turbulence enhancement and/or skin-friction drag reduction, modal and non-modal analyses are necessary.
In this regard, focusing on temporal signals of incompressible Poiseuille flow,~\cite{Massaro2023-A} explored the application of flow control techniques based on spanwise forcing.
Typically, to achieve large destabilization levels and consequently enhance turbulence intensity in isothermal wall-bounded flows, the base flow requires large Brinkman numbers.
This, however, demands high-speed flows which are unfeasible in, for example, microconfined applications. Nonetheless, at non-isothermal conditions, the system can operate at low velocities (and Brinkman numbers), while still exhibiting the destabilization benefits of strong thermodynamic gradients occurring across the pseudo-boiling region.
To this extent, following recent efforts~\citep{Bernades2024c-A}, this work is particularly focused on assessing the effects of sub-, trans- and supercritical thermodynamic regimes at isothermal and non-isothermal conditions operating at low Reynolds and Mach numbers.
In particular, with the aim of exploring the instability at different wall temperatures and bulk pressures, the analyses consider results in terms of spectrum, unstable modes, range of instability, optimal perturbations, and kinetic energy budget.
Moreover, transient growth rates are also determined to quantify energy-based transient destabilization regimes~\citep{Schmid2007-A,Schmid2014-A}. 

% Paragraph 5: Structure
The paper, thus, is organized as follows.
First, in Section~\ref{sec:flow_modeling}, the flow physics modeling is introduced.
Next, the linear stability theory, linearized equations of supercritical fluids, and discretization methods utilized are described in Section~\ref{sec:LST}.
The flow cases assessed by means of linear stability and modal \& non-modal analyses are presented and discussed in Section~\ref{sec:results}.
Finally, Section~\ref{sec:conclusions} provides concluding remarks and proposes future research directions.

\section{Flow Physics Modeling}  \label{sec:flow_modeling}	

The framework utilized in terms of (i) equations of fluid motion, (ii) real-gas thermodynamics, and (iii) high-pressure transport coefficients is briefly introduced below.

\subsection{Equations of fluid motion}  \label{sec:compressible_eq}	

% Equations of Fluid motion
The flow motion of supercritical fluids is described by the following set of dimensionless equations of mass, momentum, and total energy
\begin{align}
 \frac{\partial \rho^\star}{\partial t^\star} + \nabla^\star \cdot  \left( \rho^\star \mathbf{u}^\star \right) & = 0,    \label{eq:mass} \\
 \frac{\partial \left( \rho^\star \mathbf{u}^\star \right)}{\partial t^\star} + \nabla^\star \cdot  \left( \rho^\star \mathbf{u}^\star  \mathbf{u}^\star \right) & = -\nabla^\star P^\star + \frac{1}{{Re}} \nabla^\star \cdot \boldsymbol{\tau^\star} + \mathbf{F}^\star,  \label{eq:momentum} \\
 \frac{\partial \left( \rho^\star E^\star \right) }{\partial t^\star} + \nabla^\star \cdot  \left( \rho^\star \mathbf{u}^\star E^\star  \right) & = -\nabla^\star \cdot  (P^\star \mathbf{u}^\star)  - \frac{1}{{Re} {Br}} \nabla^\star \cdot \mathbf{q}^\star  \label{eq:energytransport} \\
 & \quad + \frac{1}{{Re}}\nabla^\star \cdot (\boldsymbol{\tau^\star} \cdot \mathbf{u}^\star) + \mathbf{F}^\star \mathbf{u}^\star, \nonumber 
\end{align}
where superscript $(\cdot)^\star$ denotes dimensionless quantities, $\rho$ is the density, $\mathbf{u}$ is the velocity vector, $P$ is the pressure, $\boldsymbol{\tau} = \mu ( \nabla \mathbf{u} + \nabla \mathbf{u}^{T} ) + \lambda(\nabla \cdot \mathbf{u})\mathbf{I}$ is the viscous stress tensor with $\mu$ the dynamic viscosity, $\lambda = - 2/3 \mu$ the bulk viscosity and $\mathbf{I}$ the identity matrix, $E = e + |\mathbf{u}|^{2}/2$ is the total energy with $e$ the internal energy, $\mathbf{q} = - {\kappa} \nabla T$ is the Fourier heat flux with $\kappa$ the thermal conductivity and $T$ the temperature, and $\mathbf{F}$ is a body force introduced to drive the flow in the streamwise direction.

The resulting set of scaled equations includes two dimensionless numbers: (i) the Reynolds number $Re = \rho_{b}U_{r}\delta/\mu_{b}$, where subscript $b$ refers to bulk quantities, $\delta$ is the channel half-height and $U_r$ is the reference streamwise velocity corresponding to its maximum value (i.e., centerline velocity for isothermal conditions), characterizing the ratio between inertial and viscous forces; and (ii) the Brinkman number $Br = \mu_b U_{r}^{2} / (\kappa T_b) = Pr Ec$, quantifying the ratio of viscous heat generation to external heating through the walls (viz. larger $Br$ values correspond to smaller heat conduction from viscous dissipation resulting in temperature increase).
The Brinkman number can also be expressed as the combination of Prandtl number $Pr =\mu_{b} {c_{P}}_{b} / \kappa_{b}$, where $c_{P}$ is the isobaric heat capacity, expressing the ratio between momentum and thermal diffusivity, and Eckert number $Ec = U_{r}^{2}/({c_{P}}_{b} T_{b})$, which accounts for the ratio between advective mass transfer and heat dissipation potential.
In this work, the Froude number, which represents the ratio between inertial and gravitational forces, is assumed to be large, and consequently buoyancy effects are not considered.
The derivation of these dimensionless equations is based on the following set of inertial-based scalings~\citep{Jofre2020b-A,Jofre2023-A}
\begin{align}
 \textbf{x}^{\star} & = \frac{\textbf{x}}{\delta}, \quad \textbf{u}^{\star} = \frac{\textbf{u}}{U_{r}}, \quad \rho^{\star} = \frac{\rho}{\rho_{b}}, \quad T^{\star} = \frac{T}{T_{b}}, \quad P^{\star} = \frac{P}{\rho_{b} U_{r}^{2}}, \nonumber \\ 
 & E^{\star} = \frac{E}{U_{r}^{2}}, \quad \mu^{\star} = \frac{\mu}{\mu_{b}}, \quad \kappa^{\star} = \frac{\kappa}{\kappa_{b}}, \quad \boldsymbol{F}^{\star} = \frac{\boldsymbol{F} U_{r} {\rho_b}^2}{\delta},\label{eq:scalings}
\end{align}
where $\textbf{x}$ is position and $H = 2 \delta$ is the total channel height.

\subsection{Equation of state}  \label{sec:Peng_Robinson}	

The CoolProp open-source library~\citep{Bell2014-A} is used in this work to describe the non-ideal thermodynamic behavior of high-pressure transcritical fluids.
In detail, the thermodynamic quantities are derived from the Helmholtz energy equation of state; their validation against NIST reference data~\citep{NIST-M} and sensitivity of the linear stability results to different thermodynamic frameworks (including ideal-gas) is covered in Appendix~\ref{sec:Appendix_A}.
In general form, it can be expressed in terms of the compressibility factor $Z = P/(\rho R^{\prime} T)$, where $R^{\prime}$ is the specific gas constant.
In dimensionless form, the equation of state reads
\begin{equation}
  P^{\star} = \frac{Z \rho^{\star} T^{\star}}{\hat{\gamma} M\!a_{b}^{2}},    \label{eq:real-gas_state_equation} 
\end{equation}
where $\hat{\gamma} \approx Z (c_{P} / c_{V}) [(Z + T (\partial Z / \partial T)_\rho)/(Z+T(\partial Z / \partial T)_P)]$ is an approximated real-gas heat capacity ratio~\citep{Firoozabadi2016-B} with $c_{V}$ the isochoric heat capacity.
As it can be noted, the dimensionless bulk Mach number $M\!a_{b} = u_b / c_b$ appears, where $c_b$ is the bulk speed of sound, which represents the ratio of flow velocity to the local speed of sound.
Real-gas equations of state need to be supplemented with the corresponding high-pressure thermodynamic variables (e.g., internal energy, heat capacities) based on departure functions~\citep{Reynolds2019-B} calculated as a difference between two states.
These functions operate by transforming the thermodynamic variables from ideal-gas conditions (low pressure - only temperature dependant) to supercritical conditions (high pressure).
The ideal-gas components are calculated by means of the NASA 7-coefficient polynomial~\citep{Burcat2005-TR}, while the analytical departure expressions to high pressures are derived from the Peng-Robinson equation of state as detailed, for example, in~\citet{Jofre2021-A}.

% High pressure coefficients
\subsection{High-pressure transport coefficients}  \label{sec:transport_coefficients}

The high pressures involved in the analyses conducted in this work prevent the use of simple relations for the calculation of dynamic viscosity $\mu$ and thermal conductivity $\kappa$.
Therefore, these quantities are also modeled by means of the CoolProp open-source library~\citep{Bell2014-A}.
Alternatively, standard methods for computing these coefficients for Newtonian fluids are based on the correlation expressions proposed by \citet{Chung1984-A,Chung1988-A}.
These correlations are mainly functions of the critical temperature $T_c$ and density $\rho_c$, molecular weight $W$, acentric factor $\omega$, association factor $\kappa_a$ and dipole moment $\mathcal{M}$, and the output from the NASA 7-coefficient polynomial~\citep{Burcat2005-TR}; details can be found in dedicated works like, for example,~\citet{Jofre2021-A} and~\citet{Poling2001-B}.
In this regard, similarly to the thermodynamic quantities, the differences in modeling the transport coefficients between CoolProp, NIST and Chung \textit{et al.} correlations are presented in Appendix~\ref{sec:Appendix_A}.

% Linear stability theory
\section{Linear stability theory}  \label{sec:LST}	

The following subsections describe the linearized stability equations resulting from the flow model presented above, and the corresponding discretization method utilized to compute the results.

\subsection{Linearized stability equations}

The flow field can be decomposed into a base state and a perturbation denoted with subscripts $(\cdot)_0$ and superscript $(\cdot)^\prime$ respectively, yielding
\begin{equation}{}
  \mathbf{q} = \mathbf{q}_0 + \mathbf{q}^\prime, \label{eq:Decomposition}
\end{equation}
where the vector $\mathbf{q}$ is composed of 5 variables: $3$ velocity components ($u$, $v$ and $w$), and $2$ independent thermodynamic variables ($\rho$ and $T$).
This decomposition yields a perturbation vector $\mathbf{q}^\prime = (\rho^\prime, u^\prime, v^\prime, w^\prime, T^\prime)^T$ superimposed to a base-flow vector, which is assumed to be parallel to the walls, and consequently only function of the wall-normal direction $y$ in the form $\mathbf{q}_0 = [\rho_0(y), u_0(y), 0, 0, T_0(y)]^T$; the derivation of the reduced set of equations is detailed in Appendix~\ref{sec:Appendix_B}.
From this point forward, the superscript $(\cdot)^\star$ is omitted and it is assumed that all equations are in dimensionless form for the sake of exposition clarity.

The selection of the two main thermodynamic variables $\rho$ and $T$ allows the perturbations of the remaining thermodynamic variables ($E, P, \mu, \kappa)$ to be expressed as a function of this pair of quantities by means of a Taylor expansion with respect to the base flow.
For example, by neglecting higher-order terms, the pressure perturbation can be approximated by expanding the base flow pressure as
\begin{equation}
P^\prime \approx \left.\frac{\partial P_0}{\partial \rho_0}\right\vert_{T_0} \rho^\prime + \left.\frac{\partial P_0}{\partial T_0}\right\vert_{\rho_0} T^\prime,
\end{equation}
with the resulting error of the approximation to be of the order of the amplitude ($\epsilon$) of the infinitesimal perturbations~\citep{Alves2016-A}.
Thus, by substituting Eq.~\ref{eq:Decomposition} into the dimensionless equations of fluid motion (Eqs.~\ref{eq:mass}-\ref{eq:energytransport}), the linear stability equations are derived as a function of the perturbation vector $\mathbf{q}^\prime$ and can be cast in compact form as
\begin{align}
     \mathbf{L_t} \frac{\partial\mathbf{q}^\prime}{{\partial t}} & + \mathbf{L_x} \frac{\partial\mathbf{q}^\prime}{{\partial x}} + \mathbf{L_y} \frac{\partial\mathbf{q}^\prime}{{\partial y}} + \mathbf{L_z} \frac{\partial\mathbf{q}^\prime}{{\partial z}} +  \mathbf{L_q} \mathbf{q}^\prime \label{eq:LST} \\
     & + \mathbf{V_{xx}} \frac{\partial^2 \mathbf{q}^\prime}{{\partial x^2}} +  \mathbf{V_{yy}} \frac{\partial^2 \mathbf{q}^\prime}{{\partial y^2}} +  \mathbf{V_{zz}} \frac{\partial^2 \mathbf{q}^\prime}{{\partial z^2}} + 
      \mathbf{V_{xy}} \frac{\partial^2 \mathbf{q}^\prime}{{\partial x \partial y}} + \mathbf{V_{xz}} \frac{\partial^2 \mathbf{q}^\prime}{{\partial x \partial z}} + \mathbf{V_{yz}} \frac{\partial^2 \mathbf{q}^\prime}{{\partial y \partial z}} = 0, \nonumber
\end{align}
where all the nonlinear terms have been neglected and $\mathbf{L_t}$, $\mathbf{L_x}$, $\mathbf{L_y}$, $\mathbf{L_z}$, $\mathbf{L_q}$, $\mathbf{V_{xx}}$, $\mathbf{V_{yy}}$, $\mathbf{V_{zz}}$, $\mathbf{V_{xy}}$, $\mathbf{V_{xz}}$ and $\mathbf{V_{yz}}$ correspond to the Jacobian matrices of the base flow and thermophysical properties.
In detail, these matrices are of size $5\times5$, corresponding to each field of the perturbation vector $\mathbf{q}^\prime$. The components of these matrices can be obtained by inspection from the resulting linear stability equations presented in Appendix~\ref{sec:Appendix_C}. In detail, Eq.~\ref{eq:LST_continuity} for continuity, Eqs.~\ref{eq:LST_momentum_X}-\ref{eq:LST_momentum_Z} for momentum, and Eq.~\ref{eq:LST_energy} for internal energy.

In linear modal stability analysis, the perturbation is assumed to have the ansatz
\begin{equation}
    \mathbf{q}^\prime(x,y,z,t) = \hat{\mathbf{q}}(y) e^{(i \alpha x + i \beta z - i \omega t)} + \textrm{c.c.}, \label{eq:LST_normalmode}
\end{equation}
where $\alpha$ and $\beta$ are the  streamwise and spanwise wavenumbers, respectively, while $\omega$ is the temporal frequency whose real and imaginary parts correspond, respectively, to the wall-normal angular temporal frequency and its local growth rate, and c.c. stands for complex conjugate.
Hence, by substituting Eq.~\ref{eq:LST_normalmode} into Eq.~\ref{eq:LST}, the following eigenvalue problem is obtained:
\begin{eqnarray}
    & (- i \omega \mathbf{L_t} + i \alpha \mathbf{L_x} + \mathbf{L_y} D + i \beta \mathbf{L_z} + \mathbf{L_q} +\label{eq:LST_eigenvalue} \\
    & - \alpha^2 \mathbf{V_{xx}} + i \alpha  \mathbf{V_{xy}} D - \alpha\beta \mathbf{V_{xz}} +  \mathbf{V_{yy}} D^2 + i \beta  \mathbf{V_{yz}} D - \beta^2  \mathbf{V_{zz}})\hat{\mathbf{q}} = 0, \nonumber
\end{eqnarray}
where $D \approx d/dy$ is the derivative operator based on the Chevyshev discretization presented in Section~\ref{sec:LST_discretization}.
Equation \ref{eq:LST_eigenvalue} can be solved either on the temporal or spatial domain as
\begin{equation}
\left. \begin{array}{ll}
\displaystyle\mathbf{L_T} \hat{\mathbf{q}} = \omega \mathbf{L_t} \hat{\mathbf{q}},\\[8pt]
\displaystyle \mathbf{L_S} \hat{\mathbf{q}} = \alpha (\beta \mathbf{V_{xz}} 
 - i  \mathbf{V_{xy}} D - i  \mathbf{L_x})\hat{\mathbf{q}} + \alpha^2  \mathbf{V_{xx}} \hat{\mathbf{q}}, \label{eq:Eigenvalue_equations}
 \end{array}\right\}
\end{equation}
where, by inspection from Eq.~\ref{eq:LST_eigenvalue}, the operator corresponding to the temporal ($\mathbf{L_T}$) and spatial ($\mathbf{L_S}$) matrices can be written as
\begin{align}
  \mathbf{L_T} & = \alpha \mathbf{L_x} - i \mathbf{L_y} D + \beta \mathbf{L_z} - i \mathbf{L_q} \\
 & + i \alpha^2 \mathbf{V_{xx}} + \alpha  \mathbf{V_{xy}} D + i \alpha\beta \mathbf{V_{xz}} - i  \mathbf{V_{yy}} D^2 + \beta  \mathbf{V_{yz}} D + i \beta^2  \mathbf{V_{zz}}, \nonumber \\
  \mathbf{L_S} & = - i \omega \mathbf{L_t} + \mathbf{L_y} D + i \beta \mathbf{L_z} + \mathbf{L_q} +  \mathbf{V_{yy}} D^2 + i \beta  \mathbf{V_{yz}} D - \beta^2  \mathbf{V_{zz}}.
\end{align}
For wall-bounded Poiseuille flow, the temporal problem is typically considered, and consequently the streamwise $\alpha$ and spanwise $\beta$ wavenumbers are prescribed, and the problem is solver for the eigenvalue $\omega$, with its real and imaginary parts corresponding, respectively, to the wall-normal wavenumber and its local growth rate.

\subsection{Discretization method} \label{sec:LST_discretization}

The discretization of the linearized equations is based on Chevyshev collocation~\citep{Trefethen2000-B} with a domain spanning the interval $0 \leq y/\delta \leq 2$ and  discretized as
\begin{equation}
    y_j = \delta \left( 1 - \textrm{cos} \frac{\pi j}{N} \right), \quad j = 0, \dots , N,
\end{equation}
where $N$ corresponds to the total number of collocation points.
In this regard, Chevyshev differentiation matrices are utilized to obtain the discretized equations and define the LST eigenvalue problem operators.
Particularly, the mesh size selected for this work is $N = 200$, which provides grid-independent results based on the convergence of the S-shaped Mack branches; the error scales with $\mathcal{O}(h^2)$ where $h = H / N$, in particular, further increasing the grid size by $50\%$ improves the accuracy by $\mathcal{O}(10^{-4})$; for brevity of exposition, the corresponding grid-convergence results are not shown in this paper. 
Moreover, the system of equations is subjected to $u^\prime = v^\prime  = w^\prime  = T^\prime = 0$ boundary conditions for both walls.

\subsection{Algebraic non-modal stability} \label{sec:algebraic_stability}

The operator $L_T$ in Eq.~\ref{eq:Eigenvalue_equations} is a non-normal matrix from which non-orthogonal eigenvectors and transient energy growth are also obtained.
The individual eigenmodes of the modal system describe the behavior of disturbances at a large asymptotic time, but they fail to capture the short-time dynamics.
As a result, the linear superposition of the non-orthogonal eigenvectors exhibits substantial energy growth for a short period of time~\citep{Thumar2024-A}.
Thus, non-modal stability theory, i.e., algebraic stability, based on the matrix exponential is used to capture the entire perturbation dynamics.
Particularly, for algebraic stability analysis, an energy norm needs to be defined.
In this regard, it is known that for compressible flows, the Chu norm~\citep{Chu1965-A,Hanifi1996-A} is a well-suited candidate, which is defined as
\begin{equation}
    E(q) = \int \left[  ({u^\prime}^\dagger {u^\prime} + {v^\prime}^\dagger {v^\prime} + {w^\prime}^\dagger {w^\prime}) + m_\rho {\rho^\prime}^\dagger {\rho^\prime} + m_T {T^\prime}^\dagger {T^\prime} \right]  dy, \label{eq:E_norm}
\end{equation}
where $(\cdot)^\dagger$ denotes complex conjugate.
For ideal-gas thermodynamics, selecting the Mack's energy norm with $m_\rho = T_0 / (\rho_0^2 \gamma {M\!a}^2)$ and $m_T = 1/(\gamma(\gamma - 1 )T_0 {M\!a}^2)$ is a common choice.
However, it has been recently proposed that for non-ideal fluids the results are more robust when $m_\rho = m_T = 1$~\citep{Ren2019b-A}.
Hence, $E(\mathbf{q})$ represents the disturbance energy subjected to the eigenvector basis obtained from modal stability.
Therefore, based on this norm, the algebraic growth rates and optimal energy amplification can be computed by performing a singular value decomposition (SVD) of the following quantity
\begin{equation}
 G(t) = \text{max}_{\mathbf{q_0}} \frac{E[\mathbf{q}(t)]}{E(\mathbf{q_0})}.
%\left. \begin{array}{ll}
%\displaystyle G(t) = \text{max}_{\mathbf{q_0}} \frac{E(\mathbf{q}(t))}{E(\mathbf{q_0})}, \\[8pt]
%\displaystyle G(x) = \text{max}_{\mathbf{q_0}} \frac{E(\mathbf{q}(x))}{E(\mathbf{q_0})}.
% \end{array}\right\}
\end{equation}
It is important to highlight that, to increase the robustness and efficiency of the computations, when performing the SVD it is beneficial to exclude growth modes, i.e., eigenvalues, that could lead to non-transient growth~\citep{Hanifi1996-A}.

\section{Results \& Discussion}  \label{sec:results}

The presentation and discussion of the results are covered in this section.
The flow cases studied are first introduced.
Next, the modal stability analyses for iso- and non-isothermal conditions are investigated.
Finally, algebraic growth analyses are provided for 2D and 3D perturbations with the corresponding optimal inputs and responses.

\subsection{Flow cases} \label{sec:flow_cases}

\begin{figure*}
	\centering
	{\includegraphics[width=0.9\linewidth]{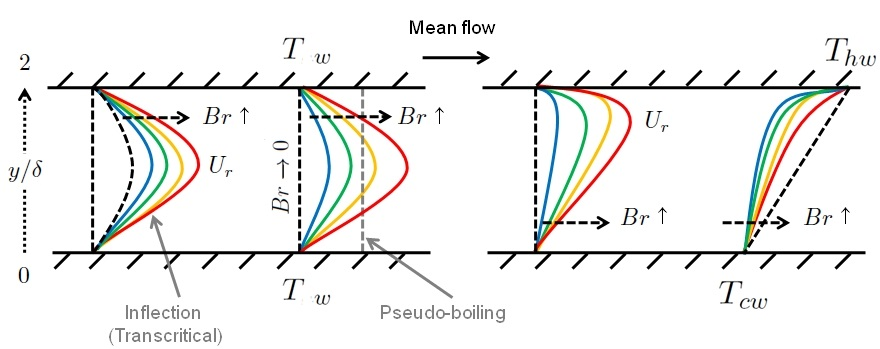}} 
	\caption{Sketch of the Pouseuille flow for iso- (left) and non-isothermal (right) cases.} 
 \label{fig:Sketch_Poiseuille_Flow}
\end{figure*}

\begin{figure*}
	\centering
	\subfloat[\vspace{-8mm}]{\includegraphics[width=0.488\linewidth]{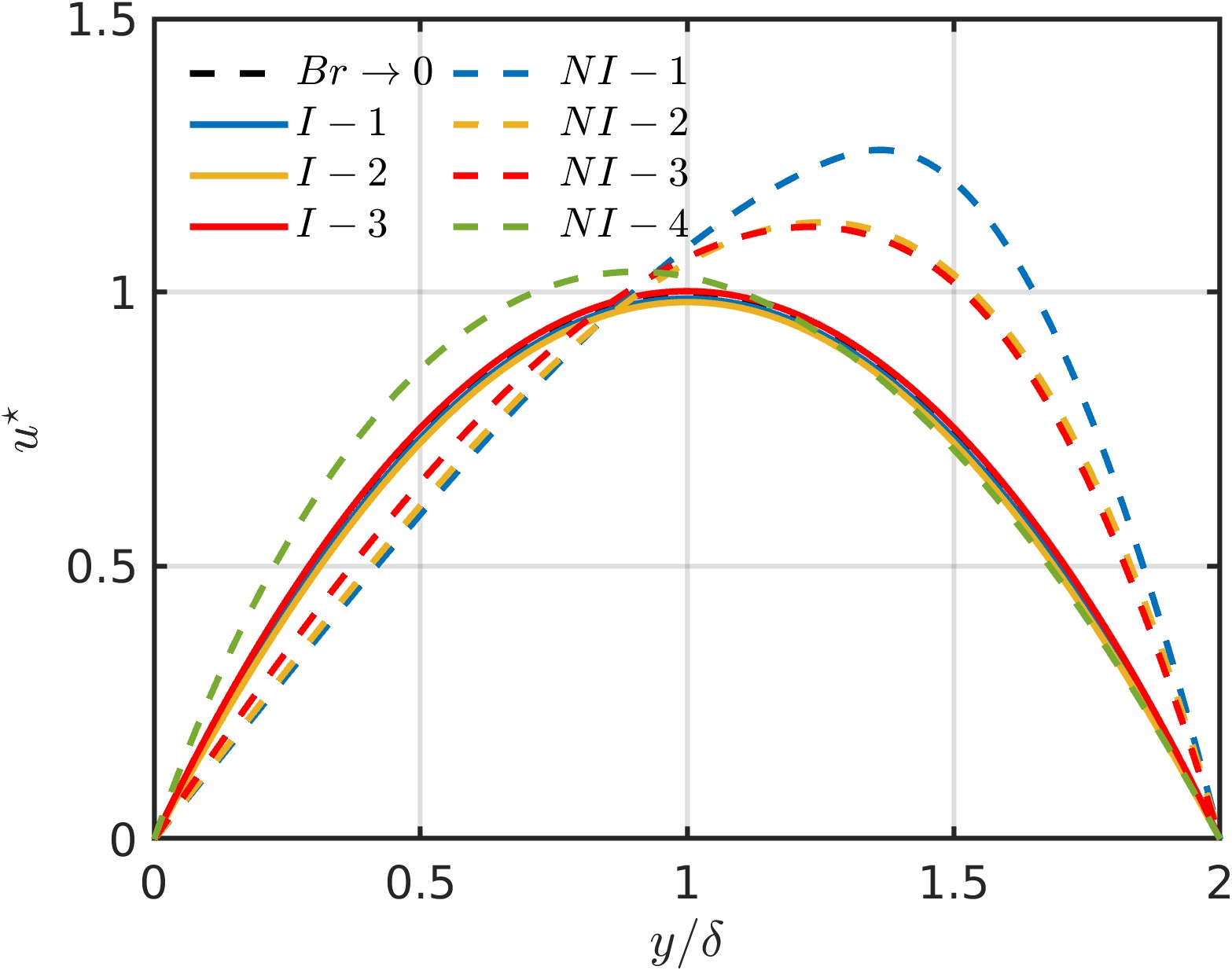}} \hspace{1mm}
    \subfloat[\vspace{-8mm}]{\includegraphics[width=0.492\linewidth]{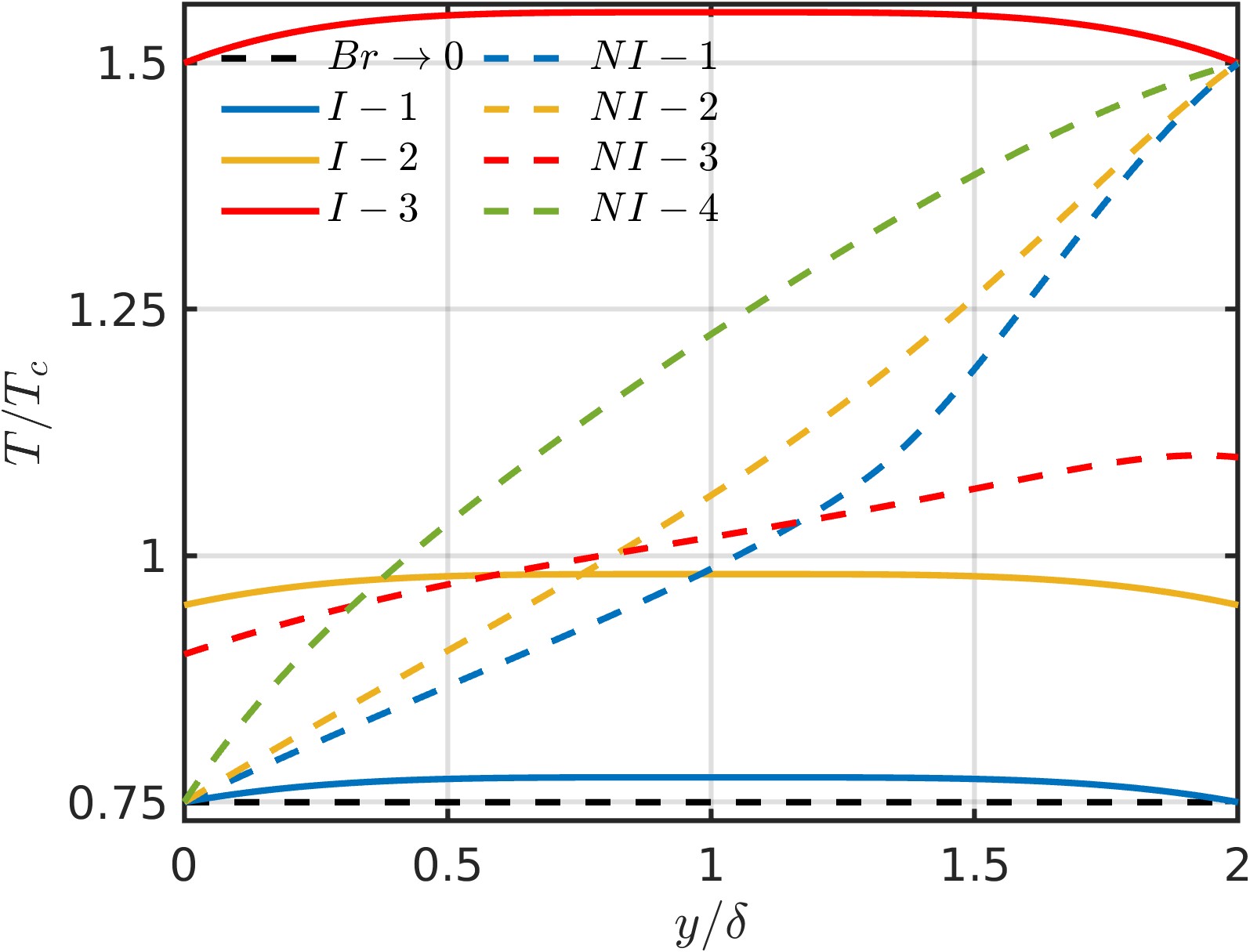}} \\ \vspace{5mm}
	\caption{Base flows profiles of the isothermal cases listed in Table~\ref{tab:summary_flowcases} for dimensionless streamwise velocity (a) and reduced temperature (b) as a function of wall-normal direction.} 
 \label{fig:baseflow_flow_cases}
\end{figure*}

As mathematically introduced in Section~\ref{sec:LST}, wall-normal instabilities are studied by means of a Poiseuille flow to isolate oblique 3D effects from the analyses.
In detail, as depicted in the sketch of Figure~\ref{fig:Sketch_Poiseuille_Flow}, the Poiseuille flow will be studied with two different base configurations: (i) isothermal conditions in which sub-, trans- or supercritical regimes are controlled by the bulk velocity and present a symmetric temperature profile with maximum at the centerline of the channel; and (ii) non-isothermal conditions by imposing a temperature difference between walls which enforces the fluid to operate within transcritical trajectories.
The complete list of cases considered and their flow parameters are provided in Table~\ref{tab:summary_flowcases}.
Particularly, Figure~\ref{fig:baseflow_flow_cases} shows the converged base flows for the iso- and non-isothermal cases studied. Details about the calculation of the base flows can be found in Appendix~\ref{sec:Appendix_B}, while a careful verification of the results at the isothermal limit is reported in Appendix~\ref{sec:Appendix_D}.
It is important to highlight that case NI-5 has been designed to mimic the operation conditions of the microconfined high-pressure transcritical flow studied utilizing DNS approaches by~\citep{Bernades2022-A}, and consequently the Brinkman number is adjusted to $Br = 5.6 \cdot 10^{-6}$ to obtain $U_r = 1\thinspace\textrm{m/s}$.
Subsequently, case NI-6 is the equivalent setup at low pressure, which is known to be steady and laminar.

\begin{table}
\renewcommand{\arraystretch}{1.5}

%\renewcommand{\arraystretch}{1.0}
%\begin{ruledtabular}
  % \centering 
\begin{tabular}{c|c|c|cc|c|c|c} \hline
\centering
Flow case &  Regime & Label & $T_{cw}/T_c$ & $T_{hw}/T_c$ & $P_b/P_c$ & $Br$ & $M\!a$\\
\hline
\multirow{1}{*}{Verification} & Superheated steam & $\text{V-1}^{(\star)}$  & 0.95 & 0.95 & 1.08 & $10^{-5}$ & $4 \cdot 10^{-3}$\\ \hline
\multirow{3}{*}{Isothermal} & High-pressure liquid-like & I-1  & 0.75 & 0.75 & 1.5 & $\le 5 \cdot 10^{-1}$ & $\le 0.45$\\
 & High-pressure transcritical & I-2  & 0.95 & 0.95 & 1.5 & $\le 5 \cdot 10^{-1}$ & $\le 1.37$\\
 & High-pressure gas-like & I-3  & 1.5 & 1.5 & 1.5 & $\le  5 \cdot 10^{-1}$  & $\le 1.35$\\ \hline
\multirow{4}{*}{Non-isothermal} & High-pressure transcritical & NI-1 & 0.75 & 1.5  & $1.5$  & $\le 10^{-1}$ & $\le 0.33$\\
 & High-pressure transcritical & NI-2  & 0.75 & 1.5 & 5  &  $\le 10^{-1}$ & $\le 0.24$\\
 & High-pressure transcritical & NI-3  & 0.9 & 1.1 & 1.5   & $\le 10^{-1}$ & $\le 0.44$ \\
 & Superheated steam & NI-4   & 0.75 & 1.5 & 0.03 & $\le 10^{-1}$ & $\le 0.58$ \\ 
 & High-pressure transcritical & NI-5  & 0.75 & 1.5 & 2.0  &  $5.6 \cdot 10^{-6}$ & $2.1 \cdot 10^{-3}$ \\ 
 & Superheated steam & NI-6  & 0.75 & 1.5 & 0.03  &  $5.6 \cdot 10^{-6}$ & $4.3 \cdot 10^{-3}$ \\ \hline
\end{tabular}
%\end{ruledtabular}
\caption{Base flow cases studied utilizing linear stability theory. The first group of cases corresponds to symmetric Poiseuille flows with isothermal walls, whereas the second group considers non-isothermal cases with different cold ($cw$) and hot ($hw$) wall temperatures. Note: $\text{V-1}^{(\star)}$ is covered for isothermal limit analysis and verification in Appendix~\ref{sec:Appendix_D}.} \label{tab:summary_flowcases}

\end{table}

\subsection{Temporal modal stability analysis}  \label{sec:Modal_stability}	

This subsection aims at quantifying flow destabilization when operating at high-pressure transcritical fluid conditions in comparison to sub- and super-critical thermodynamic regimes under 2D perturbations.
In particular, the LST studies will consider cases with different base flows obtained from varying the Reynolds and Brinkman numbers as defined in Table~\ref{tab:summary_flowcases}.

\subsubsection{Isothermal cases}  \label{sec:results_isothermal}

\begin{figure*}
	\centering
	\subfloat[\vspace{-8mm}]{\includegraphics[width=0.488\linewidth]{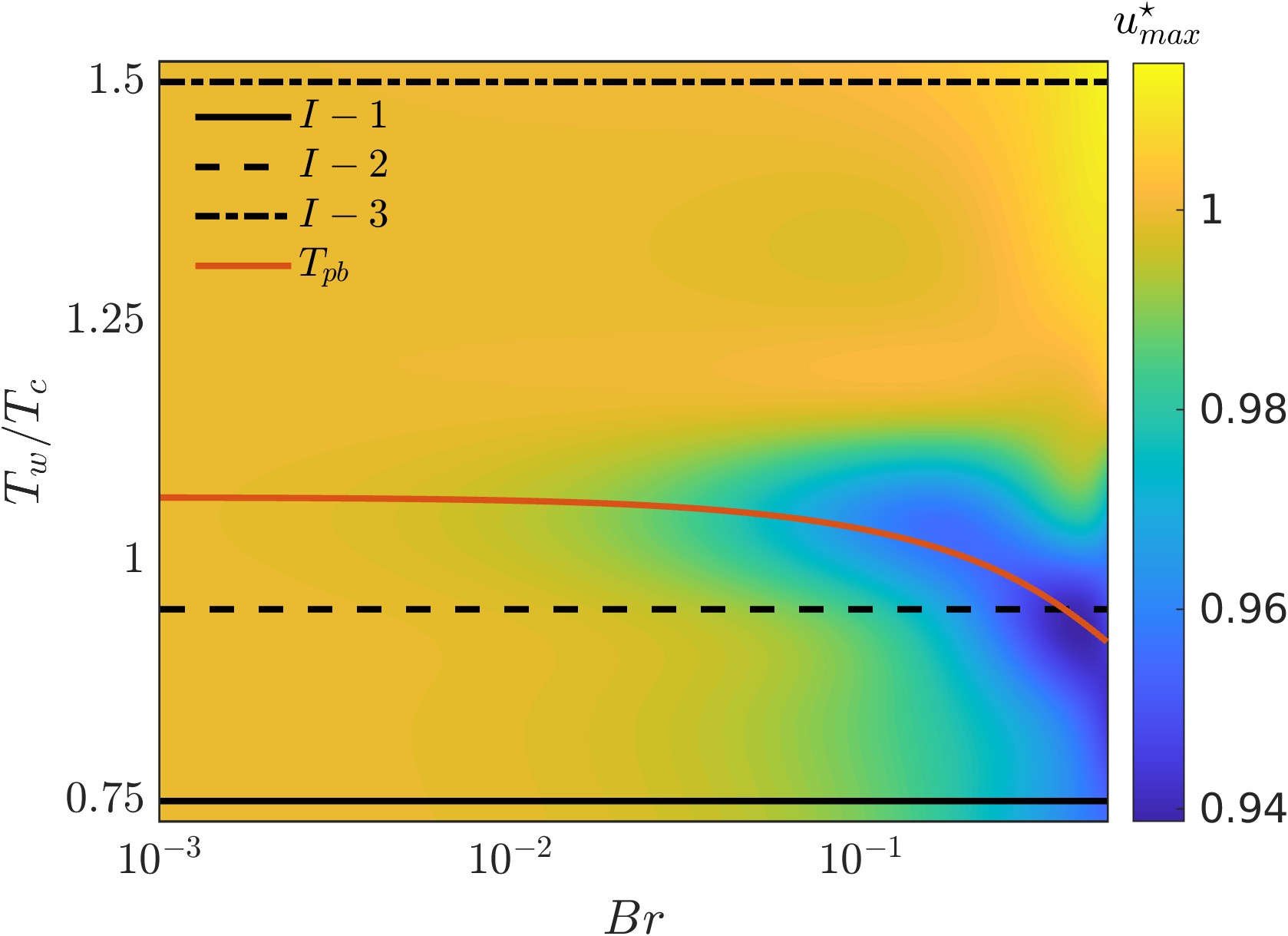}} \hspace{1mm}
    \subfloat[\vspace{-8mm}]{\includegraphics[width=0.492\linewidth]{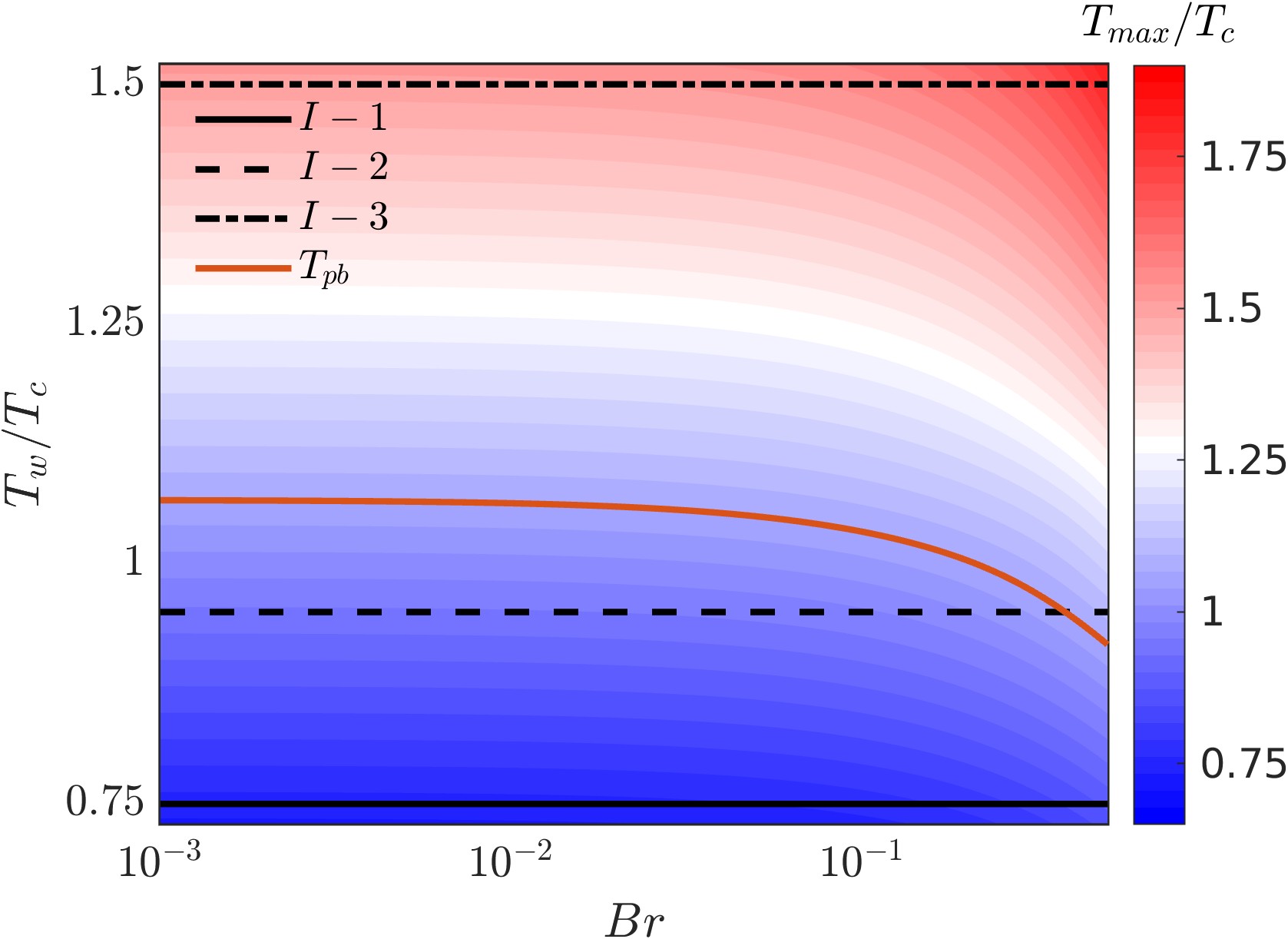}} \\ \vspace{5mm}
	\caption{Base flow in terms of maximum normalized streamwise velocity (a) and temperature (b) as a function of reduced wall temperature and Brinkman number for the cases listed in Table~\ref{tab:summary_flowcases}.} 
 \label{fig:Thermodynamic_operation_mapping_isothermal}
\end{figure*}

% Paragraph 1: Thermodynamic state
For the different isothermal cases studied (labeled I-1, I-2 and I-3 in Table~\ref{tab:summary_flowcases}), the maximum normalized streamwise velocity and temperature of the base flows considered as a function of reduced wall temperature and Brinkman number are depicted in Figure~\ref{fig:Thermodynamic_operation_mapping_isothermal} (represented by solid, dashed and dashed-dotted black curves).
The first observation to highlight is that, in the vicinity of the critical temperature at relatively large Brinkman numbers, the maximum normalized streamwise velocity is slightly reduced.
However, it is important to note that, even if the normalized streamwise velocity decreases, the absolute streamwise velocity is characterized by large values.
The second observation is that, for the isothermal setups, the only case operating across the pseudo-boiling region is the I-2 configuration.

\begin{figure*}
	\centering
	\subfloat[\vspace{-8mm}]{\includegraphics[width=0.33\linewidth]{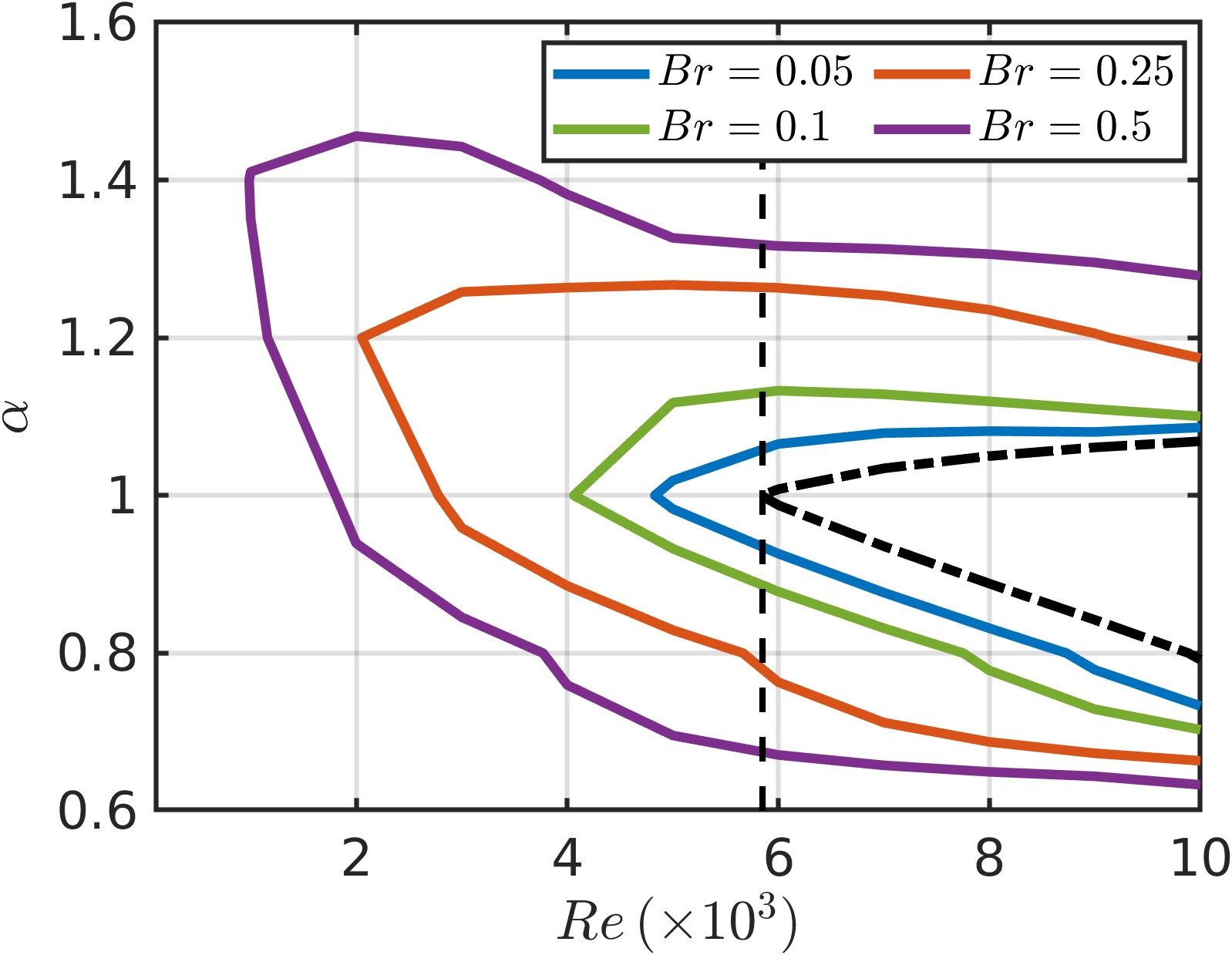}} \hfill
    \subfloat[\vspace{-8mm}]{\includegraphics[width=0.335\linewidth]{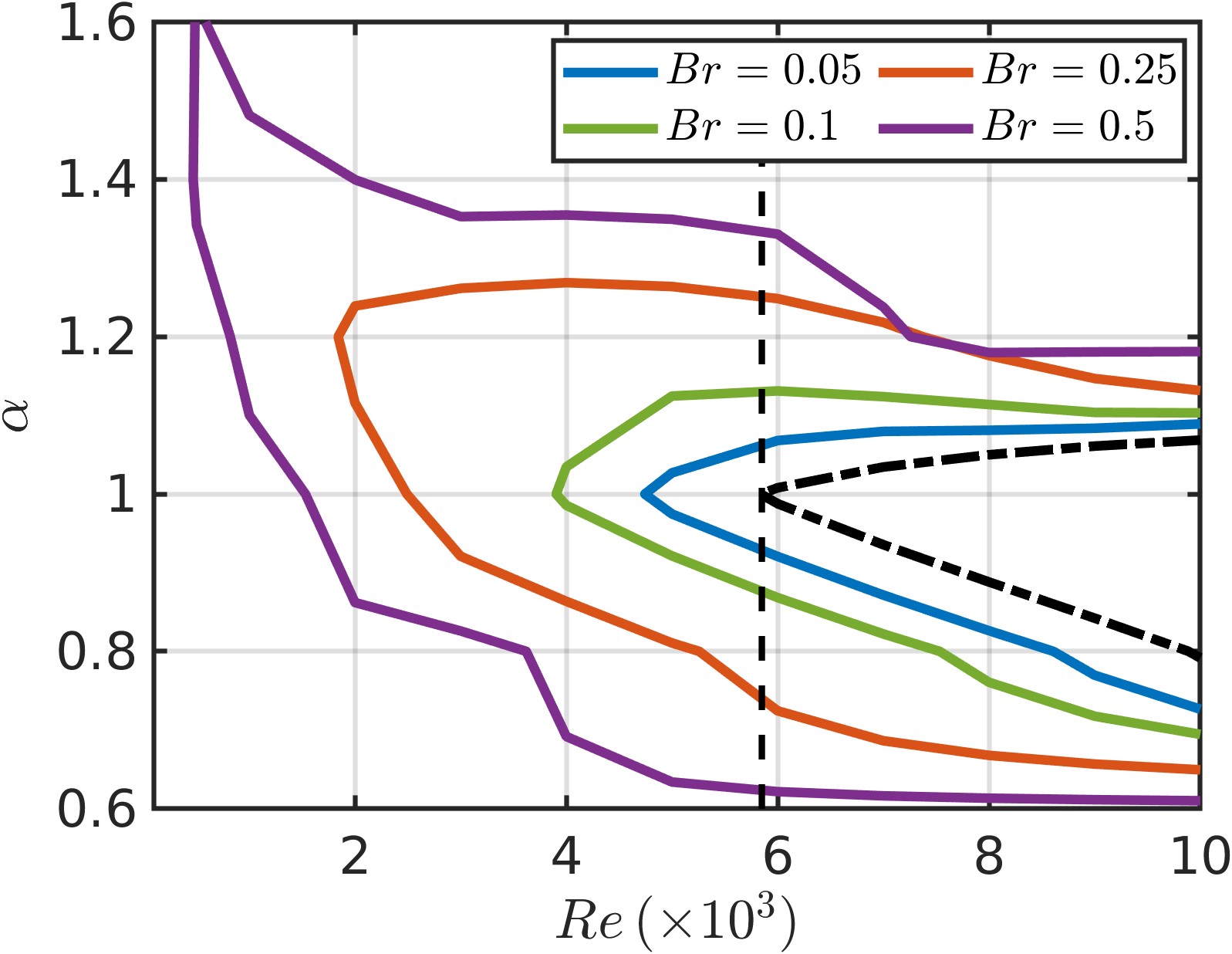}} \hfill
    \subfloat[\vspace{-8mm}]{\includegraphics[width=0.33\linewidth]{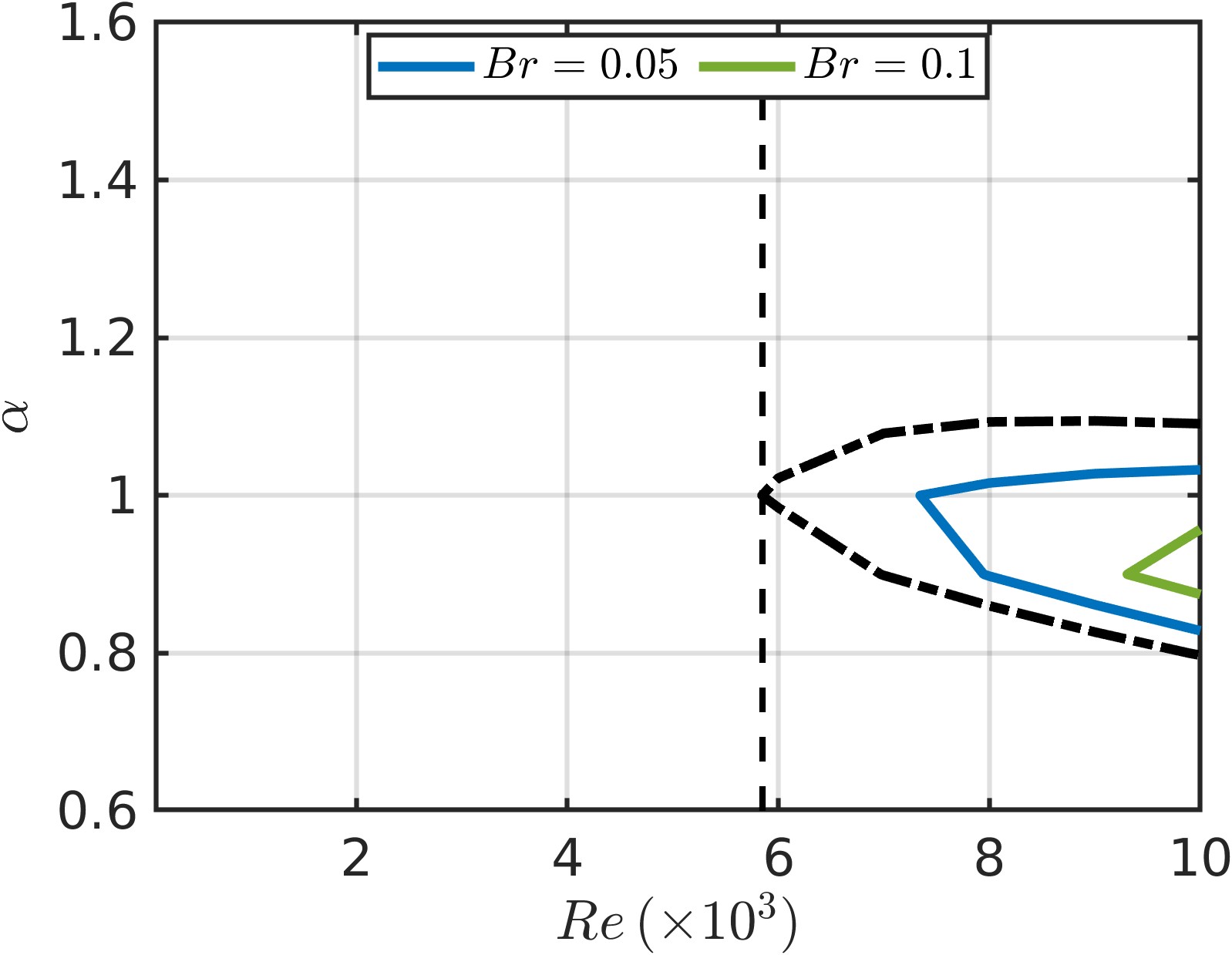}}  \\ \vspace{5mm}
	\caption{Neutral curves for various $Br$ at (a) sub-, (b) trans- and (c) supercritical regimes. The dashed-dotted line represents the isothermal limit ($Br \xrightarrow{} 0$), and the vertical dashed line indicates $Re_c = 5772$.} 
 \label{fig:isothermal_neutral_curves}
\end{figure*}

% Paragraph 2: Neutral curves comparison
Once the base flows have been characterized, the neutral curves for the different isothermal cases are shown in Figure~\ref{fig:isothermal_neutral_curves}.
First, the subcritical regime case I-1 (centerline temperature below $T_c$) yields a reduced stability region with respect to the isothermal limit equivalent setup when $Br$ increases.
In particular, the neutral curve expands to lower $Re$ values for a wider range of wavenumbers, and becomes unstable for $Re_c \approx 1000$ at $1.0 \lesssim \alpha \lesssim 1.2$; see Table~\ref{tab:summary_Re_c}.
Next, the I-2 base flow undergoes a transcritical trajectory across the pseudo-boiling region, and as a result both velocity and temperature become inflectional at $Br \gtrsim 0.35$.
This results in a lower critical Reynolds number of value $Re_c \approx 500$ (refer to Table~\ref{tab:summary_Re_c}) and a significantly wider range of wavenumbers, $1.2 \lesssim \alpha \lesssim 1.6$, where early laminar-to-turbulent transition may likely occur.
Third, the I-3 case operates at supercritical conditions behaving similarly to the ideal-gas solution~\citep{Ren2019-A}, in which increasing $Br$ enlarges the stability region. In particular, for $Br > 0.1$, the flow is stable for the entire $Re-\alpha$ parameter space considered in this work.

\begin{figure*}
	\centering
	\subfloat[\vspace{-8mm}]{\includegraphics[width=0.327\linewidth]{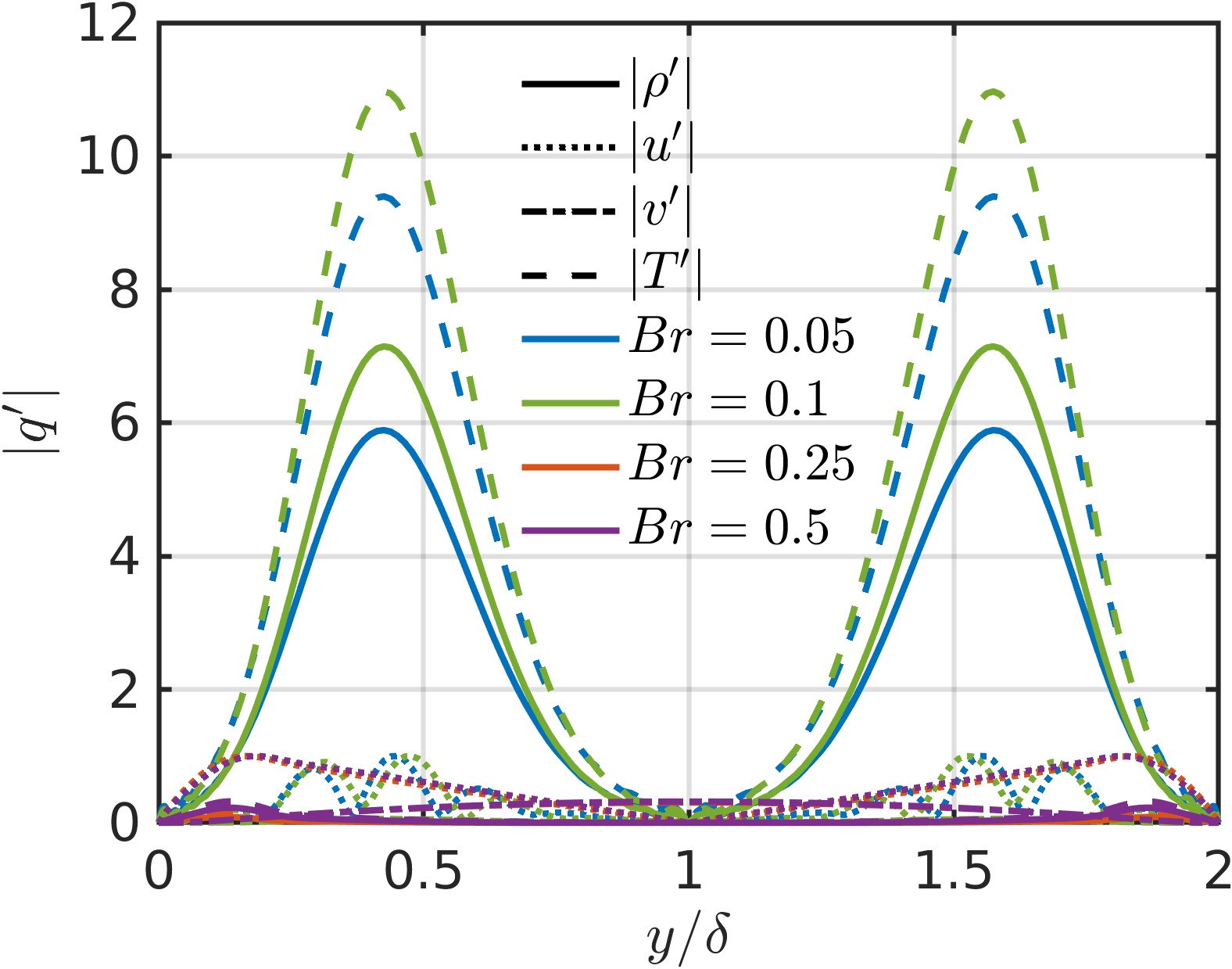}} \hfill
    \subfloat[\vspace{-8mm}]{\includegraphics[width=0.33\linewidth]{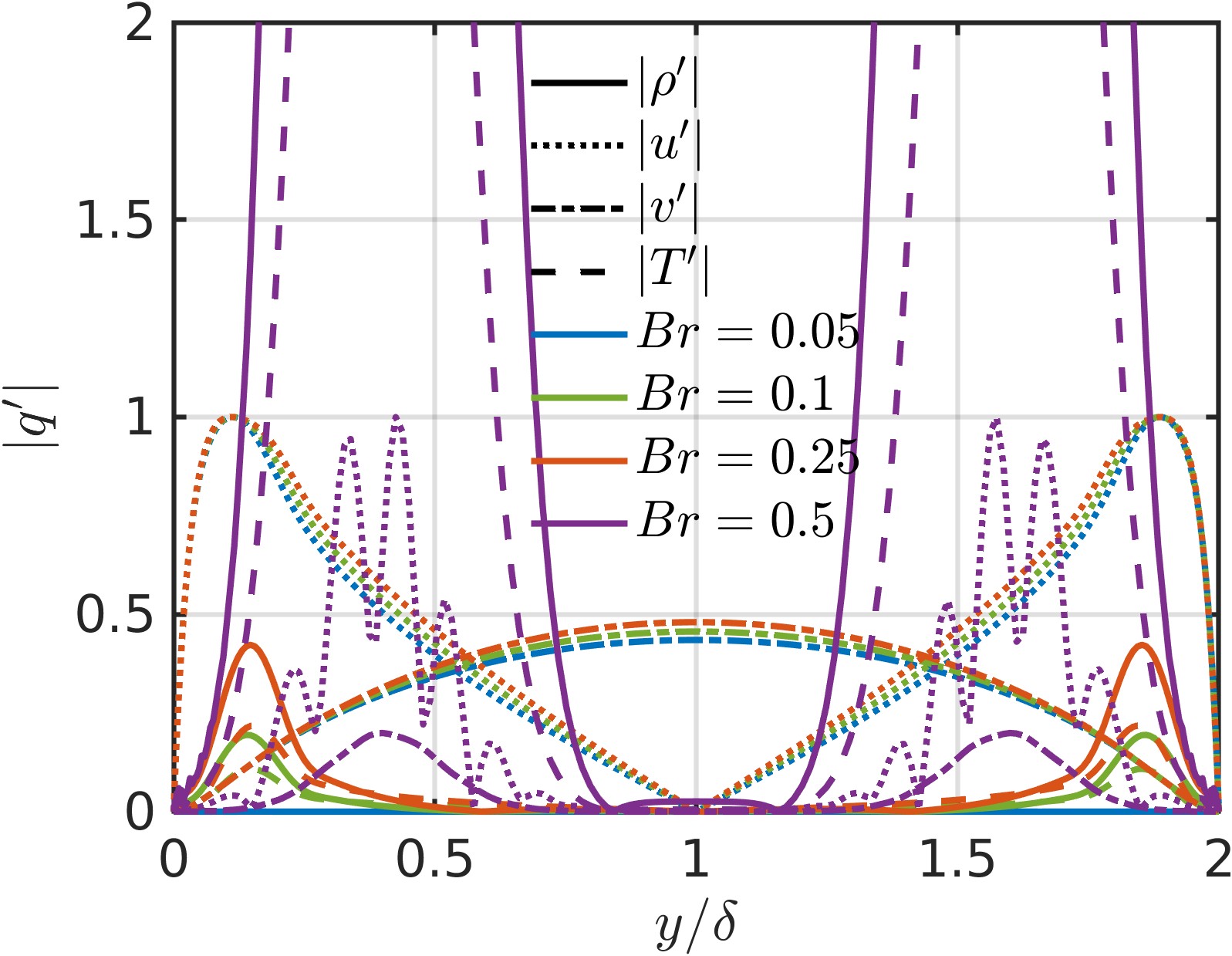}} \hfill
    \subfloat[\vspace{-8mm}]{\includegraphics[width=0.33\linewidth]{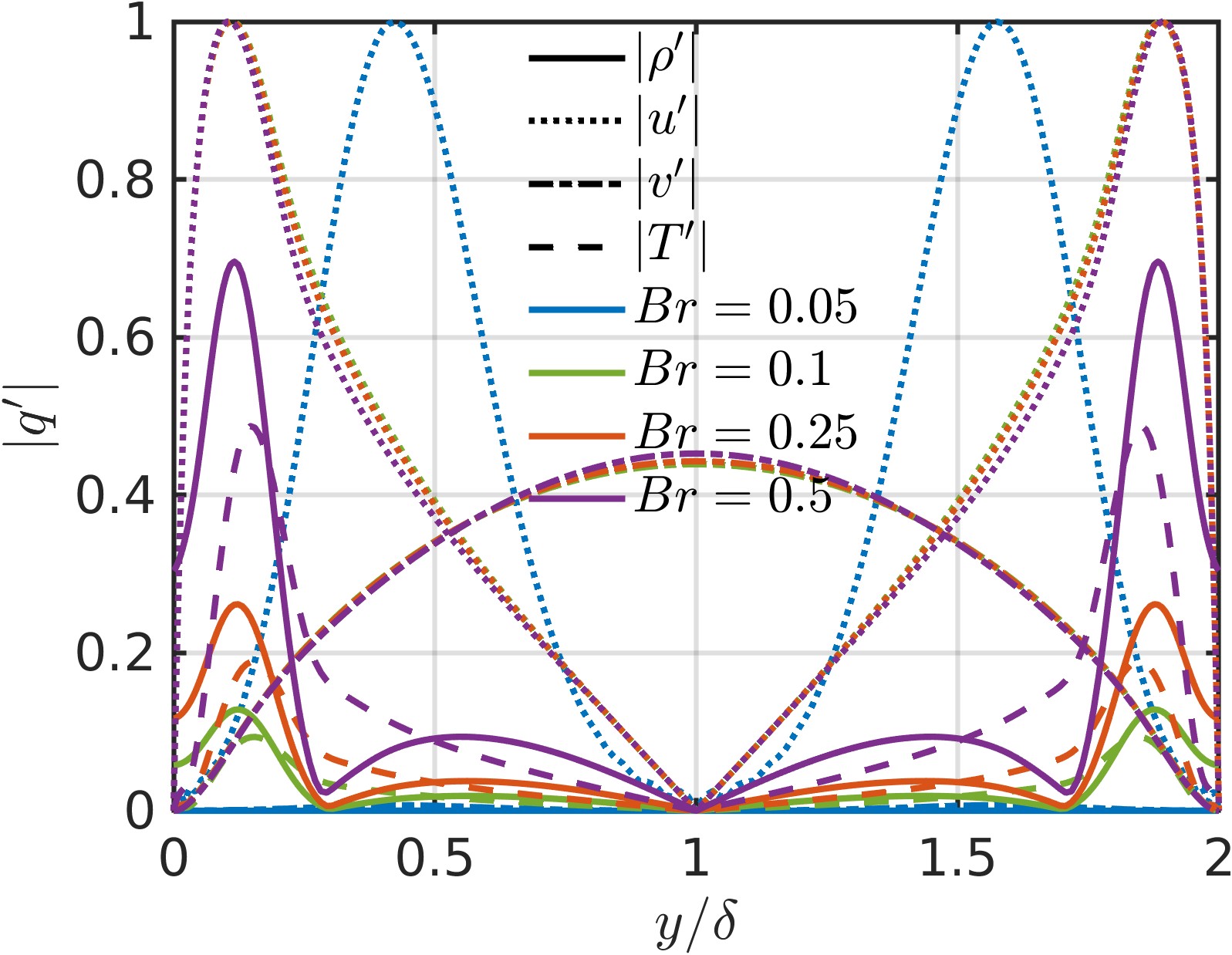}} \\ \vspace{5mm}
	\caption{Perturbation profiles of the most unstable mode at $Re = 10000$ and $\alpha = 1$ along the wall-normal direction for various Brinkman numbers at (a) sub-, (b) trans- and (c) supercritical regimes.} 
 \label{fig:isothermal_perturbations}
\end{figure*}

% Paragraph 3: Perturbation profiles
In connection to the neutral curves introduced above, Figure~\ref{fig:isothermal_perturbations} presents the perturbations of the most unstable mode for $Re = 10000$ and $\alpha = 1$ along the wall-normal direction for various Brinkman numbers at different thermodynamic regimes.
As it can be observed, unlike for the isothermal limit case, the subcritical thermodynamic regimes are dominated by density and temperature perturbations (thermophysical-driven mode).
The effects of these perturbations, however, are diminished when the Brinkman number increases and consequently the streamwise velocity dominates.
At transcritical conditions, instead, streamwise velocity perturbations dominate over the other fields the destabilization of the flow in the vicinity of walls, except for large Brinkman number where, again, the thermodynamic perturbations mandate.
However, the wall-normal velocity governs the flow instability at the centerline independently on $Br$.
Differently, for supercritical thermodynamic conditions, flow destabilization is mainly dominated throughout the entire wall-normal direction by streamwise velocity perturbations (dynamic-driven mode) with insignificant differences with Brinkman number increase regarding dynamic perturbation, but slowly rise in the vicinity of the walls for thermodynamic perturbations.

\begin{figure*}
	\centering
	\subfloat[\vspace{-8mm}]{\includegraphics[width=0.32\linewidth]{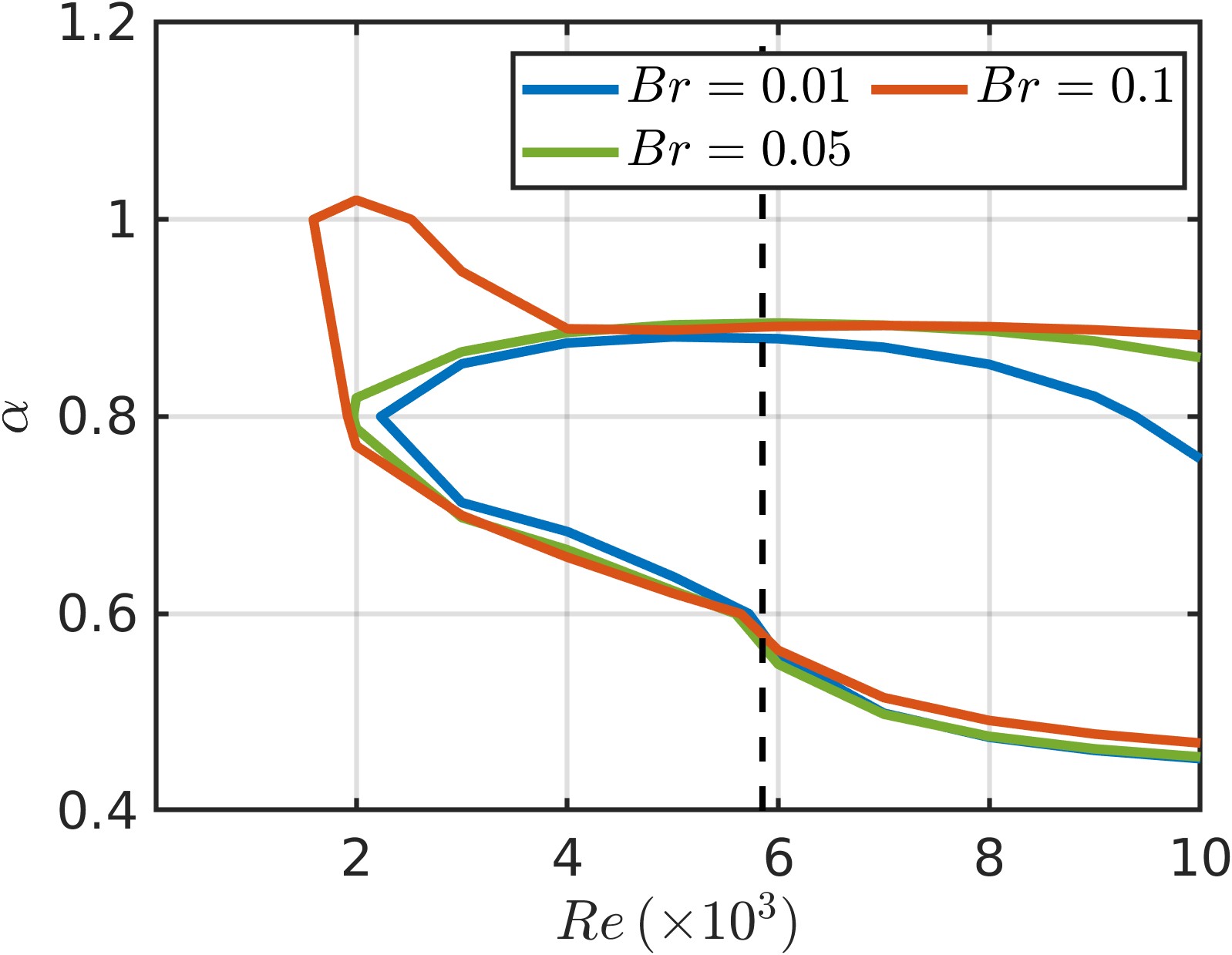}}
    \subfloat[\vspace{-8mm}]{\includegraphics[width=0.32\linewidth]{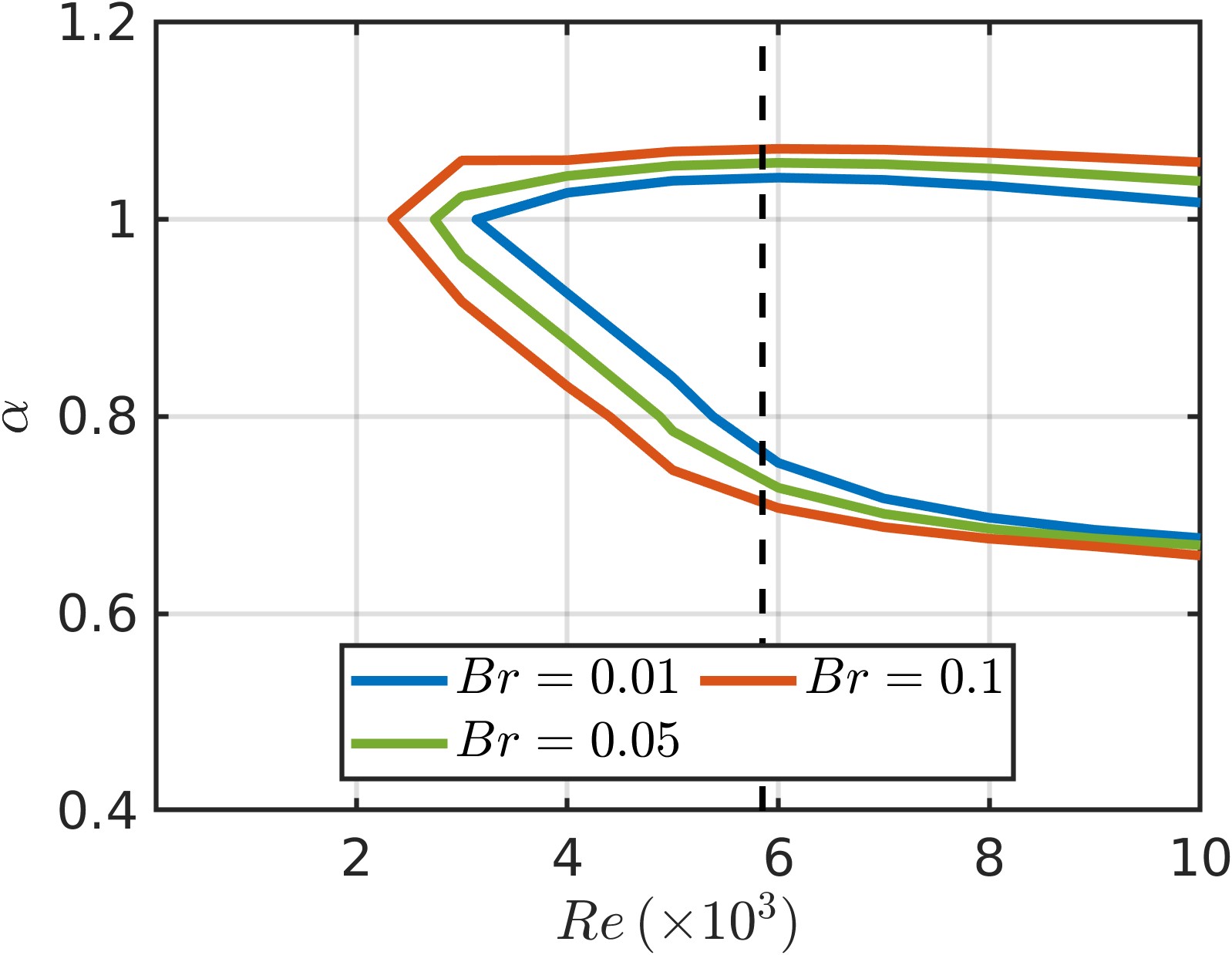}}
    \subfloat[\vspace{-8mm}]{\includegraphics[width=0.32\linewidth]{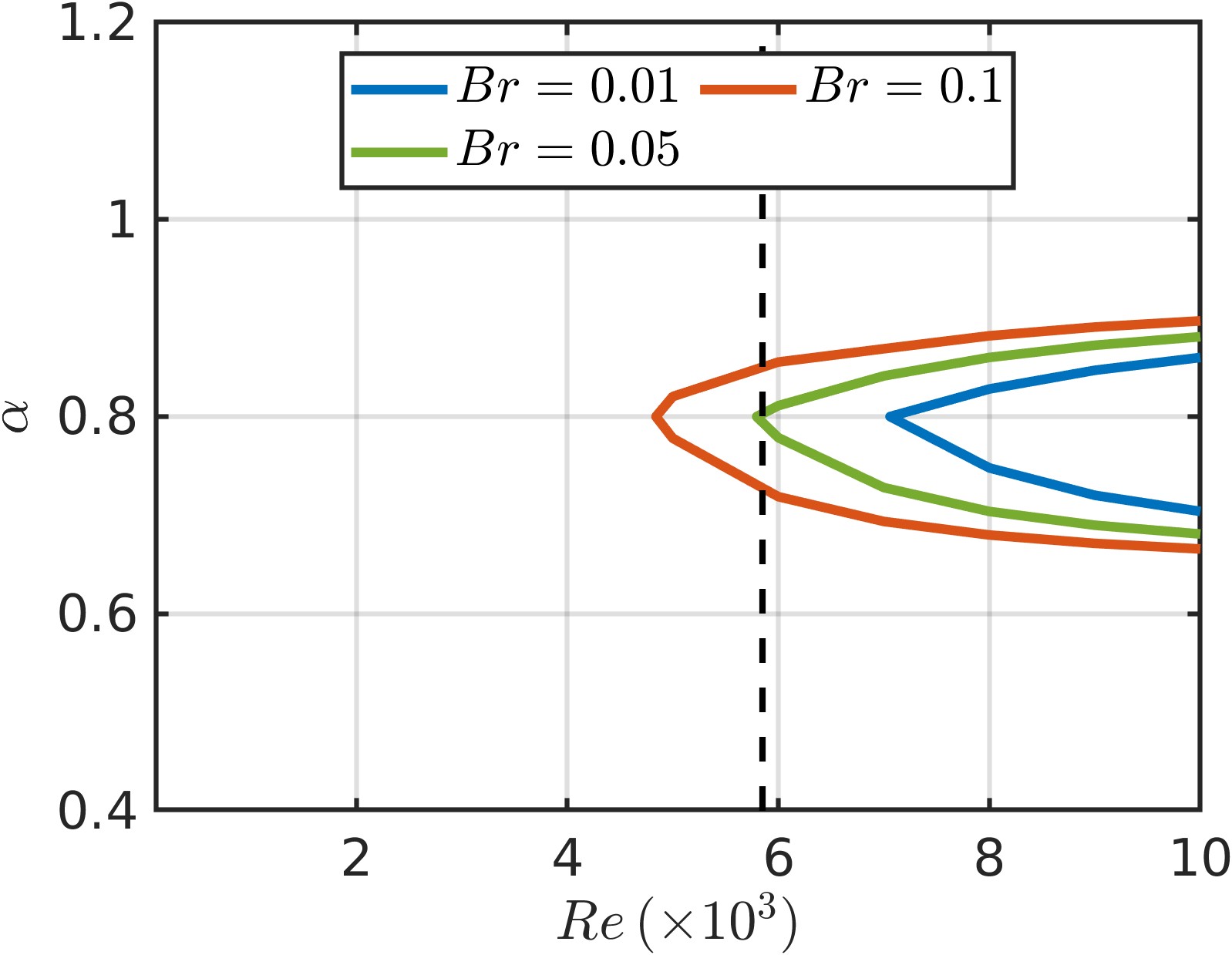}} \\  \vspace{5mm}
	\caption{Neutral curves for various Brinkman numbers for cases (a) NI-1, (b) NI-2 and (c) NI-3.} 
 \label{fig:non_isothermal_neutral_curves}
\end{figure*}

\subsubsection{Non-isothermal cases}  \label{sec:results_non_isothermal}

% Paragraph 1: Neutral curves comparison
Figure~\ref{fig:non_isothermal_neutral_curves} depicts the neutral curves for the non-isothermal cases.
It is noted that in this case the the fluid is forced to cross the pseudo-boiling region for all $Br$.
In fact, the largest value of the Brinkman number was chosen to ensure that temperature remains below the hot wall temperature, which corresponds to $Br \le 0.1$.
Particularly, while NI-1 and NI-3 cross the pseudo-boiling temperature, NI-2 is pressurized much beyond the critical pressure and consequently it operates at supercritical conditions.
The neutral curves of NI-1 are similar for all $Br$, where the instability is biased toward lower $Re$; especially, in comparison to the low-$Br$-number cases at isothermal conditions.
In detail, the wavenumber corresponding to the critical Reynolds number falls by roughly $20\%$.
Nevertheless, the NI-3 case results in a larger $Re_c$ value, which is above the isothermal limit transition when $Br < 0.05$.
Therefore, as it can be observed for case NI-1, by constraining the cold and hot temperatures closer to the pseudo-boiling temperature, the stability region is enhanced.
Finally, when increasing the pressure of the system, the neutral curves display similar envelopes as in the subcritical isothermal case.

\begin{figure*}
	\centering
	\subfloat[\vspace{-8mm}]{\includegraphics[width=0.32\linewidth]{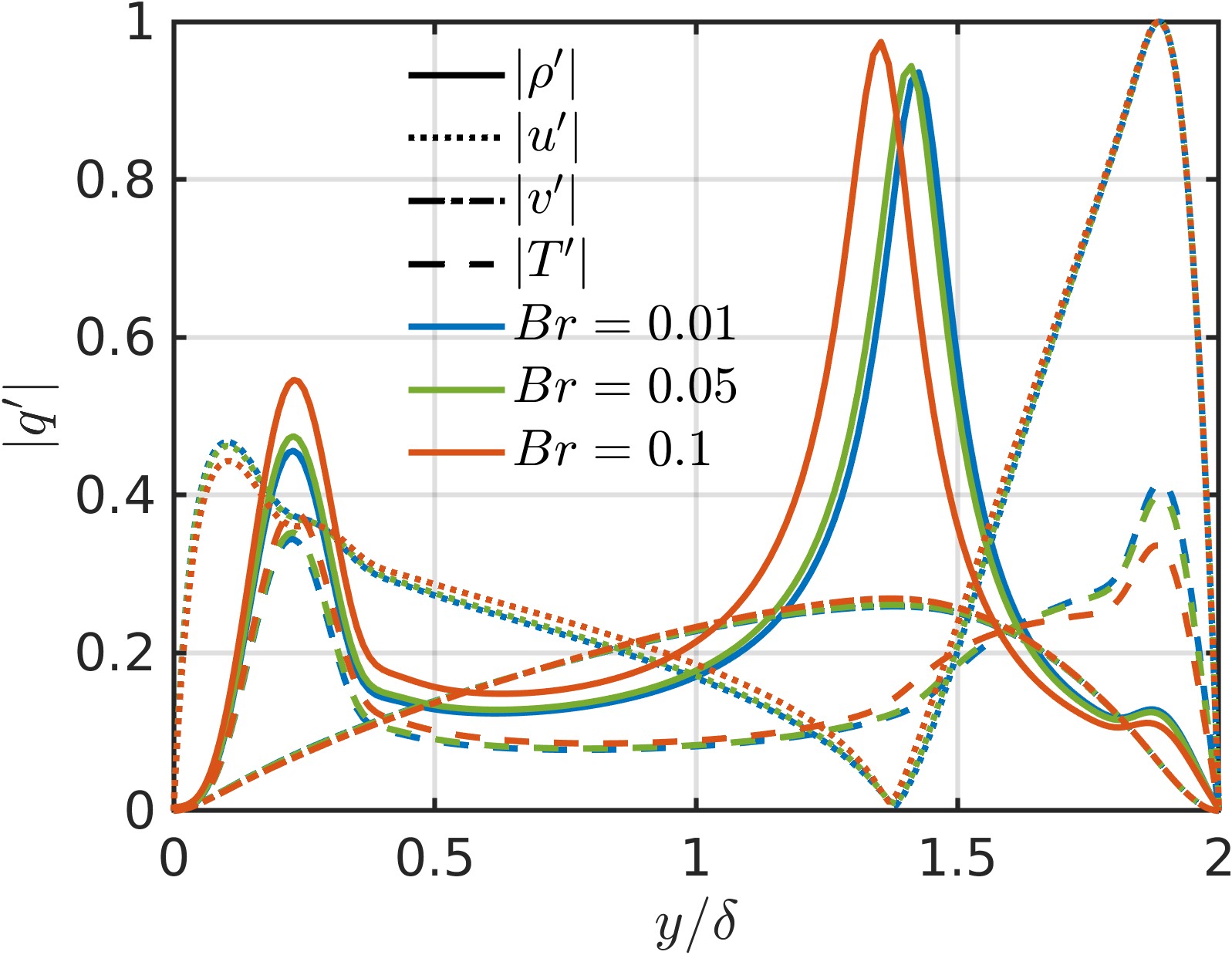}} \hfill
    \subfloat[\vspace{-8mm}] {\includegraphics[width=0.33\linewidth]{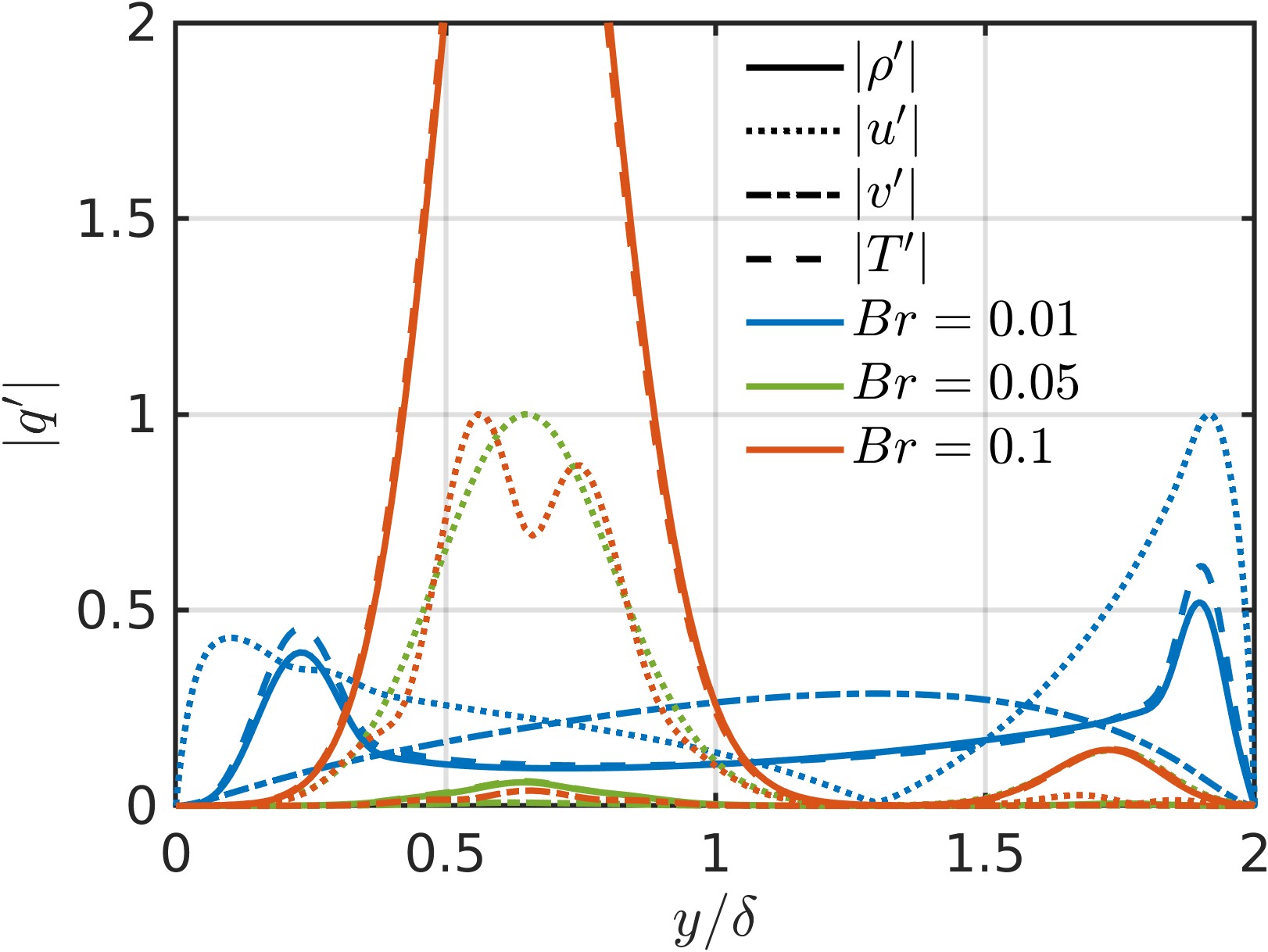}} \hfill
    \subfloat[\vspace{-8mm}]{\includegraphics[width=0.32\linewidth]{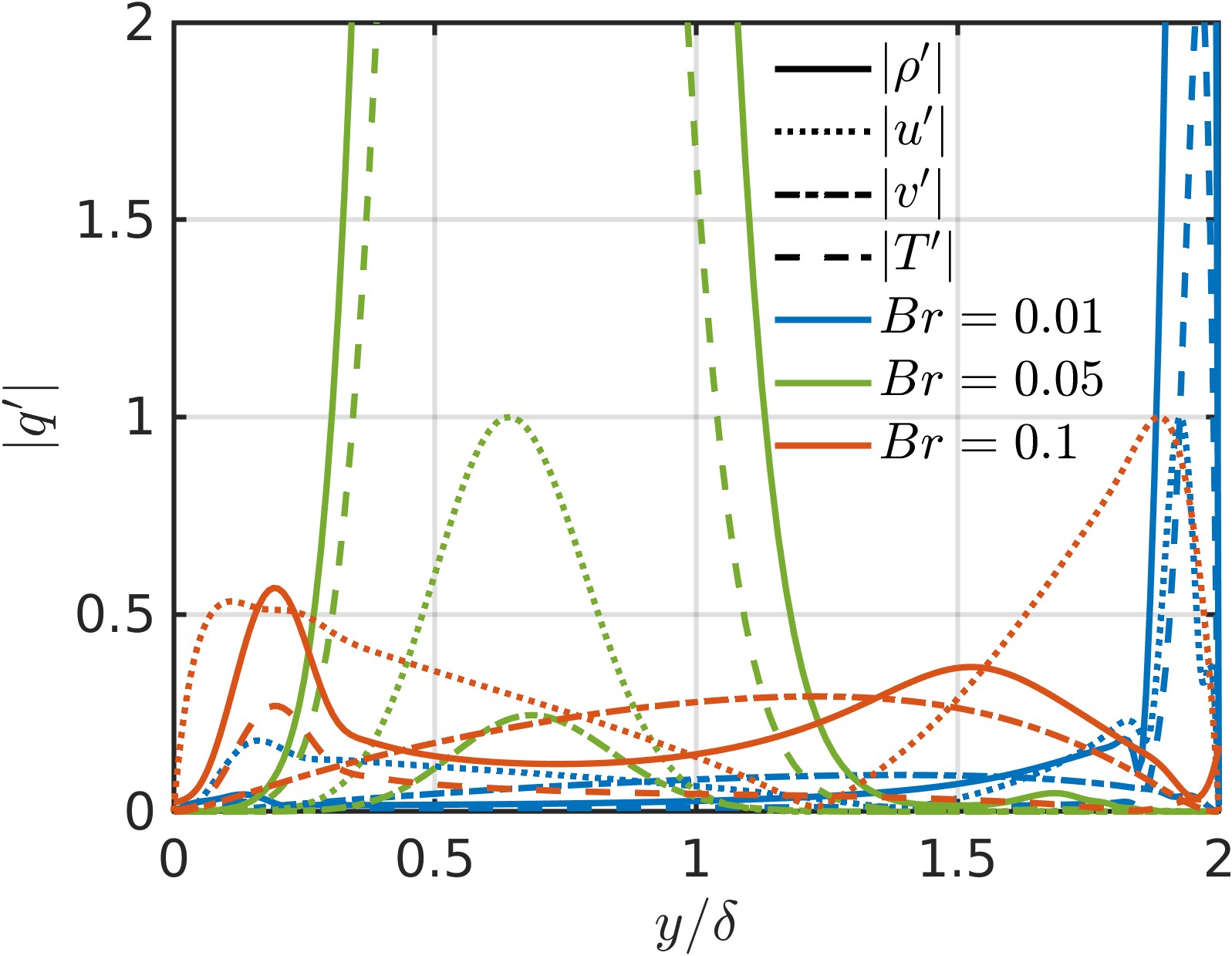}} \\ \vspace{5mm}
	\caption{Perturbation profiles of the most unstable mode at $Re = 10000$ along the wall-normal direction for various $Br$: (a) NI-1 ($\alpha = 0.8$), (b) NI-2 ($\alpha = 1$), and (c) NI-3 ($\alpha = 0.8$) cases.} 
 \label{fig:non_isothermal_perturbations}
\end{figure*}

\begin{figure*}
	\centering
	\subfloat[\vspace{-8mm}]{\includegraphics[width=0.33\linewidth]{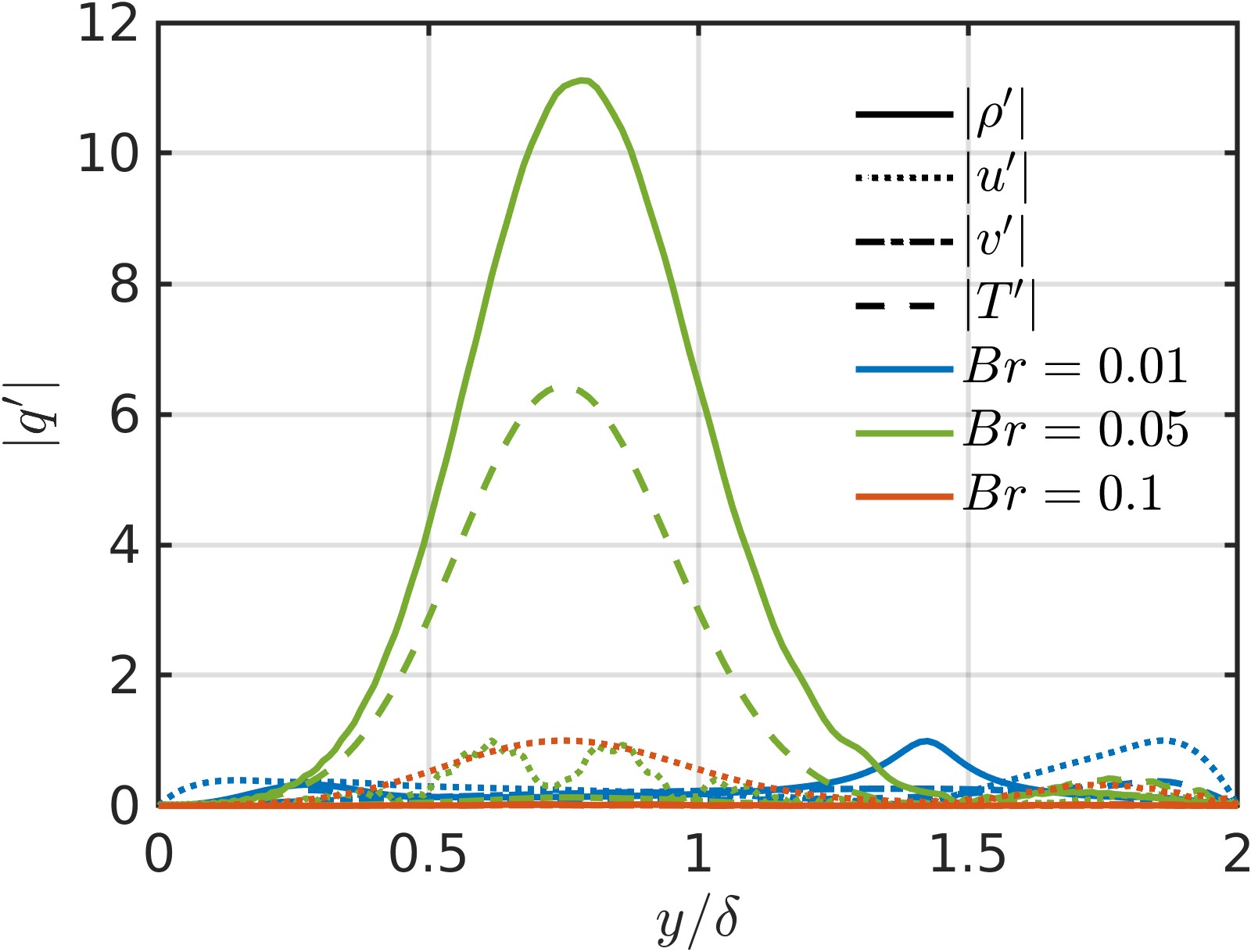}} \hfill
    \subfloat[\vspace{-8mm}]{\includegraphics[width=0.33\linewidth]{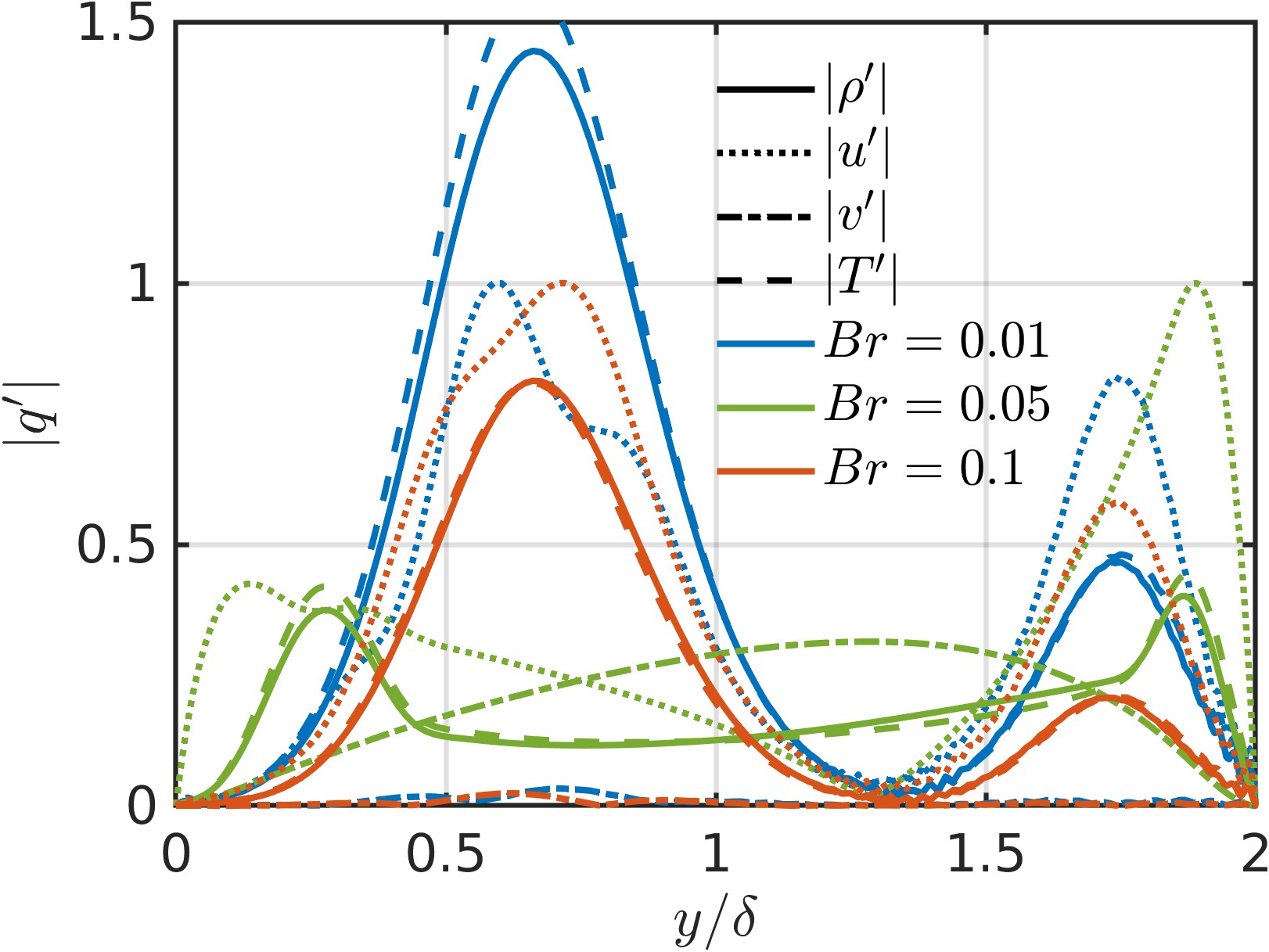}} \hfill
    \subfloat[\vspace{-8mm}] {\includegraphics[width=0.32\linewidth]{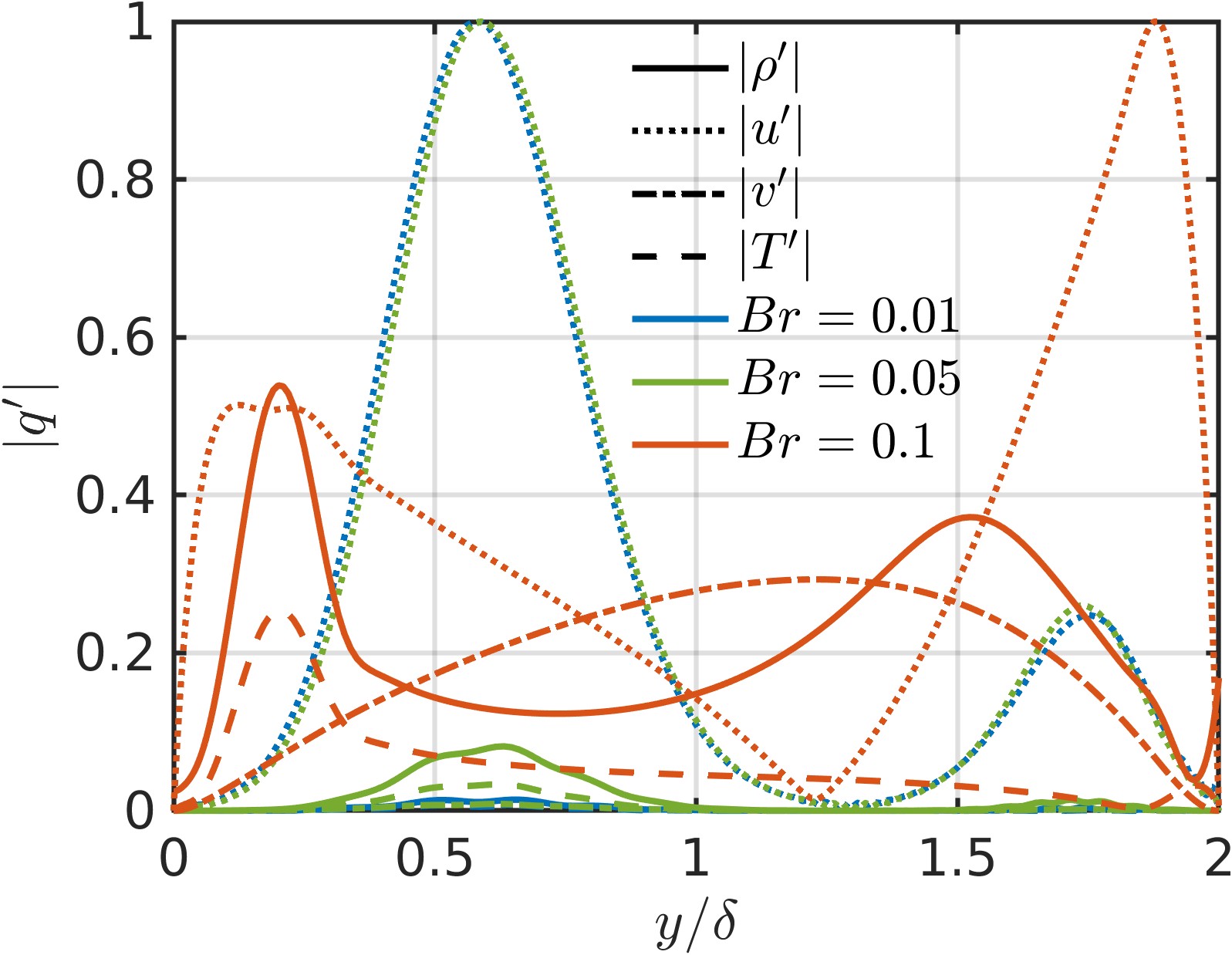}} \\ \vspace{5mm}
	\caption{Perturbation profiles of the most unstable mode for (a) NI-1 at $Re = 4000$ and $\alpha = 0.8$, (b) NI-2 at $Re = 4000$ and $\alpha = 1$, and (c) NI-3 at $Re = 8000$ and $\alpha = 0.8$.} 
 \label{fig:non_isothermal_perturbations_critical}
\end{figure*}

% Paragraph 2: Perturbation profiles
Analogously, Figure~\ref{fig:non_isothermal_perturbations} presents the perturbation profiles of the most unstable modes at the largest $Re$ analyzed.
The perturbations are dominated by density and temperature near the pseudo-boiling region, but velocity is significantly high near the hot wall. On the other hand, at the cold wall the thermodynamic and dynamic modes have similar magnitudes, although they are $50\%$ lower than at the hot wall.
This behavior is similar for both transcritical cases, NI-1 and NI-3, with small differences among $Br$ numbers for NI-1. Although NI-3 is dominated by thermodynamic perturbations for $Br = 0.01$ and $Br = 0.05$ which are roughly $10\times$ and $30\times$ greater, in magnitude, compared to velocity perturbations. Nonetheless, at largest $Br = 0.1$ the system is governed again by streamwise perturbation in the vicinity of hot wall. Instead, NI-2 is mainly dominated by the streamwise velocity near the walls and by the vertical velocity at the center.
Nevertheless, at large $Br$, the density and temperature perturbations near the  cold wall become greater than those associated with the longitudinal velocity.
Additionally, the perturbations near the value of $Re_c$ are depicted in Figure~\ref{fig:non_isothermal_perturbations_critical}.
NI-1 perturbations are similar to those obtained with larger $Re$ values for low $Br$ numbers. Instead, at $Br = 0.05$ thermodynamic modes dominate and at $Br = 0.1$ they become insignificant over streamwise perturbation.
However, for NI-2, the system is dominated by thermodynamics at $y/\delta \sim 0.75$ at low $Br$. The larger the Brinkman number, the closer dynamic and thermodynamic perturbations. Moreover, at $y/\delta \sim 1.75$ streamwise perturbation dominates; in contrast, NI-3 is clearly dominated by thermodynamic modes at both peak locations.
Finally, the NI-4 case operates at low-pressure conditions, and the linear stability results collapse to those corresponding to the ideal-gas solution.
Nonetheless, although it operates at different temperatures, the system is stable for all the Reynolds numbers and wavenumbers considered.

\subsubsection{High-pressure transcritical vs. Superheated steam non-isothermal Poiseuille flow}

\begin{figure*}
	\centering
	\subfloat[\vspace{-8mm}]{\includegraphics[width=0.47\linewidth]{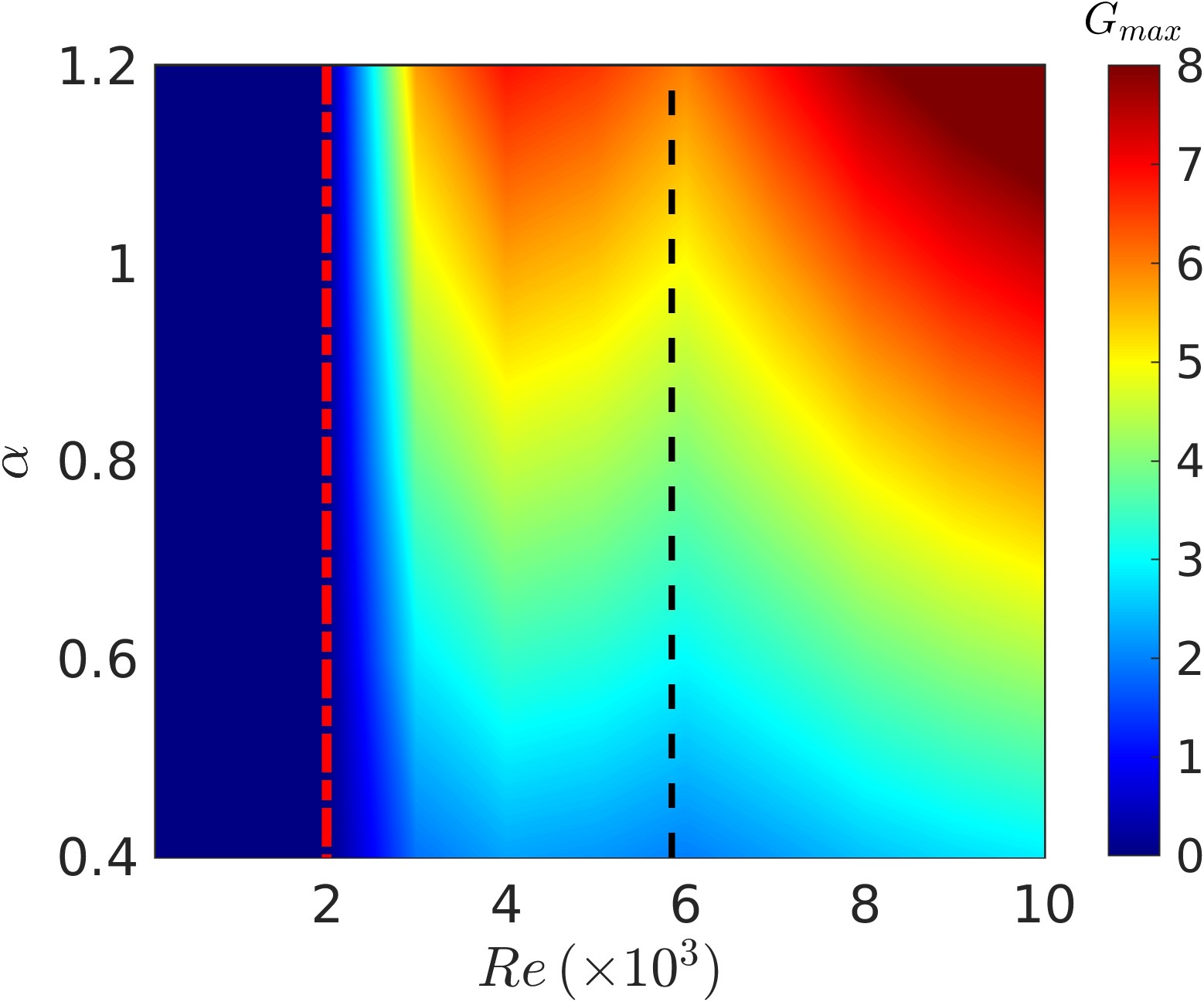}} \hfill
    \subfloat[\vspace{-8mm}]{\includegraphics[width=0.51\linewidth]{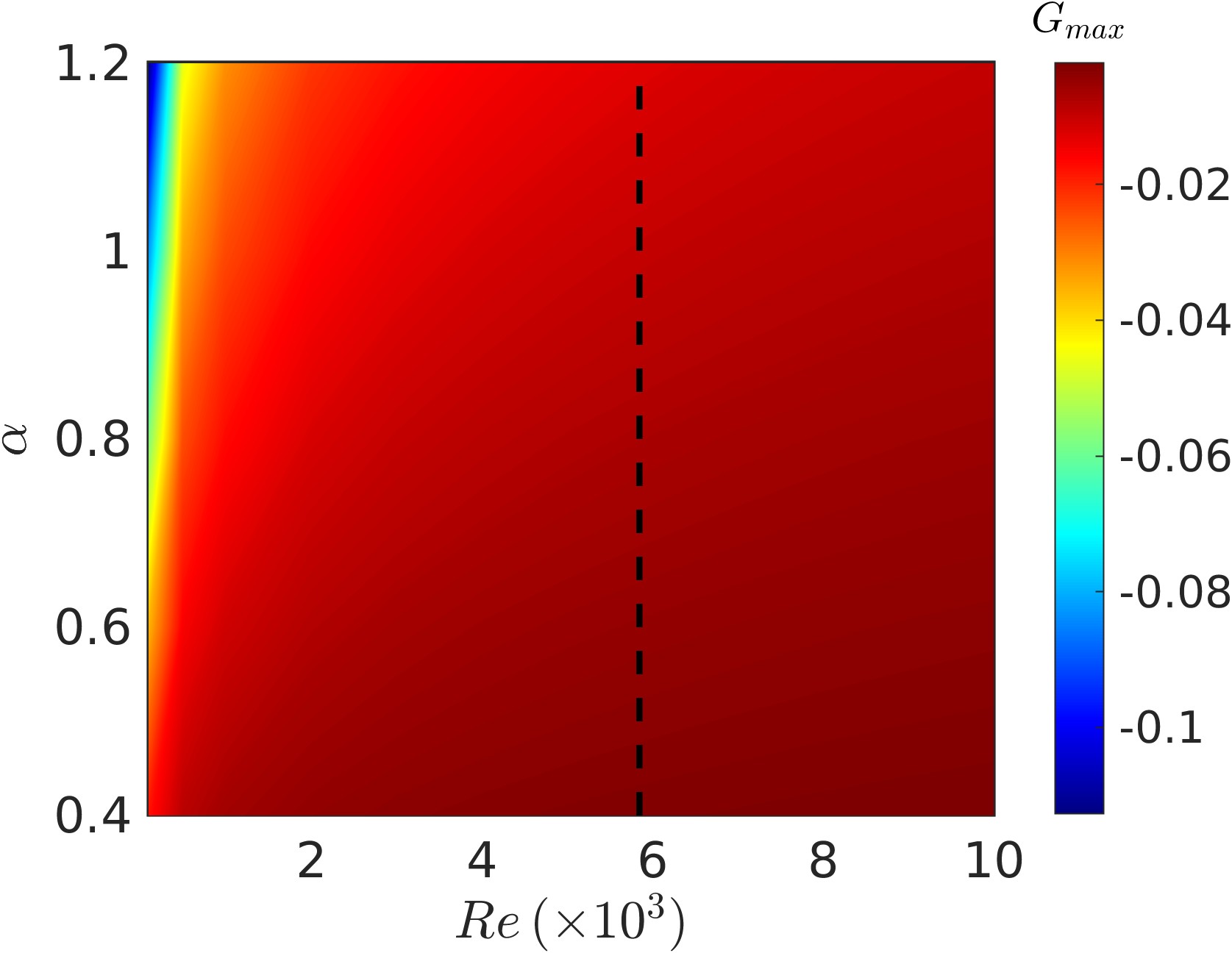}} \\  \vspace{5mm}
	\caption{Growth rate contours at $Re-\alpha$ for (a) $NI-5$ and (b) $NI-6$ cases. The neutral curve for NI-5 is highlighted in red.} 
 \label{fig:non_isothermal_growth_rate_contour}
\end{figure*}

\begin{figure*}
	\centering
	\subfloat[\vspace{-8mm}]{\includegraphics[width=0.48\linewidth]{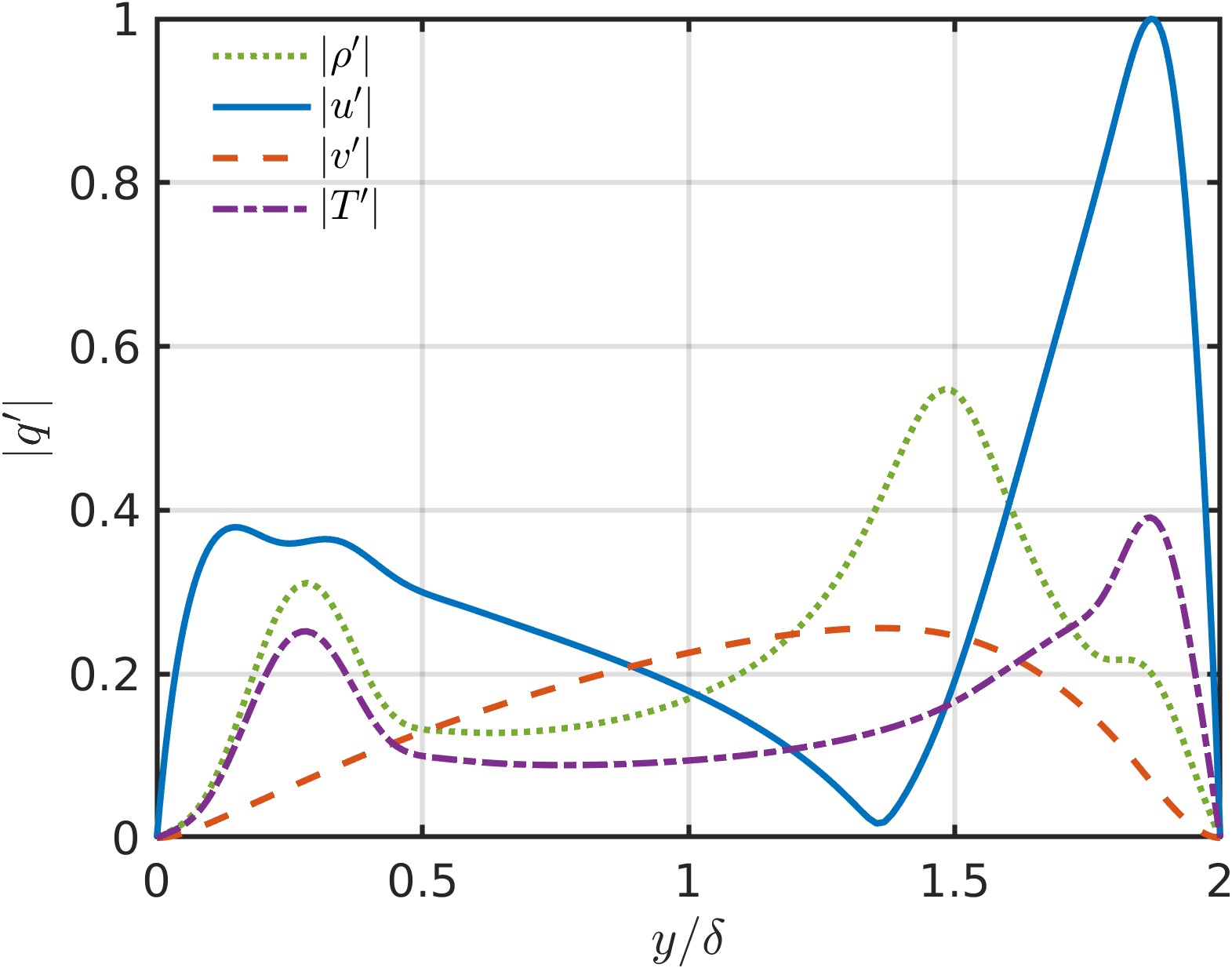}} \hfill
    \subfloat[\vspace{-8mm}]{\includegraphics[width=0.48\linewidth]{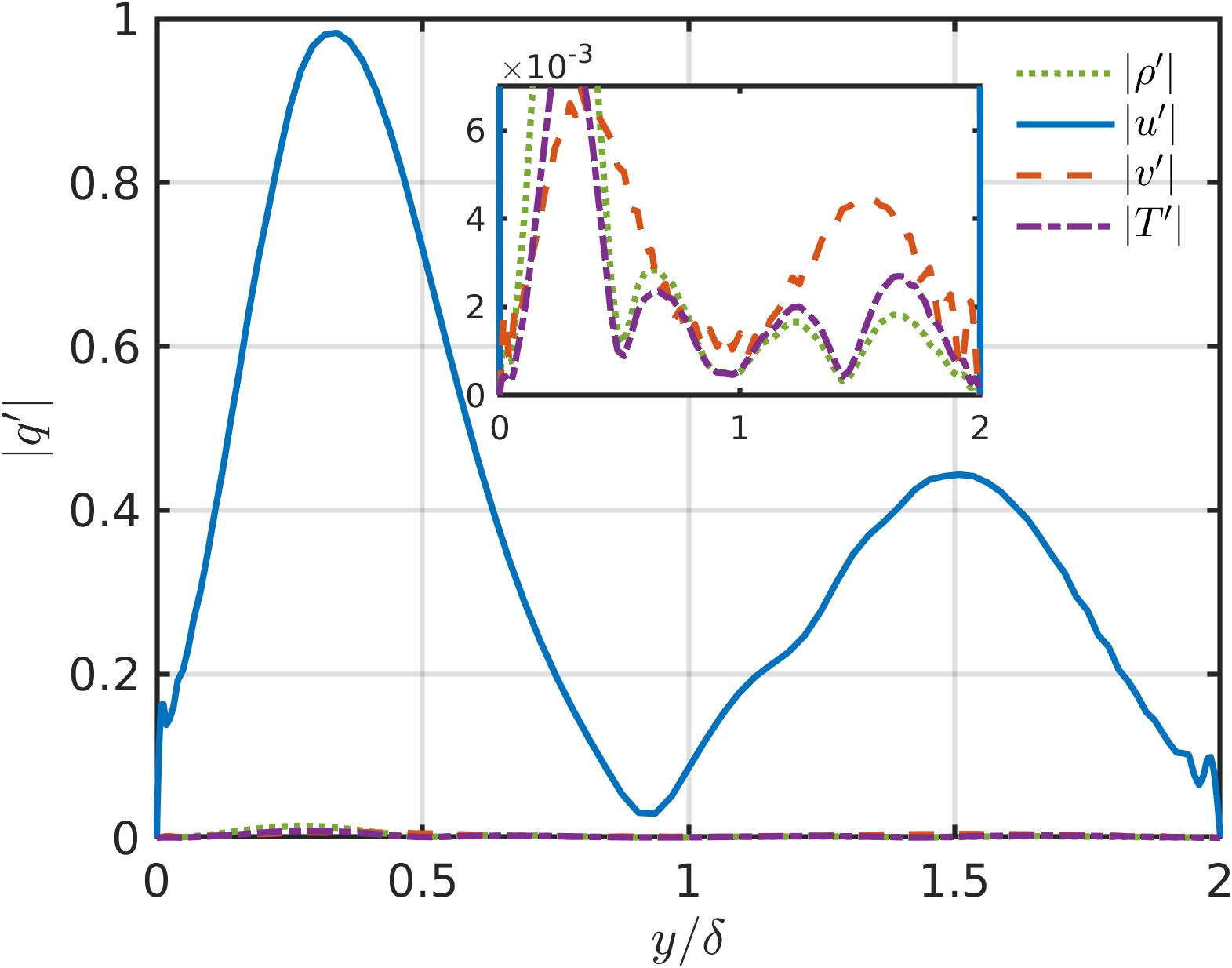}} \\ \vspace{8mm}
    \subfloat[\vspace{-8mm}]{\includegraphics[width=0.48\linewidth]{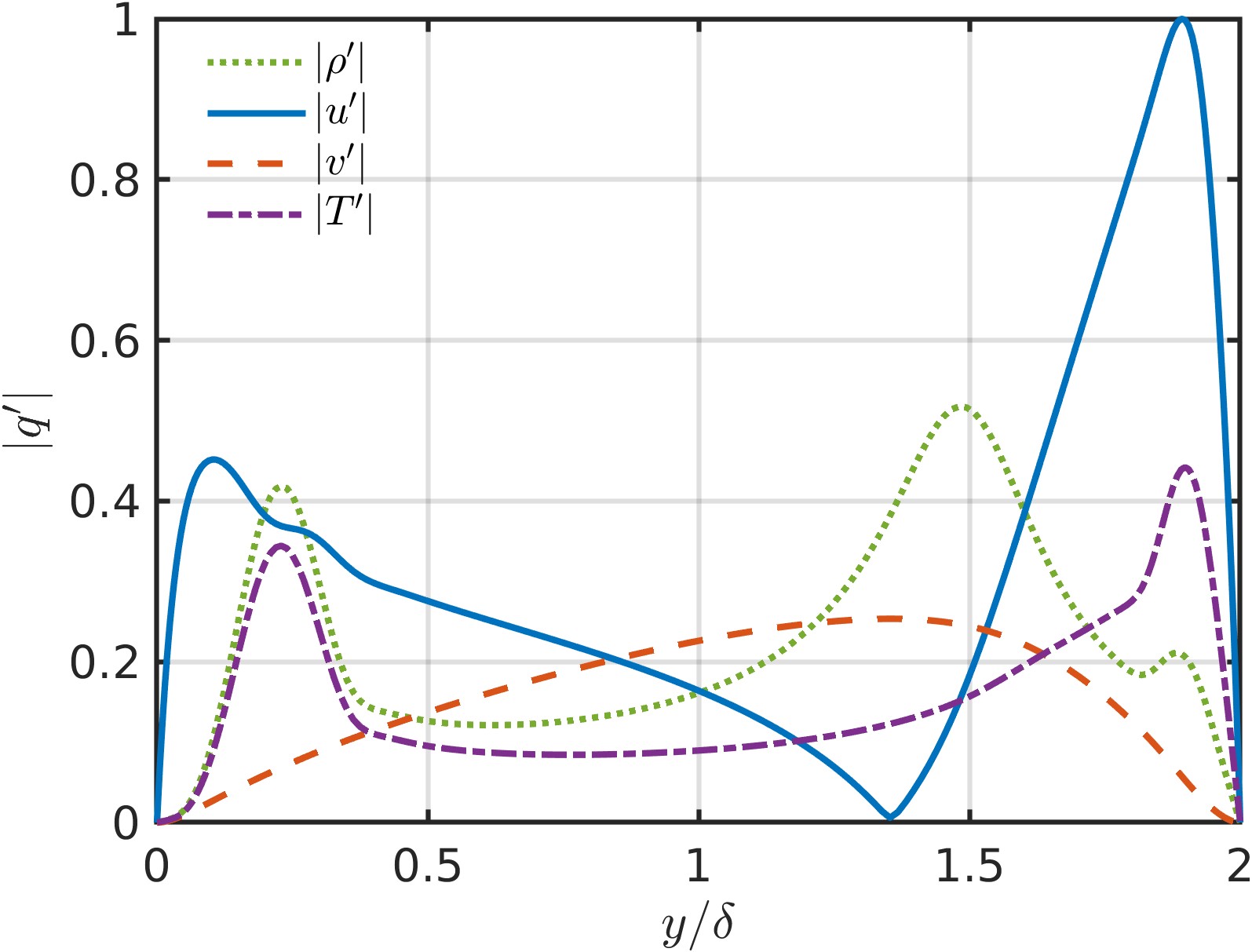}} \hfill
    \subfloat[\vspace{-8mm}]{\includegraphics[width=0.48\linewidth]{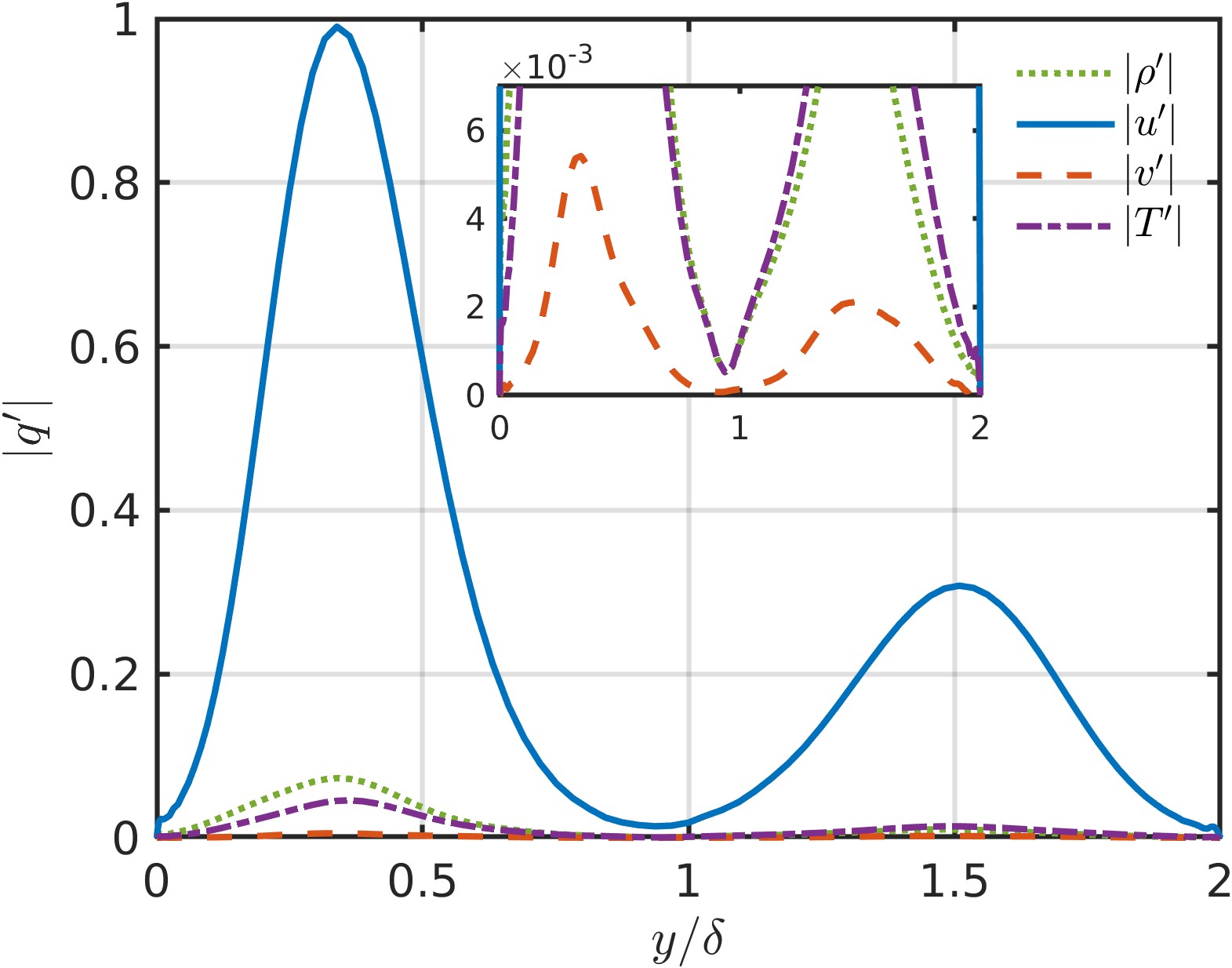}} \\ \vspace{5mm}
	\caption{Perturbation profiles of the most unstable mode at $\alpha = 0.8$ for (a) NI-5 at $Re = 4000$, (b) NI-6 at $Re = 4000$, (c) NI-5 at $Re = 10000$, and (d) NI-6 at $Re = 10000$.} 
 \label{fig:non_isothermal_perturbations_NI_5_6}
\end{figure*}

This section compares the non-isothermal flow conditions at low- and high-pressure regimes operating at low $Br$, namely the cases NI-5 and NI-6.
As concluded in previous sections, non-isothermal flows are subject to larger destabilization regions at low $Br$ numbers. In particular, NI-5 is equivalent to the transcritical channel flow setup studied via DNS by~\cite{Bernades2023c-A}, where the base flow is adjusted to obtain a reference velocity of $\mathcal{O}(1)$, as typically adopted in low Mach/Reynolds systems.
The results show that, as expected, low-Mach/Reynolds regimes are susceptible of an earlier destabilization when operating at non-isothermal conditions. Figure~\ref{fig:non_isothermal_growth_rate_contour}(a) depicts the early modal-based transition exhibited by NI-5 reaching a $Re_c = 2000$ independently of streamwise wavenumber.
In detail, the perturbations near the critical point of the solely unstable mode (i.e., $Re_b = 4000$) are mainly dominated by the streamwise velocity, although the thermodynamic variables are following a similar trajectory (at lower magnitude) as shown in Figure~\ref{fig:non_isothermal_perturbations_NI_5_6}(a). At $Re = 10000$, the dynamic components are still fully governing the destabilization process [Figure~\ref{fig:non_isothermal_perturbations_NI_5_6}(c)] specially near the hot wall. On the contrary, moving toward lower Reynolds numbers, all the analyzed modes become stable.
It is known that equivalent low-pressure setups become laminar and, therefore, stable~\citep{Bernades2024c-A}.
As seen in the previous section, the non-isothermal superheated steam system becomes stable at any $Re \le 10000$.
In particular, the NI-6 case is driven with the same $Br$ number as NI-5 and its growth rate is depicted in Figure~\ref{fig:non_isothermal_growth_rate_contour}(b) capturing the entire streamwise wavenumber range modal stability.
The perturbations are entirely dominated by the streamwise velocity for the least stable mode at $Re = 10000$ and $Re = 4000$ as depicted in Figure~\ref{fig:non_isothermal_perturbations_NI_5_6}(b,d), although thermodynamic components slowly grow their magnitudes at larger $Re$.

% Figure: Max growth rate NI-5 vs. NI-6

% Figure: Perturbation close to transition Re = 4000 and Re = 10000 NI-5 vs. NI-6

%\marc{Similarities to 2-phase flow~\citep{Fuster2009-A}. For example, when small disturbances introduced at the entrance are quickly dissipated in the bulk of the liquid due to viscous effects, they are strong enough to create some instabilities at the interface. As the perturbations grow downstream, the level of turbulence inside the gas and liquid increases, mainly in zones near the interface. The small perturbation originally induced in the liquid finally generates a strong turbulence downstream not only in the liquid, but also in the gas. In fact, in the zone where liquid is attached to the walls the instabilities are large and grow non-linearly.}

\subsubsection{Laminar-to-turbulent transition summary}

The critical Reynolds numbers for each case, extracted from the corresponding neutral curves, are summarized in Table~\ref{tab:summary_Re_c}. These results are derived from modal stability analysis, hence it is an indicator of the transition trends, and for laminar-to-turbulence thresholds the algebraic stability is needed.

\begin{table}
\caption{Critical Reynolds numbers for the flow cases described in Section~\ref{sec:flow_cases} obtained from modal stability analysis. NI-5 and NI-6 are assessed at $Br = 5.6 \cdot 10^{-6}$, i.e., low $Br$ similar to the isothermal limit ($Br \xrightarrow{}0$). \label{tab:summary_Re_c}}
\centering
%\begin{adjustbox}{width=1\textwidth}
%\renewcommand{\arraystretch}{1.0}
%\begin{ruledtabular}
  % \centering 
\begin{tabular}{c|c|ccc|cccccc} \hline
\centering
$Br$ &  V-1 & I-1 & I-2 & I-3 & NI-1 & NI-2 & NI-3 & NI-4 & NI-5 & NI-6 \\
\hline
$Br \xrightarrow{}0$ & 5844 & & 5875 &  & $-$ & $-$ & $-$ & $-$ & $2001$ & $> 10^4$ \\ \hline
$0.01$ &  & 5630 & 5609 & 6056 & 2232 & 3138 & 7058 & $> 10^4$ & $-$ & $-$\\ 
$0.05$ &  & 4829 & 4742 & 7345 & 1973 & 2742 & 5794 & $> 10^4$ & $-$ & $-$\\ 
$0.10$ & $-$ & 4062 & 3904 & 9307 & 1594 & 2340 & 4840 & $> 10^4$ & $-$ & $-$\\ 
$0.25$ &  & 2051 & 1840 & $> 10^4$& $-$ & $-$ & $-$ & $-$ & $-$ & $-$ \\ 
$0.50$ &  & 985  & 469 & $> 10^4$& $-$ & $-$ & $-$ & $-$ & $-$ & $-$\\ \hline
\end{tabular}
%\end{ruledtabular}
%\end{adjustbox}
\end{table}

\subsubsection{Energy budget} \label{sec:energy_budget}

% Introduction to energy budget
The decomposition in different terms of the kinetic-energy growth further facilitates the characterization of the destabilization mechanisms.
In particular, it identifies the main contributors to the temporal kinetic-energy growth ($K$): production ($P$), thermodynamic ($T$), and viscous dissipation ($V$).
It is important to highlight that the power of the external force introduced to drive the flow cancels out with the viscous dissipation, and it has been consequently omitted from the budget resulting the growth balance expression $K = P + T + V$~\citep{Andreolli2021-A}.
The energy balance for the 2D perturbations is derived by taking the inner product of the stream- and spanwise momentum conservation equations with their respective velocity components for the most unstable mode~\citep{Boomkamp1996-A}.
The temporal growth of density, embedded in the streamwise momentum, is obtained from the continuity equation.
The resulting equation is typically averaged over the wavelength and integrated over the height of the channel~\citep{Moeleker1998-B,Sahu2010-A,Ren2019-A}, and considering only the real part; detailed expressions are presented in Appendix~\ref{sec:Appendix_C}.

% Analysis of the budget execution
The budget is assessed at $Re = 10000$ to represent the unstable modes of the spectra at their corresponding streamwise wavenumber ranges.
It is well-known that for single-phase Poiseuille flow the instability is caused by a combination of the no-slip conditions at the boundaries and the viscous effects in the critical layer responsible for Reynolds stress destabilization~\citep{Drazin1981-A}.
In this regard, Table~\ref{tab:summary_Energy_budget} quantifies the different contribution terms for each case.
Several observations can be extracted from this table: (i) production becomes the principal contributor to the kinetic-energy growth scaling with $Br$ except for the isothermal transcritical and supercritical base flows; (ii) in particular, the I-2 velocity perturbations are limited despite the large inflection points emerging in the vicinity of walls at high-$Br$-number base flows due to compressibility effects~\citep{Xie2017-A}, and as a result the net production term decreases (the analysis of this phenomena is extended in Section~\ref{sec:instability_mechanism}); (iii) the thermodynamic term is negligible in comparison to the other ones and negative; (iv) the viscous term also decreases the temporal energy and barely changes with the Brinkman number (at low $Br$ numbers, it weights approximately $60\%$ of the production); (v) the viscous dissipation increases with $Br$ at isothermal trans- \& supercritical conditions and for non-isothermal superheated steam; (vi) the stable flow cases result in a negative growth and an increase of the thermodynamic contribution; and (vii) the energy growth for the non-isothermal cases is barely sensitive to the Brinkman number in comparison to the isothermal setups.

\begin{table}
\caption{Summary of kinetic-energy budgets ($\times 10^{-4}$) for the 2D perturbations at $Re = 10000$. Isothermal cases at $\alpha = 1$, non-isothermal NI-1-5 at $\alpha = 0.6$, and NI-2-3-4-6 at $\alpha = 0.8$. \label{tab:summary_Energy_budget}}
\centering
\begin{tabular}{c|c|ccc|cccccc} \hline
\centering
$Br$ & $E_r$ & I-1 & I-2 & I-3 & NI-1 & NI-2 & NI-3 & NI-4 & NI-5 & NI-6 \\
\hline
$Br \xrightarrow{}0$ & $K$ & $-$ & $-$ & $-$ & $-$ & $-$ & $-$ & $-$ & $4.6$ & $-17.9$ \\
                     & $P$ & $-$ & $-$ & $-$ & $-$ & $-$ & $-$ & $-$ & $9.8$ & $-11.8$ \\ 
                     & $T$ & $-$ & $-$ & $-$ & $-$ & $-$ & $-$ & $-$ & $-0.080$ & $0.29$ \\ 
                     & $V$ & $-$ & $-$ & $-$ & $-$ & $-$ & $-$ & $-$ & $-5.3$ & $-6.4$ \\ \hline
$0.01$               & $K$ & $16.5$ & $16.6$ & $20.9$ & $5.1$ & $9.9$ & $4.8$ & $-18.1$ & $-$ & $-$ \\
                     & $P$ & $41.0$ & $41.1$ & $58.9$ & $9.2$ & $17.6$ & $14.0$ & $-12.2$ & $-$ & $-$ \\
                     & $T$ & $-0.00053$ & $-0.010$ & $-0.025$ & $0.054$ & $-0.12$ & $-0.16$ & $0.31$ & $-$ & $-$ \\ 
                     & $V$ & $-24.5$ & $-24.5$ & $-38.0$ & $-4.1$ & $-7.5$ & $-9.0$ & $-6.3$ & $-$ & $-$ \\ \hline
$0.05$               & $K$ & $22.4$ & $25.0$ & $9.5$ & $5.7$ & $11.4$ & $7.2$ & $-16.4$ & $-$ & $-$ \\
                     & $P$ & $46.9$ & $52.1$ & $44.2$ & $9.9$ & $18.8$ & $15.9$ & $-11.6$ & $-$ & $-$ \\
                     & $T$ & $-0.0020$ & $-0.079$ & $-0.027$ & $0.042$ & $-0.11$ & $-0.14$ & $0.37$ & $-$ & $-$ \\ 
                     & $V$ & $-24.5$ & $-27.0$ & $-34.6$ & $-4.2$ & $-7.3$ & $-8.6$ & $-16.4$ & $-$ & $-$ \\ \hline
$0.10$               & $K$ & $32.4$ & $38.5$ & $-1.8$ & $11.0$ & $13.2$ & $10.5$ & $-14.3$ & $-$ & $-$ \\
                     & $P$ & $59.2$ & $70.0$ & $23.4$ & $18.8$ & $20.4$ & $20.0$ & $-10.7$ & $-$ & $-$ \\
                     & $T$ & $-0.02$ & $-0.26$ & $0.12$ & $0.049$ & $-0.11$ & $-0.20$ & $0.42$ & $-$ & $-$ \\ 
                     & $V$ & $-26.7$ & $-31.2$ & $-25.2$ & $-7.8$ & $-7.1$ & $-9.3$ & $-4.1$ & $-$ & $-$ \\ \hline
$0.25$               & $K$ & $61.8$ & $32.1$ & $-25.7$ & $-$ & $-$ & $-$ & $-$ & $-$ & $-$ \\
                     & $P$ & $91.1$ & $47.6$ & $-4.2$ & $-$ & $-$ & $-$ & $-$ & $-$ & $-$ \\
                     & $T$ & $-0.15$ & $-0.66$ & $1.4$ & $-$ & $-$ & $-$ & $-$ & $-$ & $-$ \\ 
                     & $V$ & $-29.1$ & $-14.9$ & $-22.9$ & $-$ & $-$ & $-$ & $-$ & $-$ & $-$ \\ \hline
$0.50$               & $K$ & $124.4$ & $4.0$ & $-27.9$ & $-$ & $-$ & $-$ & $-$ & $-$ & $-$ \\
                     & $P$ & $159.3$ & $5.2$ & $-21.1$ & $-$ & $-$ & $-$ & $-$ & $-$ & $-$ \\
                     & $T$ & $-0.91$ & $-0.24$ & $2.8$ & $-$ & $-$ & $-$ & $-$ & $-$ & $-$ \\ 
                     & $V$ & $-34.0$ & $-0.9$ & $-9.6$ & $-$ & $-$ & $-$ & $-$ & $-$ & $-$ \\ \hline

\end{tabular}
%\end{ruledtabular}
%\end{adjustbox}
\end{table}

\begin{figure*}
	\centering
	\subfloat[\vspace{-8mm}]{\includegraphics[width=0.495\linewidth]{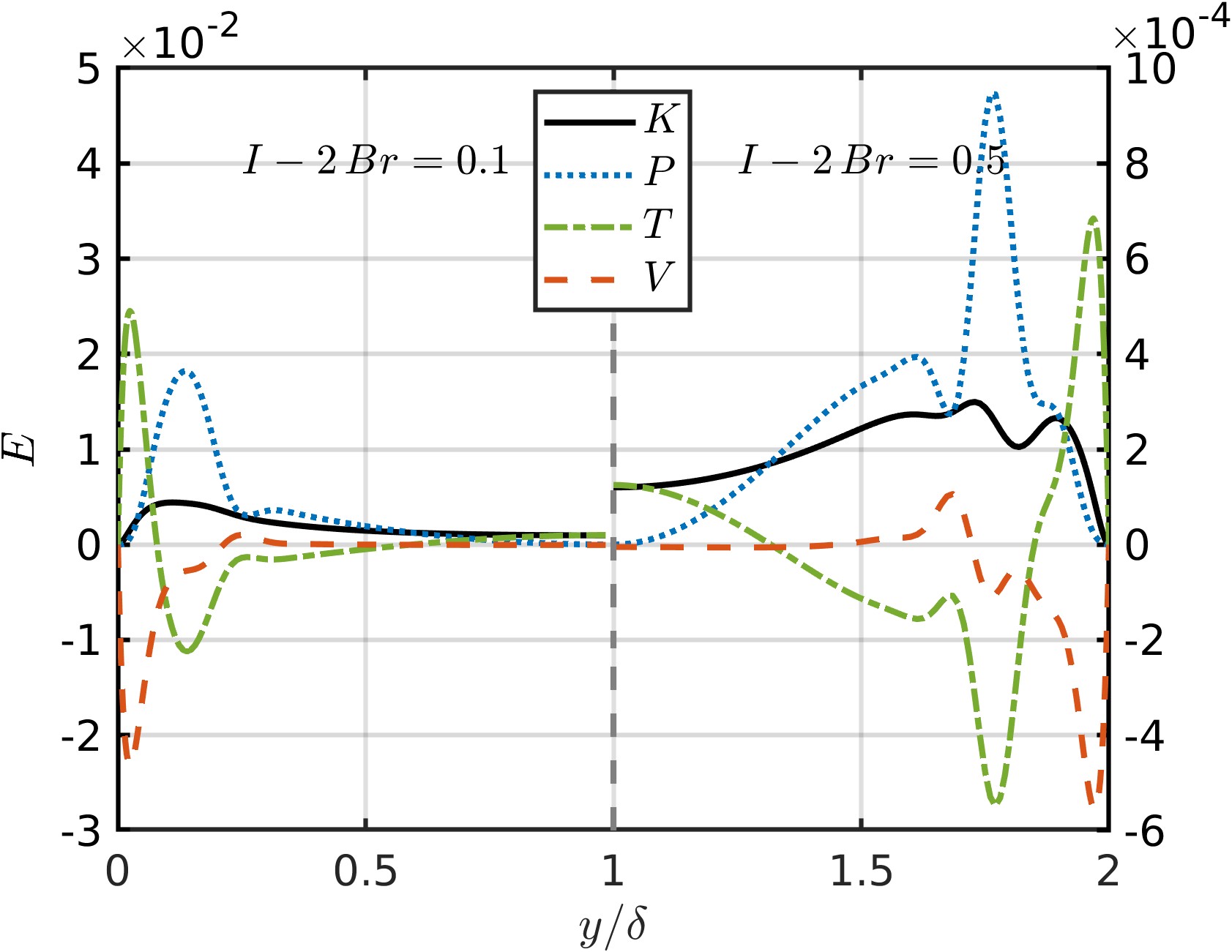}} \hfill
    \subfloat[\vspace{-8mm}]{\includegraphics[width=0.495\linewidth]{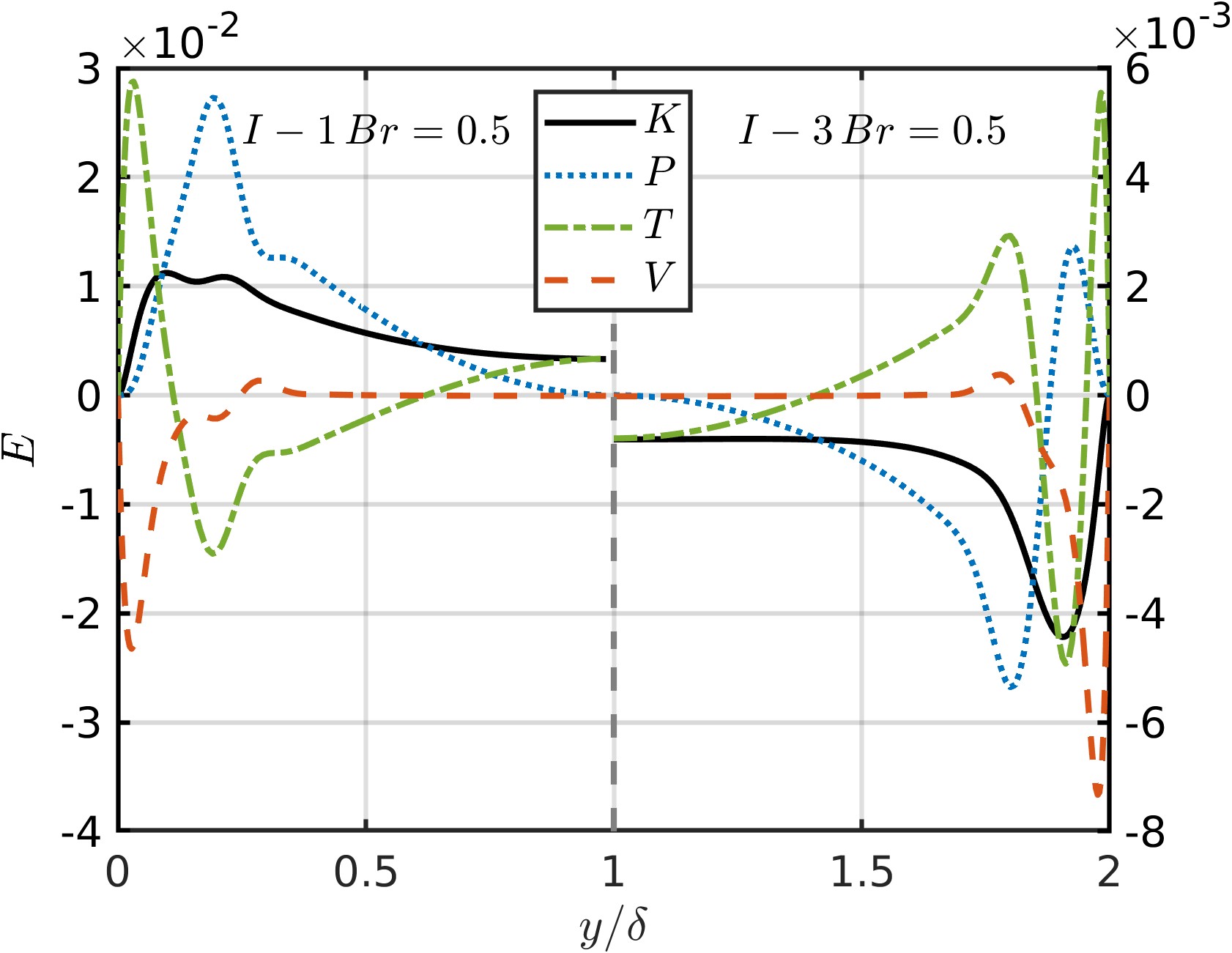}} \\ \vspace{8mm}
    \subfloat[\vspace{-8mm}]{\includegraphics[width=0.485\linewidth]{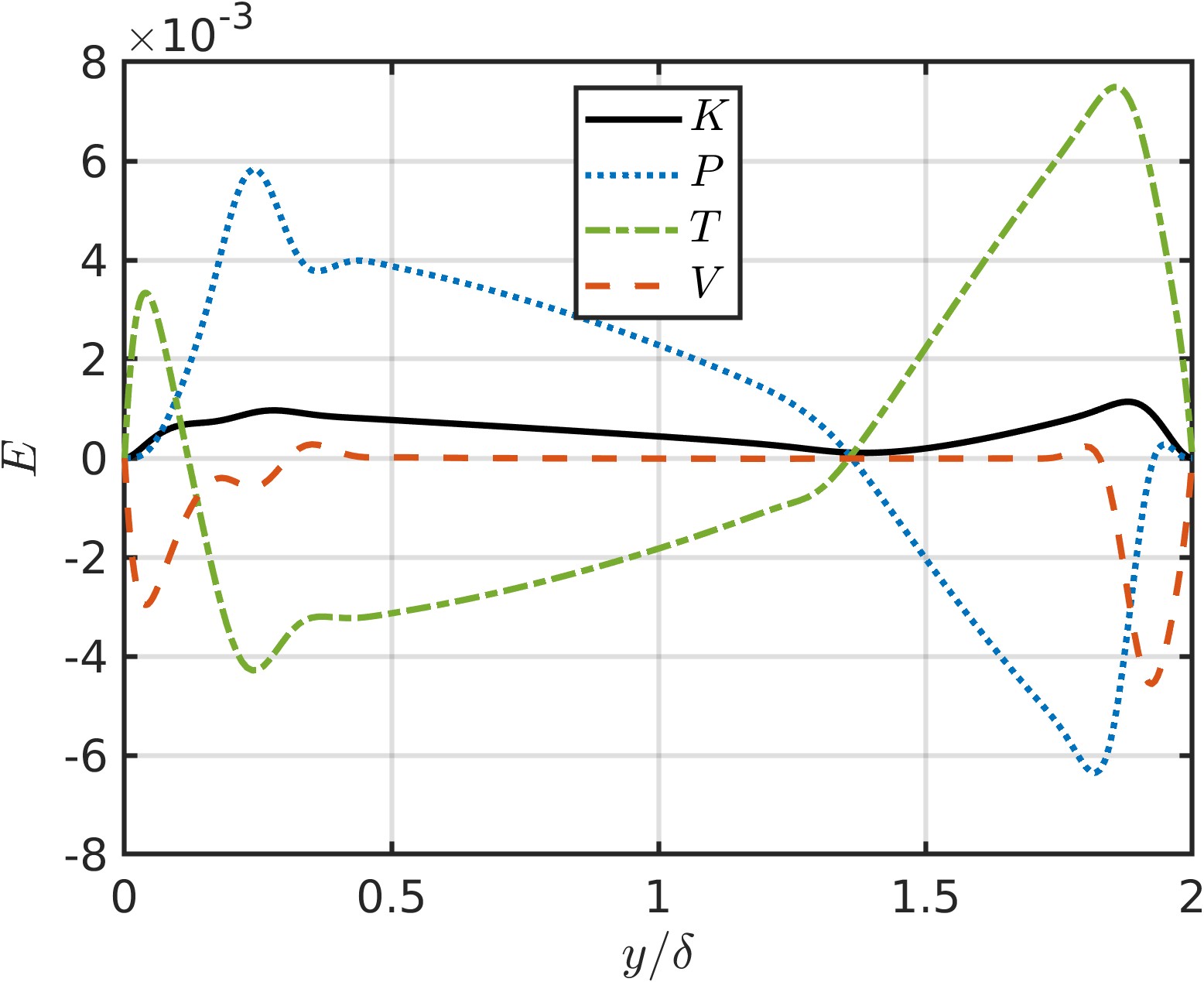}} \hfill
    \subfloat[\vspace{-8mm}]{\includegraphics[width=0.475\linewidth]{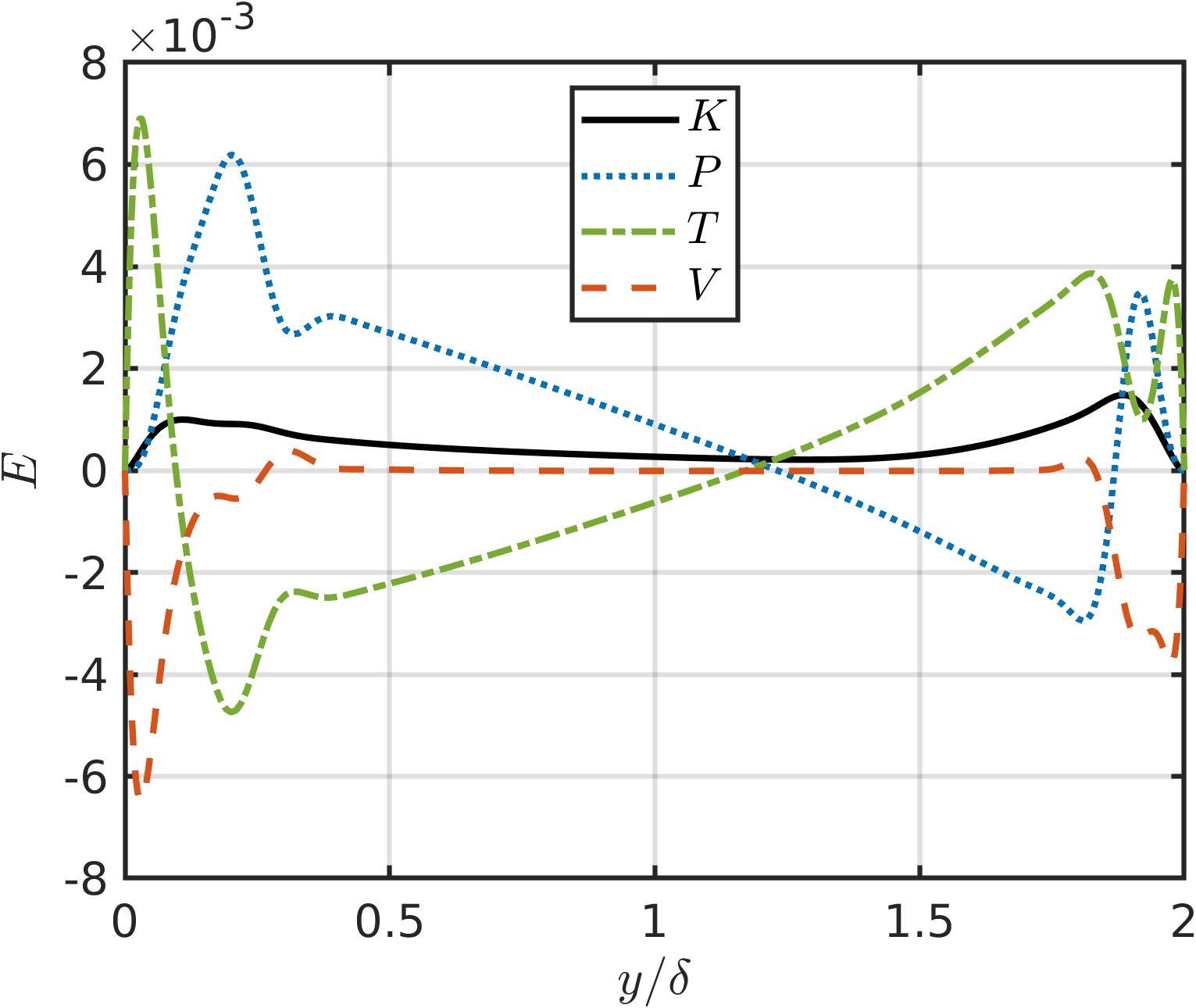}} \\ \vspace{5mm}
	\caption{Energy budget terms along the wall-normal direction at $Re = 10000$ for (a) I-2 at $Br = 0.1$ (left) and $Br = 0.5$ (right), (b) I-1 (left) and I-3 (right) both at $Br = 0.5$, (c) NI-1 at $Br = 0.1$, and (d) NI-3 at $Br = 0.1$.} 
 \label{fig:Energy_budget_spatial}
\end{figure*}

% Paragraph: spatial contribution
Figure~\ref{fig:Energy_budget_spatial} characterizes the spatial growth along the wall-normal direction.
First, the differences across the isothermal setups are depicted in the top row up to the half-plane symmetry.
In particular, Figure~\ref{fig:Energy_budget_spatial}(a) shows the I-2 case at $Br = 0.1$ (below the pseudo-boiling region) and $Br = 0.5$ (crossing the pseudo-boiling region).
As anticipated, it is found that the production term decreases (even in non-dimensionless form) due to the supersonic regimes subjected at large $Br$.
Nonetheless, unlike at lower $Br$ values, the production term rapidly grows away from the centerline achieving its maximum at the inflection point, where, in contrast, the thermodynamic term reaches its minimum.
In general, the thermodynamic contribution dominates at the centerline and in the vicinity of the walls. Moreover, regardless of the $Br$ value, the thermodynamic term introduces flow destabilization near the walls, while viscous dissipation stabilises the flow at similar rates resulting in the cancellation of each other to provide a null net kinetic-energy growth.
In terms of sub- versus supercritical behavior,  Figure~\ref{fig:Energy_budget_spatial}(b) compares I-1 against I-3 at large $Br$ values.
It can be clearly observed that cases at subcritical thermodynamic conditions tend to destabilize the flow, whereas the opposite is obtained at supercritical regimes.
In detail, at subcritical conditions production is important toward the walls and decrease toward the center, whereas the supercritical cases present an opposite behavior changing sign at the centerline and reaching its maximum (positive) in the vicinity of the walls.
Moreover, the thermodynamic contribution is largest at the channel half-height for the subcritical setups, yielding negative values where production peaks and becoming once again the largest destabilization contributor near the walls.
Instead, for the I-3 case the thermodynamic term is negative at the centerline and increases toward the boundaries becoming the main instability source.
This behavior, however, is modified in the wall-normal position where production peaks and the thermodynamic contribution reverses its effect following a double valley-peak trajectory.
However, this change of behavior is not important enough to obtain a positive kinetic-energy growth.
On the other hand, viscous dissipation stabilises the flow near walls in both scenarios.
Finally, the second row of the figure shows the behavior of the non-isothermal cases.
As it can be observed, no important differences can be observed as the Brinkman number changes between cases.
In this regard, Figure~\ref{fig:Energy_budget_spatial}(c-d) depicts the energy budget contributions at $Br = 0.1$ for cases NI-1 and NI-3.
Interestingly, production becomes the key mechanism in regions of relatively large density near the cold wall and achieves its negative peak near the pseudo-boiling region.
This stable region is wider for NI-2, which presents larger temperature difference across walls.
Moreover, case NI-3 (characterized by a smaller temperature difference between walls) yields lower production and thermodynamic peaks, but even so the production term is responsible for destabilizing the flow near the hot wall.
Thus, the thermodynamic term becomes the principal budget contributor in the vicinity of walls and peaking at the pseudo-boiling region, where it changes sign similar to the production term.
Oppositely, viscous dissipation stabilises the flow in the vicinity of walls and vanishes elsewhere.
Hence, it is clear that for isothermal flow setups the production term dominates the kinetic-energy budget, which is similar to what happens in the cold region (near the wall) of the non-isothermal cases.
Nonetheless, in the vicinity of the hot wall and pseudo-boiling region, flow destabilization is dominated by the thermodynamic term.

% Figure: Budget spatial distribution 

\subsubsection{Instability mechanism} \label{sec:instability_mechanism}

% Introduction of potential mechanisms
The growth in kinetic  energy is mostly governed by the production term~\citep{Boomkamp1996-A,Xie2017-A,Samanta2020-A,Andreolli2021-A}.
It plays a crucial role in controlling disturbances by supplying energy to them by means of the Reynolds stresses.
In this regard, it is helpful to examine the vorticity transport equation~\citep{Xie2017-A}, and in particular the components dominating the spanwise vorticity perturbations, which are relevant for shear-flow instabilities.
Thus, this section aims at carefully studying these effects.

\begin{figure*}
	\centering
    \subfloat[\vspace{-8mm}]{\includegraphics[width=0.48\linewidth]{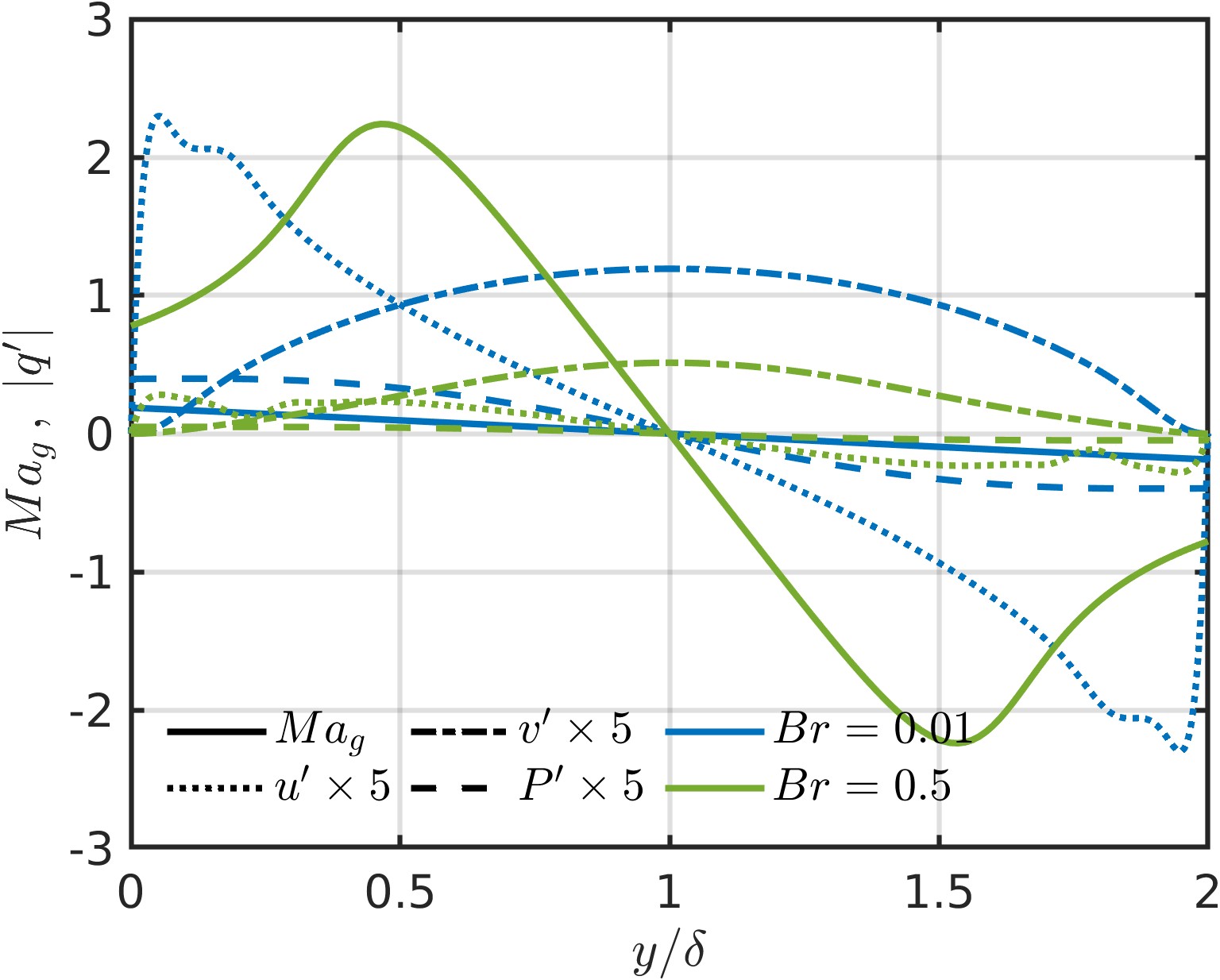}} \hfill
    \subfloat[\vspace{-8mm}]{\includegraphics[width=0.49\linewidth]{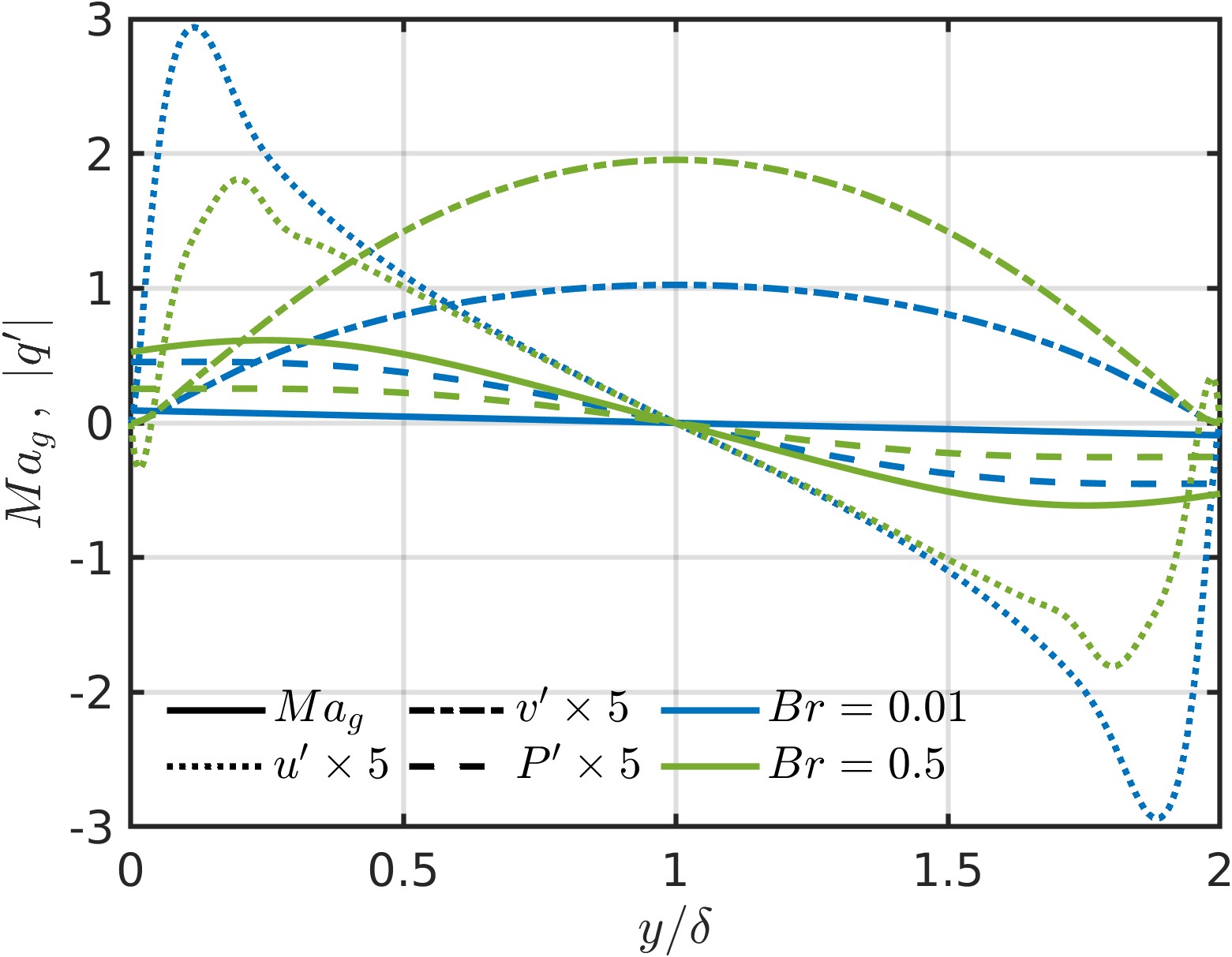}} \\\vspace{5mm}
	\caption{Energy budget compressibility effects for (a) I-2 case at $Br = 0.01$ (${M\!a}_b = 0.1$ and $U_r = 42.3 m/s$) and $Br = 0.5$ (${M\!a}_b = 1.4$ and $U_r = 351 m/s$) depicting the evolution of ${M\!a}_g$, $u^\prime$, $v^\prime$ and $P^\prime$ along the wall-normal direction. The perturbations are scaled by a factor of $5\times$ for visualization. For comparison subcritical (b) I-1 is also reported.} 
 \label{fig:Energy_budget_compressibility}
\end{figure*}

% Detail analysis of the observed compressibility effects
The singular behavior of the production term in the case of isothermal transcritical high-speed flows is briefly characterized, which is beyond the scope of this work as it focuses mainly on the stability of high-pressure fluids at low $M\!a$ and $Re$ conditions.
In this regard, it is known that for incompressible flows the instability mechanisms related to streamwise perturbations is driven by the Tollmien-Schlichting (T-S) waves.
However, for high-speed flows, the T-S waves are typically subjected to the effects of compressibility phenomena.
This type of effects can be quantified by means of the gradient Mach number~\citep{Sarkar1995-A} ${M\!a}_g = [\partial u_0 / \partial y] / [c_0 \sqrt(\alpha^2 + \beta^2)]$.
In detail, given that obliqueness effects are null for the current 2D modal assessment ($\beta = 0$), ${M\!a}_g$ is the dominant factor controlling the production rate.
Generally, under such circumstances, the gradient Mach number exceeds unity near the walls, and consequently these regions are subjected to strong pressure waves and dilatational velocity fluctuations, and exhibit different flow behavior compared to low-speed regimes.
In this context, Figure~\ref{fig:Energy_budget_compressibility} depicts the behavior of case I-2 varying from $Br = 0.01$ to $Br = 0.5$.
As it can be observed, large ${M\!a}_g$ values are encountered near the boundaries, and as a result the fluctuations in streamwise and wall-normal directions are significantly attenuated.
Therefore, the production term of the energy budget is notably reduced yielding a decay of kinetic energy growth.

% Figure: Pressure effects (Mg, qu, qv, qp) low vs. high speed flows isothermal 

\begin{figure*}
	\centering
	\subfloat[\vspace{-8mm}]{\includegraphics[width=0.48\linewidth]{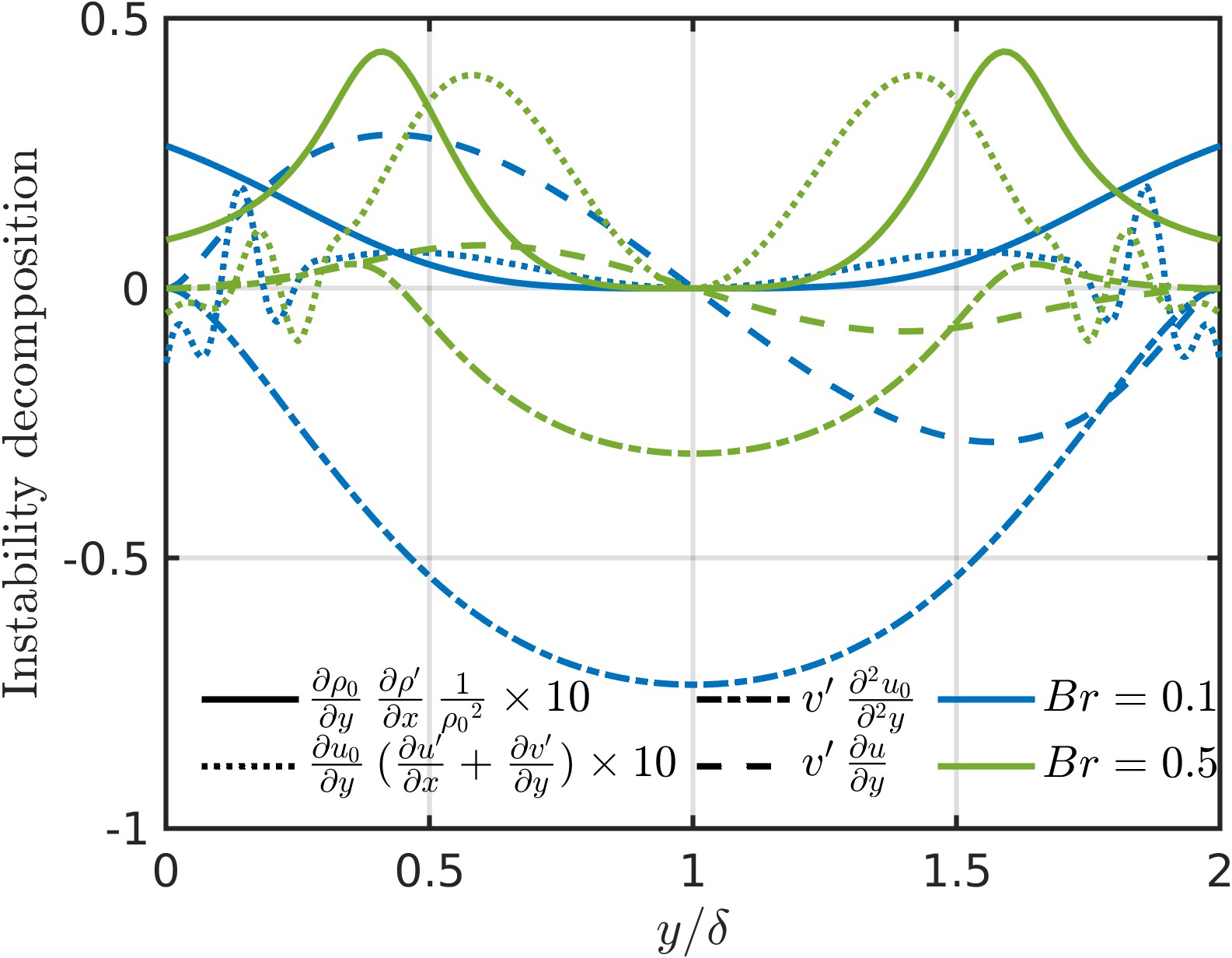}} \hfill
    \subfloat[\vspace{-8mm}]{\includegraphics[width=0.48\linewidth]{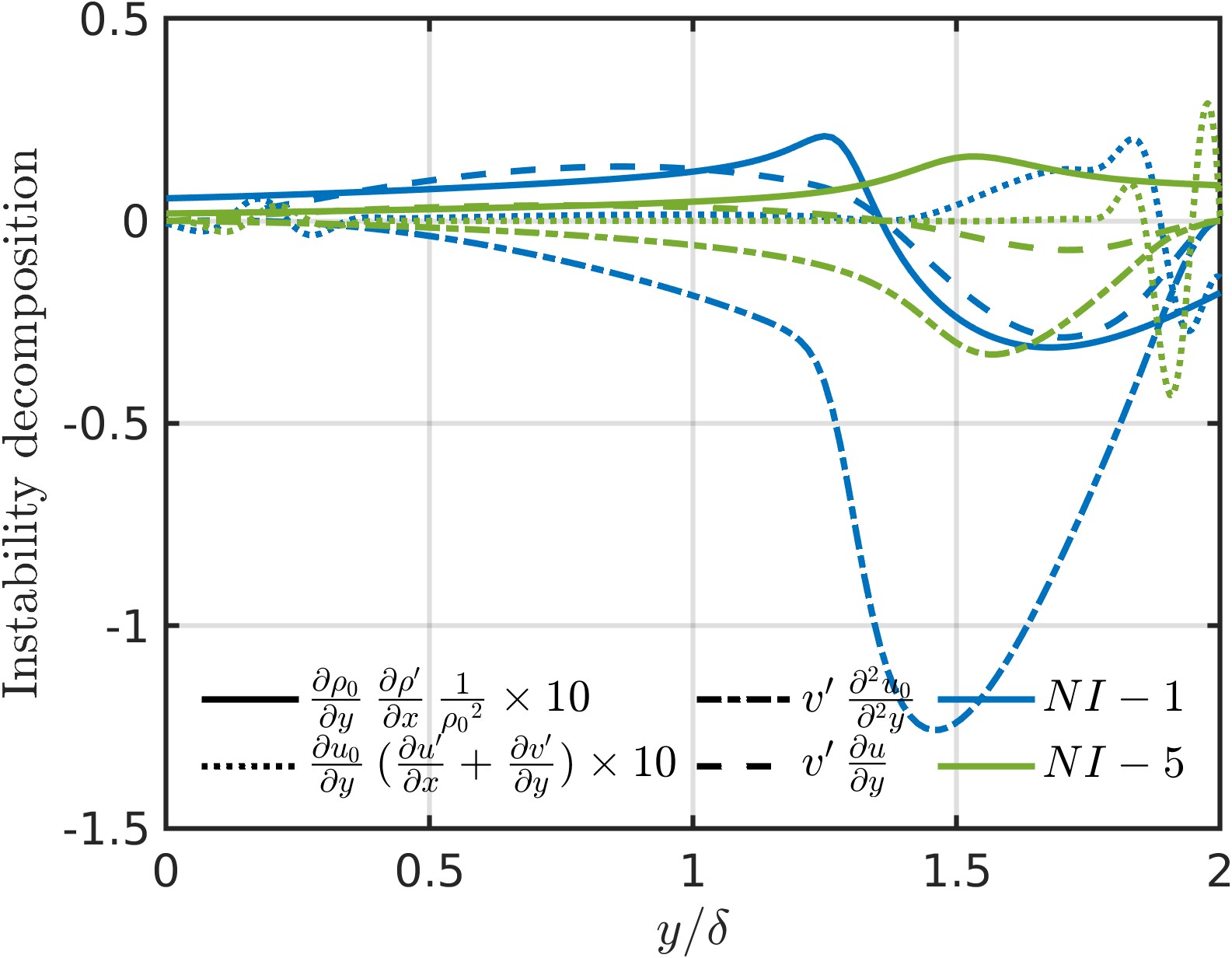}} \\ \vspace{5mm}
	\caption{Energy instability mechanisms along the wall-normal direction at $Re = 10000$ for (a) $I-2$ at $Br = 0.1$ and $Br = 0.5$, and (b) $NI-1$ at $Br = 0.1$ and $NI-5$. The baroclinic effect and compressible vortex production are scaled by a factor of $10\times$ for visualization.} 
 \label{fig:Energy_budget_instability}
\end{figure*}

% Vorticity asessment
The spanwise vorticity is decomposed into three main components \citep{Xie2017-A}: (i) baroclinic effect $(\partial \rho_0 / \partial y)(\partial \rho^\prime / \partial x) / {\rho_0}^2$, (ii) compressible vortex production $(\partial u_0 / \partial y)(\partial u^\prime / \partial x + \partial v^\prime / \partial y)$, and (iii) second derivative effect $v^\prime (\partial^2 u_0 / \partial^2 y)$.
Among these contributions, the shear flow instability is typically critical for second derivative and the production field from the velocity equation $v^\prime \partial u / \partial y$.
These multiple mechanisms can be identified in Figure~\ref{fig:Energy_budget_instability}.
On the one hand, the isothermal transcritical case is depicted in Figure~\ref{fig:Energy_budget_instability}(a) to clarify the importance of both vorticity second derivative (especially at the channel centerline) and production terms.
The baroclinic effect and compressible vortex production are one order of magnitude smaller.
Interestingly, the larger the $Br$ the more important these two terms become in detriment of the dominant ones.
On the other hand, Figure~\ref{fig:Energy_budget_instability}(b) compares NI-1 at $Br = 0.1$ and NI-5.
For both cases, the second derivative of vorticity emerges as the principal destabilization mechanism, specially near the pseudo-boiling region.
In particular, in the case of low-speed conditions, the baroclinic effect achieves similar magnitude as for larger velocity flows at the pseudo-boiling region.
It is important to note that when the temperature difference between walls is reduced (case NI-3) the magnitude of all mechanisms become significantly smaller, specially the velocity production.
The baroclinic effect, however, preserves similar magnitudes at the pseudo-boiling region.

% Figure: Compare growing mechanisms (baroclinic effect, compressible vortex production, second derivative) isothermal trans- vs. non-isothermal trans

\subsection{Transient energy growth analysis}  \label{sec:Algebraic_growth}	

Linear modal stability is limited to the asymptotic behavior of disturbances, omitting large transient amplifications at short time.
This short-time energy growth can be significant compared to the initial perturbation level.
Therefore, to capture more insights into the mechanisms of transition to turbulence, the non-normality of the linearized Navier-Stokes operator needs to be taken into account~\citep{Schmid2007-A,Ghosh2019-A,Nastro2020-B}.
In this regard, based on 3-D perturbations ($\beta > 0$), algebraic instabilities are quantified for the various flow cases studied.
Particularly, this section characterizes (i) the maximum growth rate and transient amplifications, and (ii) the optimal perturbation profiles which generate the maximum growth rate; for completeness, validation against incompressible~\citep{Reddy1993-A} and isothermal~\citep{Ren2019b-A} high-pressure Poiseuille flow cases are reported in Appendix~\ref{sec:Appendix_E}.

\subsubsection{Maximum growth rate and transient amplification}

\begin{figure*}
	\centering
    \subfloat[\vspace{-6mm}]{\includegraphics[width=0.32\linewidth]{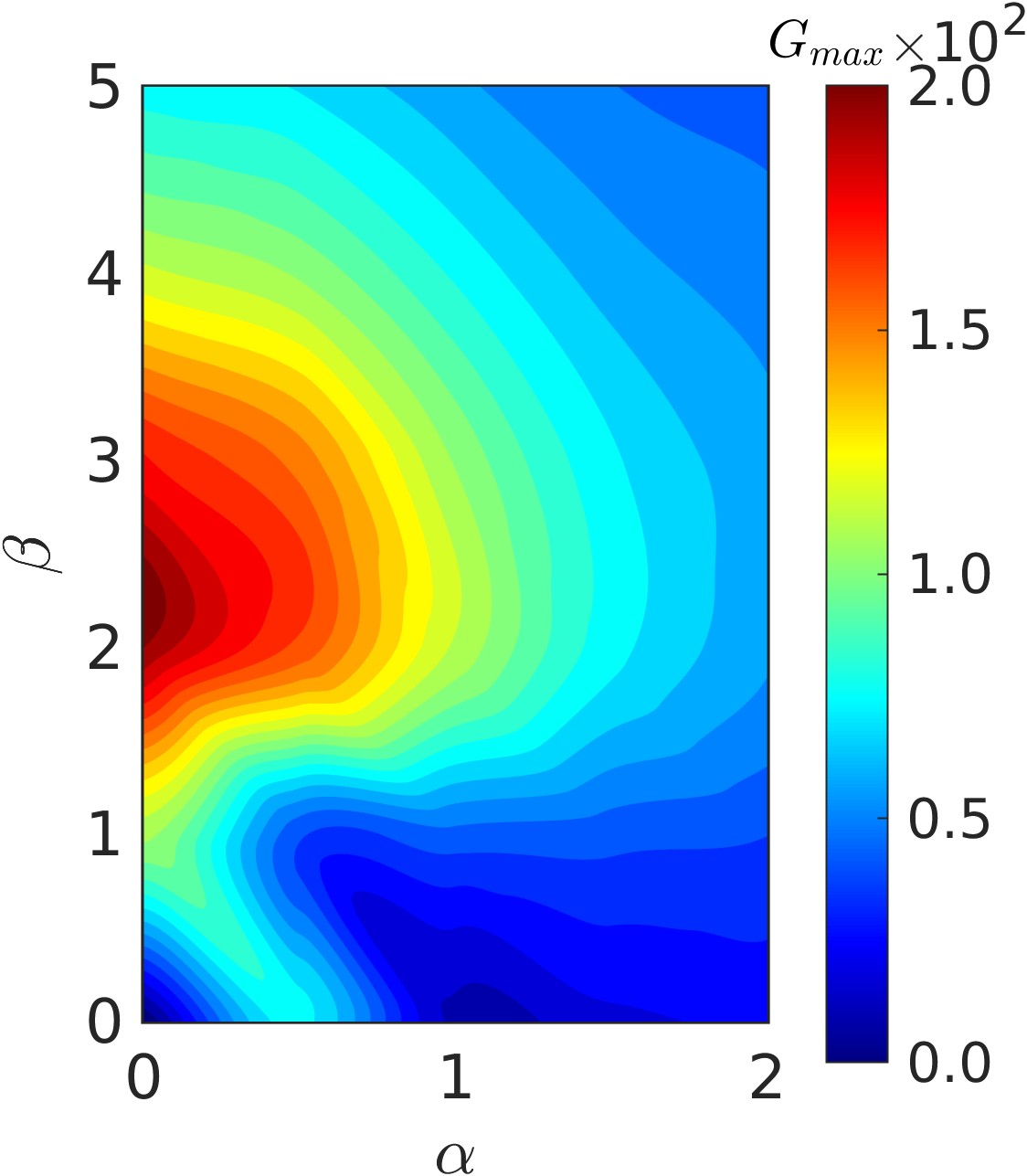}} \hspace{0.5mm}
    \subfloat[\vspace{-6mm}]{\includegraphics[width=0.32\linewidth]{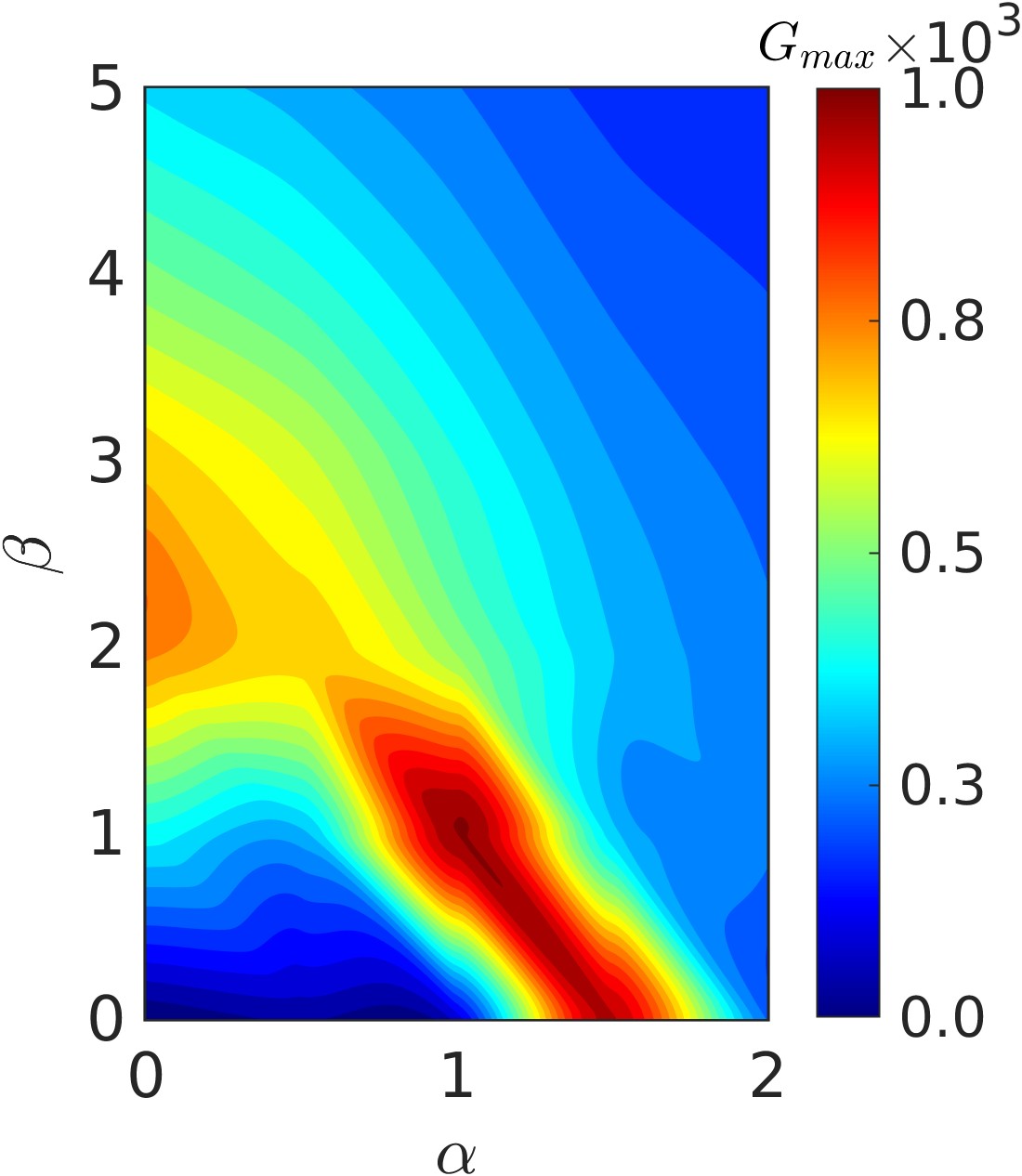}} \hspace{0.5mm} 
    \subfloat[\vspace{-6mm}]{\includegraphics[width=0.32\linewidth]{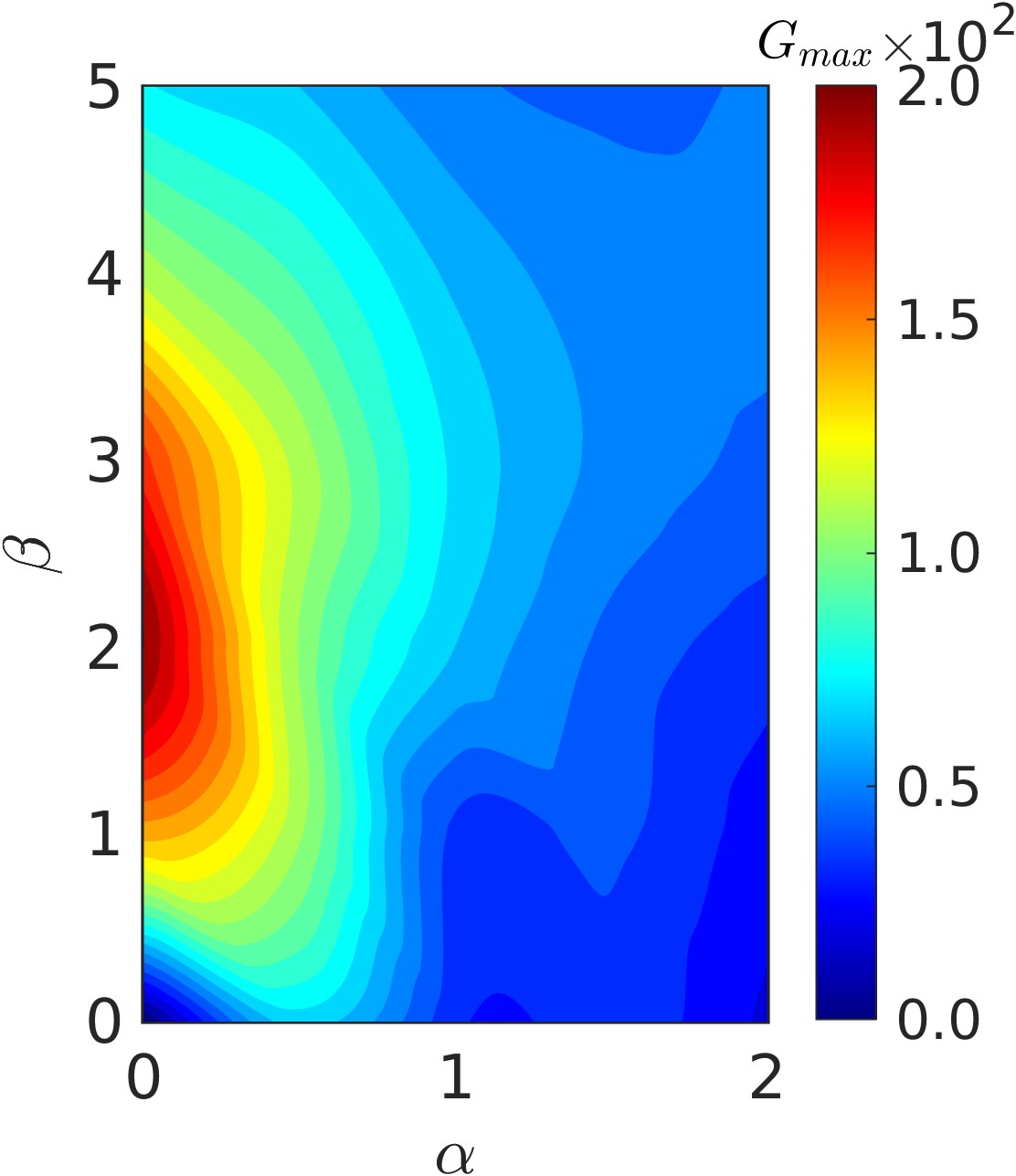}} \\ \vspace{2mm}
	\caption{Transient growth envelopes at $Re = 1000$ and $Br = 0.5$ for (a) I-1, (b) I-2 and (c) I-3 cases. The resolution range for case I-2 has been limited to $G_{max} = 1000$.} 
 \label{fig:Transient_growth_maps_Iso}
\end{figure*}

\begin{figure*}
	\centering
    \subfloat[\vspace{-6mm}]{\includegraphics[width=0.32\linewidth]{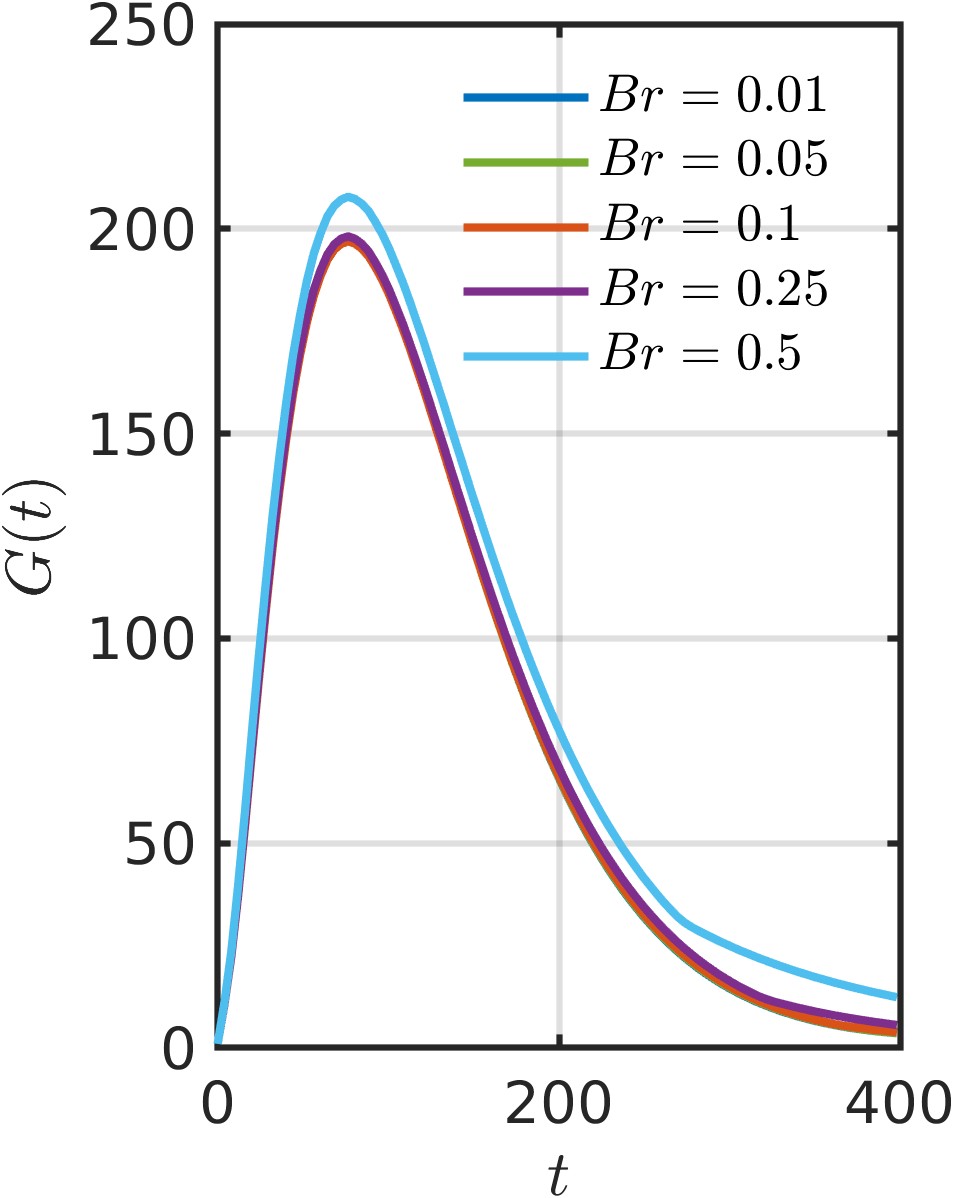}} \hspace{0.5mm}
    \subfloat[\vspace{-6mm}]{\includegraphics[width=0.32\linewidth]{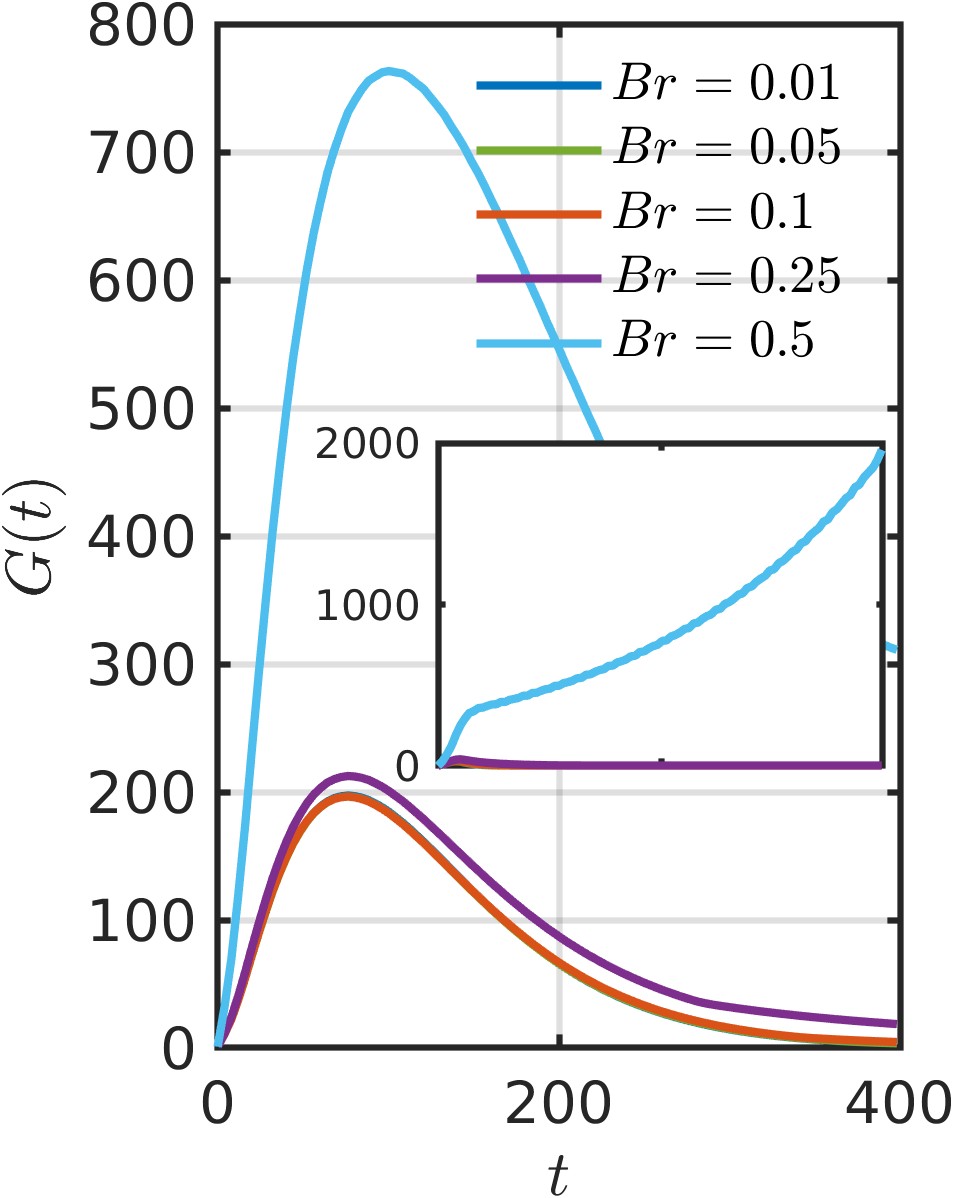}} \hspace{0.5mm}
    \subfloat[\vspace{-6mm}]{\includegraphics[width=0.32\linewidth]{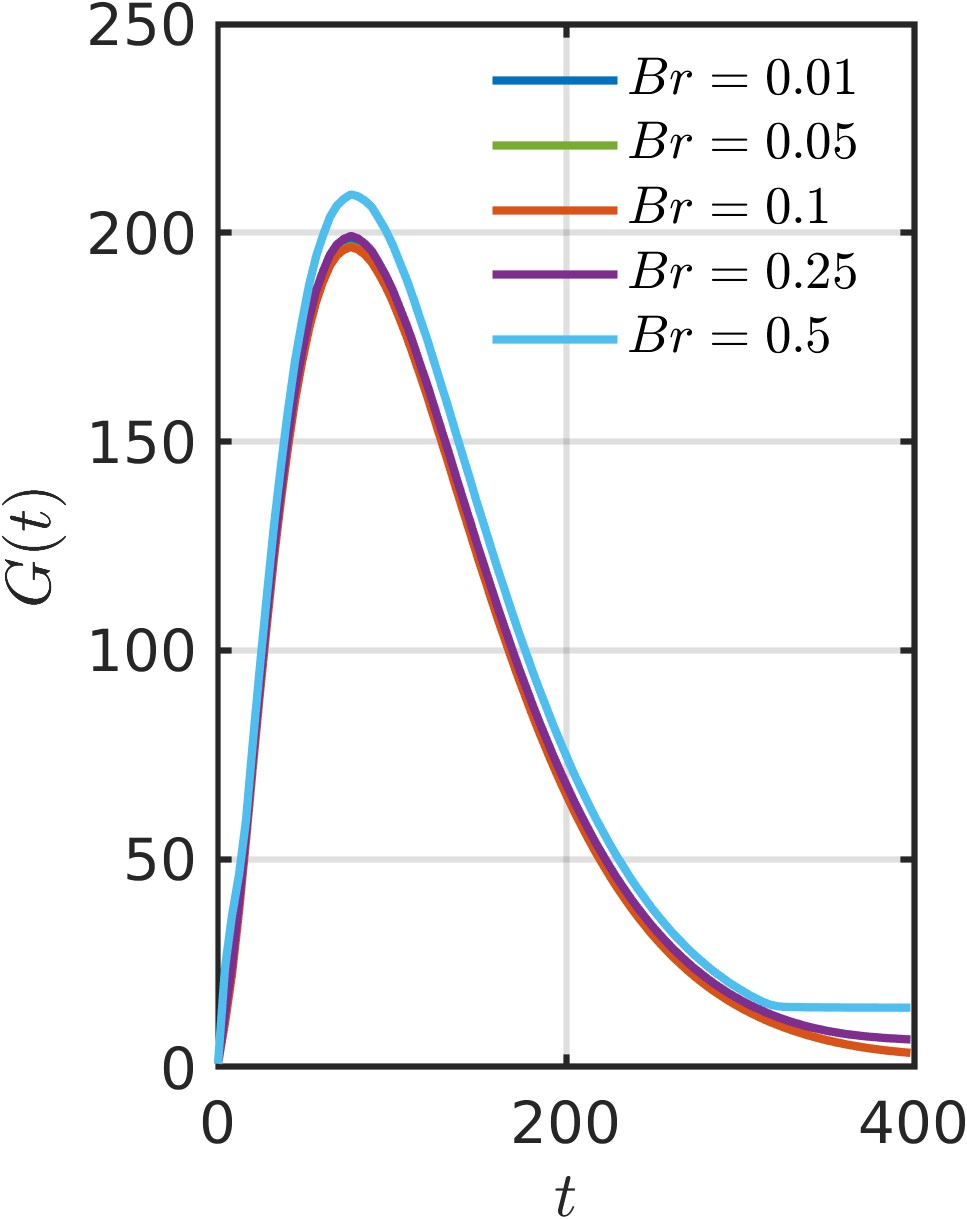}} \\ \vspace{2mm}
	\caption{Amplification rates for $Re = 1000$ at $\alpha = 0$ and $\beta = 2$ for (a) I-1, (b) I-2, and (c) I-3 cases. Inset of (b) corresponds to the exponential amplification rate at $\alpha = \beta =1$ for case I-2.} 
 \label{fig:Amplification_rates_Iso}
\end{figure*}

% Isothermal cases
The effects of the Brinkman number and thermodynamic regime at iso- and non-isothermal flow conditions on the non-modal growth of the disturbance energy is investigated.
The Reynolds number $Re = 1000$ is chosen, far below the transition point based on the linear stability analyses summarized in Table~\ref{tab:summary_flowcases}.
First, the transient growth envelopes on the $\alpha$-$\beta$ space for the isothermal setups are detailed in Figure~\ref{fig:Transient_growth_maps_Iso}.
They reveal that the kinetic energy of the perturbations grows by a factor of $200$ when operating at sub- and super-critical regimes.
Furthermore, the kinetic energy can be enhanced by a factor of $800\times$ when applying an invariant streamwise perturbation ($\alpha = 0$), and even become inviscid unstable at $\alpha = \beta = 1$.
In fact, similar behavior was observed by~\citet{Schmid2007-A} for Poiseuille flow under 3D perturbations at $Re = 10000$.
Particularly, the largest (asymptotic exponential) transient growth occurs for streamwise independent perturbations at $\alpha = 0$ and $\beta = 2$.
Nonetheless, it is noted that case I-2 deviates from this trend at $Br = 0.5$, showcasing a region in which the maximum growth tends to infinity for $\alpha = \beta = 1$.
In detail, the spectra yield an unstable mode responsible for exponential growth rate over time, and as a consequence the critical Reynolds number is notably reduced. At $Br = 0.25$ this behavior is no longer captured and smaller $Br$ values result in maximum growth localized only at the invariant streamwise perturbation.
Moreover, the colormaps of cases I-1 and I-3 at low $Br$ values behave similarly to the results previously presented.
In particular, they have indistinguishable shape and maxima with slight reduction of the maximum growth region, which is localized at $\alpha = 0$ and $\beta = 2$.
Hence, for the sake of brevity, only results for the largest Brinkman number $Br = 0.5$ are depicted. 
Figure~\ref{fig:Amplification_rates_Iso} depicts the evolution of the growth rate over time at the maximum growth rate on the $\alpha-\beta$ space.
As it can be observed, the resulting growth rate presents a monotonically increasing trend reaching the maximum destabilization energy after it starts attenuating.
In detail, cases I-1 and I-3 develop similar energy instability behavior for the different base flows analyzed.
However, the transcritical case evolves to larger amplifications and it is exacerbated for $\alpha = \beta = 1$.

% Figure Growth rate: I-1-2-3 for Br = 0.5

% Figure Growth rate I-1-2-3 for Br = 0.5, alpha = 0, beta = 2

\begin{figure*}
	\centering
    \subfloat[\vspace{-6mm}]{\includegraphics[width=0.32\linewidth]{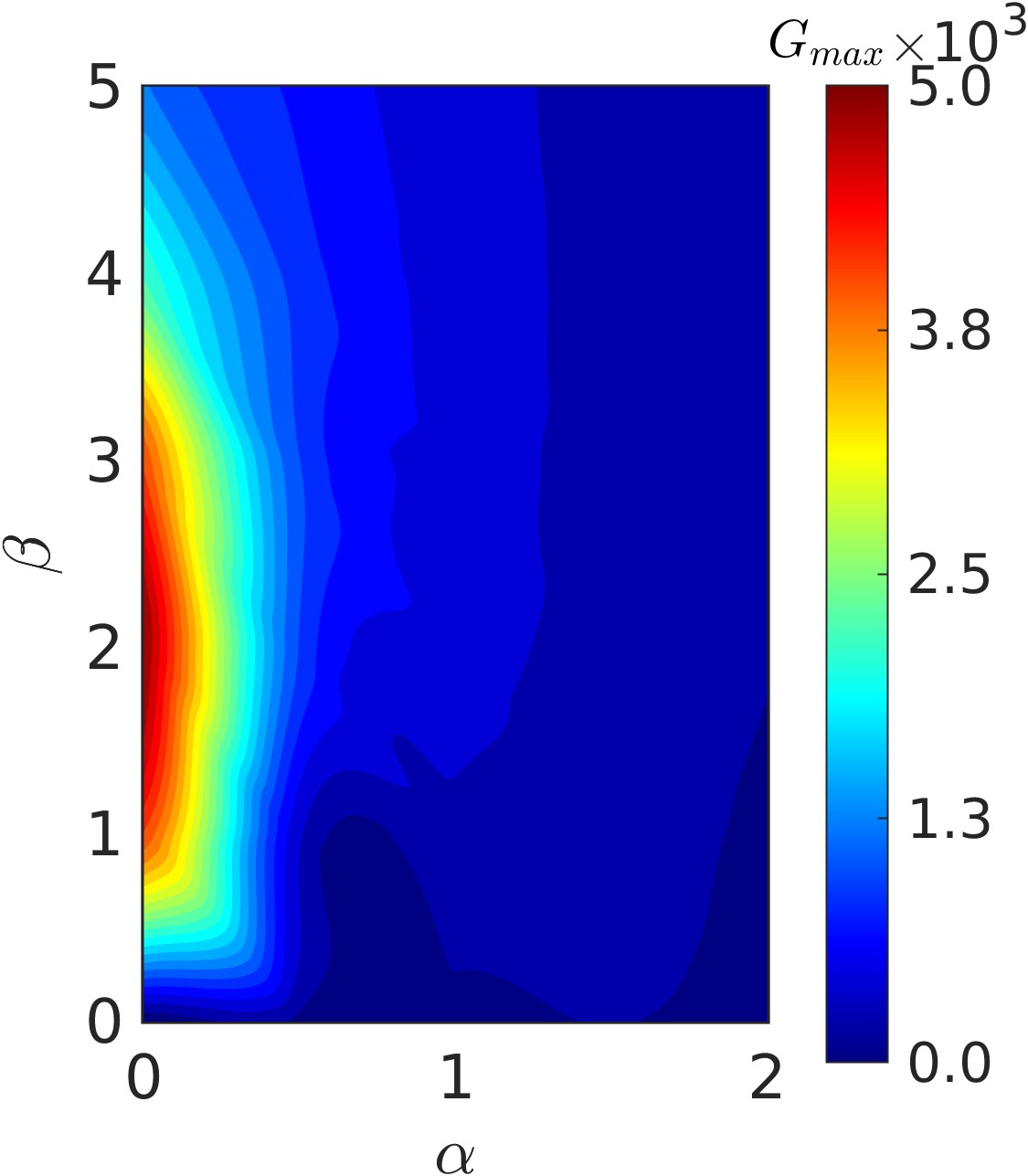}} \hspace{0.5mm}
    \subfloat[\vspace{-6mm}]{\includegraphics[width=0.32\linewidth]{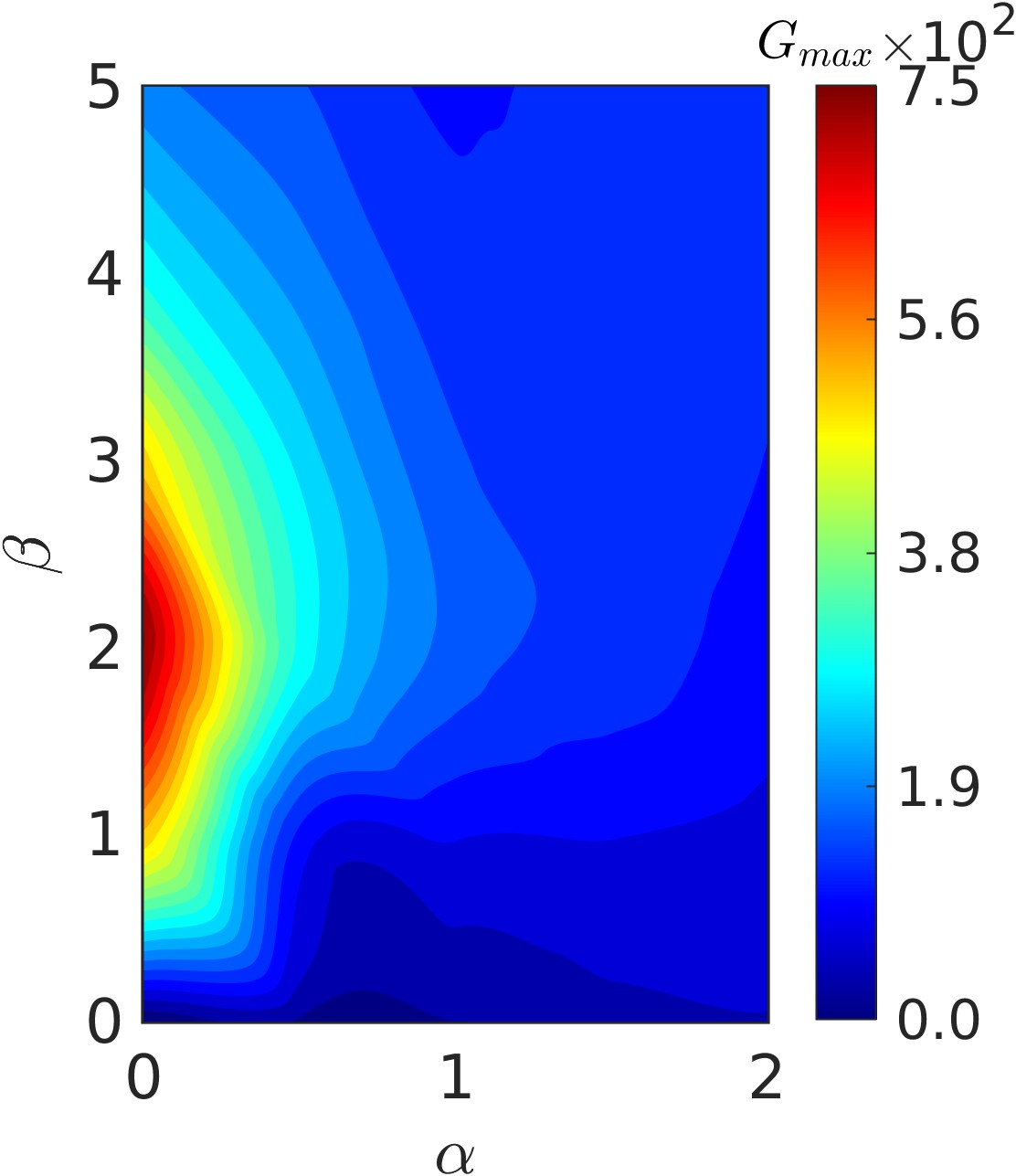}} \hspace{0.5mm} 
    \subfloat[\vspace{-6mm}]{\includegraphics[width=0.32\linewidth]{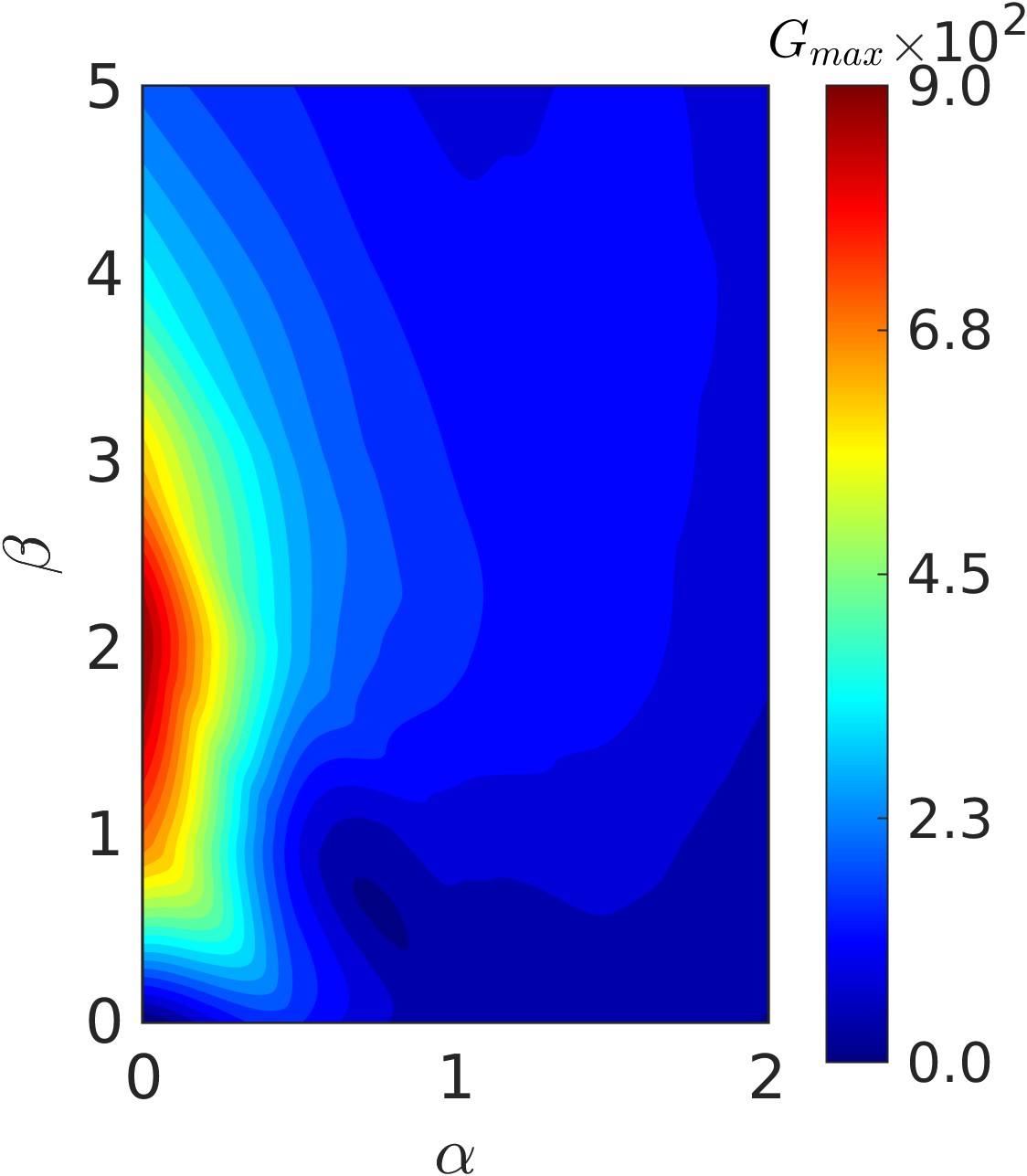}} \\ \vspace{2mm}
	\caption{Transient growth envelopes at $Re = 1000$ and $Br = 0.1$ for (a) NI-1, (b) NI-2, and (c) NI-3 cases.} 
 \label{fig:Transient_growth_maps_NonIso}
\end{figure*}

\begin{figure*}
	\centering
    \subfloat[\vspace{-6mm}]{\includegraphics[width=0.32\linewidth]{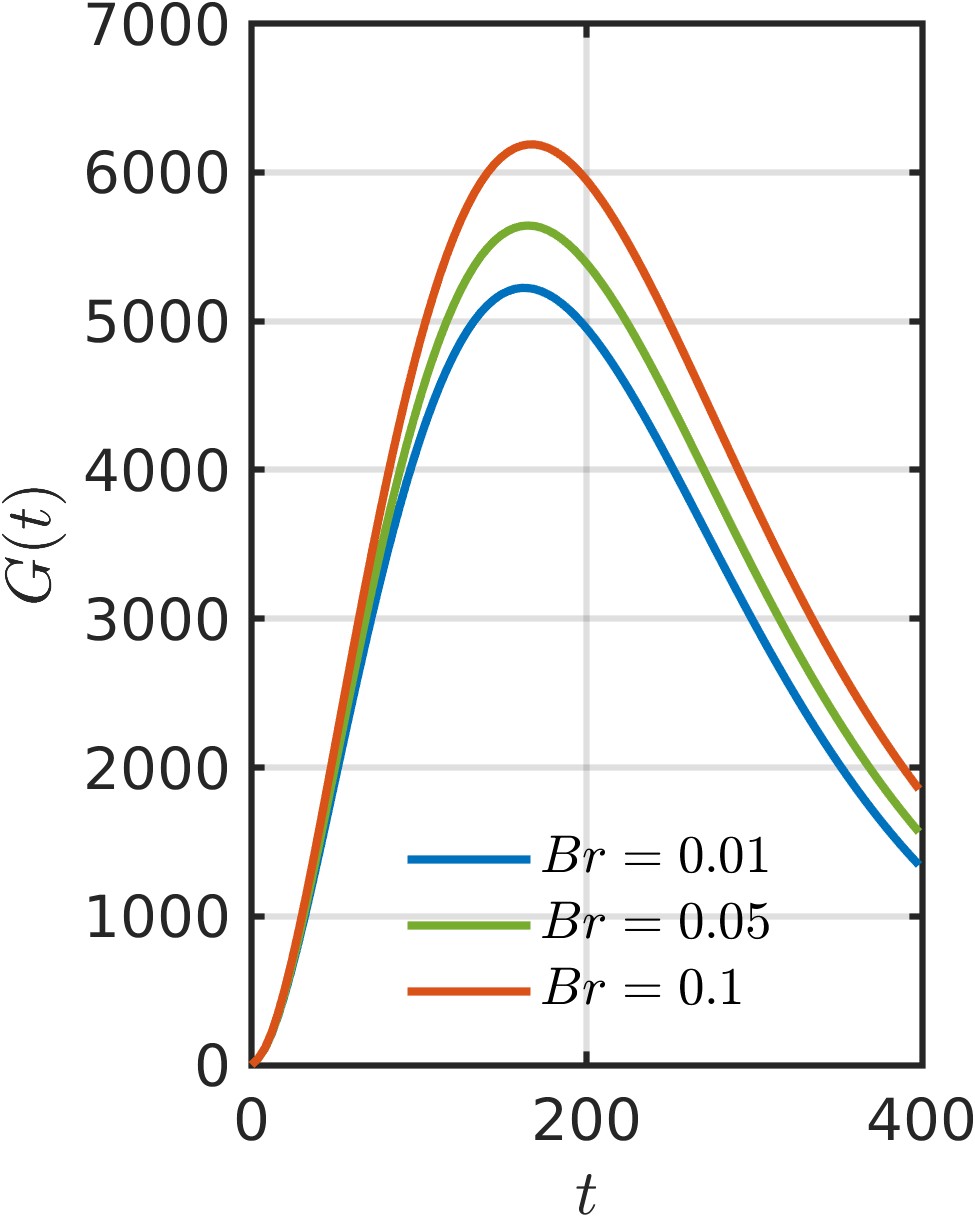}} \hspace{0.5mm}
    \subfloat[\vspace{-6mm}]{\includegraphics[width=0.315\linewidth]{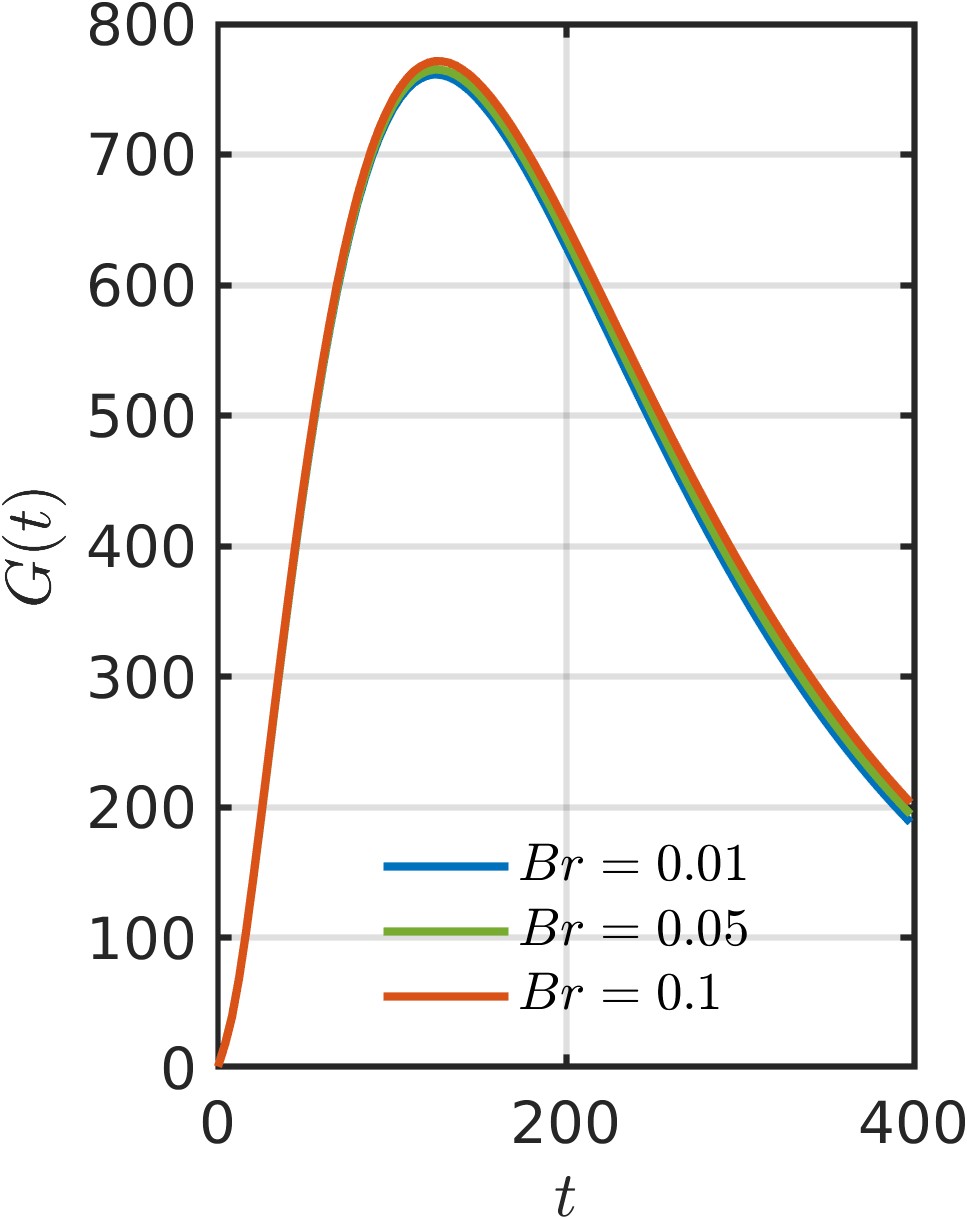}} \hspace{0.5mm}
    \subfloat[\vspace{-6mm}]{\includegraphics[width=0.325\linewidth]{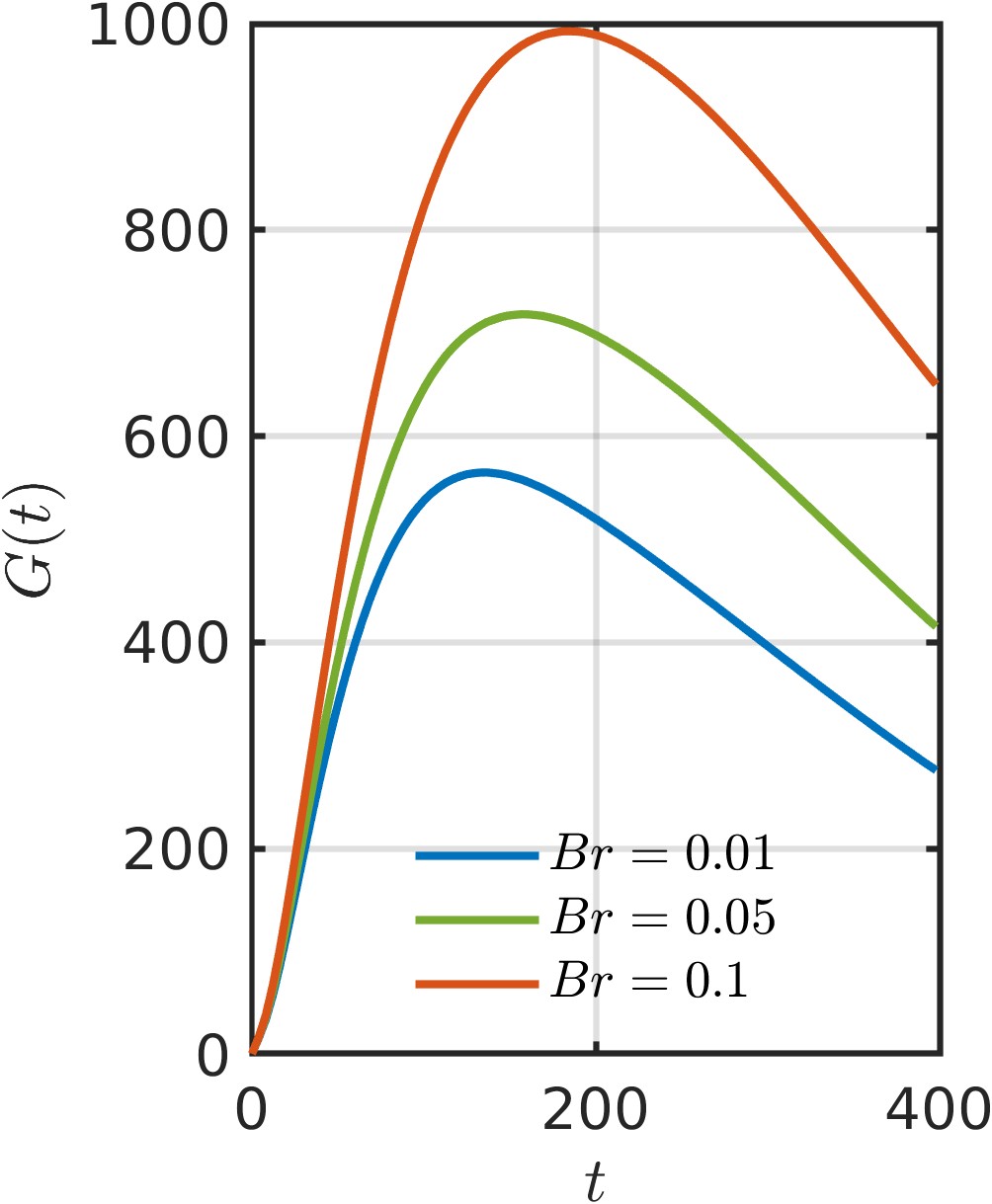}} \\ \vspace{2mm}
	\caption{Amplification rates for $Re = 1000$ at $\alpha = 0$ and $\beta = 2$ for (a) NI-1, (b) NI-2, and (c) NI-3 cases.} 
 \label{fig:Amplification_rates_NonIso}
\end{figure*}

% Non-isothermal cases (NI-1,2,3)
Analogously to the discussion above focused on the isothermal cases, the transient growths of the non-isothermal NI-1, NI-2 and NI-3 cases are shown in Figure~\ref{fig:Transient_growth_maps_NonIso}.
It can be observed that there is a clear range at $\alpha = 0$ and $1.0 \le \beta \le 3.0$ where amplification is large despite  corresponding to fully stable spectra.
Nonetheless, this range is notably reduced when operating at $P_b/P_c = 5$ (case NI-2) for the optimum wavenumber $\alpha = 0$ and $\beta = 2$ in comparison to the isothermal case.
In particular, the amplification is significantly reduced by a factor of roughly $8\times$.
Moreover, narrowing the temperature operating window reduces the amplification by approximately $5\times$, but the energy boost from the non-modal instability is longer sustained within the system as quantified by the amplification rate evolution in Figure~\ref{fig:Amplification_rates_NonIso}.
As previously introduced, Brinkman numbers are limited to $Br \le 0.1$ so that temperatures are constrained within the range imposed by the wall temperatures.
It is important to note, therefore, that when $Br$ increases the transient growth significantly increases along a large region of peak locations for $\beta = 2$ and $\alpha \ge 0$.

% Figure Growth rate: NI-1-2-3 for Br = 0.1

% Figure Amplifications I-1-2-3 for Br = 0.5, alpha = 0, beta = 2

\begin{figure*}
	\centering
    \subfloat[\vspace{-6mm}]{\includegraphics[width=0.35\linewidth]{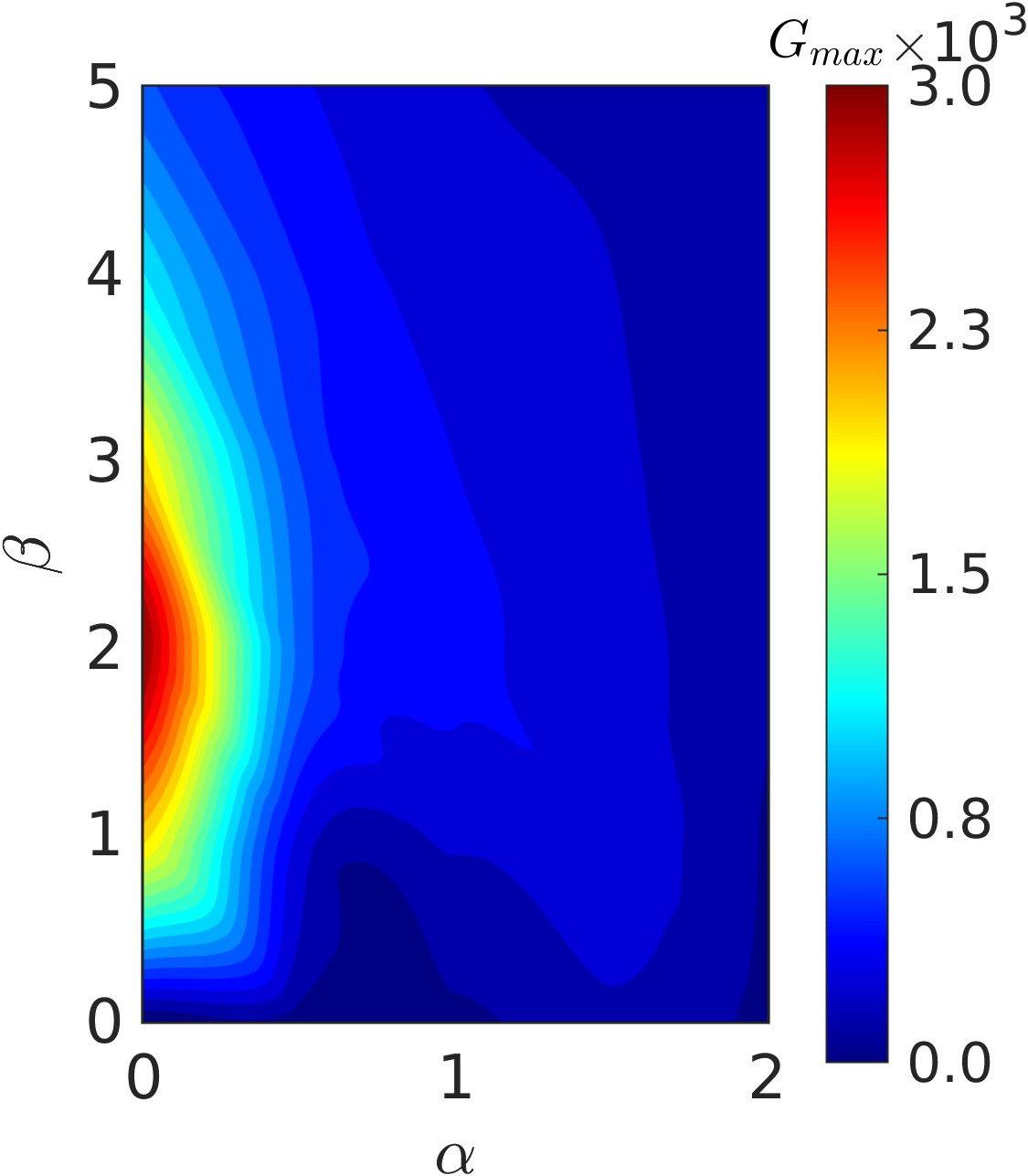}} \hspace{5mm}
    \subfloat[\vspace{-6mm}]{\includegraphics[width=0.35\linewidth]{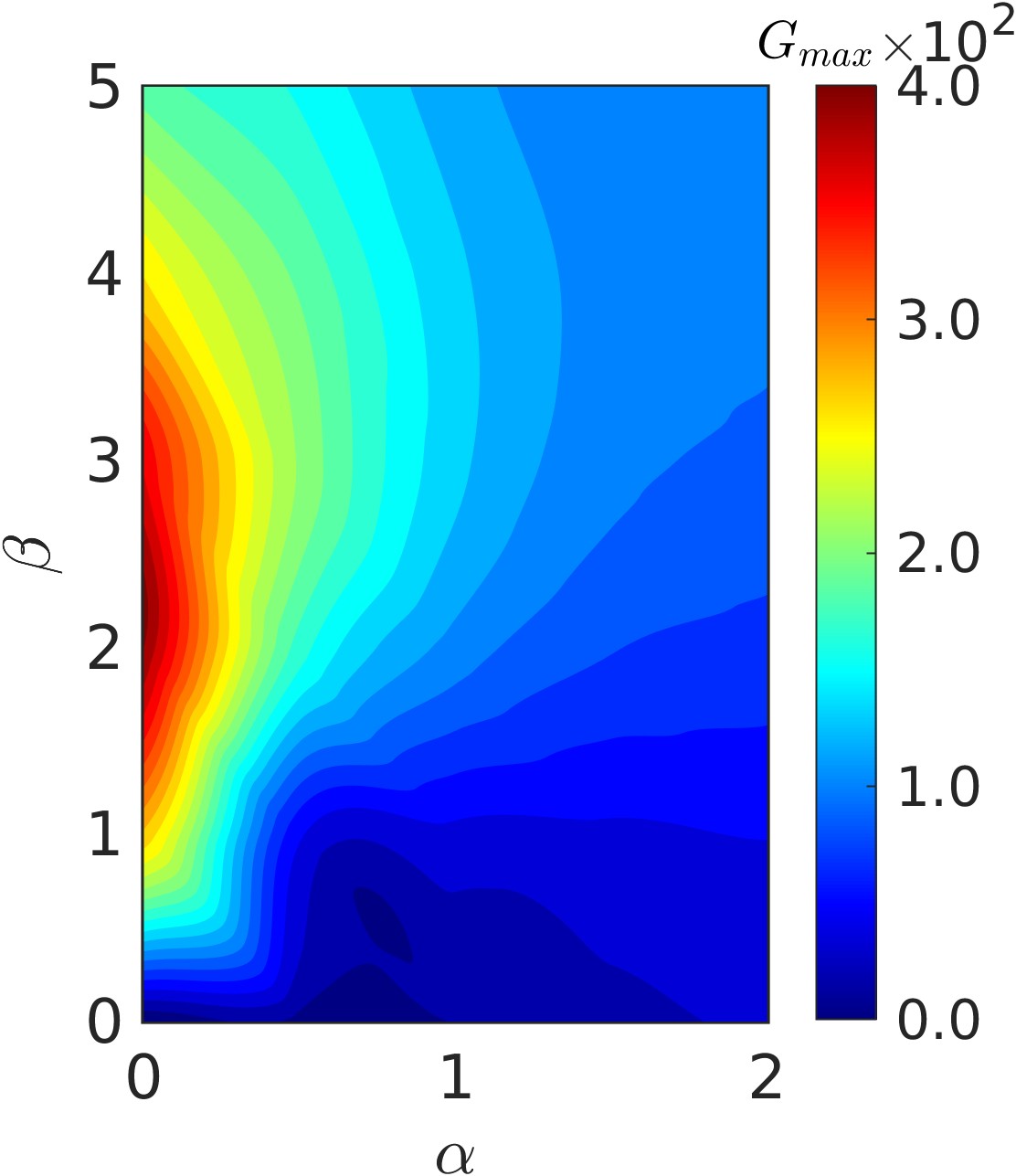}} \\ \vspace{2mm}
	\caption{Transient growth envelopes at $Re = 1000$ and $Br = 5.6 \cdot 10^{-6}$ for (a) NI-5 and (b) NI-6 cases.} 
 \label{fig:Amplification_rates_NonIso_Ur_1}
\end{figure*}

% Low Mach cases (NI-5-6)
The different transition properties of the non-isothermal cases with respect to the isothermal setups for low Reynolds and Mach number conditions has been discussed in Section~\ref{sec:Modal_stability}.
Therefore, non-modal analyses are carried out only for cases NI-5 ($P_b/P_c = 2$) and NI-6 ($P_b/P_c = 0.03$).
In this regard, Figure~\ref{fig:Amplification_rates_NonIso_Ur_1} depicts the transient growth envelopes of these two cases.
As it can be observed, although the $Br$ values are relatively small, the energy growth presents large rates when operating at transcritical conditions (NI-5) with only an approximate $50\%$ reduction in maximum growth with respect to larger $Br$ cases.
Moreover, the optimum remains located at a similar wavelength region.
Instead, if the system operates at atmospheric pressure conditions, the growth rate is notably reduced.

% Figure Transient Growth Rate NI-5 and NI-6

\begin{figure*}
	\centering
    \subfloat[\vspace{-6mm}]{\includegraphics[width=0.32\linewidth]{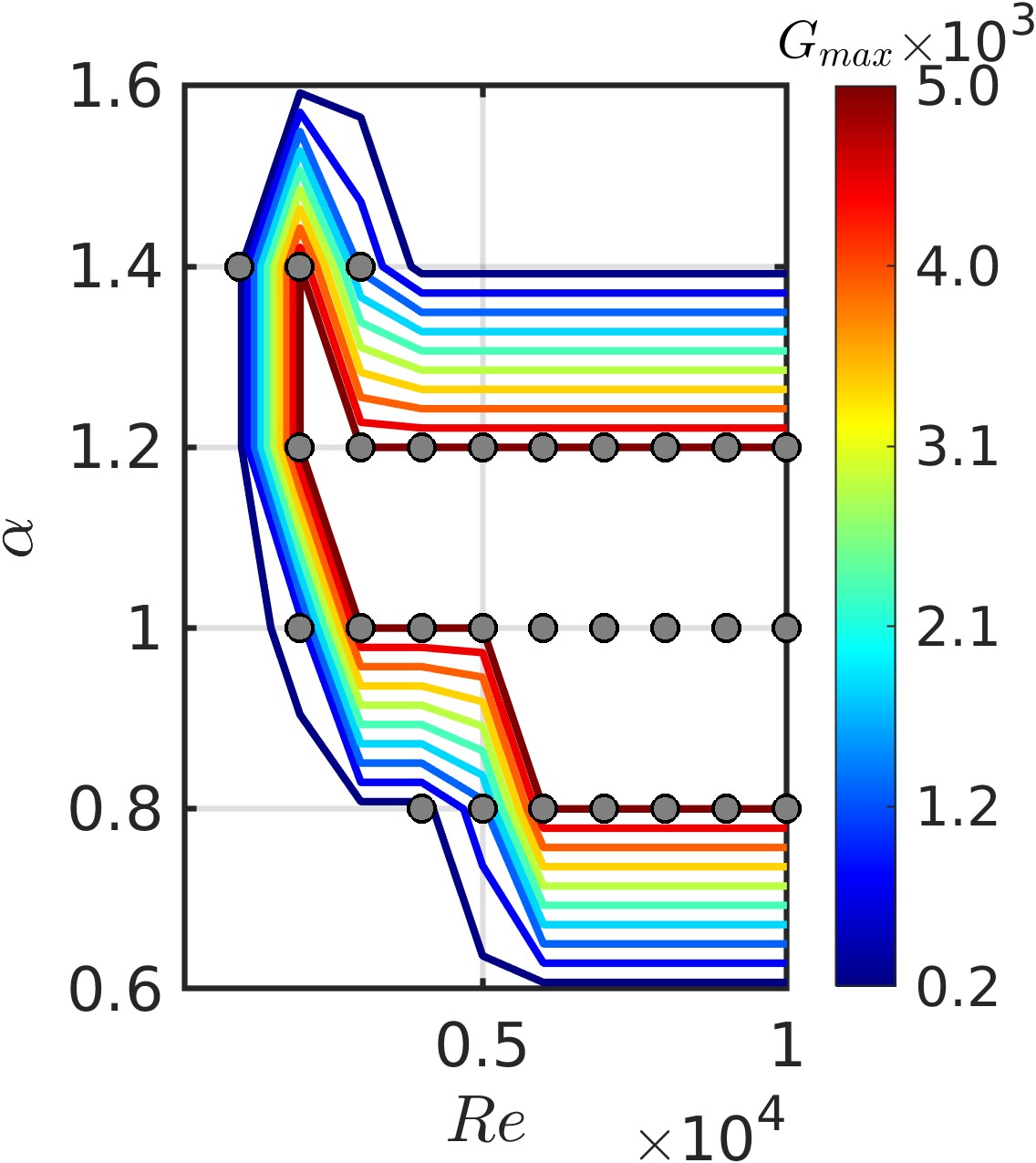}} \hspace{0.5mm}
    \subfloat[\vspace{-6mm}]{\includegraphics[width=0.32\linewidth]{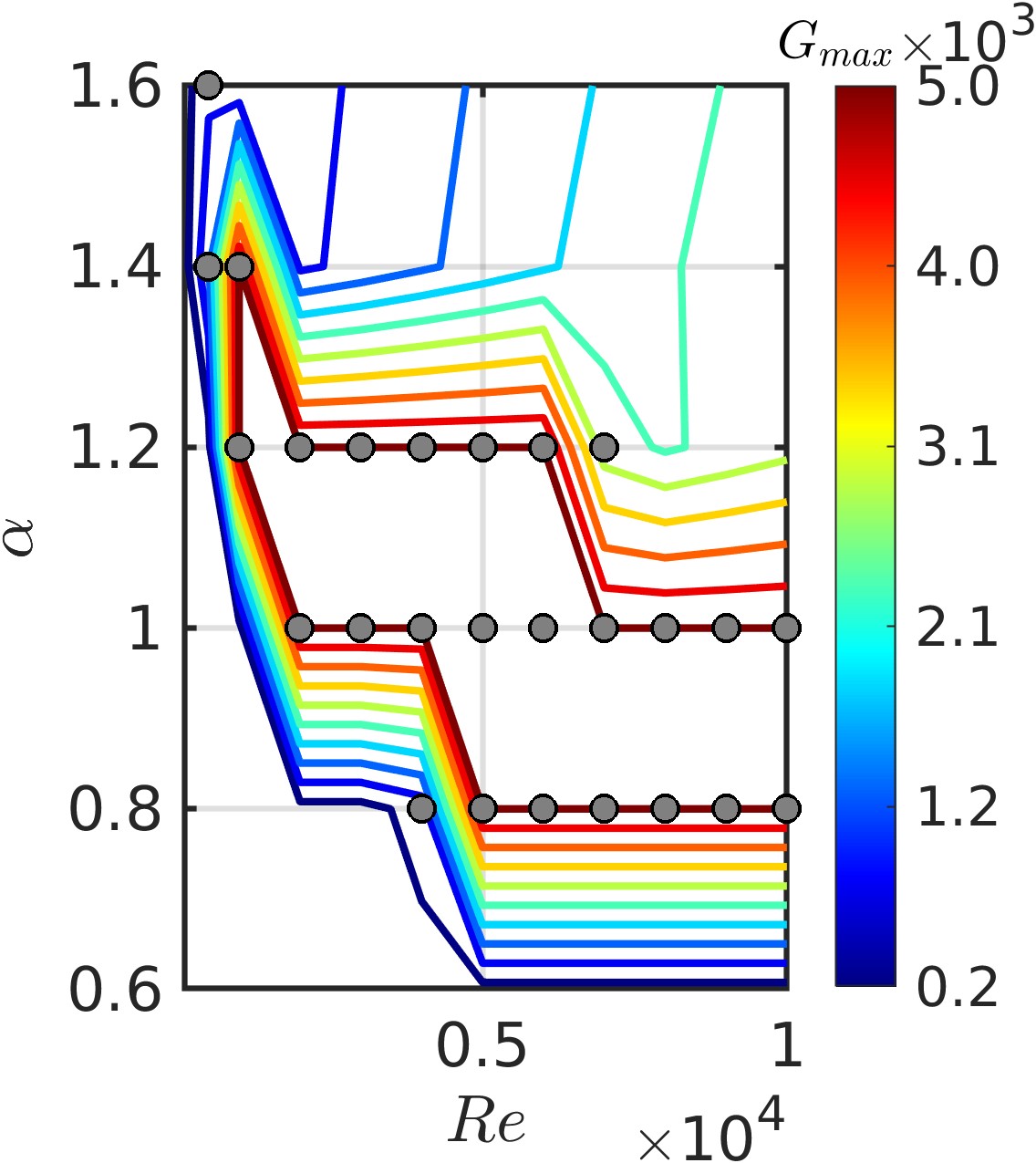}} \hspace{0.5mm} 
    \subfloat[\vspace{-6mm}]{\includegraphics[width=0.32\linewidth]{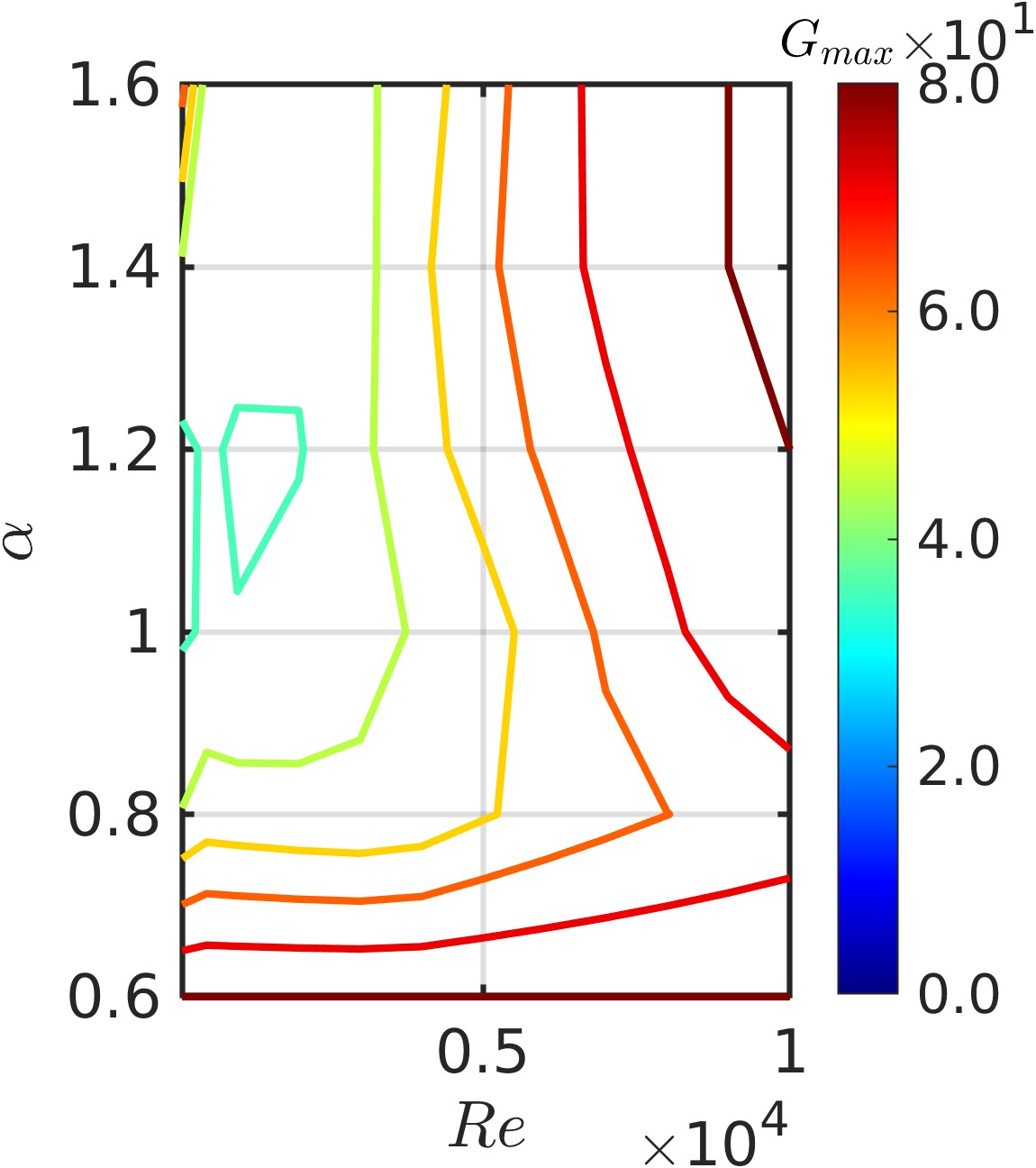}} \\ \vspace{2mm}
	\caption{Transient growth envelopes at $\beta = 0$ and $Br = 0.5$ for (a) I-1, (b) I-2, and (c) I-3 cases with $10$ spaced contour levels. Grey-filled circles denote infinite energy growth. The resolution range for cases I-1 and I-2 has been limited to $G_{max} = 5000$.} 
 \label{fig:Transient_growth_maps_Iso_beta0}
\end{figure*}

\begin{figure*}
	\centering
    \subfloat[\vspace{-6mm}]{\includegraphics[width=0.40\linewidth]{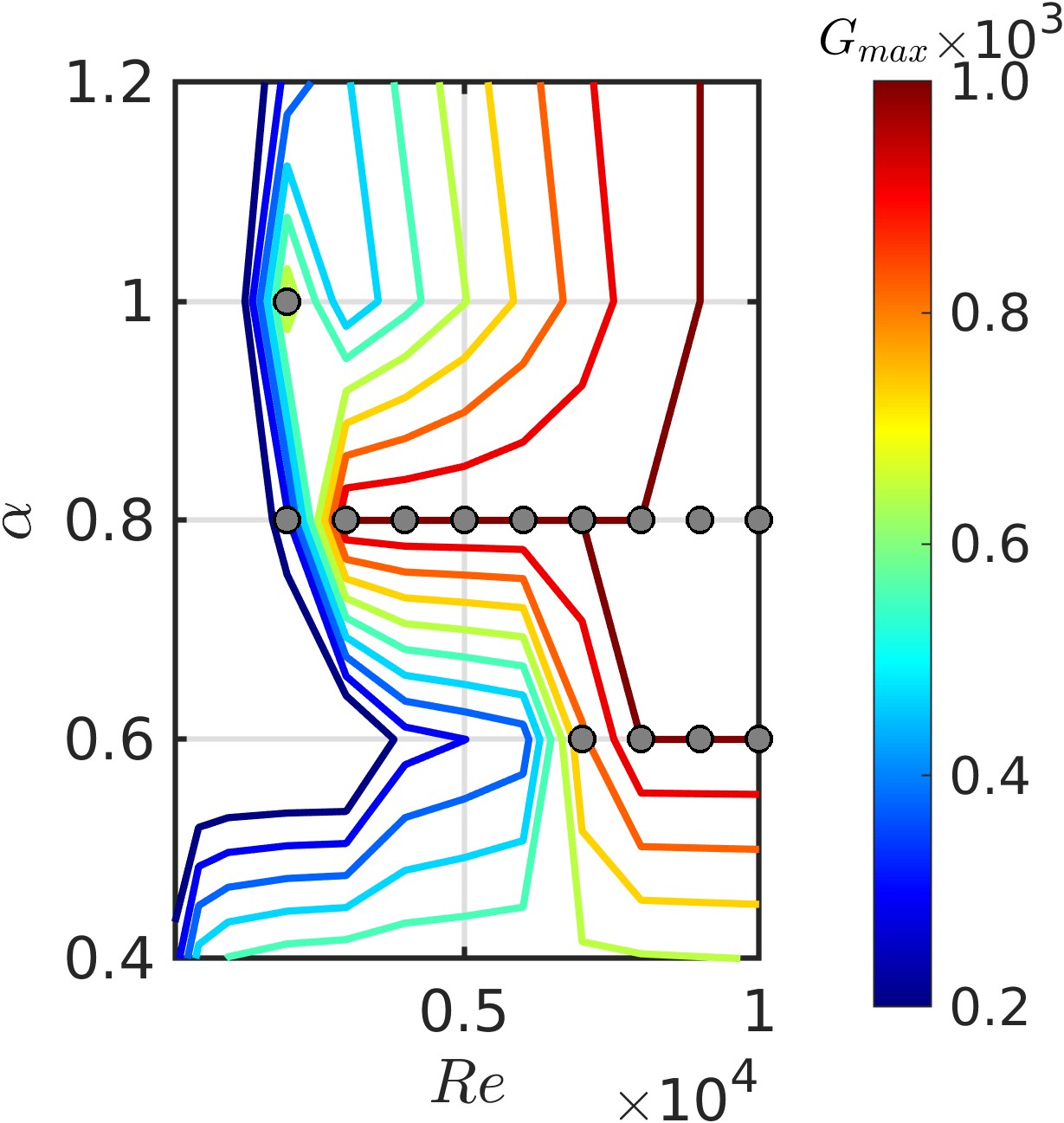}} \hspace{5mm}
    \subfloat[\vspace{-6mm}]{\includegraphics[width=0.40\linewidth]{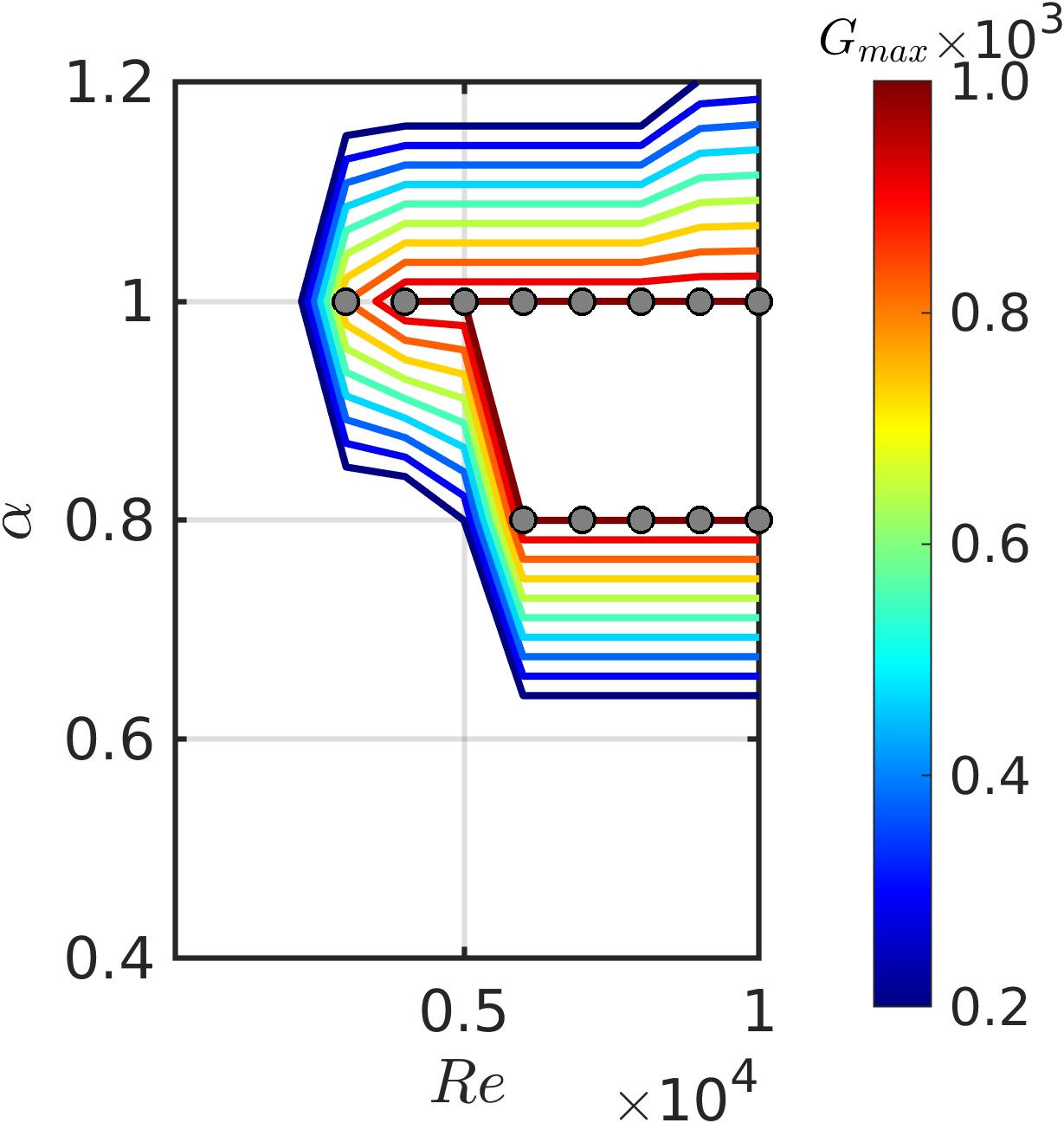}} \\ \vspace{6mm}
    \subfloat[\vspace{-6mm}]{\includegraphics[width=0.40\linewidth]{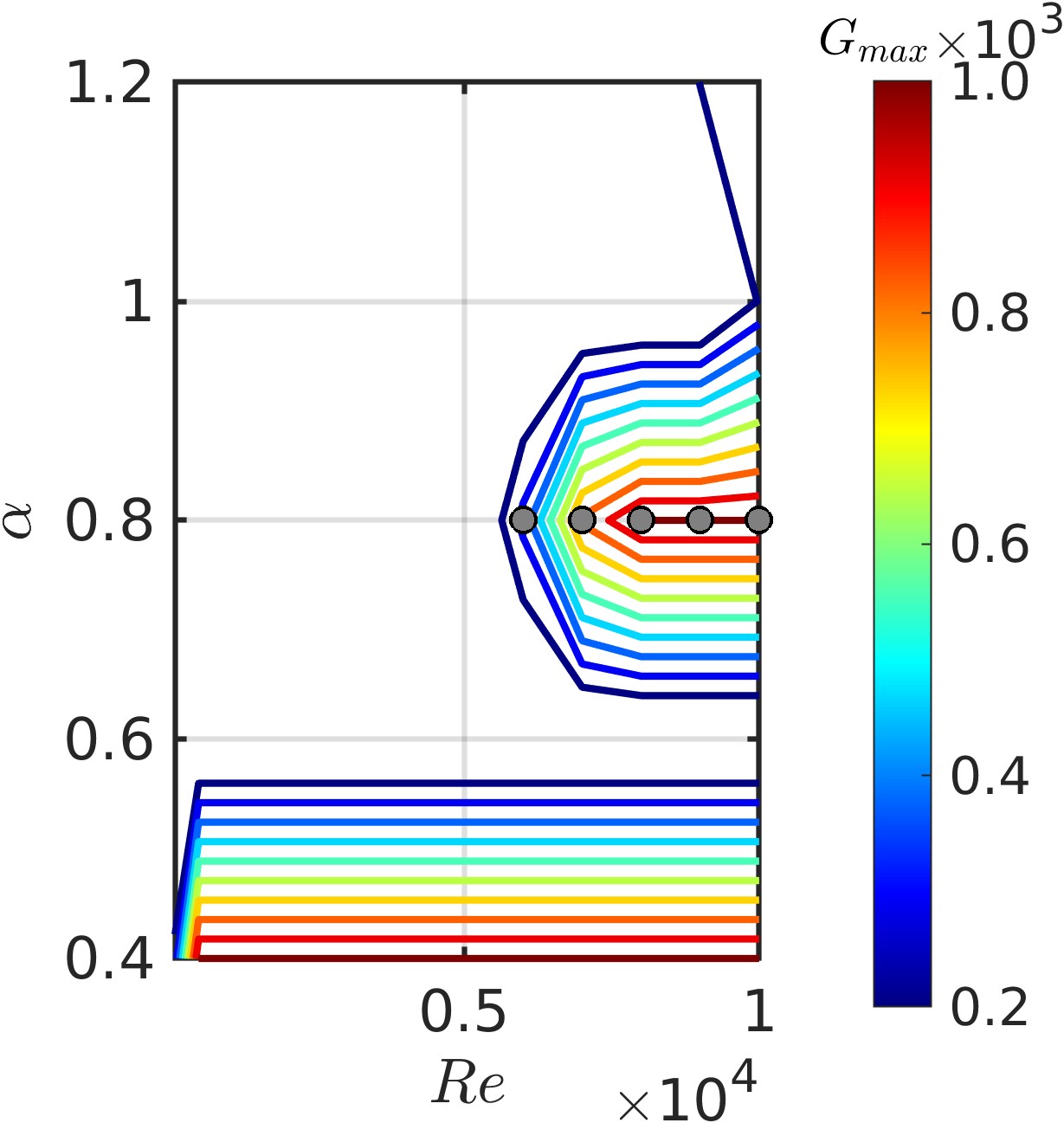}} \hspace{5mm}
    \subfloat[\vspace{-6mm}]{\includegraphics[width=0.40\linewidth]{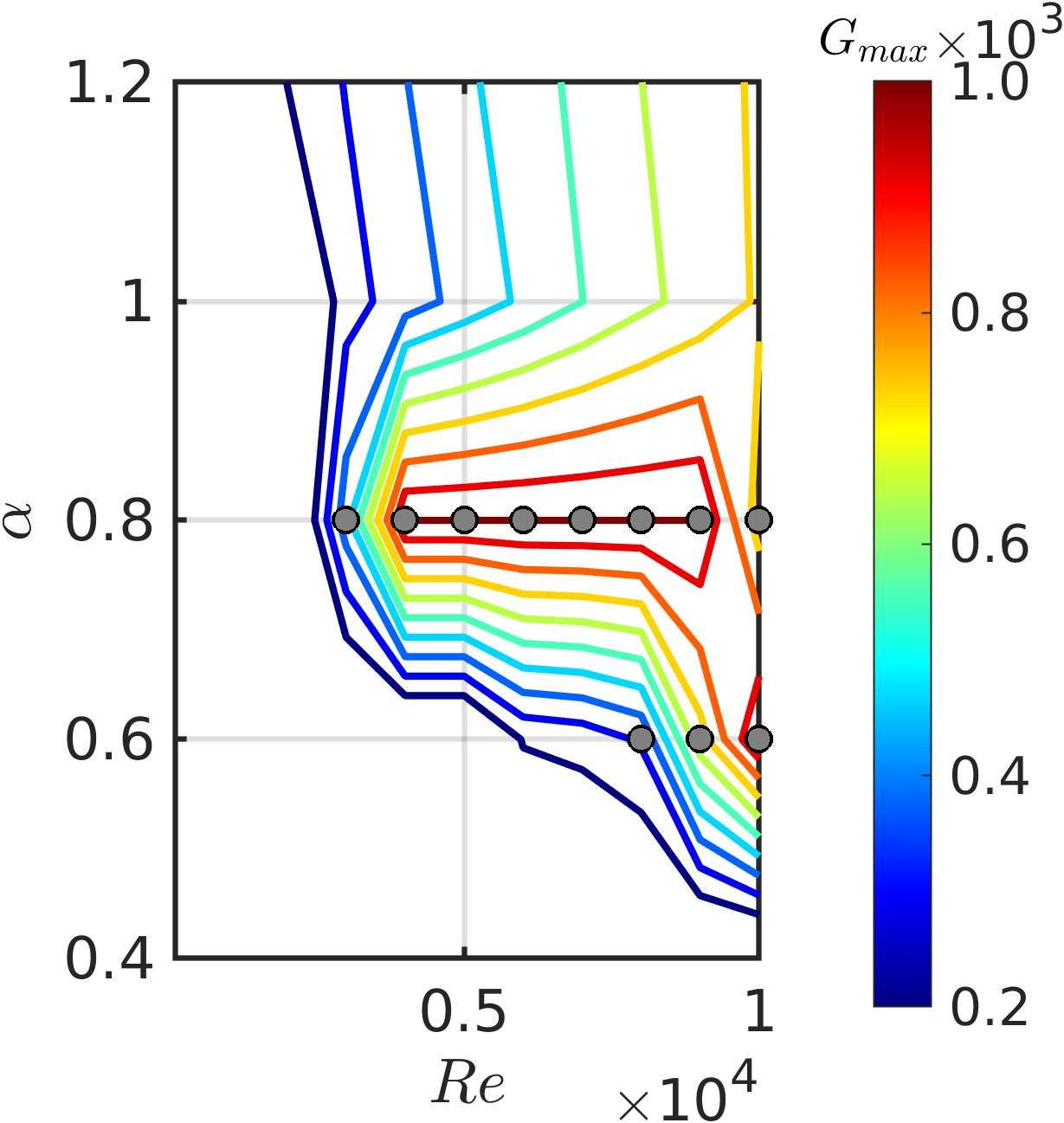}} \\ \vspace{2mm}
	\caption{Transient growth envelopes at $\beta = 0$ and $Br = 0.1$ for (a) NI-1, (b) NI-2, and (c) NI-3 cases, and $Br = 5.6 \cdot 10^{-6}$ for (d) NI-5. The resolution range has been limited to $G_{max} = 1000$.} 
 \label{fig:Transient_growth_maps_NonIso_beta0}
\end{figure*}

% Transient growth at 2D - Re sweep at beta = 0 to detail the growth rate
The discussion below aims at detailing the stability maps obtained from the modal analyses.
In particular, the focus is placed on the non-orthogonal effects of the operator.
The neutral curves and the corresponding critical Reynolds numbers can be directly compared to the exponential energy growth.
In this regard, Figures~\ref{fig:Transient_growth_maps_Iso_beta0}-\ref{fig:Transient_growth_maps_NonIso_beta0} depict (i) the growth rate maps on the $\alpha-Re$ space under 2D perturbations ($\beta = 0$) at different contour levels, and (ii) the limit beyond which energy grows to infinity.
Similar to the transient growth maps, equivalent behavior is exhibited at low $Br$ values as in the results from modal analysis.
Consequently, colormaps for large $Br$ cases are only depicted.
As it can be observed, since the plots present similar unstable regions, the results from energy growth validate the modal analysis results in the case of isothermal conditions.
Furthermore, Figure~\ref{fig:isothermal_neutral_curves}(c) confirms that transient growth becomes also stable at large $Br$ values as a result of lower energy levels.
The non-isothermal cases are also well captured, as well as the extended instability region at low Reynolds number for case NI-1 at $\alpha \approx 1$.
It is important to note that increasing the bulk pressure to $P_b/P_c = 5$ results in a shift of the instability to larger streamwise wavenumbers, and consequently a narrower envelope is obtained with $G_{max} \ge 200$ as depicted in Figure~\ref{fig:Transient_growth_maps_NonIso_beta0}(b).
In addition, case NI-3 results in an even smaller region despite of operating at transcritical conditions as shown in Figure~\ref{fig:Transient_growth_maps_NonIso_beta0}(c).
It is worth noting that an additional large growth rate area is magnified at low streamwise wavenumber. Although the energy remains stable over the time, this narrower temperature difference setup yields an important operating window of high energy growth at $\alpha = 0.4$. In particular, it confirms the trend observed in the energy growth rate map at $Re = 1000$ [Figure~\ref{fig:Transient_growth_maps_NonIso}(c)] which also tailed large enough energy growths toward this wavenumber space. In particular, though, this growth is within the same magnitude compared to unstable wavenumber $\alpha = 0.8$ at largest $Re = 10000$. Unlike $\alpha = 0.8$, the $\alpha = 0.4$ region does not achieve an exponentially amplification growth but the energy level remains steady its quantity.
Moreover, Figure~\ref{fig:Transient_growth_maps_NonIso_beta0}(d) confirms that the low-velocity flow case NI-5 obtains a similar envelope as NI-1.
Thus, this case is the most efficient candidate to enhance flow destabilization.
For brevity, cases NI-4 and NI-6 are not shown as they behave as the supercritical flow case I-3 [Figure~\ref{fig:isothermal_neutral_curves}(c)]. 

% Summary table comments
Finally, Table~\ref{tab:summary_Re_c_energy} summarizes the critical Reynolds number and maximum growth rate for the optimal $\alpha-\beta$ parameter space found in the energy growth maps for the largest $Br$ values studied.
In detail, $\alpha = 0$ and $\beta=2$ results in a common region with maximum growth rate for all flow cases, except for case I-2 (and I-1) in which a significantly large growth rate is also reported at $\alpha = 1$-$\beta=1$. In detail, at higher $Re$ this growth rises exponentially.
These conditions result in early transition for case NI-1-3 at $Re \approx 2500$, while case NI-5 at low-Mach conditions achieves the earliest transition at similar regime.
However, despite the energy increase, the behavior is not exponential, and consequently the energy may eventually decay at larger times.
%Instead, case NI-2 provided early transition at $Re = 1000$, but for $Re > 4000$ the energy growth decays.
Interestingly, the sub- and transcritical cases I-1 and I-2 provide an unstable energy region only within the range $900 \le Re \le 4000$ for I-2 and move toward large values $1500 \le Re \le 5100$ for I-2. Nonetheless, I-1 presents rates that are roughly $8\times$ smaller in comparison to the transcritical case. Of note, this amplification rates are achieved with non-sothermal cases at lower Reynolds regimes.

% Figure Growth rate beta = 0: I-1-2-3 for Br = 0.5

% Figure Growth rate: NI-1-2-3 at Br = 0.1 and NI-5 at Br = 5.6 x10^-6

% Critical Reynolds summary table
\begin{table}
\caption{Energy-based critical Reynolds numbers and energy growths for the flow cases listed in Section~\ref{sec:flow_cases} at a two-parameter space $(\alpha, \beta)$ levels. Cases I-1, I-2 and I-3 are assessed at $Br = 0.5$, NI-1, NI-2, NI-3 and NI-4 at $Br = 0.1$, and NI-5 and NI-6 at $Br = 5.6 \cdot 10^{-6}$. Transition criteria defined if energy grows beyond elapsed time ($t \le 400$) highlighting with superscript $(\cdot)^\star$ the exponentially algebraic growth.}\label{tab:summary_Re_c_energy}
\centering
%\begin{adjustbox}{width=1\textwidth}
%\renewcommand{\arraystretch}{1.0}
%\begin{ruledtabular}
  % \centering 
\begin{tabular}{c|c|ccc|cccccc} \hline
\centering
 & $(\alpha, \beta)$ &  I-1 & I-2 & I-3 & NI-1 & NI-2 & NI-3 & NI-4 & NI-5 & NI-6 \\
\hline
$Re_c$ & $(0,2)$ &  $5300$ & $3900$ & $5100$ &  $2500$ & $3100$ & $2300$ & $4500$ & $2100$ & $4300$ \\
 & $(1,1)$ &  $1500-5100^\star$ & $900-4000^\star$ & $-$   &  $-$     &  $-$    & $-$    & $-$   & $-$ &  $-$ \\ \hline
 $G_{max}$ & $(0,2)$ &  $5800$&  $11300$ & $5400$ &  $38500$ & $7400$ & $5200$ & $8800$ & $16200$ & $8300$ \\
 & $(1,1)$ &  $180-6950$ & $800-54500$ & $-$   &  $-$     &  $-$    & $-$    & $-$   & $-$ &  $-$ \\ \hline
\end{tabular}
%\end{ruledtabular}
%\end{adjustbox}
\end{table}

\subsubsection{Optimal input profile and output responses}

\begin{figure*}
	\centering
    \subfloat[\vspace{-6mm}]{\includegraphics[width=0.305\linewidth]{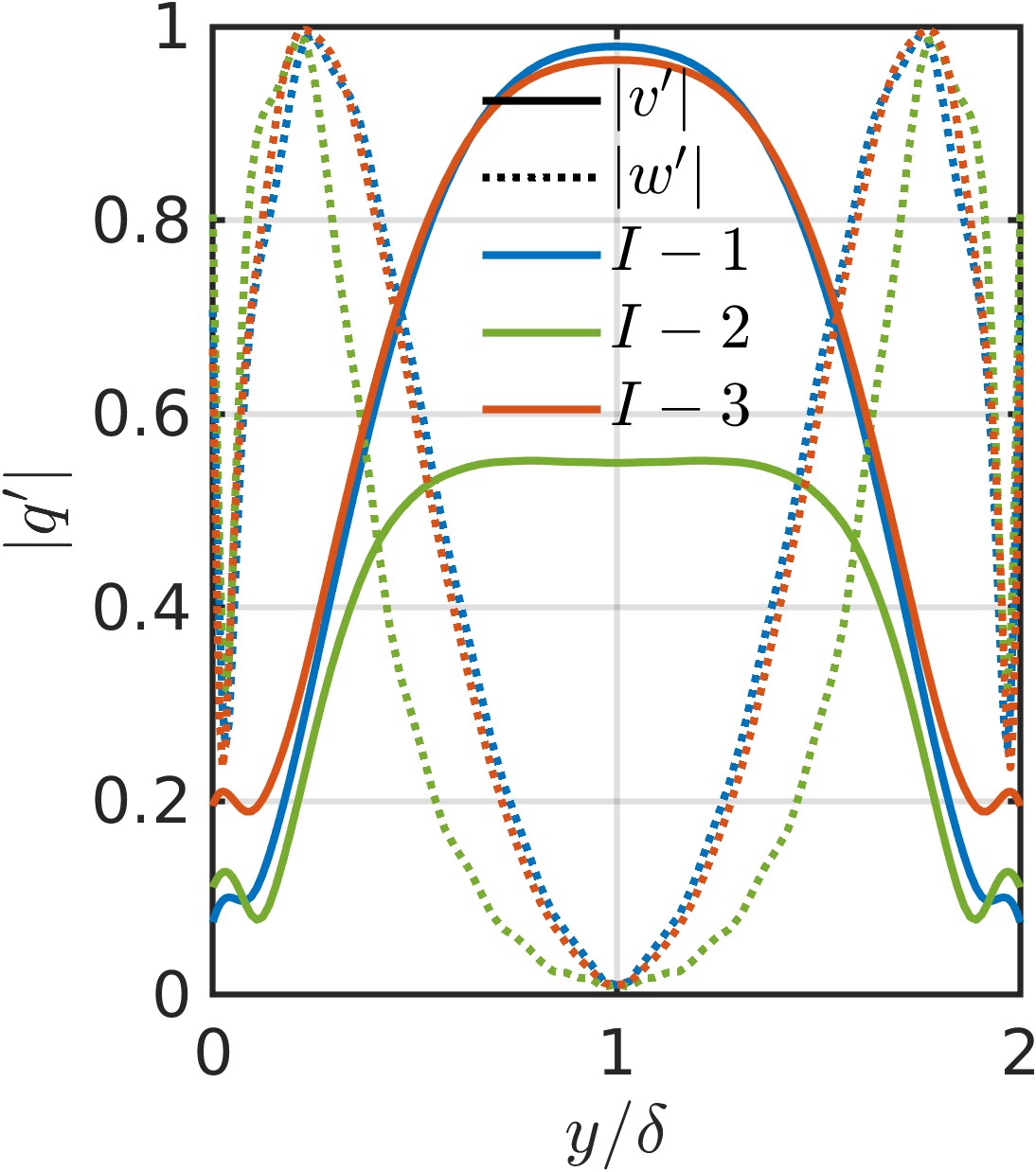}} \hspace{2mm}
    \subfloat[\vspace{-6mm}]{\includegraphics[width=0.305\linewidth]{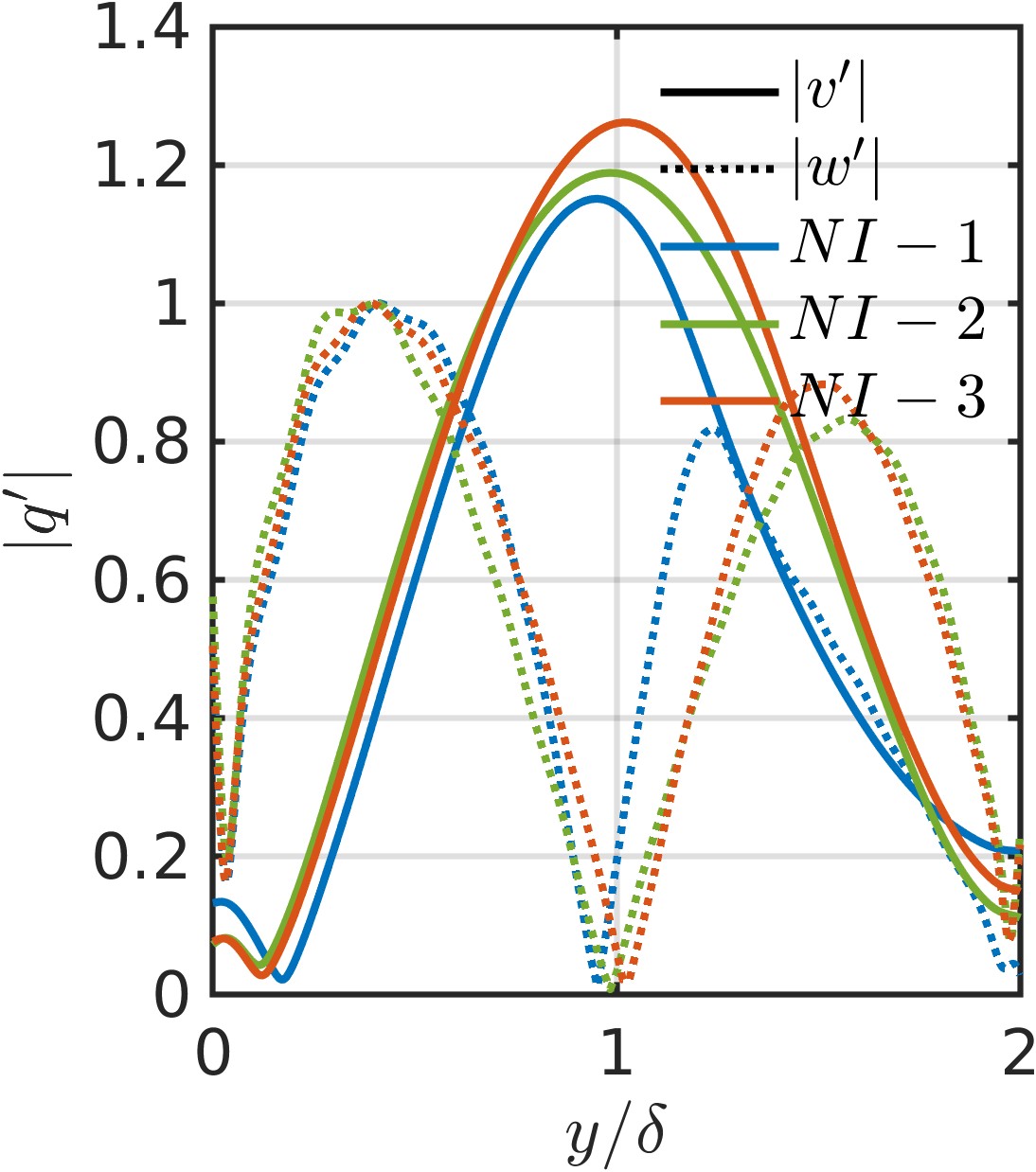}} \hspace{2mm}
    \subfloat[\vspace{-6mm}]{\includegraphics[width=0.305\linewidth]{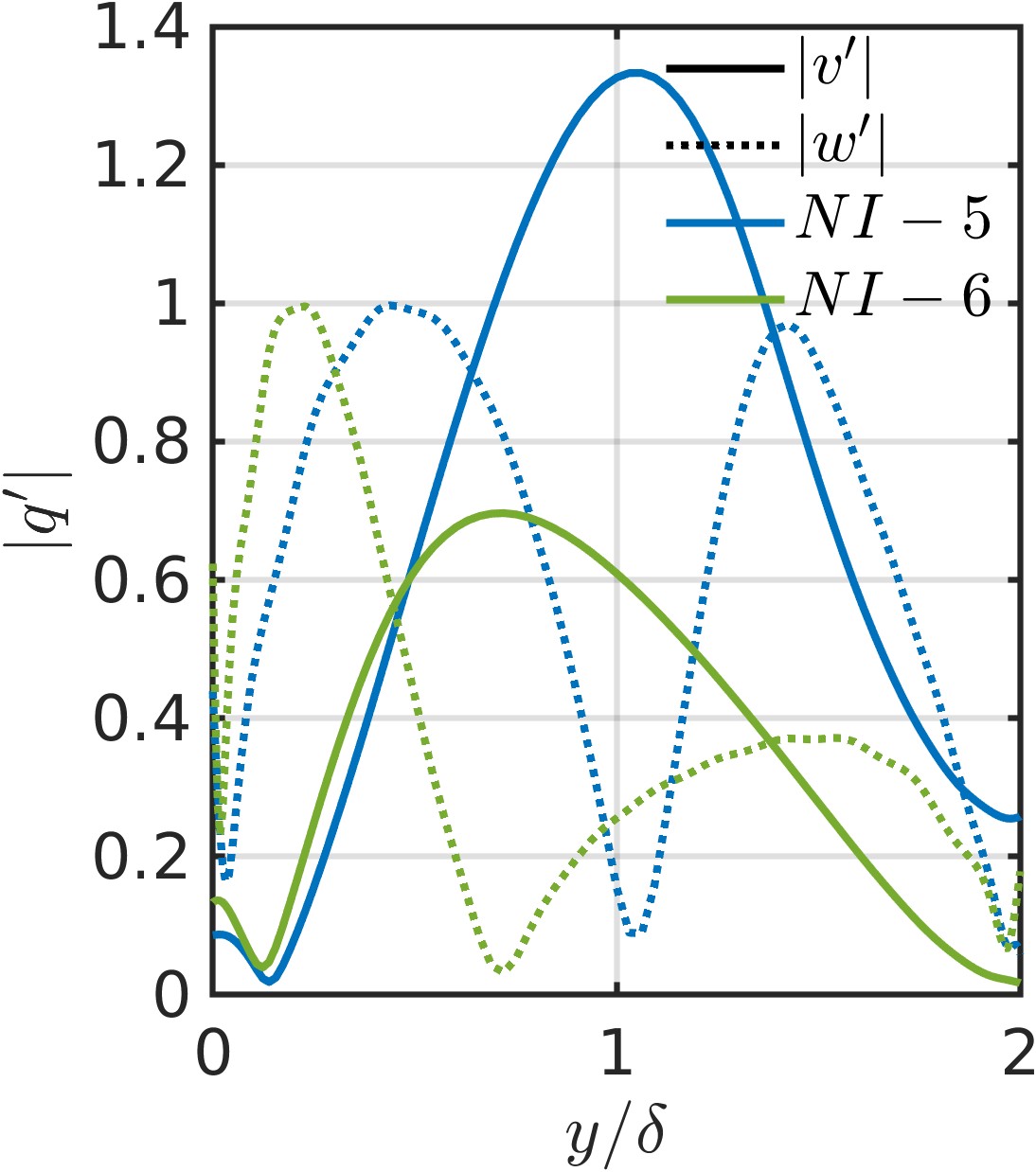}} \\ \vspace{2mm}
	\caption{Optimum input perturbation profiles at $Re = 1000$, $\alpha = 0$ and $\beta = 2$ for the isothermal flow cases (a) I-1, I-2 and I-3 at $Br = 0.5$, (b) NI-1, NI-2 and NI-3 at $Br = 0.1$, and (c) NI-4 and NI-5 at $Br = 5.6 \cdot 10^{-6}$. Results are normalized by $w^\prime$ for all cases.} 
 \label{fig:Optimum_profile_INPUT}
\end{figure*}

\begin{figure*}
	\centering
    \subfloat[\vspace{-6mm}]{\includegraphics[width=0.305\linewidth]{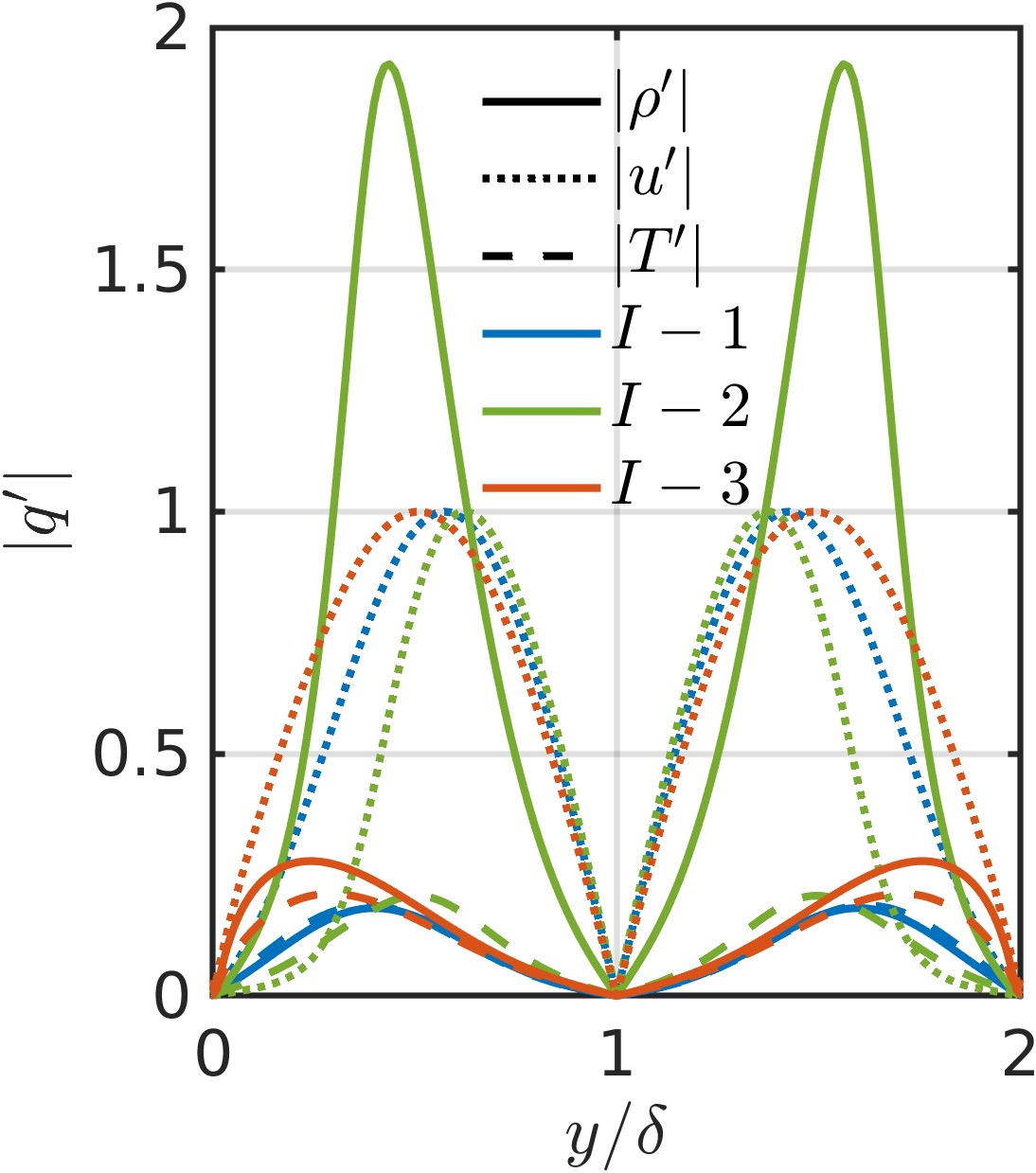}} \hspace{2mm}
    \subfloat[\vspace{-6mm}]{\includegraphics[width=0.305\linewidth]{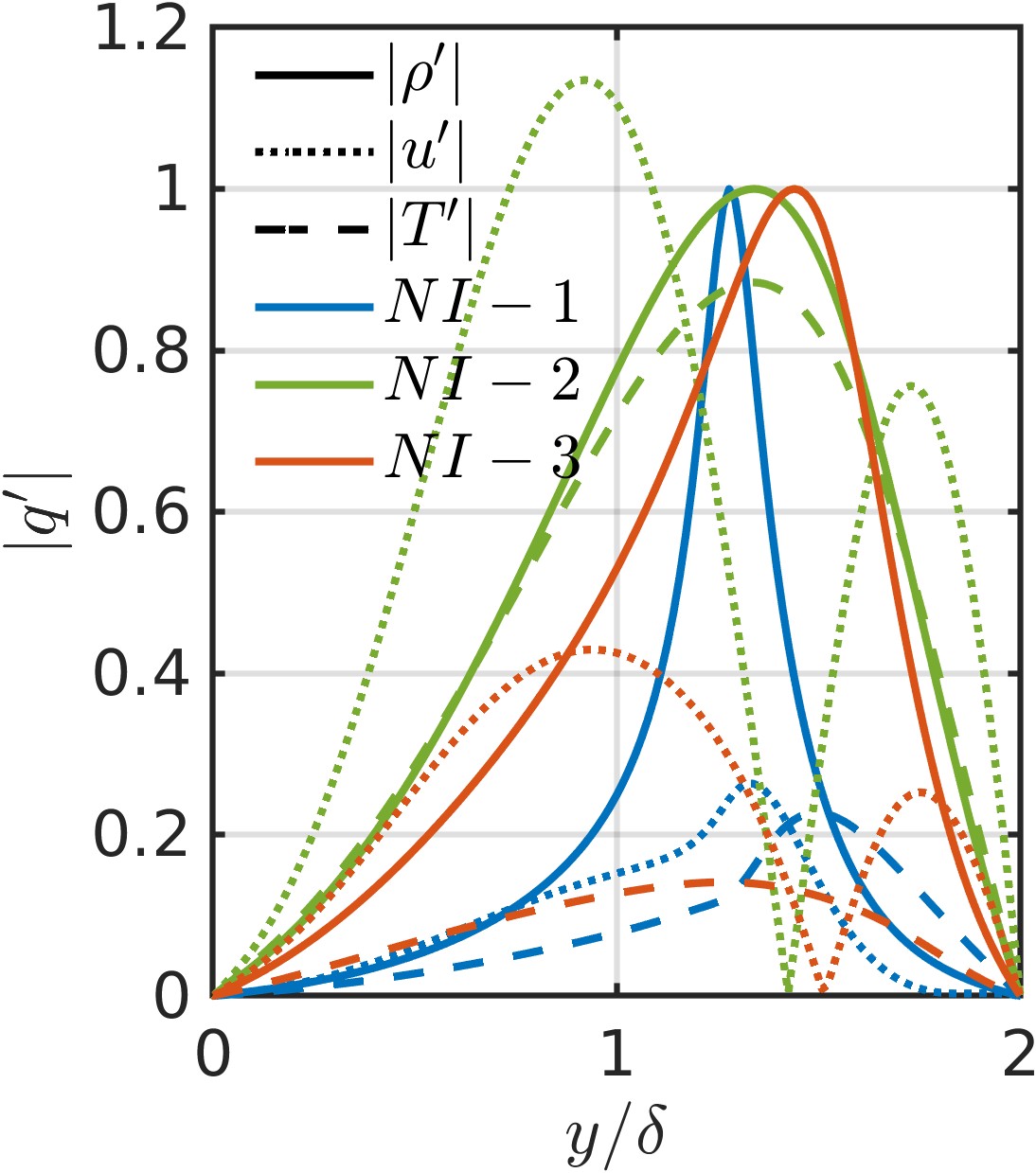}} \hspace{2mm}
    \subfloat[\vspace{-6mm}]{\includegraphics[width=0.305\linewidth]{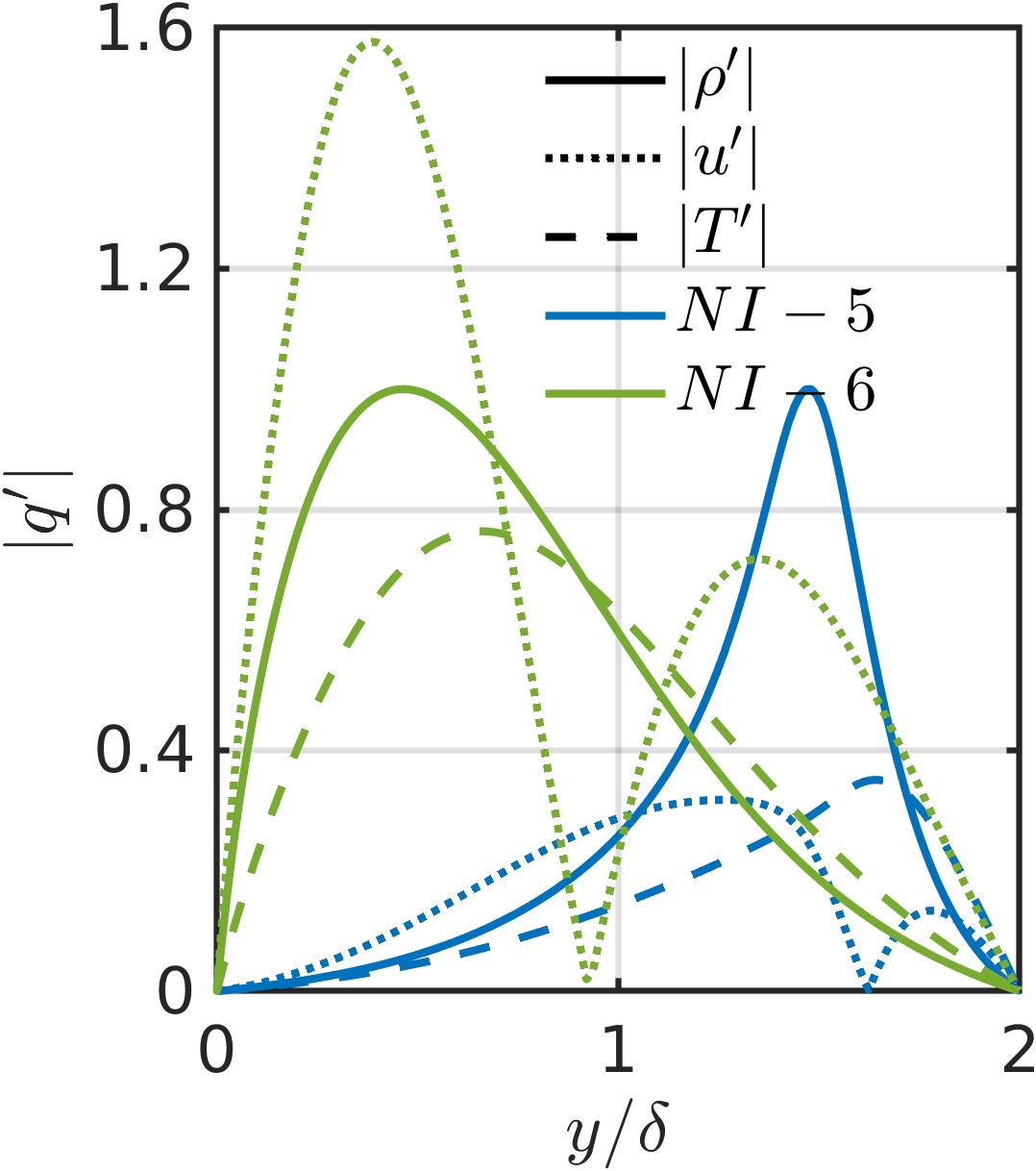}} \\ \vspace{2mm}
	\caption{Optimum output response at $Re = 1000$, $\alpha = 0$ and $\beta = 2$ for the isothermal flow cases (a) I-1, I-2 and I-3 at $Br = 0.5$, (b) NI-1, NI-2 and NI-3 at $Br = 0.1$, and (c) NI-5 and NI-6 at $Br = 5.6 \cdot 10^{-6}$. Results are normalized by $u^\prime$ for (a) and $\rho^\prime$ for (b,c).} 
 \label{fig:Optimum_profile_OUTPUT}
\end{figure*}

\begin{figure*}
	\centering
    \subfloat[\vspace{-6mm}]{\includegraphics[width=0.34\linewidth]{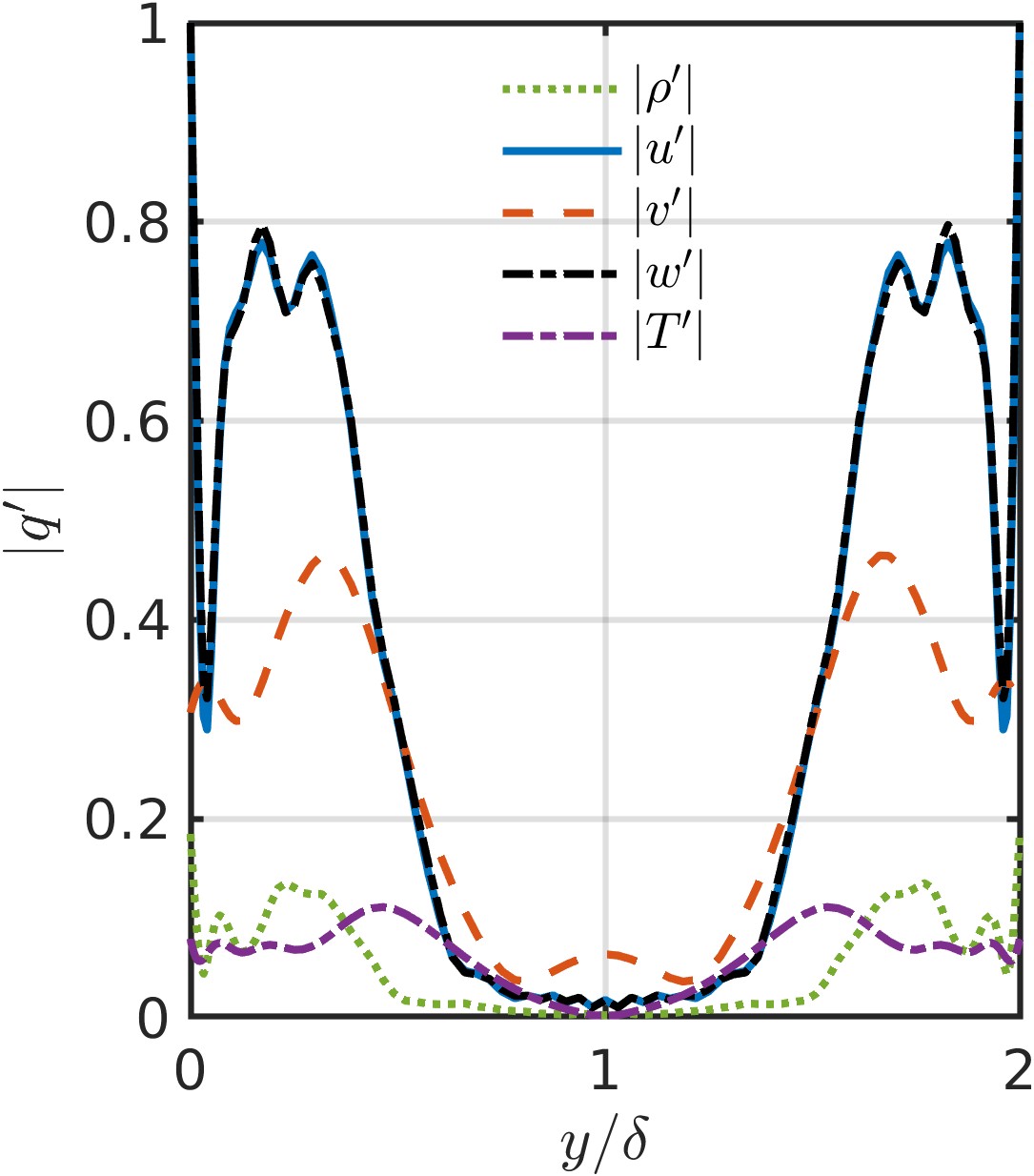}} \hspace{2mm}
    \subfloat[\vspace{-6mm}]{\includegraphics[width=0.33\linewidth]{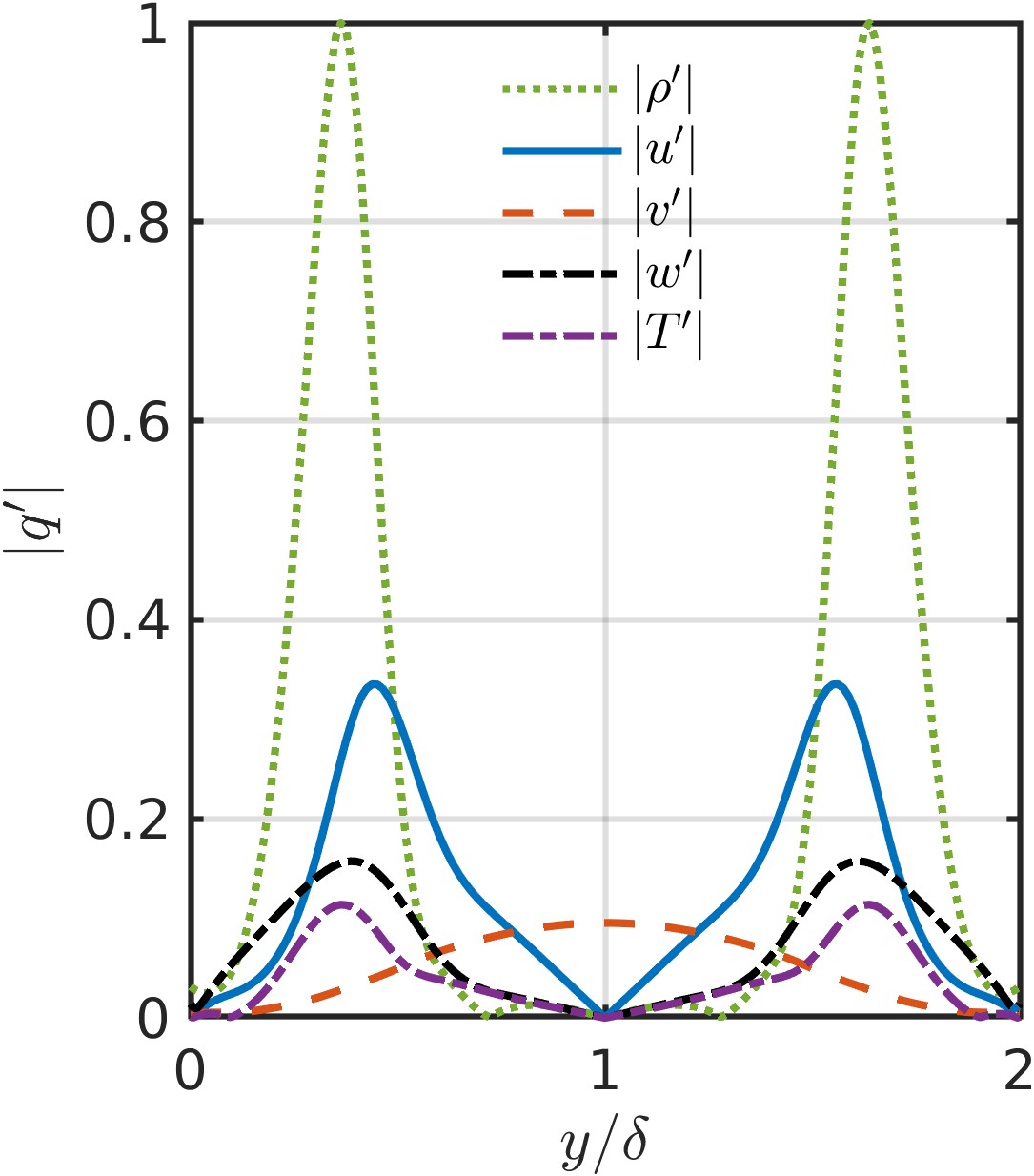}} \\ \vspace{2mm}
	\caption{Optimum (a) perturbation and (b) response at $Re = 1000$, $\alpha = \beta = 1$ for the isothermal flow case I-2 at $Br = 0.5$. Results are normalized by $w^\prime$ for (a) and $\rho^\prime$ for (b).} 
 \label{fig:Optimum_profile_IN_OUT_I-2}
\end{figure*}

% Optimal perturbations
Given that the growth rate over time describes the maximum energy ampliﬁcation over all possible initial conditions, it is interesting to quantify which speciﬁc initial perturbation is responsible for the maximum amplification energy output at a given time.
This initial condition is easily recovered by means of a SVD of the matrix exponential~\citep{Schmid2007-A}.
In this regard, the optimal initial condition and the corresponding response are, respectively, shown in Figure~\ref{fig:Optimum_profile_INPUT} and Figure~\ref{fig:Optimum_profile_OUTPUT} at wavelengths $\alpha = 0$ and $\beta = 2$ providing the maximum amplification.
On the one hand, for the isothermal cases, the streamwise vortices and velocity streaks are recovered for the optimal perturbation and its response, similarly to the case of incompressible flows.
However, the thermal streaks are also important due to the crucial role of density and temperature at high-pressure transcritical conditions, whereas the dynamic streaks remain of the same importance as the streamwise velocity.
On the other hand, the non-isothermal cases yield a distinct behavior.
The optimal profile results in a parabolic wall-normal velocity perturbation.
The response is prominent for thermal streaks located within the the pseudo-boiling region (near the hot wall) and dominated by density over the temperature variations.
Moreover, for case NI-5, similar optimal input and output responses are exhibited.
Instead, when operating at atmospheric pressure conditions (case NI-6), the velocity streaks are again enhanced and more prominent in the vicinity the cold wall.
In particular, thermal streaks are significant within this region and more prominent than the isothermal sub- and supercritical flow setups.
It is important to note that the the outstanding energy growth obtained for case I-2 at $\alpha = \beta = 1$ is quantified in Figure~\ref{fig:Optimum_profile_IN_OUT_I-2}.
Thus, the optimal input results in 3D-like vortices with similar streamwise and spanwise components.
Nonetheless, the wall-normal velocity is relatively lower for this case, and the response is enlarged for the density streaks whereas the streamwise velocity vortex becomes asymmetric and gradually decreases its magnitude toward the channel centerline.

% Figure Optimum profiles INPUT

% Figure Optimum profiles OUTPUT

% Figure Optimum INPUT AND OUTPUT for I-2 Alpha = Beta = 1 (3D)

\begin{figure*}
	\centering
    \subfloat[\vspace{-6mm}]{\includegraphics[width=0.32\linewidth]{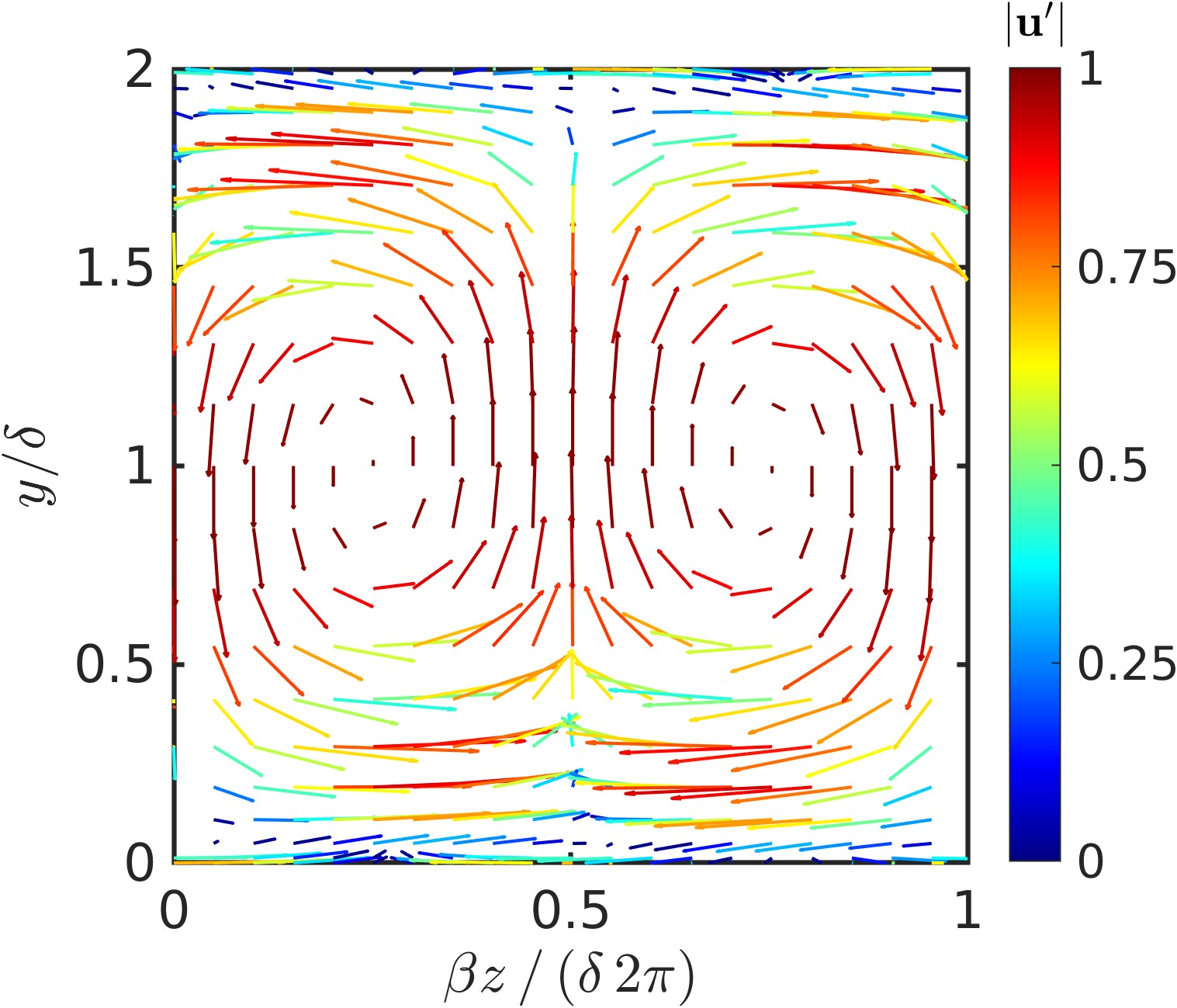}} \hfill
    \subfloat[\vspace{-6mm}]{\includegraphics[width=0.32\linewidth]{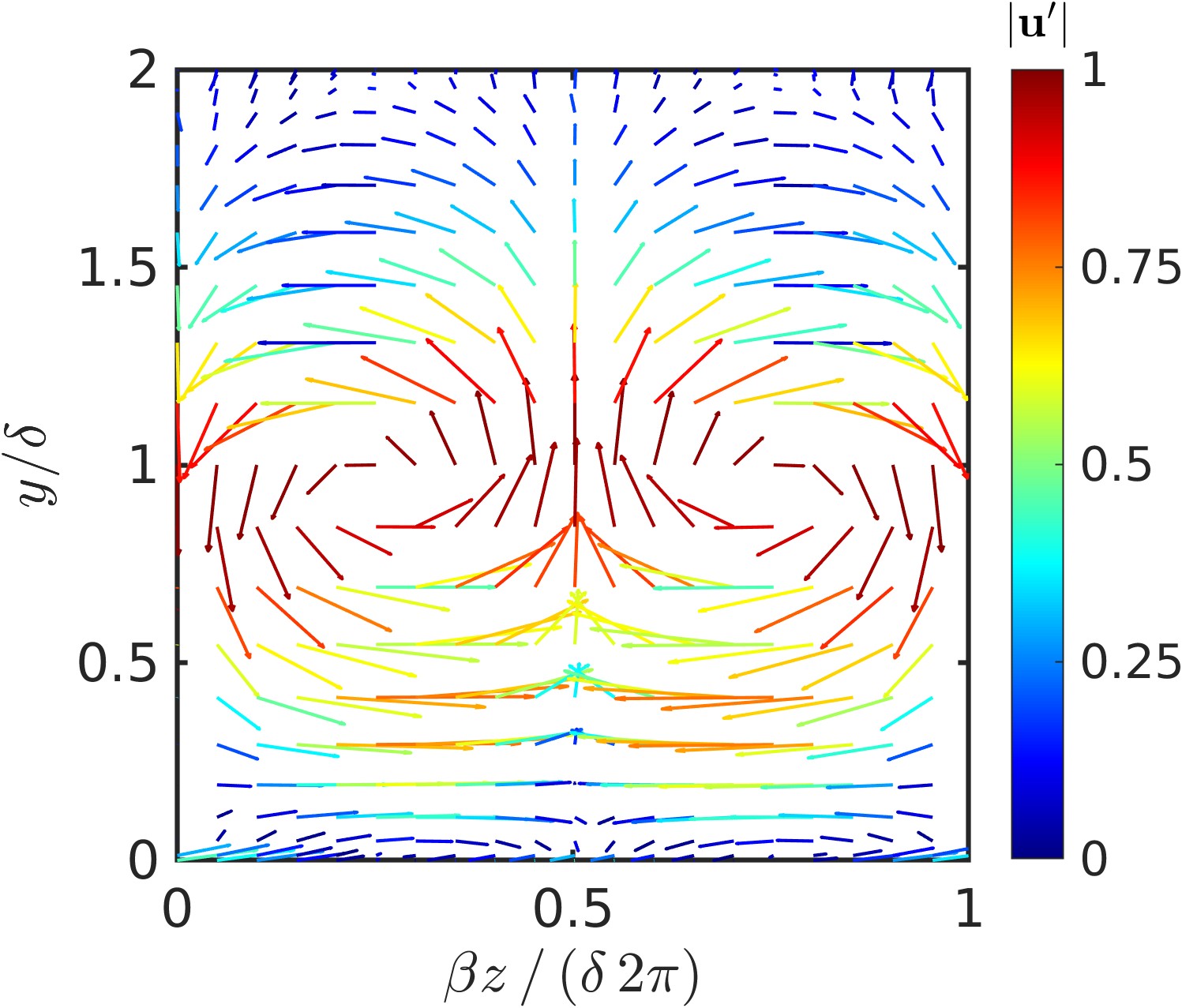}} \hfill
    \subfloat[\vspace{-6mm}]{\includegraphics[width=0.32\linewidth]{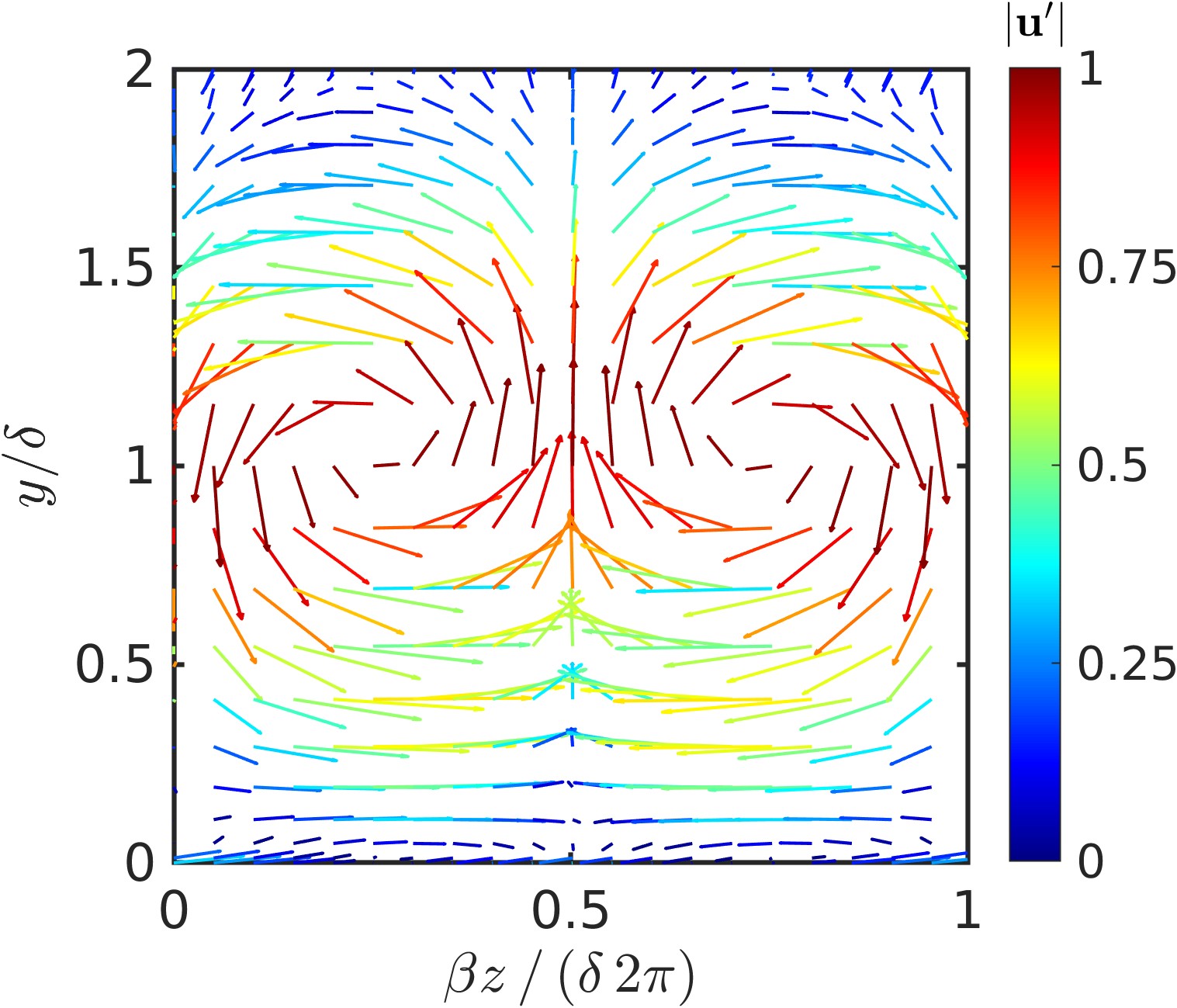}} \\ \vspace{2mm}
	\caption{Optimum velocity perturbation at $Re = 1000$, $\alpha = 0$ and $\beta = 2$ for cases (a) I-1 at $Br = 0.5$, (b) NI-1 at $Br = 0.1$, and (c) NI-5 at $Br = 5.6 \cdot 10^{-6}$.} 
 \label{fig:Optimum_profile_INPUT_Pattern}
\end{figure*}

% Optimal patterns
Figure~\ref{fig:Optimum_profile_INPUT_Pattern} displays the optimal velocity patterns across the $(z-y)$-plane at wavenumbers $\alpha = 0$ and $\beta = 2$; in this case, invariant along streamwise direction.
The pattern resembles the streamwise vortices typical of shear flows~\citep{Malik2006-A}.
These counter-rotating vortices tend to be compacted within the half-plane region at non-isothermal conditions for both Brinkman numbers.
Nonetheless, at low velocity, the vortices are slightly biased to the hot wall.
Likewise, the optimal responses recover the spanwise alternating streamwise velocities.
The non-isothermal flow cases, however, report an interesting phenomenon.
At large $Br$ values for case NI-1, there is a single alternating streamwise velocity at wall-normal position $0.5 \le y/\delta \le 1.5$ presenting an elongated ellipsoid-like shape.
The temperature field presents a similar behavior, but its maximum is shifted to $y/\delta \sim 1.6$, which corresponds to the pseudo-boiling region.
Interestingly, the density field presents a well-defined ellipsoid at $y/\delta \sim 1.3$.
Furthermore, at lower Reynolds and Mach numbers (case NI-5), the counter-rotating vortices are clearly apparent and interacting with the pseudo-boiling region that splits their rotation in two.
Instead, the thermodynamic field behaves similarly to case NI-1, but they are moderately stretched to the cold wall.
It is important to highlight that DNS results of an equivalent 3D turbulent flow setup~\citep{Bernades2022c-A} indicated a similar behavior of flow structures.

%Figure INPUT

%Figure OUTPUT
\begin{figure*}
	\centering
    \subfloat[\vspace{-6mm}]{\includegraphics[width=0.32\linewidth]{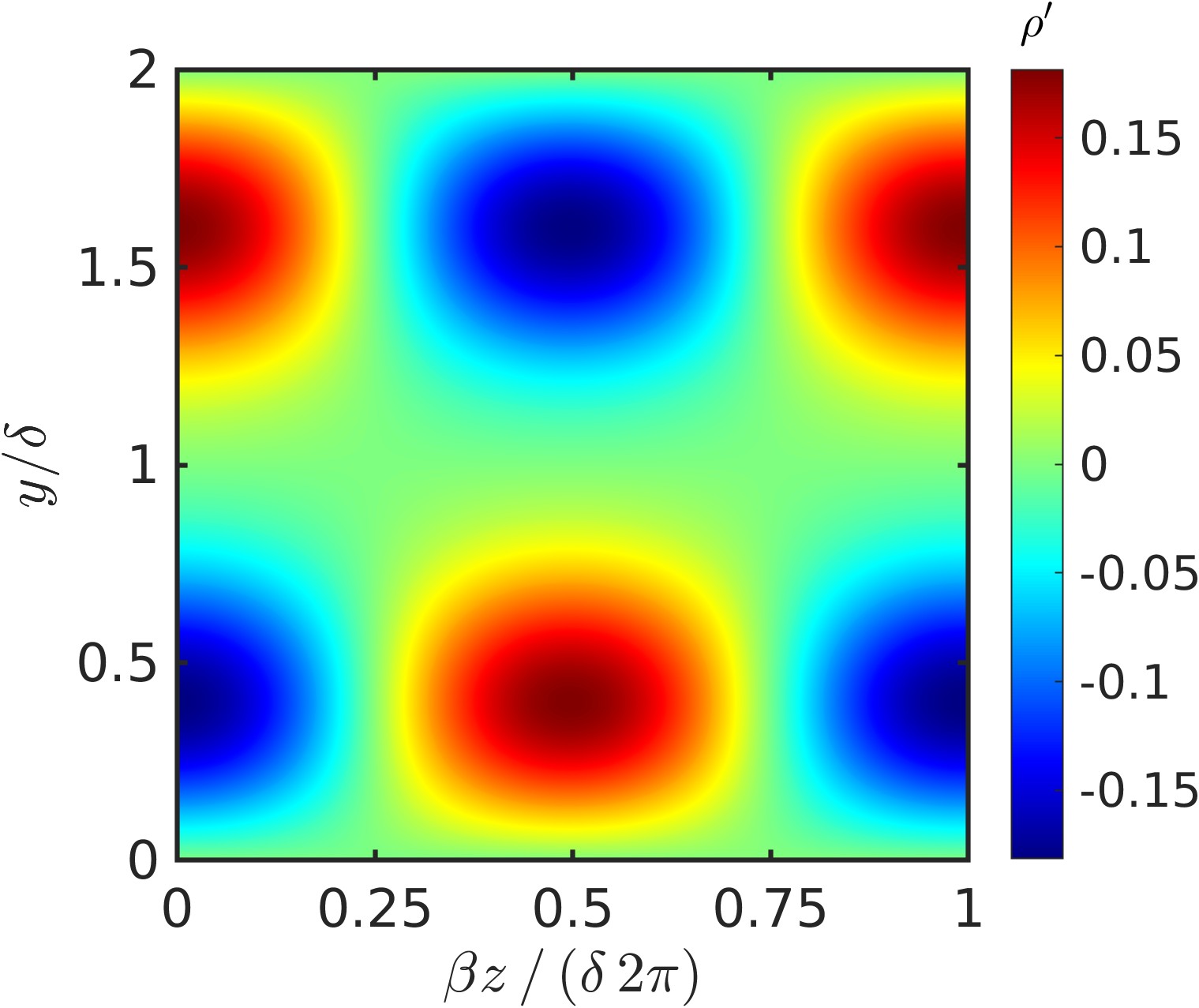}} \hfill
    \subfloat[\vspace{-6mm}]{\includegraphics[width=0.32\linewidth]{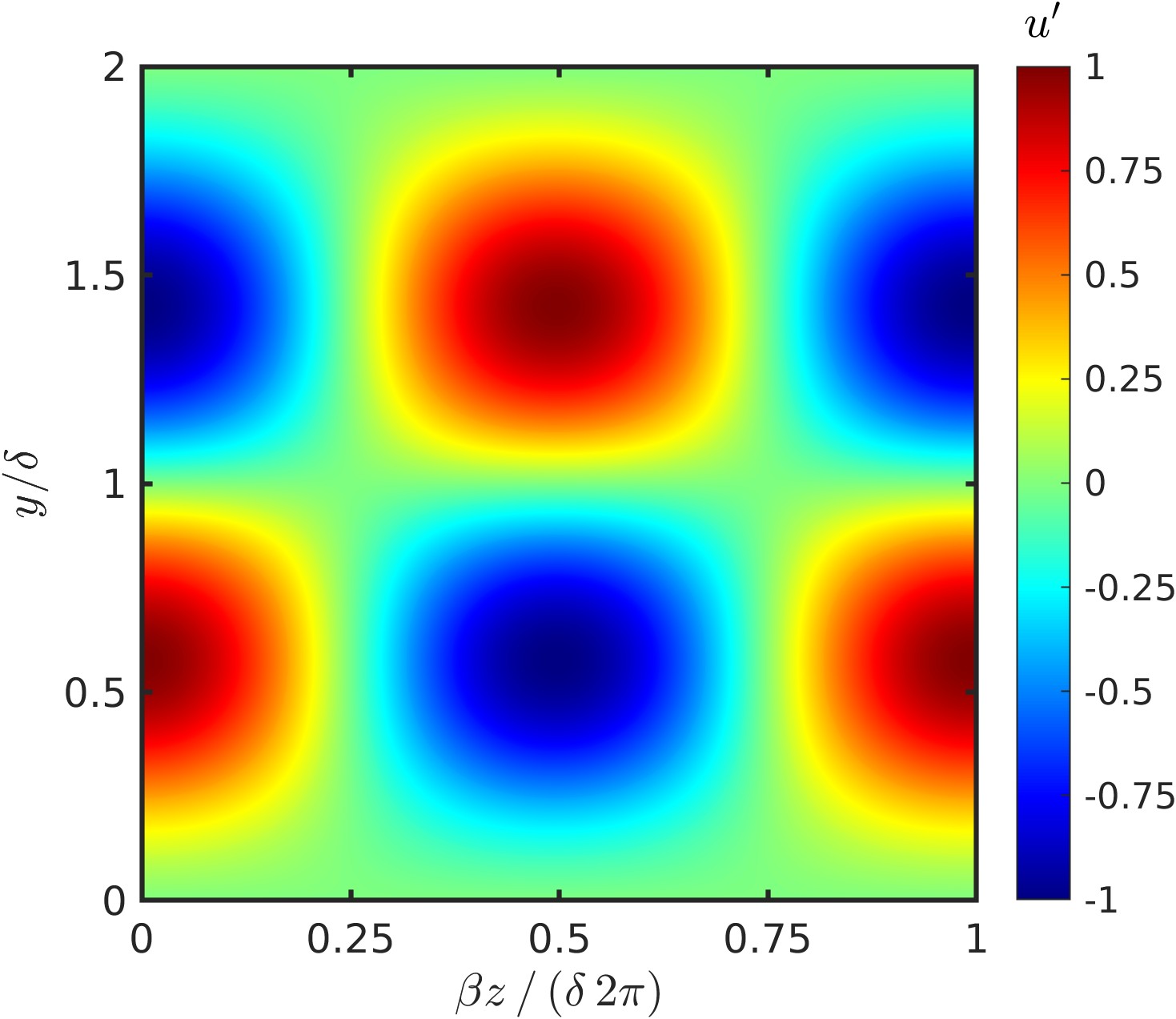}} \hfill
    \subfloat[\vspace{-6mm}]{\includegraphics[width=0.32\linewidth]{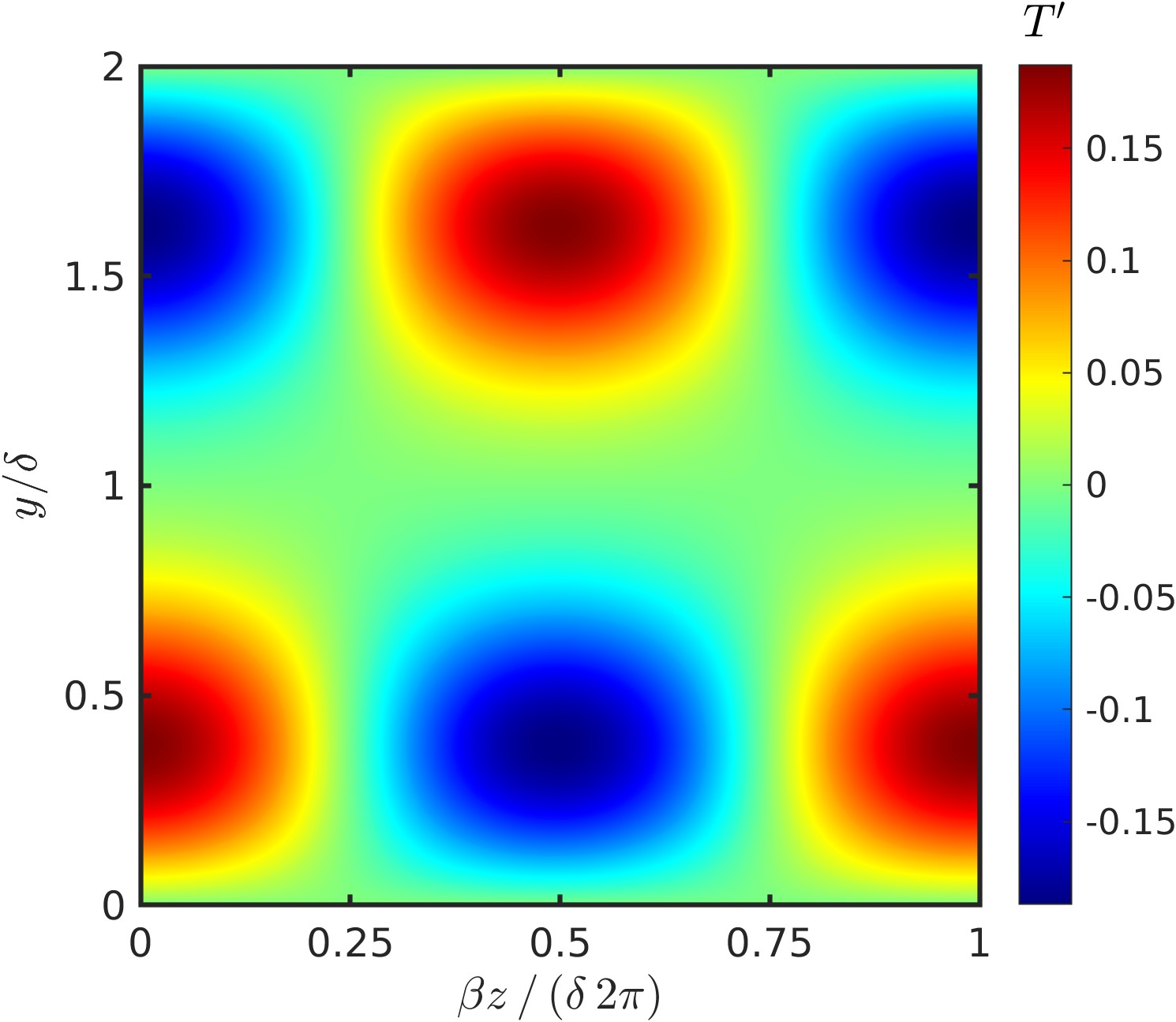}} \\  \vspace{4mm}
    \subfloat[\vspace{-6mm}]{\includegraphics[width=0.32\linewidth]{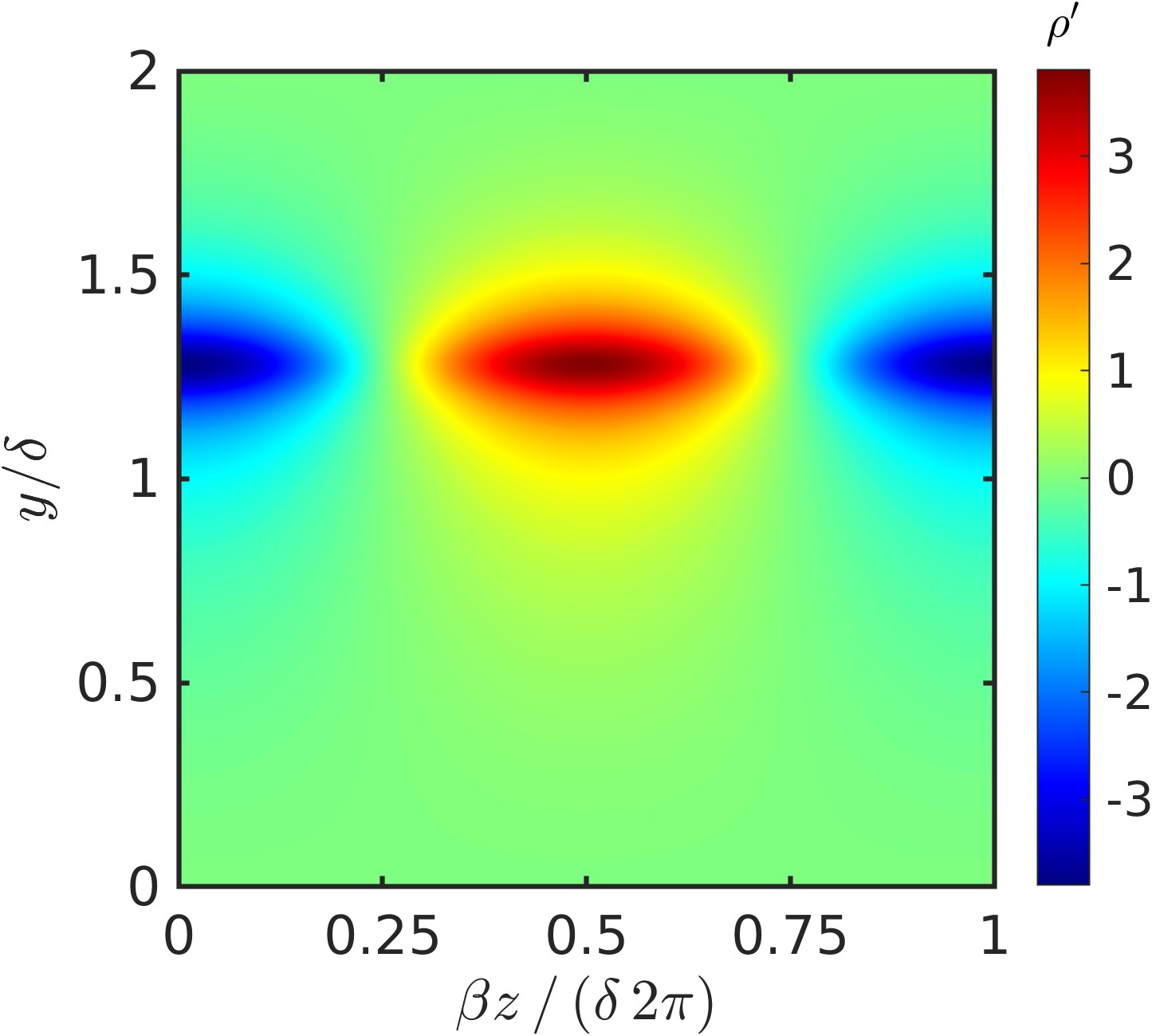}} \hfill
    \subfloat[\vspace{-6mm}]{\includegraphics[width=0.32\linewidth]{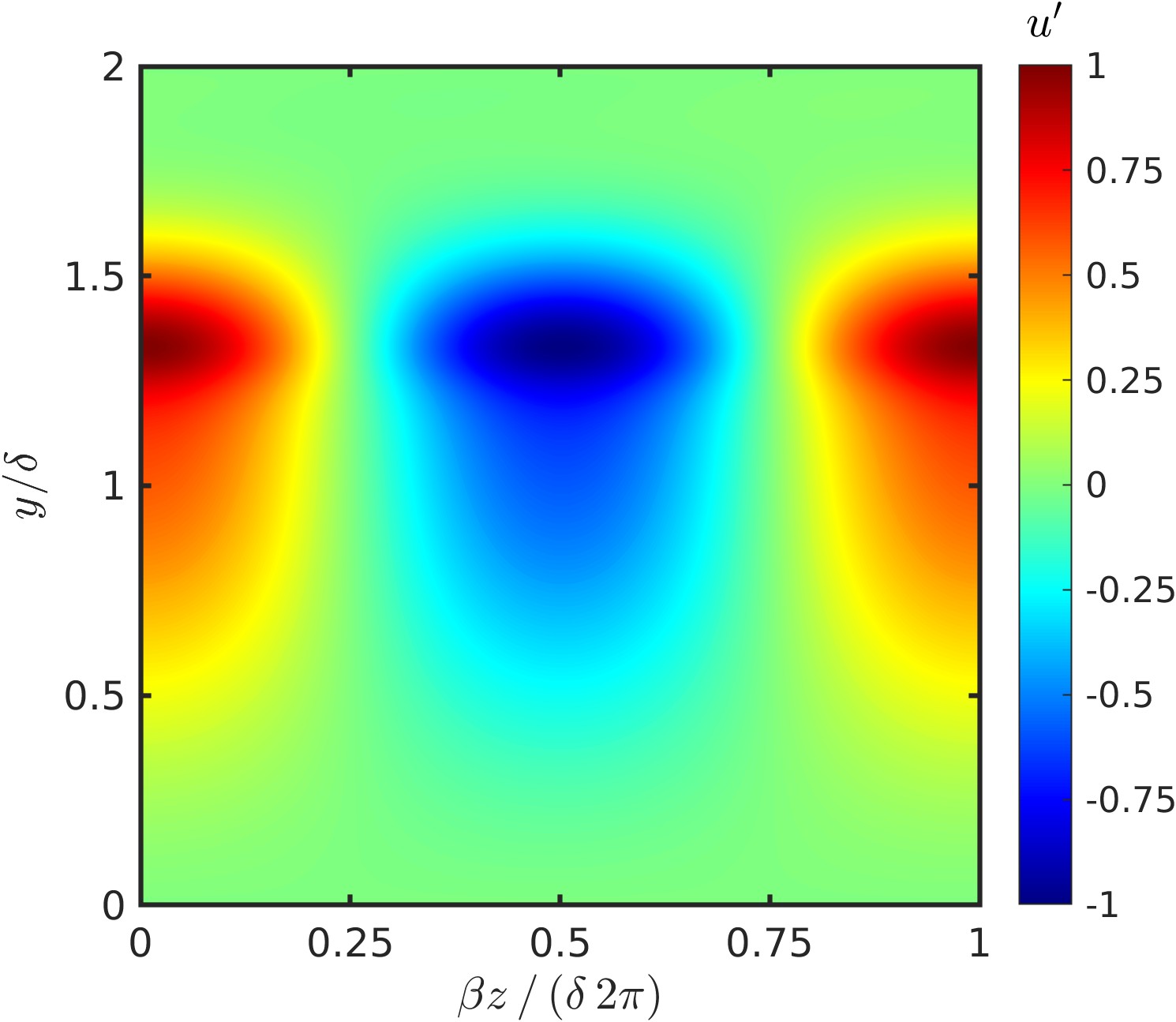}} \hfill
    \subfloat[\vspace{-6mm}]{\includegraphics[width=0.32\linewidth]{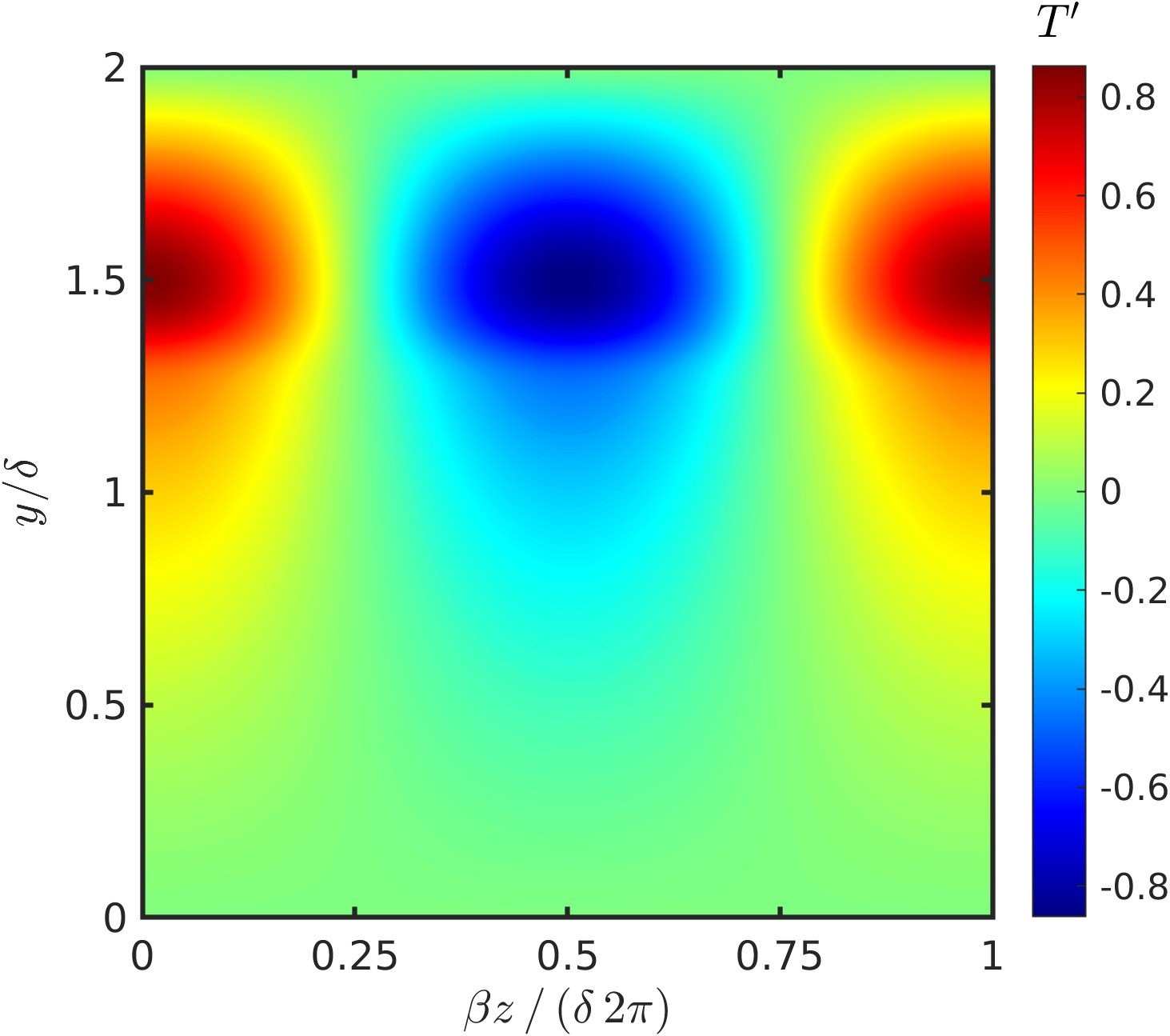}} \\  \vspace{4mm}
    \subfloat[\vspace{-6mm}]{\includegraphics[width=0.32\linewidth]{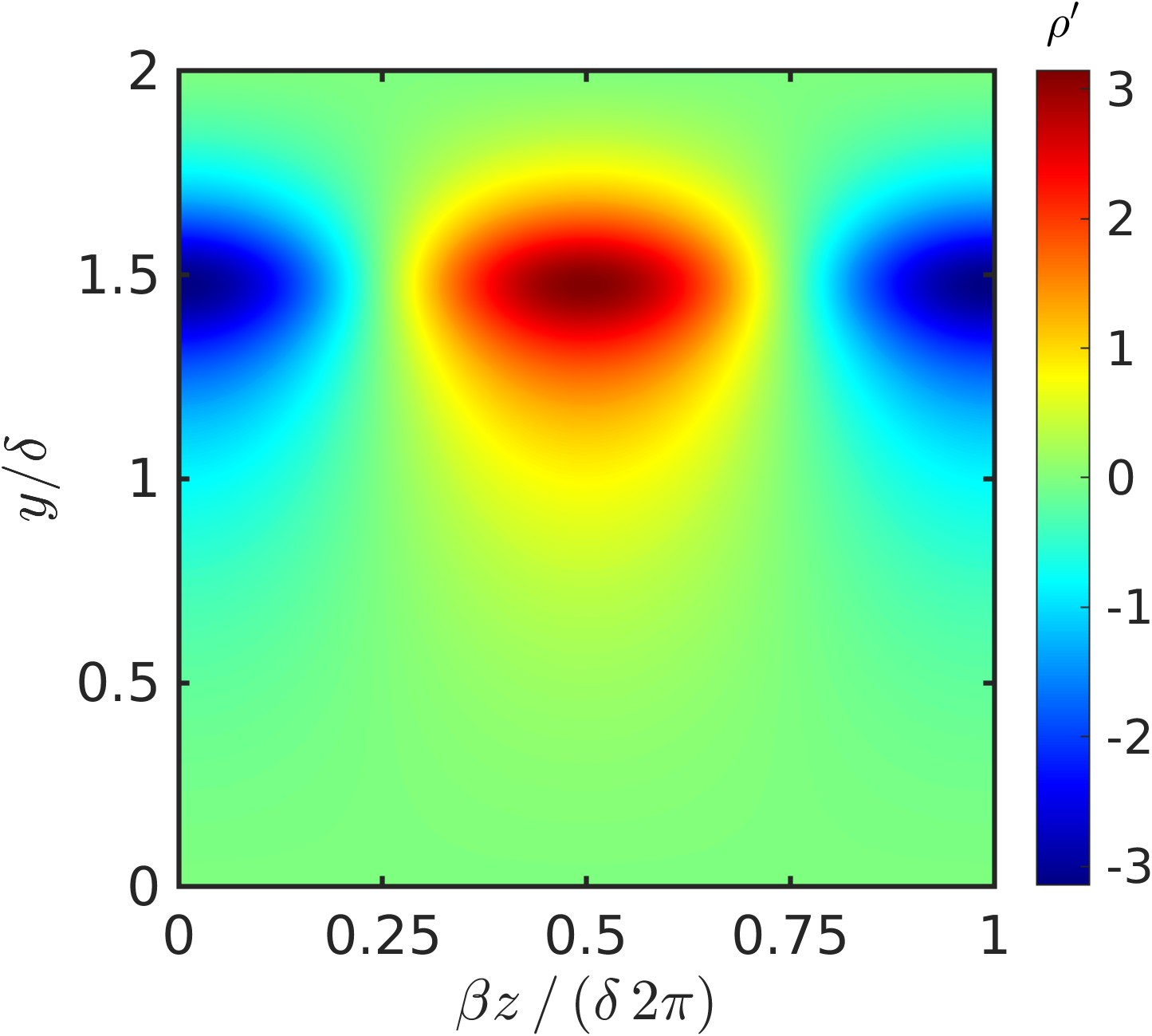}} \hfill
    \subfloat[\vspace{-6mm}]{\includegraphics[width=0.32\linewidth]{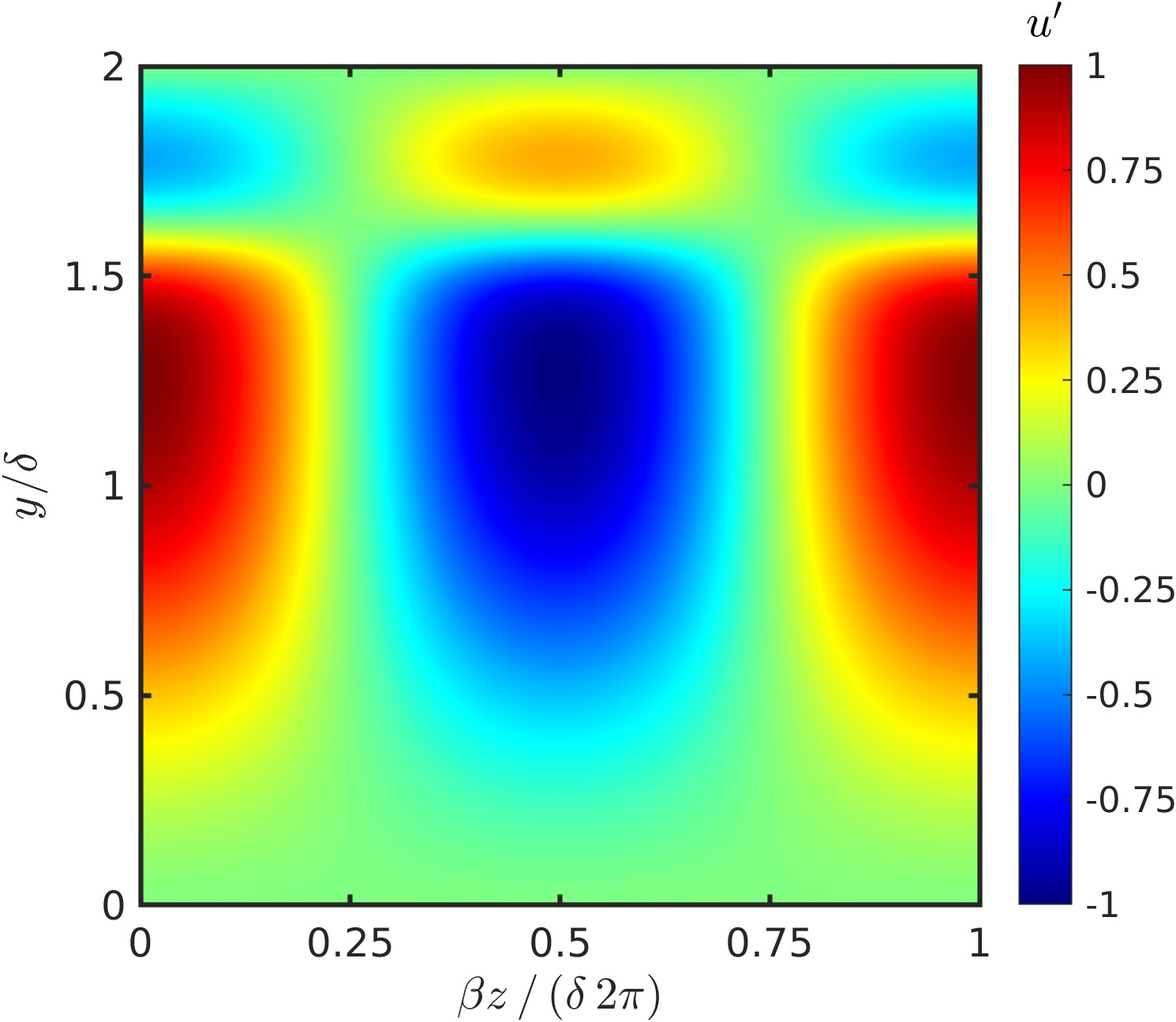}} \hfill
    \subfloat[\vspace{-6mm}]{\includegraphics[width=0.32\linewidth]{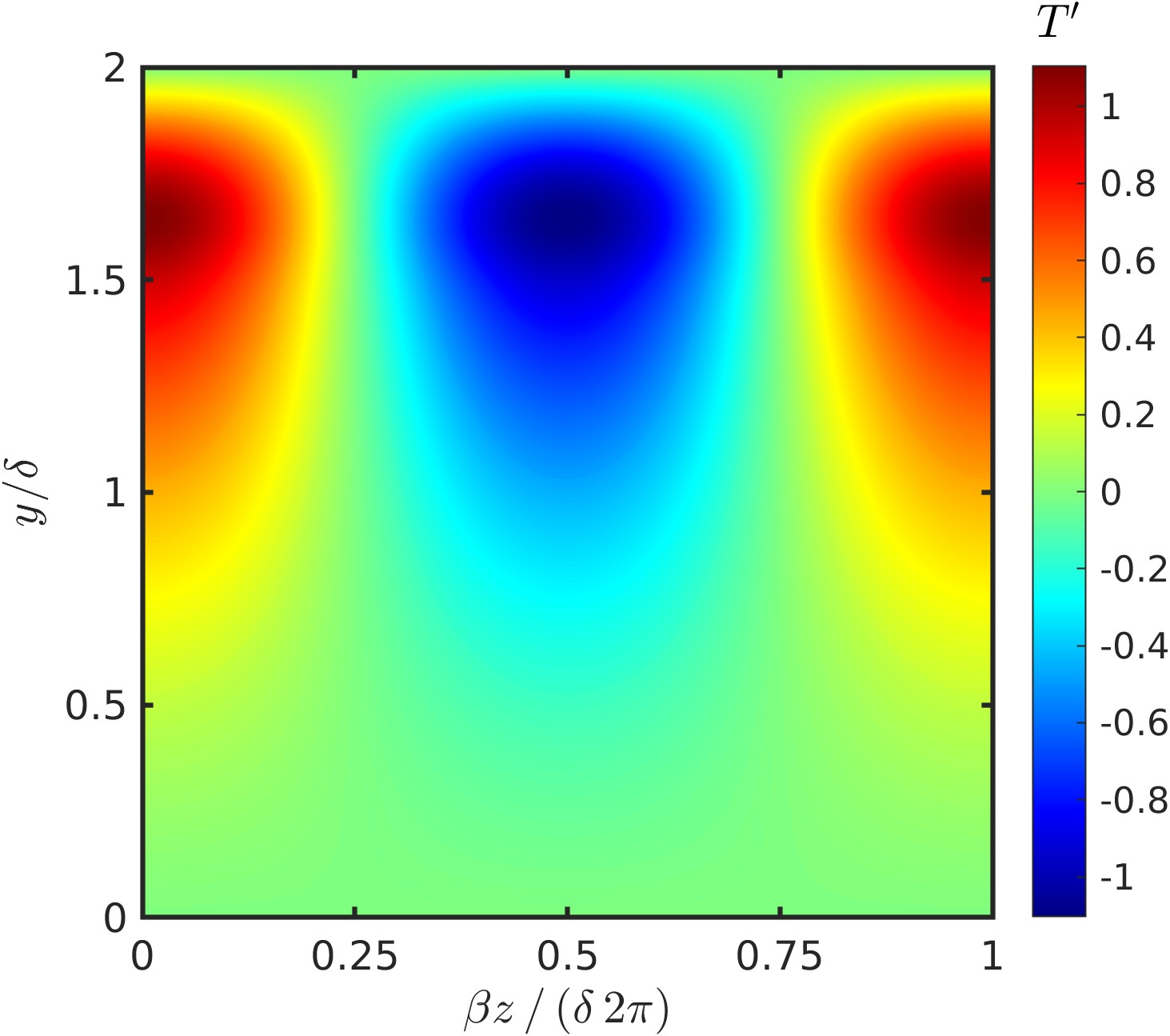}} \\ \vspace{2mm}
	\caption{Optimum response at $Re = 1000$, $\alpha = 0$ and $\beta = 2$ for (a,b,c) I-1 at $Br = 0.5$, (d,e,f) NI-1 at $Br = 0.1$, and (g,h,i) NI-5 at $Br = 5.6 \cdot 10^{-6}$.} 
 \label{fig:Optimum_profile_OUTPUT_Pattern}
\end{figure*}

%\textcolor{red}{Compared with Q-criteria vortices from DNS with the objective to see if the optimum response is similar in structure as the vortices highlighted. Liase with the CTRSP Flow Physics Proceedings...}

% Figure Optimal Perturbation at alpha = 0 beta = 2 for I-2 and NI-1

%\subsection{DNS results}  \label{sec:DNS}	
%DNS will be used to validate the LST. If any misscorrelation, DNS should point out the mode difference with respect to LST such as dual-mode coexistence not captured by LST.
%To this extent, the section covers (i) stability map, (ii) profiles of the perturbation and (iii) DNS validation of growth rate and phase velocity across streamwise direction on non-isothermal low-pressure, viz. laminar channel flow.
%quantify any difference on large density gradients and validate dual-mode coexistence.
%First, the base flow from DNS laminar case is compared to LST framework. The validation results are covered in Appendix~\ref{sec:Appendix_B}. Therefore, based on these results

%\subsubsection{DNS stability maps}
%The stability map of the studied case with the perturbations from LST at different $Re$ levels

%\subsubsection{DNS perturbation profile}
%Similarly as previous section this compares the perturbation profile at chosen $Re-\alpha$ with LST.

%\subsubsection{DNS growth rate}
%The growth rates at several $Re$ are compared against LST

\section{Conclusions}   \label{sec:conclusions}

% Paragraph 1 - Summary
A real-gas framework for linear stability analysis of plane Poiseuille flows has been developed and verified at the isothermal limit to study the stability of high-pressure transcritical flows.
The subsequent modal analysis has focused on characterizing the effects of streamwise perturbations for iso- and non-isothermal flow setups at various Brinkman numbers.
On the one hand, it has been found that for isothermal cases the destabilization occurs at lower Reynolds numbers when the Brinkman number increases.
In particular, at transcritical conditions, the modal-based laminar-to-turbulence transition is triggered at smaller Reynolds numbers than at subcritical thermodynamic regimes.
Instead, similarly to ideal-gas cases, the supercritical cases result in stability enhancement when the Brinkman number increases.
On the other hand, non-isothermal setups have been studied by imposing temperature differences between wall such that the fluid undergoes a transcritical trajectory across the pseudo-boiling region.
For these cases, it has been found that destabilization effects arise at relatively lower Brinkman numbers.
Moreover, component-wise energy budget analyses have revealed that their redistribution through pressure-velocity interactions plays a crucial role on driving energy growth/decay.
In detail, kinetic-energy balances have indicated that the production term is responsible for the destabilization phenomena observed under isothermal subcritical conditions, whereas the transcritical cases are limited only to compressibility effects.
Instead, the non-isothermal cases at low velocities are dominated by both the production and thermodynamic terms, specially the latter which mainly drives the instability mechanism.
From a non-modal perspective, similar patterns are observed when treating unforced algebraic growth compared to modal analysis, where exponential energy growths are captured at similar $\alpha-Re$ parameter values.
This confirms that amplification rates are significantly large for the non-isothermal cases, even for low-Brinkman-number base flows.
In addition, the algebraic growth of the 3D perturbations provides a maximum amplification within a region similar to incompressible flows.
Nonetheless, this region is significantly enlarged in the case of non-isothermal flows with amplification rates on the order of $10\times$ in comparison to isothermal transcritical and equivalent sub/super-critical non-isothermal setups.

% Paragraph 2 - Detailed conclusions
The modal stability analysis has also highlighted that for the isothermal transcritical cases the enhancement of destabilization effects results in a critical Reynolds number that is $5\times$ smaller than that of the isothermal limit case.
Instead, at supercritical conditions at the same Brinkman number the neutral curve is positively shifted a factor of $2\times$ with respect to the transcritical case.
In particular, density and temperature perturbations are mostly responsible for such destabilization for both regimes.
However, supercritical flows enhance the stability if the Brinkman number increases.
Alternatively, the non-isothermal setups exhibit flow destabilization at even lower Brinkman numbers for wavenumbers that are roughly $20\%$ smaller than for the isothermal cases.
It is important to note that by increasing the pressure of the system by $5\times$ similar neutral curves as for the isothermal subcritical case are recovered, although the critical Reynolds number remains slightly lower.
Nonetheless, if the asymmetric wall temperature difference is reduced, the laminar-to-turbulent transition is significantly delayed and comparable to the isothermal subcritical cases.
Moreover, the perturbations are dominated by the thermodynamic modes, especially near the pseudo-boiling region.
The precise mechanism driving and suppressing energy growth is explored by considering the energy budget of 2D perturbations.
The additional stability provided by large Brinkman numbers is shown to be due to a combination of decreased energy production and a more active velocity-pressure gradient term, which forces a component redistribution of energy in a way that significantly dampens the wall-normal fluctuations.
For isothermal flows, the velocity production and second derivative of the vorticity transport equation dominate, whereas the former is more important for the non-isothermal flows except near the pseudo-boiling region where the baroclinic effect becomes dominant and independent of the Brinkman number.
Furthermore, the results of algebraic growth have been found to perform similar to the ones of the modal analysis.
Next, under 3D perturbations it has been found that at $Re = 1000$ a co-existing destabilization region is obtained where amplifications tend to infinity.
This region corresponds to $\alpha = 0$ and $\beta \sim 2$, which for the non-isothermal transcritical flow expands to a wider spanwise region at $1 \le \beta \le 3$.
Interestingly, the isothermal transcritical case presents a co-existing region at $\alpha = \beta \approx 1$ where amplifications grow exponentially.
In addition, it is noted that the maximum amplifications are exacerbated for the non-isothermal cases by a factor of roughly $10\times$ in comparison to the isothermal transcritical cases.
Nonetheless, this energy enhancement is lost when operating at atmospheric pressure conditions or far beyond the critical point.

% Paragraph 3 - Future work
Future work will consider extension of these analyses for the non-isothermal cases.
In particular, energy budgets and growth rates comparing the iso- and non-isothermal setups with 3D perturbations will be further characterized.
Moreover, the unstable modes resulting from the 2D perturbations will also be investigated to evaluate the presence of additional modes beyond the transcritical cases covered. 
Additionally, focusing on the destabilizing mechanisms, the optimum profiles for skin-friction reduction will be explored, and DNS will be used to further validate the LST results, by calculating the time evolution of base flows superimposed with the most unstable modes.
Finally, adaptive resolvent analyses will be carried out to capture the sensitivity to the input responses of the operator and to explore data-driven forcings and their corresponding responses. 

%\backsection[Supplementary data]{\label{SupMat}Supplementary material and movies are available at \\https://doi.org/10.1017/jfm.2019...}

\backsection[Acknowledgements]{The authors acknowledge support from the \textit{Formaci\'o de Professorat Universitari} scholarship (FPU-UPC R.D 103/2019), and the \textit{Serra H\'unter} and SGR (2021-SGR-01045) programs of the Generalitat de Catalunya (Spain).}

\backsection[Funding]{This work is funded by the European Union (ERC, SCRAMBLE, 101040379). Views and opinions expressed are however those of the authors only and do not necessarily reflect those of the European Union or the European Research Council. Neither the European Union nor the granting authority can be held responsible for them.}

\backsection[Declaration of interests]{The authors report no conflict of interest.}

\backsection[Data availability statement] %{Data will be made available upon reasonable request.}
{The data reported in this paper has been generated by means of the in-house MATLAB code named High-Pressure Linear Stability Analysis (HPLSA).
The source code is openly accessible at: \url{https://github.com/marc-bernades/HPLSA}.
HPLSA was designed to serve as a flexible tool to develop linear stability methods for ideal-gas and real-gas thermodynamic frameworks, including wrappers for CoolProp and RefProp libraries. The code is equipped with several comments for readability. 
Additionally, the repository includes a description of the code, instructions and a guide for users.
Therefore, the linear stability modal and non-modal analysis can be fully reproduced by the interested reader.
Although the code is suitable for Poiseuille flows, it can be easily adapted to other wall-bounded cases, such as Couette flow, by adjusting the initial and boundary conditions.
Furthermore, the operator is built for temporal eigenproblems (prescribing streamwise and spanwise wavenumbers and solving the eigenproblem for the angular frequency and growth rate of the perturbation), but it is prepared to be expanded to solve spatial problems also, which is typical in the case of external flows.
This solver requires the previous installation of the high-pressure compressible flow solver (HPCFS) available at \url{https://github.com/marc-bernades/HPCFS}, which embeds some functionalities and thermodynamic models necessary for the complete usage of HPLSA.}

\backsection[Author ORCIDs]{Marc Bernades, https://orcid.org/0000-0003-3761-2038; Francesco Capuano, https://orcid.org/0000-0003-0274-5260; Llu\'is Jofre, https://orcid.org/0000-0003-2437-259X.}

\backsection[Author contributions]{Marc Bernades: Conceptualization, Formal analysis, Investigation, Software, Writing – original draft; Francesco Capuano: Conceptualization, Investigation, Writing – review \& editing; Lluís Jofre: Conceptualization, Funding acquisition, Investigation, Writing – review \& editing.}

%%%%%%%%%%%

\appendix

\section{Thermophysical properties validation} \label{sec:Appendix_A}

\begin{figure*}
	\centering
	\subfloat[]{\includegraphics[width=0.49\linewidth]{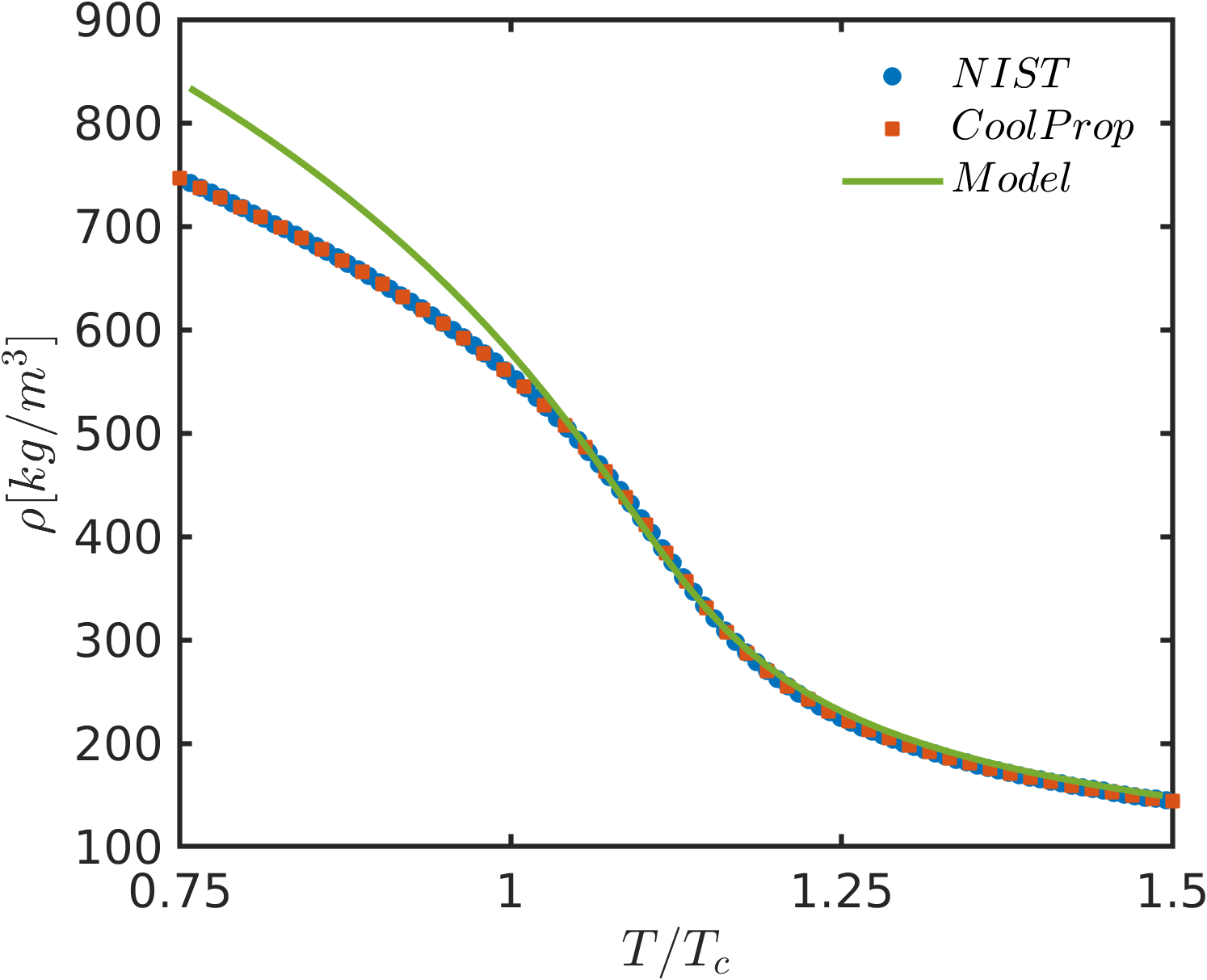}}
    \subfloat[]{\includegraphics[width=0.49\linewidth]{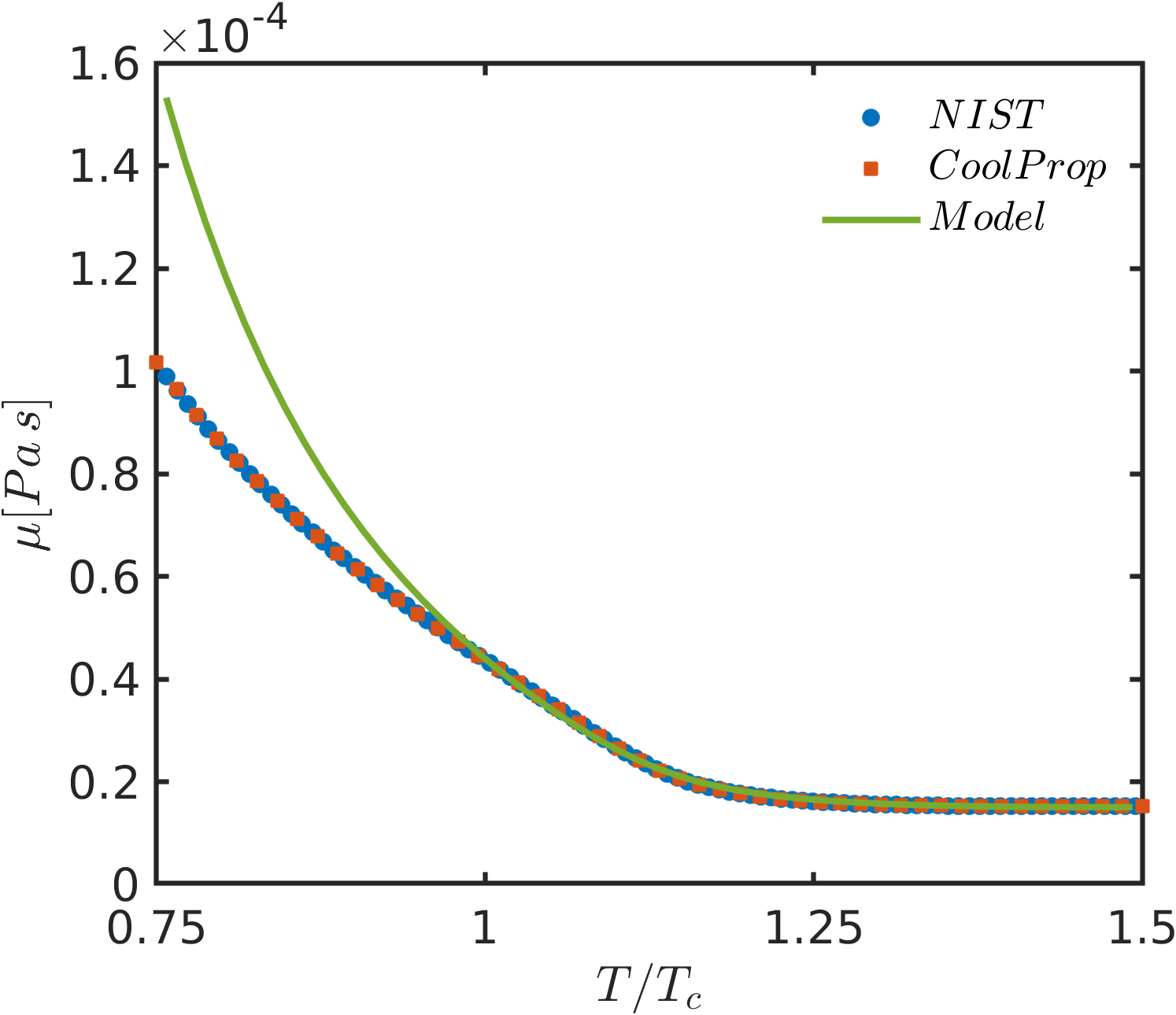}} \\
    \subfloat[]{\includegraphics[width=0.49\linewidth]{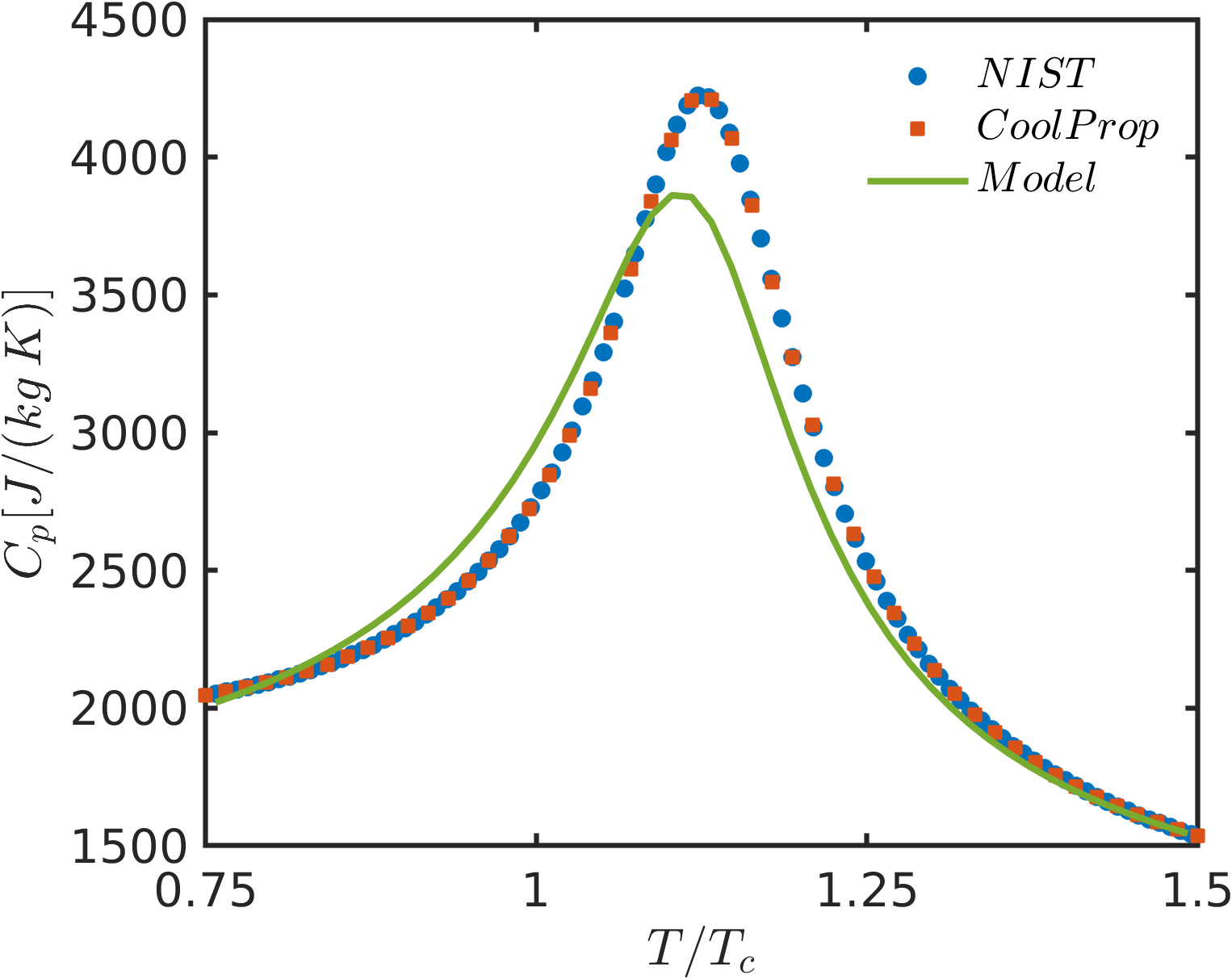}}
    \subfloat[]{\includegraphics[width=0.49\linewidth]{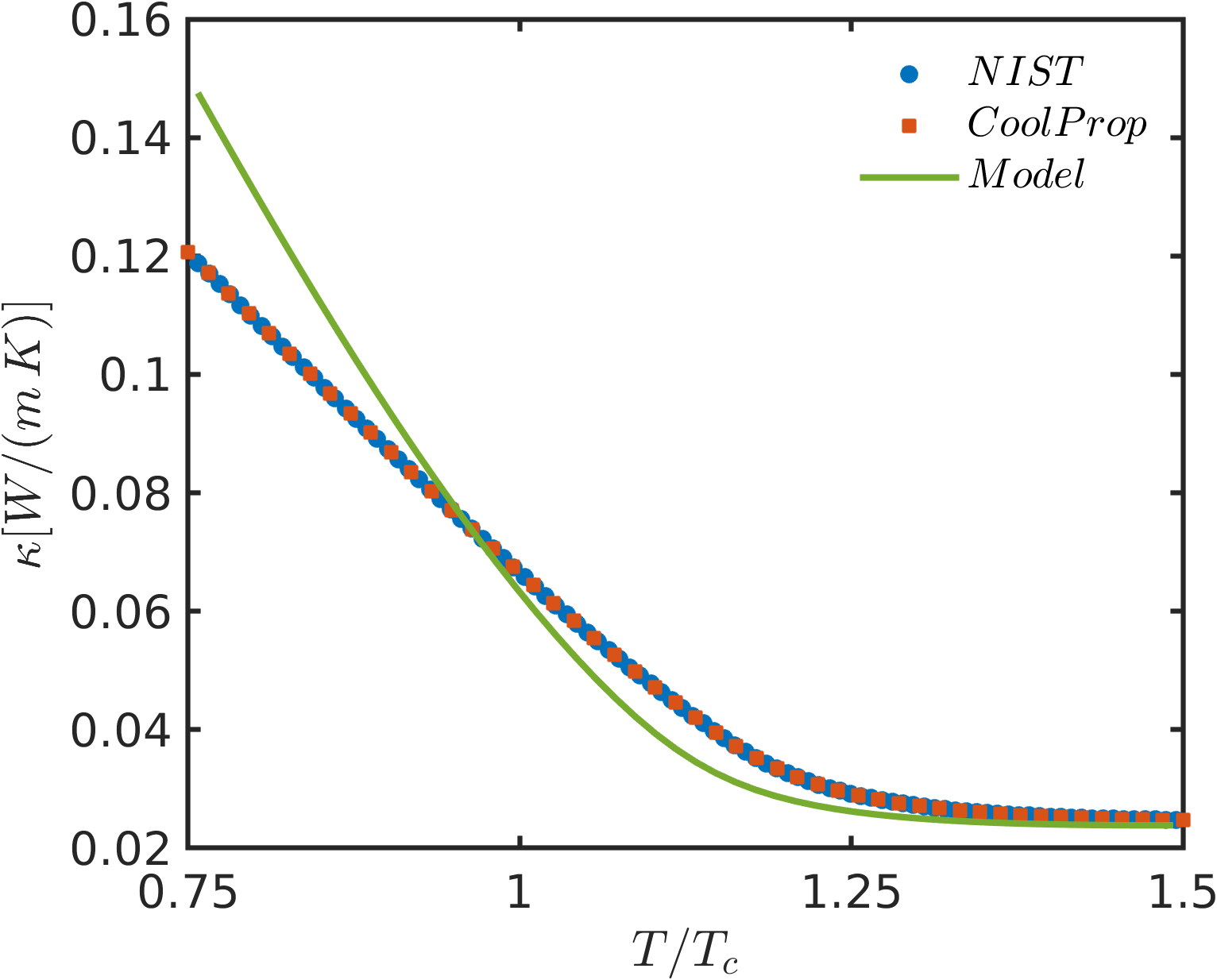}} \\
	\caption{Thermodynamic and transport properties of Nitrogen at $P/ P_c = 2$ with temperature range from $T / T_c = 0.75$ to $T / T_c = 1.5$ for NIST, CoolProp and \textit{Model} (Peng-Robinson equation of state with Chung \textit{et al.} high-pressure coefficients) for (a) density $\rho$, (b) dynamic viscosity $\mu$, (c) isobaric heat capacity $c_p$, and (d) thermal conductivity $\kappa$.} 
 \label{fig:Thermodynamic_validation}
\end{figure*}

The thermodynamic quantities and transport coefficients obtained from the CoolProp library~\citep{Bell2014-A} are validated against NIST data.
They are also compared against the Peng-Robinson equation of state coupled with the high-pressure coefficients from Chung \textit{et al.} labelled as \textit{Model}.
In this regard, Figure~\ref{fig:Thermodynamic_validation} depicts the density, viscosity, isobaric heat capacity and thermal conductivity for Nitrogen at $P / P_c = 2$ and temperature range from $T / T_c = 0.75$ to $T / T_c = 1.5$.
It can be observed that the CoolProp results matches with the NIST reference data.
However, the \textit{Model} solution results in differences in the vicinity of the pseudo-boiling line and at subcritical temperatures at liquid-like state.
The accuracy of such model has been extensively analyzed by~\citet{Bernades2022-A} at different bulk pressures for different substances.
Additionally, the results from the NIST-based library RefProp has also been validated obtaining similar results to the open-source CoolProp library (results not shown to facilitate visualization).

\section{Base flow equations of fluid motion}   \label{sec:Appendix_B}

The base flow is driven by a body force in the streamwise direction and is obtained by solving the Navier-Stokes equations assuming that the flow is fully developed in a laminar state, spanwise and streamwise independent, steady and parallel, i.e., $\partial(\cdot)/\partial x = \partial(\cdot)/\partial z = 0 $, $\partial(\cdot)/\partial t = 0 $, $ v = w = 0$. The dimensionless compressible equations of fluid motion (Eq.~\ref{eq:mass}-\ref{eq:energytransport}) are thus simplified to
\begin{align}
    \overbrace{\frac{d \mu}{d y} \frac{d u}{d y} + \mu \frac{d^2 u}{d y^2} }^{\frac{d}{d y} \left( \mu \frac{d u}{d y} \right)} & = - F,  \label{eq:momentum_baseflow_u} \\
    \frac{d P}{d y} & = 0, \label{eq:momentum_baseflow_v} \\
    \underbrace{u \left( \frac{d \mu}{d y}  \frac{d u}{d y} + \mu \frac{d^2 u}{d y^2} \right) + \mu \left( \frac{d u}{d y} \right)^2 +  \frac{d \kappa}{d y}  \frac{d T}{d y} + \kappa  \frac{d^2 T}{d y^2}}_{\frac{d}{d y} \left( u \mu  \frac{d u}{d y} \right) + \frac{d}{d y} \left( \kappa  \frac{d T}{d y} \right)} & = - F u, \label{eq:momentum_baseflow_E}
\end{align}
where density depends on the equation of state.
In this regard, the streamwise velocity is calculated from Eq.~\ref{eq:momentum_baseflow_u} and temperature from Eq.~\ref{eq:momentum_baseflow_E}.
At this stage, density, viscosity and thermal conductivity can be determined with the updated temperature.
Based on the initial conditions, the body force $F$ can be calculated from Eq.~\ref{eq:momentum_baseflow_u} and controls the desired velocity profile to achieve the desired Reynolds number.

\subsection*{Base flow verification}

\begin{figure*}
	\centering
	\subfloat[\vspace{-6mm}]{\includegraphics[width=0.49\linewidth]{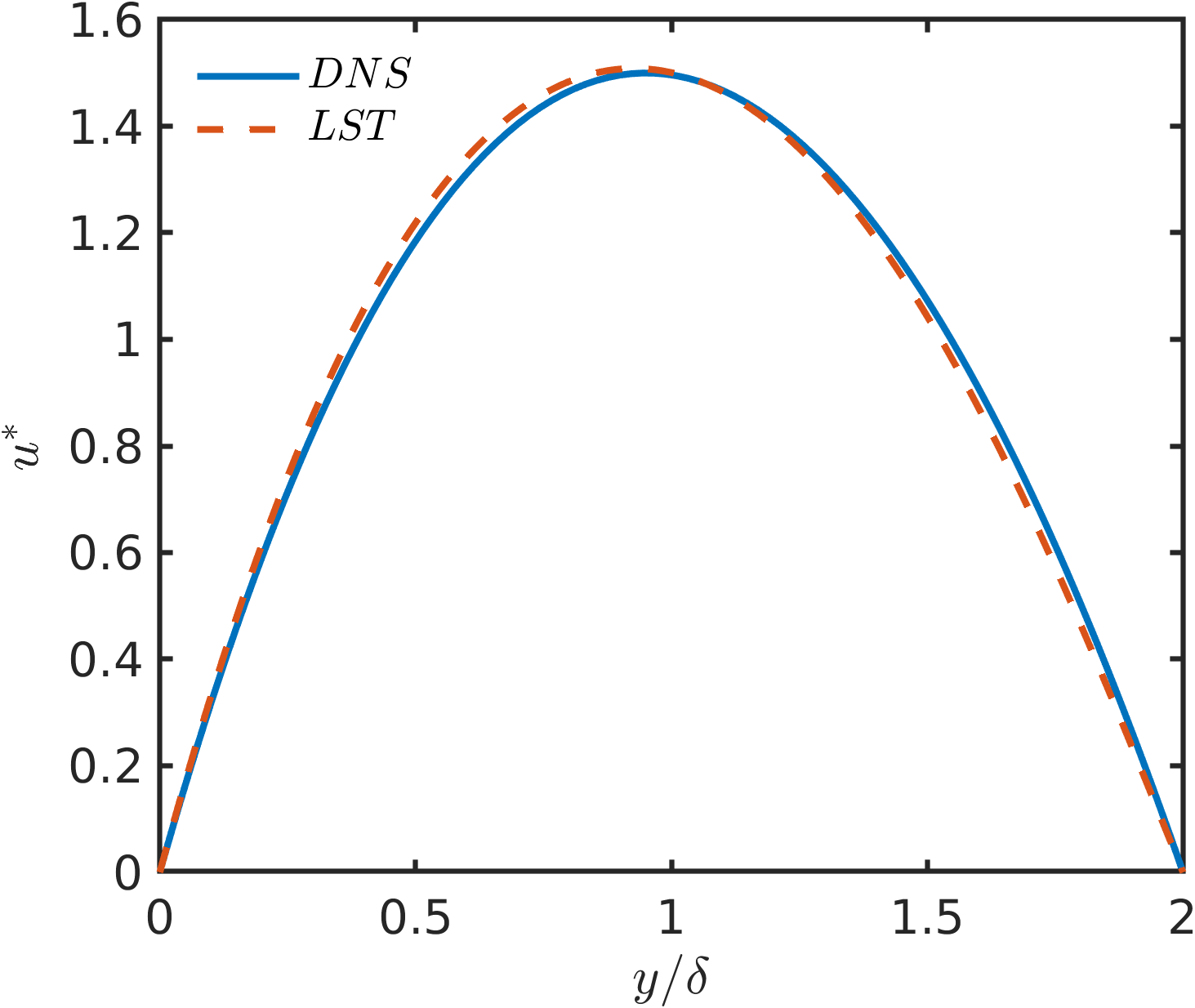}}
    \subfloat[\vspace{-6mm}]{\includegraphics[width=0.49\linewidth]{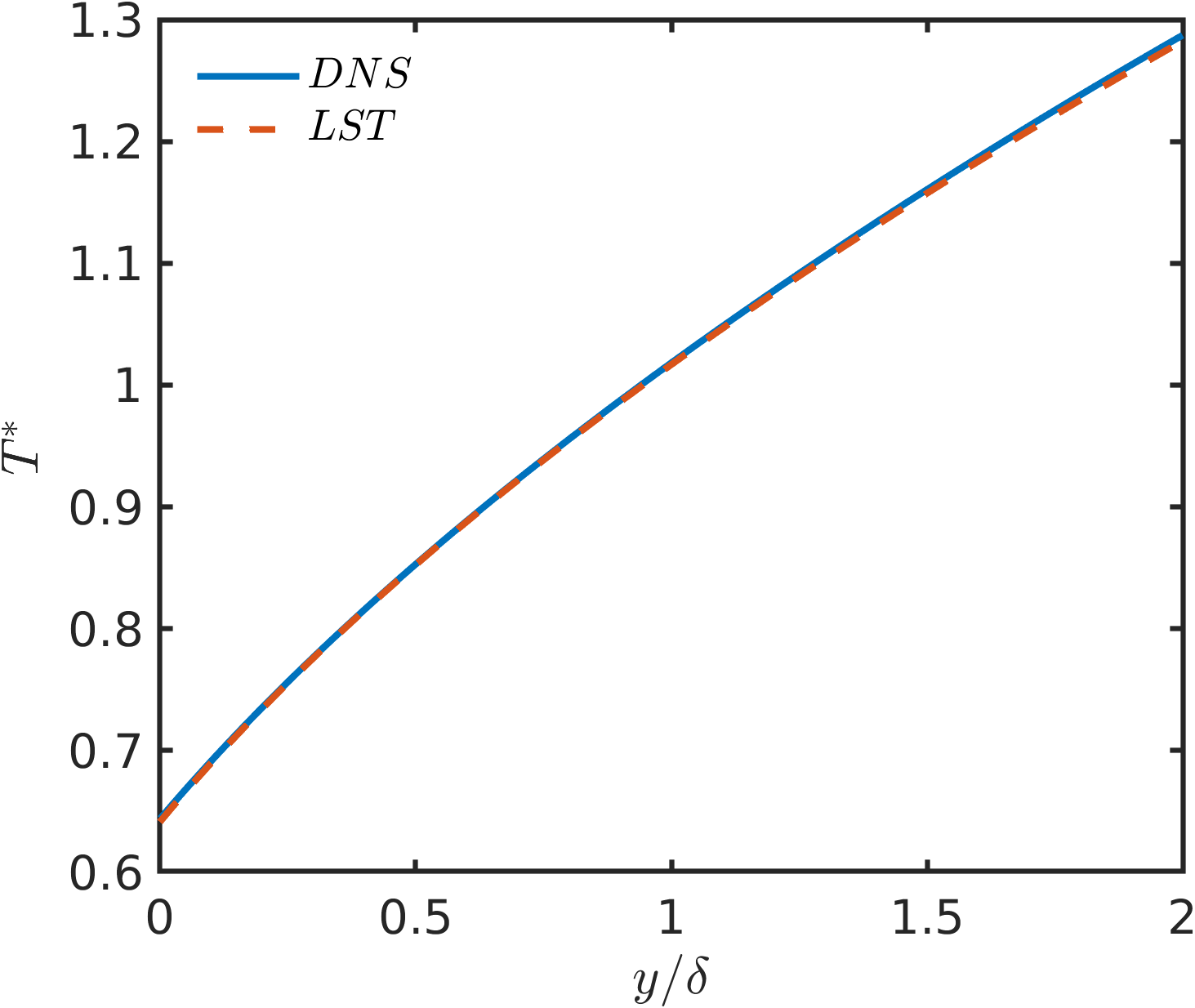}} \\
    \subfloat[\vspace{-6mm}]{\includegraphics[width=0.49\linewidth]{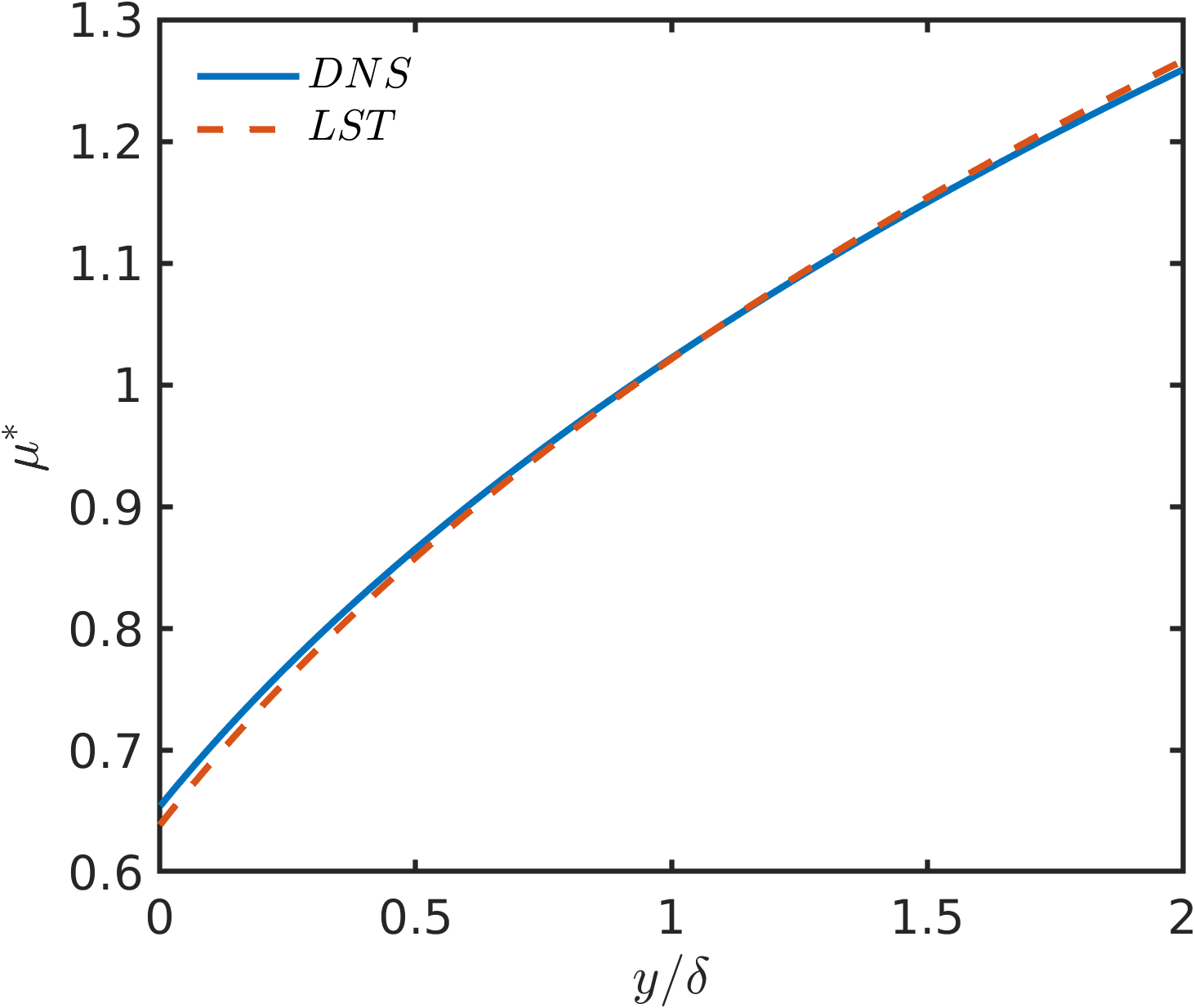}}
    \subfloat[\vspace{-6mm}]{\includegraphics[width=0.49\linewidth]{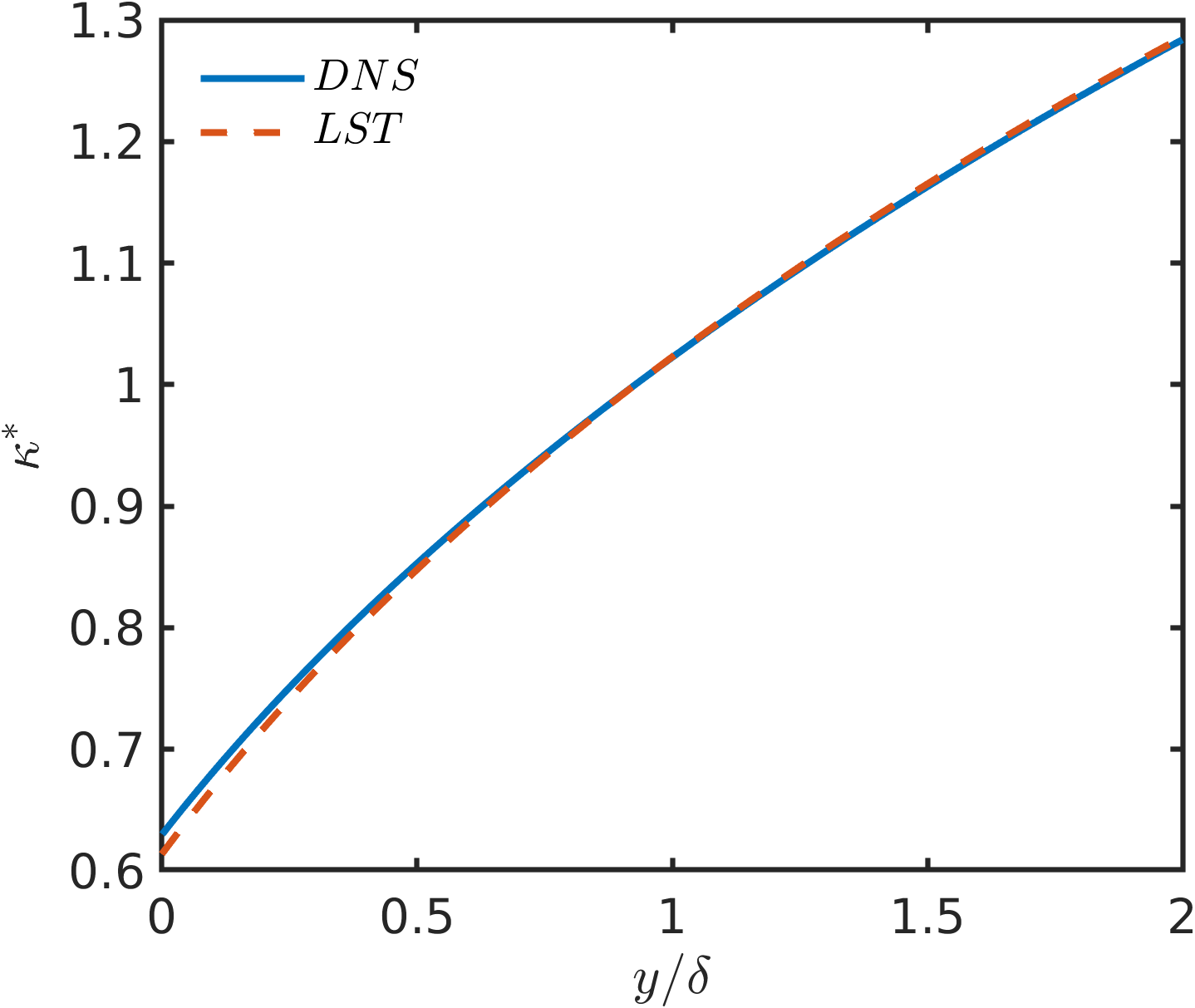}} \\ \vspace{2mm}
	\caption{Low-pressure $P/P_c = 0.03$ non-isothermal base flow with temperature range from $T / T_c = 0.75$ to $T / T_c = 1.5$ utilizing Nitrogen from (i) ensemble-averaged DNS high-pressure transcritical channel flow, and (ii) linear stability solver for (a) velocity, (b) temperature, (c) dynamic viscosity, and (d) thermal conductivity. Both frameworks utilize the thermodynamic model of Peng-Robinson equation of state and Chung \textit{et al.} high-pressure transport coefficients.} 
 \label{fig:baseflow_validation}
\end{figure*}

The base flow of the linear stability solver is verified against the solution of a laminar steady-state channel flow.
This case operates with Nitrogen at low-pressure $P/P_c = 0.03$ and the setup is such that it matches the same input pressure as the transcritical equivalent case at $P/P_c = 2$~\citep{Bernades2023-A}.
The resulting dimensionless numbers and setup for the iterative LST base flow are imposed similar to the DNS solution as $Br = 7.7 \cdot 10^{-4}$, where the reference velocity scaling is based on the bulk velocity.
For consistency, instead of imposing a constant normalized body force, the LST incorporates a feedback control that adjusts the value of $\hat{F}$.
Moreover for this verification, the LST uses the same thermodynamic model used for the DNS computation, i.e., the Peng-Robinson equation of state and high-pressure coefficients from Chung \textit{et al.}~\citep{Chung1984-A,Chung1988-A}
The base flow iterative process for the linear stability is independent of Reynolds number.
In this regard, Figure~\ref{fig:baseflow_validation} depicts the velocity, temperature and transport properties comparing the LST and DNS frameworks.
As it can be observed, good agreement between both approaches is found, which consequently verifies the LST algorithm.

\section{Linear stability equations}    \label{sec:Appendix_C}

This section presents the linearized equations of fluid motion. They are derived by substituting Eq.~\ref{eq:Decomposition} into the dimensionless Navier-Stokes equations (Eq.~\ref{eq:mass}-\ref{eq:energytransport}) with appropriate algebra developments.
The derived discrete equations can then be rewritten in matrix form as formulated in Eq.~\ref{eq:LST}. Therefore, by inspection each of the $5x5$ terms of $\mathbf{L_t}$, $\mathbf{L_x}$, $\mathbf{L_y}$, $\mathbf{L_z}$, $\mathbf{L_q}$, $\mathbf{V_{xx}}$, $\mathbf{V_{yy}}$, $\mathbf{V_{zz}}$, $\mathbf{V_{xy}}$, $\mathbf{V_{xz}}$ and $\mathbf{V_{yz}}$, the corresponding matrices can be obtained.

\subsubsection{Continuity equation}

Substituting the decomposition $\rho = \rho_0 + \rho^\prime$ into Eq.~\ref{eq:mass} reads
\begin{align}
    & \frac{\partial (\rho_0 + \rho^\prime)}{\partial t} = -   \nabla \cdot  [ (\rho_0 + \rho^\prime) (\mathbf{u_0} + \mathbf{u^\prime}) ] = \nonumber \\
    & -\rho_0 \nabla \cdot  \mathbf{u^\prime} -  \underbrace{\rho_0 \nabla \cdot  \mathbf{u_0}}_{=0} -  \rho^\prime \nabla \cdot  \mathbf{u^\prime} -   \underbrace{\rho^\prime \nabla \cdot  \mathbf{u_0}}_{=0}  - \underbrace{\mathbf{u_0} \nabla\rho_0}_{=0}  -  \underbrace{\mathbf{u_0} \nabla\rho^\prime}_{u_0 \frac{d \rho^\prime}{dx}}  -  \underbrace{\mathbf{u^\prime} \nabla \rho_0}_{v^\prime \frac{d \rho_0}{dy}}  -  \underbrace{\mathbf{u^\prime} \nabla \rho^\prime}_{= 0}, 
\end{align}
which, by further manipulation, results in
\begin{align}
\frac{\partial \rho^\prime}{\partial t} = -  \rho_0 \left( \frac{du^\prime}{dx} + \frac{dv^\prime}{dy} + \frac{dw^\prime}{dz}\right) - u_0 \frac{d \rho^\prime}{dx} - v^\prime \frac{d \rho_0}{dy}. \label{eq:LST_continuity}
\end{align}

\subsubsection{Momentum equations}

Analogously to the continuity equation, substituting the decomposition of variables into Eq.~\ref{eq:momentum} yields
\begin{align}
     & \frac{\partial [ (\rho_0 + \rho^\prime) \mathbf{(u_0 + u^\prime)} ]}{\partial t} + \nabla \cdot  [ (\rho_0 + \rho^\prime) \mathbf{(u_0 + u^\prime)}  \mathbf{(u_0 + u^\prime)}] + \nabla ((P_0 + P^\prime)) =  \nonumber \\
     & \frac{1}{{Re}} \nabla \cdot \{ (\mu_0 + \mu^\prime) [\nabla (\mathbf{u_0 + u^\prime}) + \nabla  (\mathbf{u_0 + u^\prime})^{T} ] + (\lambda_0 + \lambda^\prime)[\nabla \cdot (\mathbf{u_0 + u^\prime})]\boldsymbol{I} \} + \textbf{F}. 
\end{align}

Based on this expression, the linearized momentum in each direction is detailed as follows.

Momentum streamwise direction:
\begin{eqnarray}
    &&  \rho_0 \frac{\partial u^\prime}{\partial t} + u_0 \frac{\partial \rho^\prime}{\partial t} + u_0 u_0 \frac{\partial \rho^\prime}{\partial x} + \frac{\partial P_0}{\partial \rho_0} \frac{\partial \rho^\prime}{\partial x} + 2 \rho_0 u_0 \frac{\partial u^\prime}{\partial x} + \rho_0 u_0 \frac{\partial v^\prime}{\partial y} + \rho_0 u_0 \frac{\partial w^\prime}{\partial z} \nonumber \\
    &&  + \rho_0 \frac{\partial u_0}{\partial y} v^\prime +  u_0 \frac{\partial \rho_0}{\partial y} v^\prime = \frac{1}{Re} \frac{\partial \mu_0}{\partial y}\frac{\partial v^\prime}{\partial x} + \frac{1}{Re} \frac{\partial \mu_0}{\partial y}\frac{\partial u^\prime}{\partial y} \nonumber \\
    &&  + \frac{2 \mu_0 - \lambda_0}{Re} \frac{\partial^2 u^\prime}{\partial^2 x} + \frac{\mu_0}{Re} \frac{\partial^2 u^\prime}{\partial^2 y} + \frac{\mu_0}{Re} \frac{\partial^2 u^\prime}{\partial^2 z} + \frac{\mu_0 + \lambda_0}{Re} \frac{\partial^2 v^\prime}{\partial x \partial y} + \frac{\mu_0 + \lambda_0}{Re} \frac{\partial^2 w^\prime}{\partial x \partial z} \nonumber \\
    &&  + \frac{\partial P_0}{\partial T_0}\frac{\partial T^\prime}{\partial x} + \frac{1}{Re} \frac{\partial \mu_0}{\partial \rho_0}\frac{\partial u_0}{\partial y}\frac{\partial \rho^\prime}{\partial y} + \frac{1}{Re} \frac{\partial \mu_0}{\partial T_0}\frac{\partial u_0}{\partial y}\frac{\partial T^\prime}{\partial y} + \frac{1}{Re} \frac{\partial \mu_0}{\partial \rho_0}\frac{\partial^2 u_0}{\partial^2 y} \rho^\prime \nonumber \\
    &&  + \frac{1}{Re} \frac{\partial \mu_0}{\partial T_0}\frac{\partial^2 u_0}{\partial^2 y} T^\prime  + \frac{1}{Re} \left( \frac{\partial^2 \mu_0}{\partial^2 \rho_0} \frac{\partial \rho_0}{\partial y} + \frac{\partial^2 \mu_0}{\partial \rho_0 \partial T_0} \frac{\partial T_0}{\partial y} \right) \frac{\partial u_0}{\partial y} \rho^\prime  \nonumber \\
    && + \frac{1}{Re} \left( \frac{\partial^2 \mu_0}{\partial^2 T_0} \frac{\partial T_0}{\partial y} + \frac{\partial^2 \mu_0}{\partial T_0 \partial \rho_0} \frac{\partial \rho_0}{\partial y} \right) \frac{\partial u_0}{\partial y} T^\prime + \frac{\hat{F}}{Re}. \label{eq:LST_momentum_X}  
\end{eqnarray}

Momentum wall-normal direction:
\begin{eqnarray}
    && \rho_0 \frac{\partial v^\prime}{\partial t} + \rho_0 u_0 \frac{\partial v^\prime}{\partial x} = \frac{1}{Re} \frac{\partial \lambda_0}{\partial y}\frac{\partial u^\prime}{\partial x} + \frac{1}{Re} \frac{\partial \lambda_0}{\partial y}\frac{\partial v^\prime}{\partial y} + \frac{2}{Re} \frac{\partial \mu_0}{\partial y}\frac{\partial v^\prime}{\partial y} + \frac{1}{Re} \frac{\partial \lambda_0}{\partial y}\frac{\partial w^\prime}{\partial z}  \nonumber \\
    && + \frac{\mu_0}{Re} \frac{\partial^2 v^\prime}{\partial^2 x} + \frac{2 \mu_0 + \lambda_0}{Re} \frac{\partial^2 v^\prime}{\partial^2 y} + \frac{\mu_0}{Re} \frac{\partial^2 v^\prime}{\partial^2 z} + \frac{\mu_0 + \lambda_0}{Re} \frac{\partial^2 u^\prime}{\partial x \partial y} + \frac{\mu_0 + \lambda_0}{Re} \frac{\partial^2 w^\prime}{\partial y \partial z} \nonumber \\
    && + \frac{1}{Re} \frac{\partial \mu_0}{\partial \rho_0}\frac{\partial u_0}{\partial y}\frac{\partial \rho^\prime}{\partial x} + \frac{1}{Re} \frac{\partial \mu_0}{\partial T_0}\frac{\partial u_0}{\partial y}\frac{\partial T^\prime}{\partial x} + \frac{\partial P_0}{\partial \rho_0}\frac{\partial \rho^\prime}{\partial y} + \frac{\partial P_0}{\partial T_0}\frac{\partial T^\prime}{\partial y} + \frac{\partial^2 P_0}{\partial^2 \rho_0} \frac{\partial \rho_0}{\partial y} \rho^\prime \nonumber \\
    && + \frac{\partial^2 P_0}{\partial T_0 \partial \rho_0} \frac{\partial T_0}{\partial y} \rho^\prime + \frac{\partial^2 P_0}{\partial^2 T_0} \frac{\partial T_0}{\partial y} T^\prime + \frac{\partial^2 P_0}{\partial \rho_0 \partial T_0} \frac{\partial \rho_0}{\partial y} T^\prime. \label{eq:LST_momentum_Y}
\end{eqnarray}

Momentum spanwise direction:
\begin{eqnarray}  
    &&\rho_0 \frac{\partial w^\prime}{\partial t} + \rho_0 u_0 \frac{\partial w^\prime}{\partial x} = \frac{1}{Re} \frac{\partial \mu_0}{\partial y}\frac{\partial w^\prime}{\partial y} + \frac{1}{Re} \frac{\partial \mu_0}{\partial y}\frac{\partial v^\prime}{\partial z} +
      \frac{\mu_0}{Re} \frac{\partial^2 w^\prime}{\partial^2 x} + \frac{\mu_0}{Re} \frac{\partial^2 w^\prime}{\partial^2 y} \nonumber \\
     && + \frac{2 \mu_0 + \lambda_0}{Re} \frac{\partial^2 w^\prime}{\partial^2 z} + \frac{\mu_0 + \lambda_0}{Re} \frac{\partial^2 u^\prime}{\partial x \partial z} + \frac{\mu_0 + \lambda_0}{Re} \frac{\partial^2 v^\prime}{\partial y \partial z} + \frac{\partial P_0}{\partial \rho_0}\frac{\partial \rho^\prime}{\partial z} + \frac{\partial P_0}{\partial T_0}\frac{\partial T^\prime}{\partial z}. \label{eq:LST_momentum_Z}
\end{eqnarray}

\subsubsection{Internal energy equation}

Expressing Eq.~\ref{eq:energytransport} in terms of internal energy by removing the kinetic-energy part, the corresponding transport equation reads as
\begin{equation}
    \frac{\partial \left( \rho e \right) }{\partial t} + \nabla \cdot  \left( \rho \mathbf{u} e  \right) = -P \thinspace \nabla \cdot  ( \mathbf{u})  - \frac{1}{{Re} {Br}} \nabla \cdot \mathbf{q} + \frac{1}{{Re}}\nabla \cdot (\boldsymbol{\tau} \cdot \mathbf{u}) +  F \mathbf{u}.\label{eq:internal_energy_transport}
\end{equation}

Similar to the subsections above, by substituting the expressions $\rho = \rho_0 + \rho^\prime$, $\mathbf{u} = \mathbf{u_0} + \mathbf{u^\prime}$ and $e = e_0 + e^\prime$ into Eq.~\ref{eq:internal_energy_transport}, one reads
\begin{eqnarray}
    && \frac{\partial \left(  (\rho_0 + \rho^\prime) (e_0 + e^\prime)\right) }{\partial t^\star} + \nabla^\star \cdot  \left( (\rho_0 + \rho^\prime) (\mathbf{u_0} + \mathbf{u^\prime}) (e_0 + e^\prime)  \right) + (P_0 + P^\prime)  \nabla \cdot   (\mathbf{u_0} + \mathbf{u^\prime}) = \nonumber \\
    && +\frac{1}{{Re} {Br}} \nabla \cdot \left((\kappa_0 + \kappa^\prime) \nabla (T_0 + T^\prime) \right) \nonumber \\
    && + \frac{1}{{Re}}\nabla \cdot \left[\left( (\mu_0 + \mu^\prime) \left( \nabla (\mathbf{u_0 + u^\prime}) + \nabla  (\mathbf{u_0 + u^\prime})^{T}  \right) \right. \right.  \nonumber \\
    && + \left. \left. (\lambda_0 + \lambda^\prime)(\nabla \cdot (\mathbf{u_0 + u^\prime}))\boldsymbol{I} \right) \cdot  (\mathbf{u_0} + \mathbf{u^\prime})\right] + F (\mathbf{u_0} + \mathbf{u^\prime}), 
\end{eqnarray}
which, by (i) substituting the terms $e^\prime$, $P^\prime$, $\kappa^\prime$ and $\mu^\prime$ with their corresponding Taylor series expansion approximation as a function of the independent thermodynamic pair of variables selected ($\rho^\prime$-$T^\prime$), and (ii) linearizing the equations, the following expression is obtained
\begin{eqnarray}
    && \left( e_0 + \rho_0 \frac{\partial e_0}{\partial \rho_0} \right)  \frac{\partial \rho^\prime}{\partial t} + \rho_0 \frac{\partial e_0}{\partial T_0} \frac{\partial T^\prime}{\partial t} + \left( e_0 u_0 + \rho_0 u_0 \frac{\partial e_0}{\partial \rho_0} \right) \frac{\partial \rho^\prime}{\partial x} \nonumber \\
    && + \left( \rho_0 e_0 + P_0 \right) \left(\frac{\partial u^\prime}{\partial x} + \frac{\partial v^\prime}{\partial y} + \frac{\partial w^\prime}{\partial z} \right) + \rho_0 u_0 \frac{\partial e_0}{\partial T_0} \frac{\partial T^\prime}{\partial x} = \frac{2}{Re} \mu_0 \frac{\partial u_0}{\partial y} \frac{\partial v^\prime}{\partial x} + \frac{2}{Re} \mu_0 \frac{\partial u_0}{\partial y} \frac{\partial u^\prime}{\partial y} \nonumber \\
    && +\frac{1}{Re Br} \frac{\partial \kappa_0}{\partial \rho_0} \frac{\partial T_0}{\partial y} \frac{\partial \rho^\prime }{\partial y} + \frac{1}{Re Br} \left(\frac{\partial \kappa_0}{\partial y} + \frac{\partial \kappa_0}{\partial T_0} \frac{\partial T_0}{\partial y} \right) \frac{\partial T^\prime }{\partial y} + \frac{\kappa_0}{Re Br} \left( \frac{\partial^2 T^\prime}{\partial^2 x} + \frac{\partial^2 T^\prime}{\partial^2 y} + \frac{\partial^2 T^\prime}{\partial^2 z}\right) \nonumber \\
    && + \frac{1}{Re Br} \left( \frac{\partial^2 T_0}{\partial y^2} \frac{\partial \kappa_0}{\partial \rho_0} +  \left( \frac{\partial^2 \kappa_0}{\partial {\rho_0}^2} \frac{\partial \rho_0}{\partial y} + \frac{\partial^2 \kappa_0}{\partial \rho_0 \partial T_0} \frac{\partial T_0}{\partial y} \right) \frac{\partial T_0}{\partial y} \right) \rho^\prime + \frac{1}{Re}  \frac{\partial \mu_0}{\partial \rho_0} \left( \frac{\partial u_0}{\partial y} \right)^2 \rho^\prime \nonumber \\
    && + \frac{1}{Re Br} \left( \frac{\partial^2 T_0}{\partial y^2} \frac{\partial \kappa_0}{\partial T_0} +  \left( \frac{\partial^2 \kappa_0}{\partial {T_0}^2} \frac{\partial T_0}{\partial y} + \frac{\partial^2 \kappa_0}{\partial T_0 \partial \rho_0} \frac{\partial \rho_0}{\partial y} \right) \frac{\partial T_0}{\partial y} \right) T^\prime  + \frac{1}{Re}  \frac{\partial \mu_0}{\partial T_0} \left( \frac{\partial u_0}{\partial y} \right)^2 T^\prime \nonumber \\
    && + \frac{\hat{F}}{Re} u^\prime. \label{eq:LST_energy}
\end{eqnarray}

It is important to note that the Jacobian terms of the equation of state fields are obtained from the Coolprop library.
However the derivatives of the high-pressure coefficients $\mu$ and $\kappa$ are computed numerically by means of second-order finite differences with a derivation step of $\Delta T = 10^{-2} K$ and $\Delta \rho = 10^{-2} kg \thinspace m^3$.
This relatively small step is robust enough to accurately calculate the gradients of the thermodynamic and transport coefficients fields at each point.

\subsection{Energy balance equations}

The kinetic energy can be expressed as the difference between the averaged Reynolds stresses term, which determines the production of energy rate due to transfer of energy from the base flow to the disturbances, and the dissipation energy.
Therefore, the equation isolates the mechanisms by which energy is transferred from the base flow to the disturbances.
In this regard, the contributors of the kinetic energy balance equation described in Section~\ref{sec:energy_budget} are detailed below.
For ease of exposition, the notation $\hat{\cdot}$ denoting perturbation vectors is dropped from the equations, yielding
\begin{eqnarray}
   && K = - i \omega \int \rho_0 (u u^\dagger + v v^\dagger) dy, \\
   && \Theta = - i \alpha \int \rho_0 u_0 (u u^\dagger + v v^\dagger) dy, \\
   && P = - \int \rho_0 \frac{\partial u_0}{\partial y}  v u^\dagger dy, \\
   && T = - \int \left[ i \alpha \left( \frac{\partial P_0}{\partial \rho_0} \rho  +  \frac{\partial P_0}{\partial T_0} T \right) u^\dagger + \left(\frac{\partial P_0}{\partial \rho_0} \frac{\partial \rho}{\partial y} + \frac{\partial P_0}{\partial T_0} \frac{\partial T}{\partial y} \right) v^\dagger \right. \nonumber \\
   && + \left. \left( \frac{\partial^2 P_0}{\partial {\rho_0}^2} \frac{\partial \rho_0}{\partial y} +  \frac{\partial^2 P_0}{\partial {\rho_0} \partial T_0} \frac{\partial T_0}{\partial y} \right) \rho v^\dagger + \left( \frac{\partial^2 P_0}{\partial {T_0}^2} \frac{\partial T_0}{\partial y} +  \frac{\partial^2 P_0}{\partial {T_0} \partial T_0} \frac{\partial \rho_0}{\partial y} \right) T v^\dagger \right] dy, \\
   && V = \frac{1}{Re} \int \left[ - \alpha^2 (2 \mu_0 + \lambda_0) u u^\dagger + \mu_0 \frac{\partial^2}{\partial y^2} u^\dagger + i \alpha (\mu_0 + \lambda_0) \frac{\partial v}{\partial y} u^\dagger \right. \nonumber \\
   && + i \alpha \frac{\partial \mu_0}{\partial y} v u^\dagger + \frac{\partial \mu_0}{\partial \rho_0} \frac{\partial u_0}{\partial y} \frac{\partial \rho}{\partial y} u^\dagger + \frac{\partial mu_0}{\partial y} \frac{\partial u}{\partial y} u^\dagger + \frac{\partial \mu_0}{\partial T_0} \frac{\partial u_0}{\partial y} \frac{\partial T}{\partial y} u^\dagger \nonumber \\
   && + \frac{\partial \mu_0}{\partial \rho_0} \frac{\partial^2 u_0}{\partial y^2} \rho u^\dagger + \frac{\partial u_0}{\partial y} \left( \frac{\partial^2 \mu_0}{\partial {\rho_0}^2} \frac{\partial \rho_0}{\partial y} + \frac{\partial^2 \mu_0}{\partial \rho_0 \partial T_0} \frac{\partial T_0}{\partial y} \right) \rho u^\dagger \nonumber \\
   && + \frac{\partial \mu_0}{\partial T_0} \frac{\partial^2 u_0}{\partial y^2} T u^\dagger + \frac{\partial u_0}{\partial y} \left( \frac{\partial^2 \mu_0}{\partial {T_0}^2} \frac{\partial T_0}{\partial y} + \frac{\partial^2 \mu_0}{\partial T_0 \partial \rho_0} \frac{\partial \rho_0}{\partial y} \right) T u^\dagger \nonumber \\
   && - \alpha^2 \mu_0 v v^\dagger + (2 \mu_0 + \lambda_0) \frac{\partial^2 v}{\partial y^2} v^\dagger + i \alpha (\mu_0 + \lambda_0) \frac{\partial u}{\partial y} v^\dagger \nonumber \\
   && \left. + i \alpha \frac{\partial \mu_0}{\partial \rho_0} \frac{\partial u_0}{\partial y} \rho v^\dagger + i \alpha \frac{\partial \lambda_0}{\partial y} u v^\dagger + i \alpha \frac{\partial \mu_0}{\partial T_0} \frac{\partial u_0}{\partial y} T v^\dagger + \left( 2 \frac{\partial \mu_0}{\partial y} + \frac{\partial \lambda_0}{\partial y} \right) \frac{\partial v}{\partial y} v^\dagger \right] dy.
\end{eqnarray}
\section{Linear stability verification at the isothermal limit} \label{sec:Appendix_D}

\begin{figure*}
	\centering
	\subfloat[\vspace{-6mm}]{\includegraphics[width=0.49\linewidth]{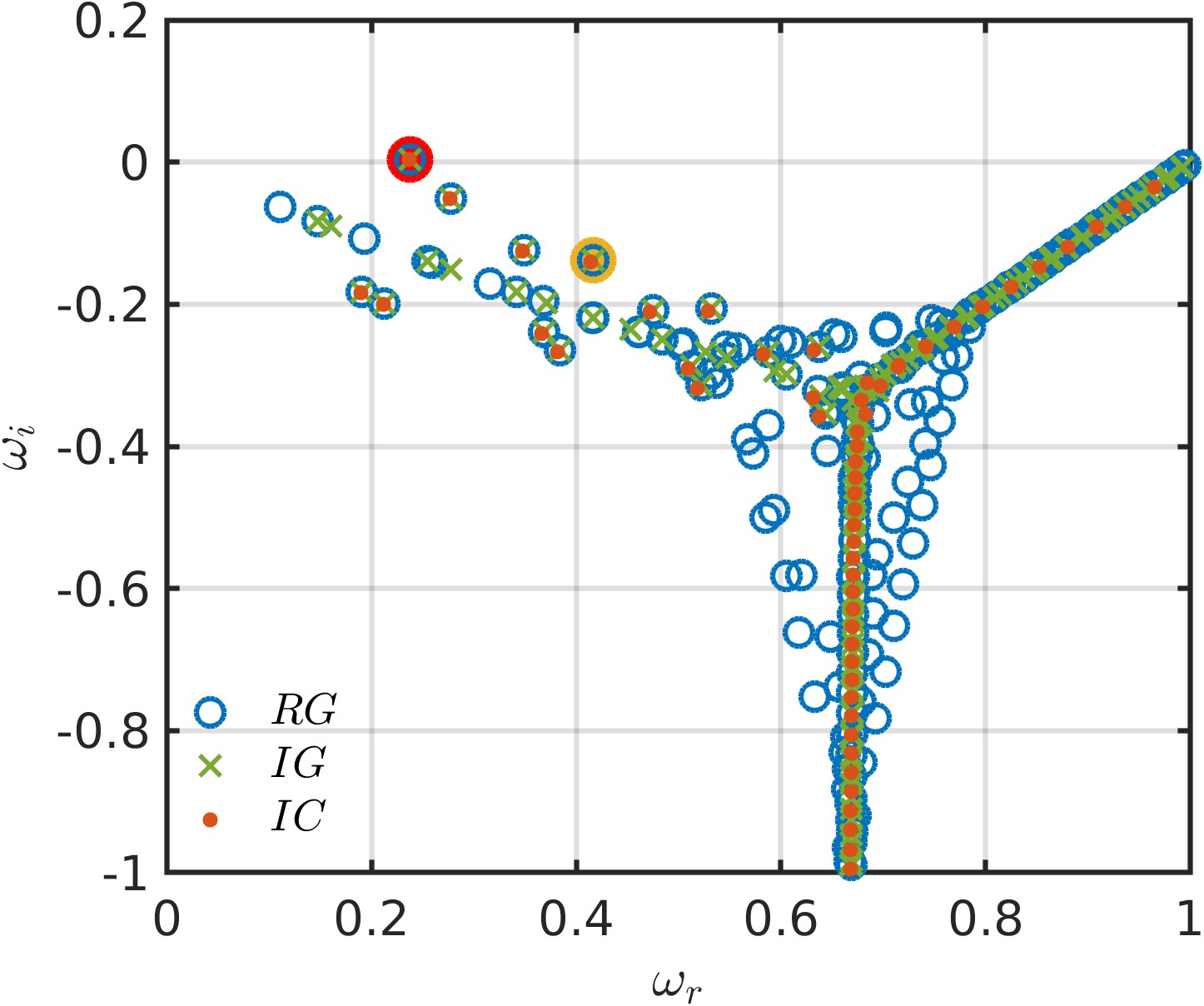}}
    \subfloat[\vspace{-6mm}]{\includegraphics[width=0.49\linewidth]{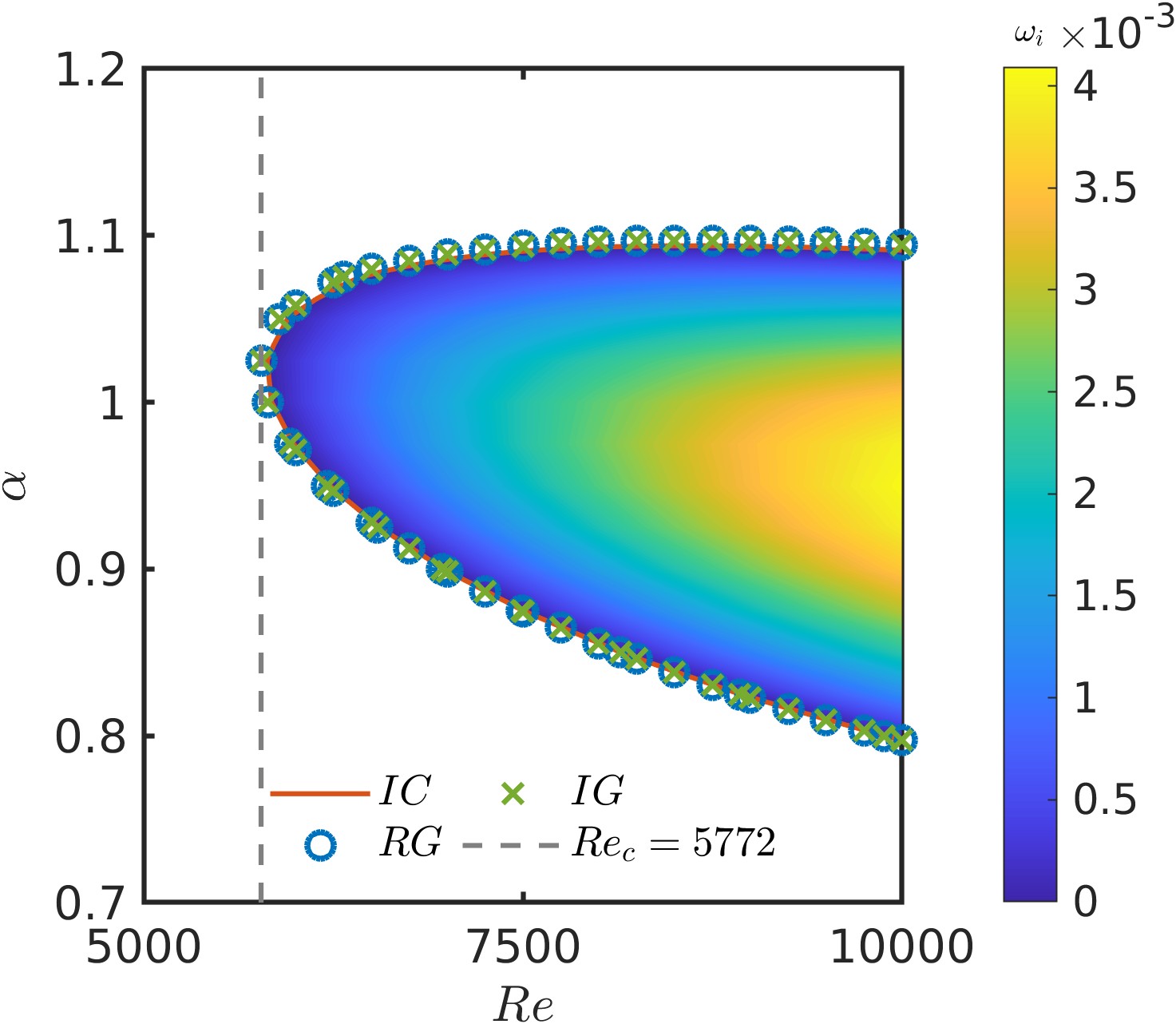}} \\ \vspace{2mm}
	\caption{(a) Eigenspectrum at $Re = 10000$ and wavenumber $\alpha = 1$ and (b) neutral curve. Real-gas framework with CoolProp thermodynamic and transport properties model (RG), ideal-gas with power law (IG), and incompressible framework (IC). In (a), the red highlighted eigenvalue corresponds to an unstable mode ($\omega = 0.2375 + 0.0037i$), whereas dark yellow corresponds to a stable mode ($\omega = 0.4164 - 0.1382i$) whose perturbations are depicted in Figure~\ref{fig:Isothermal_limit_verification_modes}.} 
 \label{fig:Isothermal_limit_verification_spectrum}
\end{figure*}

\begin{figure*}
	\centering
    \subfloat[\vspace{-6mm}]{\includegraphics[width=0.49\linewidth]{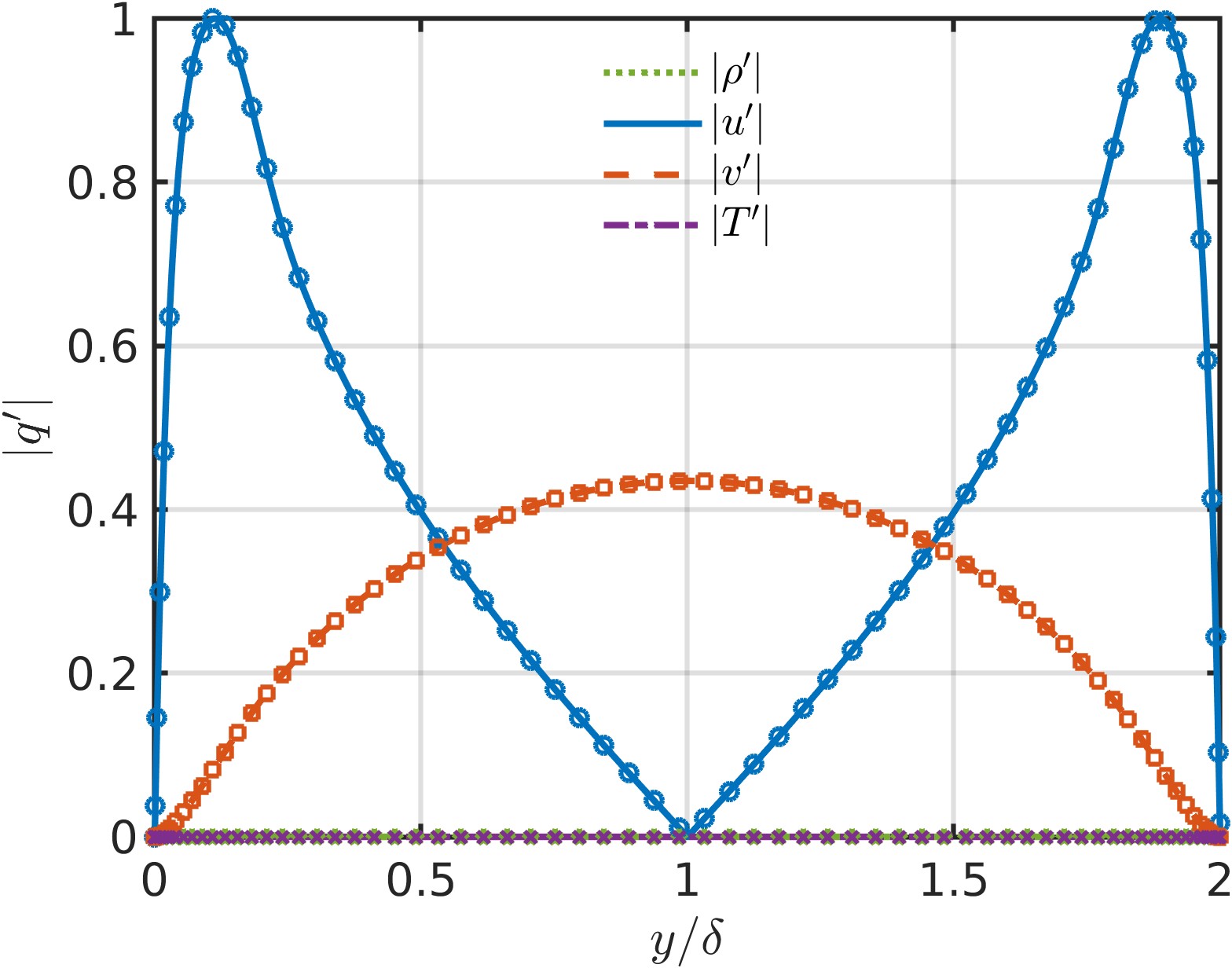}}
    \subfloat[\vspace{-6mm}]{\includegraphics[width=0.49\linewidth]{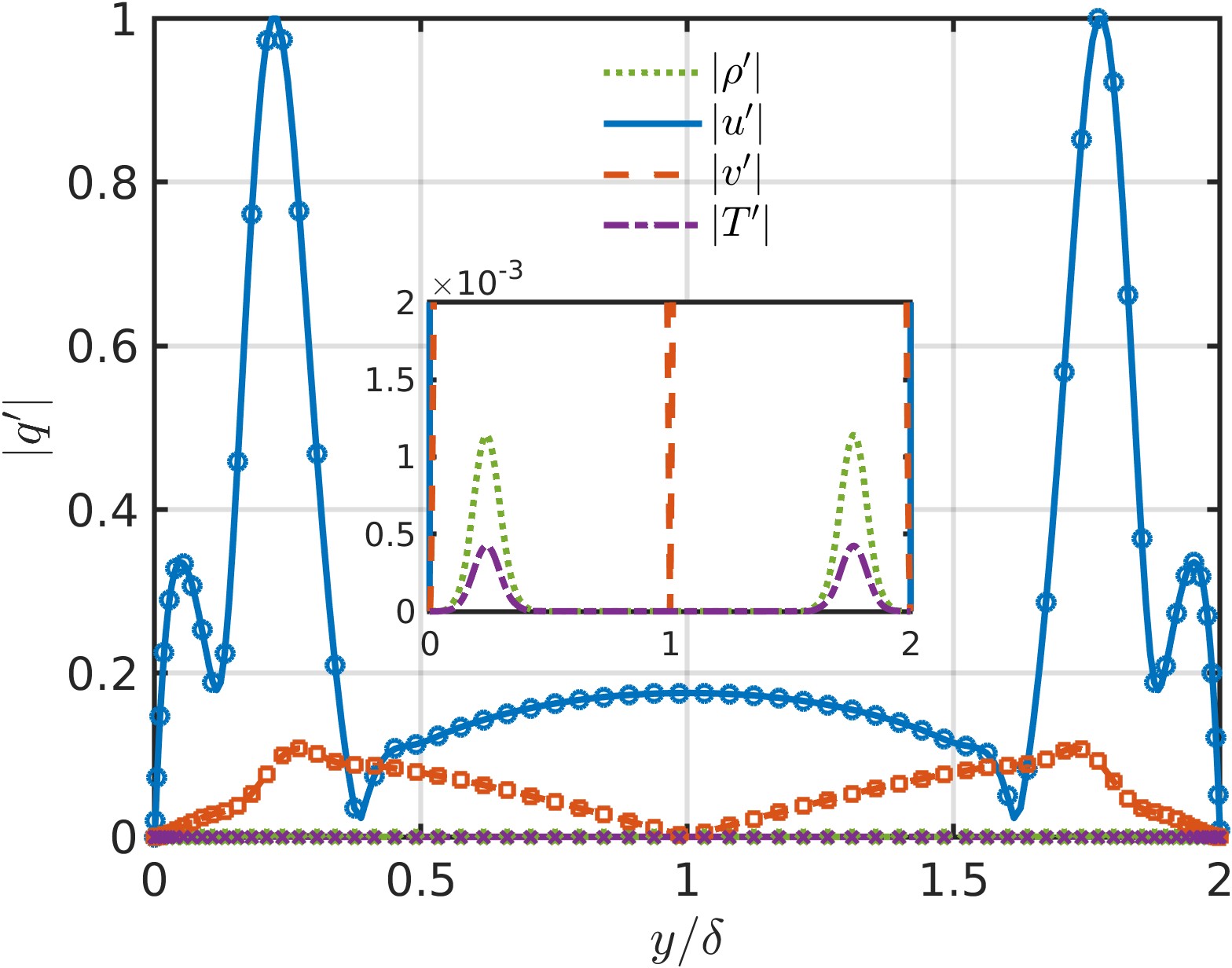}} \\ \vspace{2mm}
	\caption{Perturbation profiles of (a) unstable ($\omega = 0.2375 + 0.0037i$) and (b) stable mode ($\omega = 0.4164 - 0.1382i$) normalized by $| u^{'}|$. Real-gas framework with CoolProp thermodynamic and transport properties model (RG) depicted by solid lines, and ideal-gas with power law (IG) with markers.} 
 \label{fig:Isothermal_limit_verification_modes}
\end{figure*}

The linear stability solver is verified with respect to the incompressible reference~\citet{Trefethen1993-A} and extended replications~\citet{Trefethen2000-B}.
In this regard, the base flow is solved from the stability equations at the isothermal limit with (i) the CoolProp real-gas and transport coefficients framework, and (ii) the ideal-gas equation of state and power law for the transport coefficients.
For this comparison, wall-scalings have been used with $CO_2$ as operating fluid to match the Poiseuille flow conditions from~\citet{Ren2019-A} at $T_{cw} = T_{hw} = 290 K$ and $P = 8 MPa$.
The spectrum is computed at $Re = 10000$ and wavenumber $\alpha = 1$ solved for 2D perturbations, i.e., $\beta = 0$.
Figure~\ref{fig:Isothermal_limit_verification_spectrum} depicts the corresponding (a) eigenspectrum and (b) neutral curve.
It can be observed that the~\citet{Mack1976-A} branches named as A-,P- and S are reproduced by the incompressible case.
At the isothermal limit, the ideal-gas result collapses to these branches.
However, the real-gas solution adds additional modes due to thermodynamic effects, i.e., modes driven by thermodynamics.
The same resulting unstable mode is captured by the studied frameworks as it is a dynamic mode independent of thermodynamic effects.
Hence, the thermodynamic and transport properties models do not influence the dynamic modes in the isothermal limit.
Moreover,  the neutral curve shows good agreement for the different models considered and, in fact, they all provide the same critical Reynolds number $Re = 5772$ determined for Poiseuille flow by~\citet{Thomas1953-A} and~\citet{Orzag1971-A}.
Next, Figure~\ref{fig:Isothermal_limit_verification_modes} depicts the perturbations of the (a) unstable mode and (b) stable mode [dark yellow highlighted eigenvalue in Figure~\ref{fig:Isothermal_limit_verification_spectrum} (a)].
Thus, the statement claimed above can be verified since the unstable modes of the ideal- and real-gas frameworks are the same.
Instead, some small differences emerge for the stable mode since the compressibility effects on the density perturbation are represented by the real-gas framework as shown in the inset of Figure~\ref{fig:Isothermal_limit_verification_modes}(b).

\section{Transient growth verification} \label{sec:Appendix_E}

\begin{figure*}
	\centering
	{\includegraphics[width=0.6\linewidth]{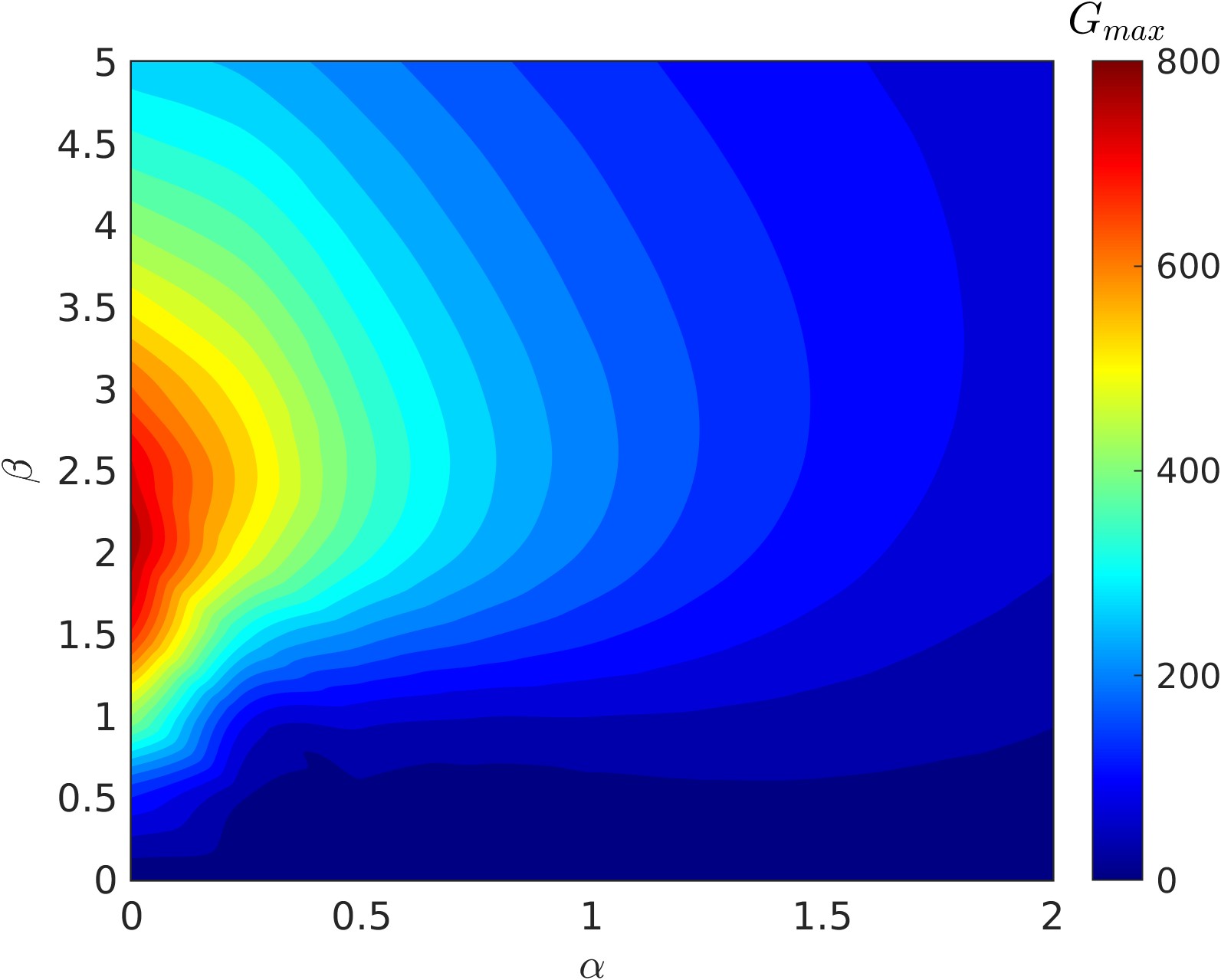}} \\
	\caption{Transient growth map at incompressible conditions ($Br \sim 0$) for $Re = 2000$.} 
 \label{fig:Transient_growth_incompressible}
\end{figure*}

\begin{figure*}
	\centering
	    \subfloat[\vspace{-6mm}]{\includegraphics[width=0.45\linewidth]{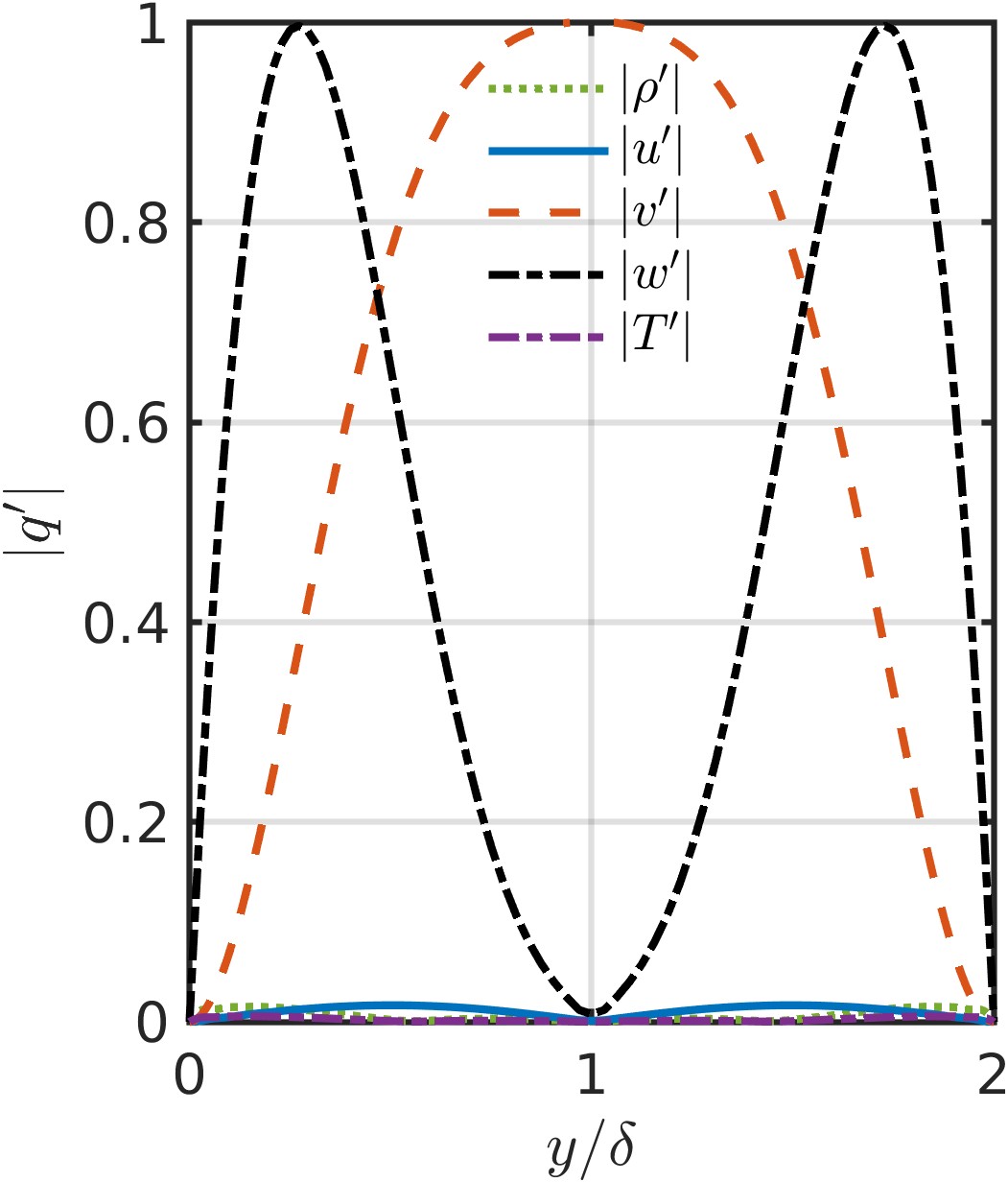}} \hspace{5mm}
    \subfloat[\vspace{-6mm}]{\includegraphics[width=0.46\linewidth]{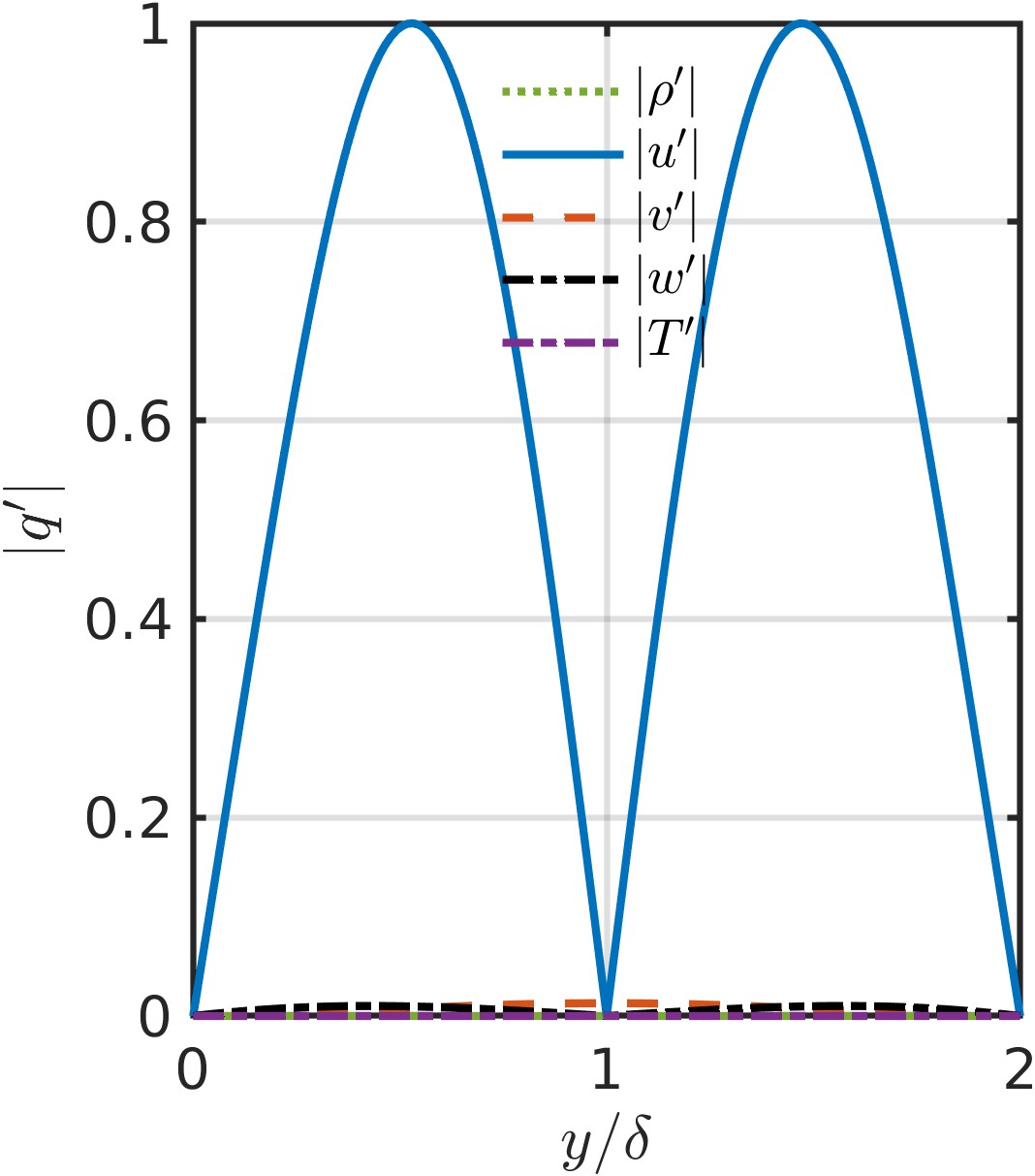}} \\ \vspace{3mm}
	\caption{Optimum eigenvector profiles at incompressible conditions ($Br \sim 0$) for $Re = 2000$ at maximum growth ($\alpha = 0$ and $\beta = 2$) for (a) input and (b) output. Results are normalized by (a) $w^\prime$ and (b) ) $u^\prime$.} 
 \label{fig:Optimum_profile_incompressible}
\end{figure*}

The transient growth framework is firstly verified against the incompressible reference~\citep{Reddy1993-A} (refer to Figure 15).
Figure~\ref{fig:Transient_growth_incompressible} highlights that the largest transient growth is achieved at $\alpha = 0$ and $\beta \sim 2$.
The classic optimal perturbation with streamwise vortices and the corresponding outputs with streaks are also recovered as depicted in Figure~\ref{fig:Optimum_profile_incompressible}.
Hence, no significant differences are observed when operating at the isothermal limit using the high-pressure framework with respect to the incompressible flow reference solution.

% Figure E.1 - Transient map

% Figure E.2 - Optimum profile

Finally, the transient growth framework is also verified against the high-pressure transcritical results presented by~\cite{Ren2019-A} at isothermal conditions with $CO_2$ as working fluid (refer to Figures 12, 13 and 14).
Despite the utilization of different thermodynamic frameworks, the largest amplifications and their corresponding wavenumbers are properly recovered.
It is important to note that the optimum perturbation profiles are also accurately obtained.
The small differences observed can be attributed to the large sensitivity of the operator to the thermodynamic framework utilized.

% Figure E.3 - Transient growth at sub-, trans- and super-critical conditions at Br = 0.07 (reference)
\begin{figure*}
	\centering
	\subfloat[\vspace{-6mm}]{\includegraphics[width=0.33\linewidth]{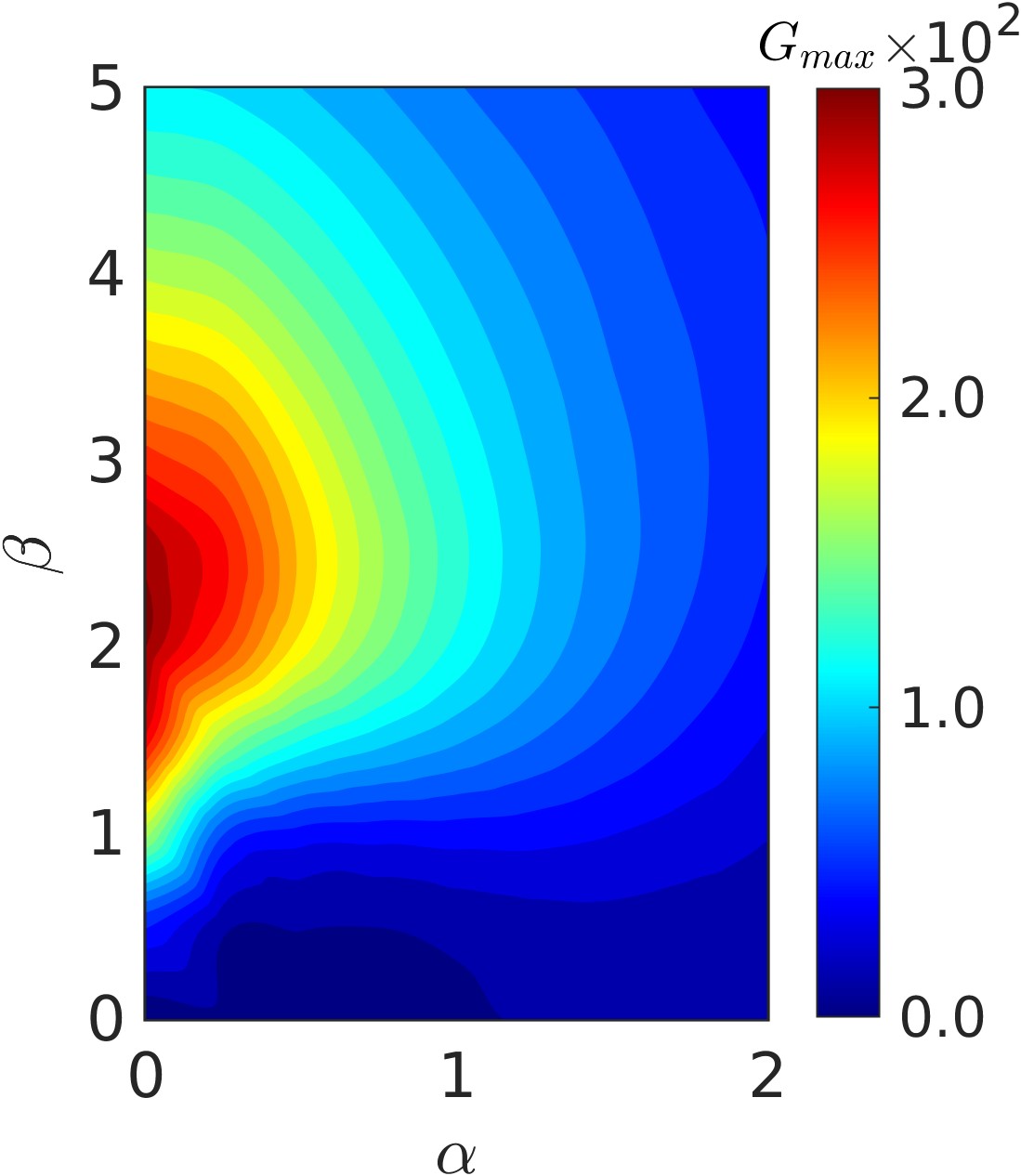}} \hfill
    \subfloat[\vspace{-6mm}]{\includegraphics[width=0.33\linewidth]{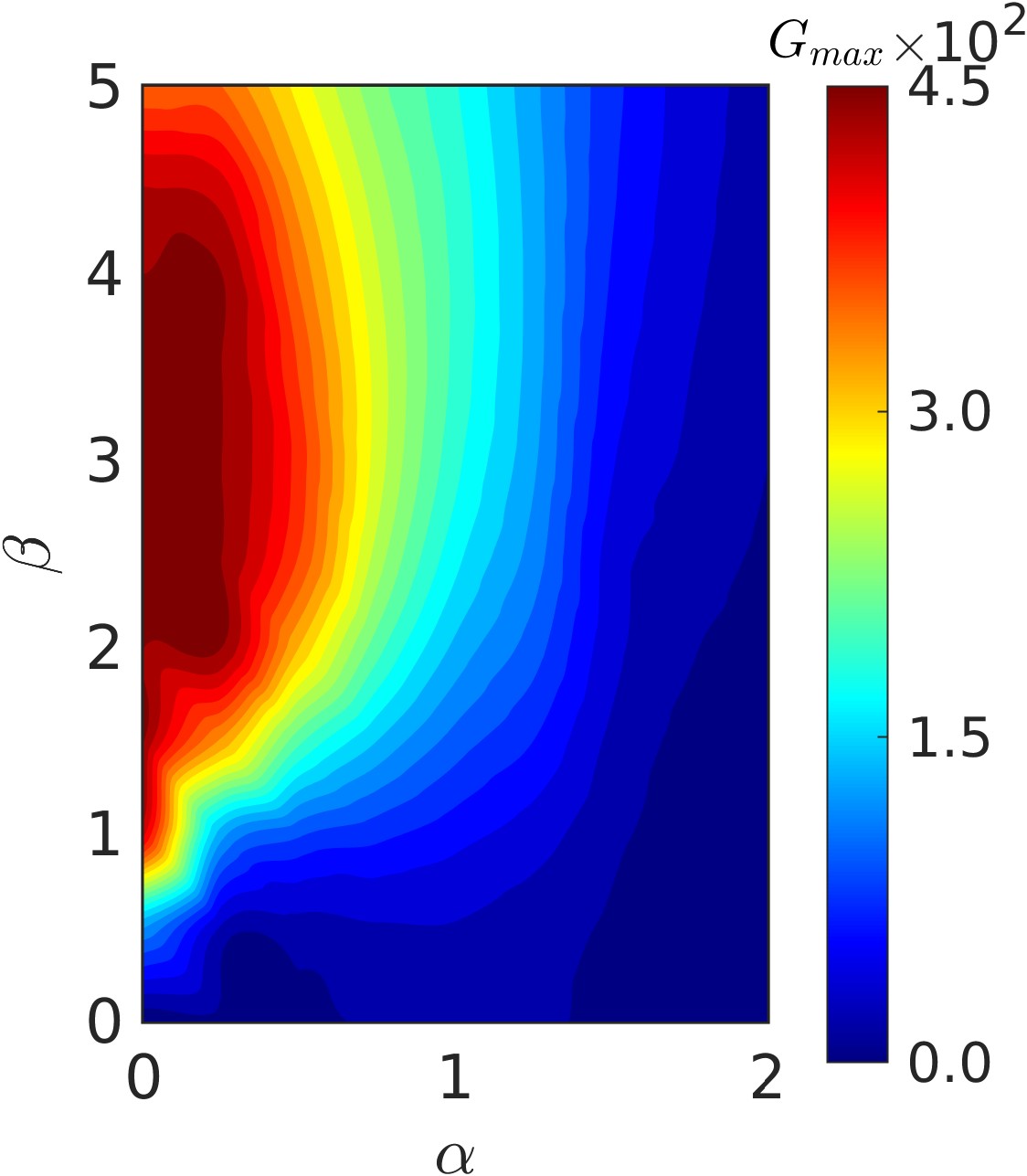}} \hfill
    \subfloat[\vspace{-6mm}]{\includegraphics[width=0.33\linewidth]{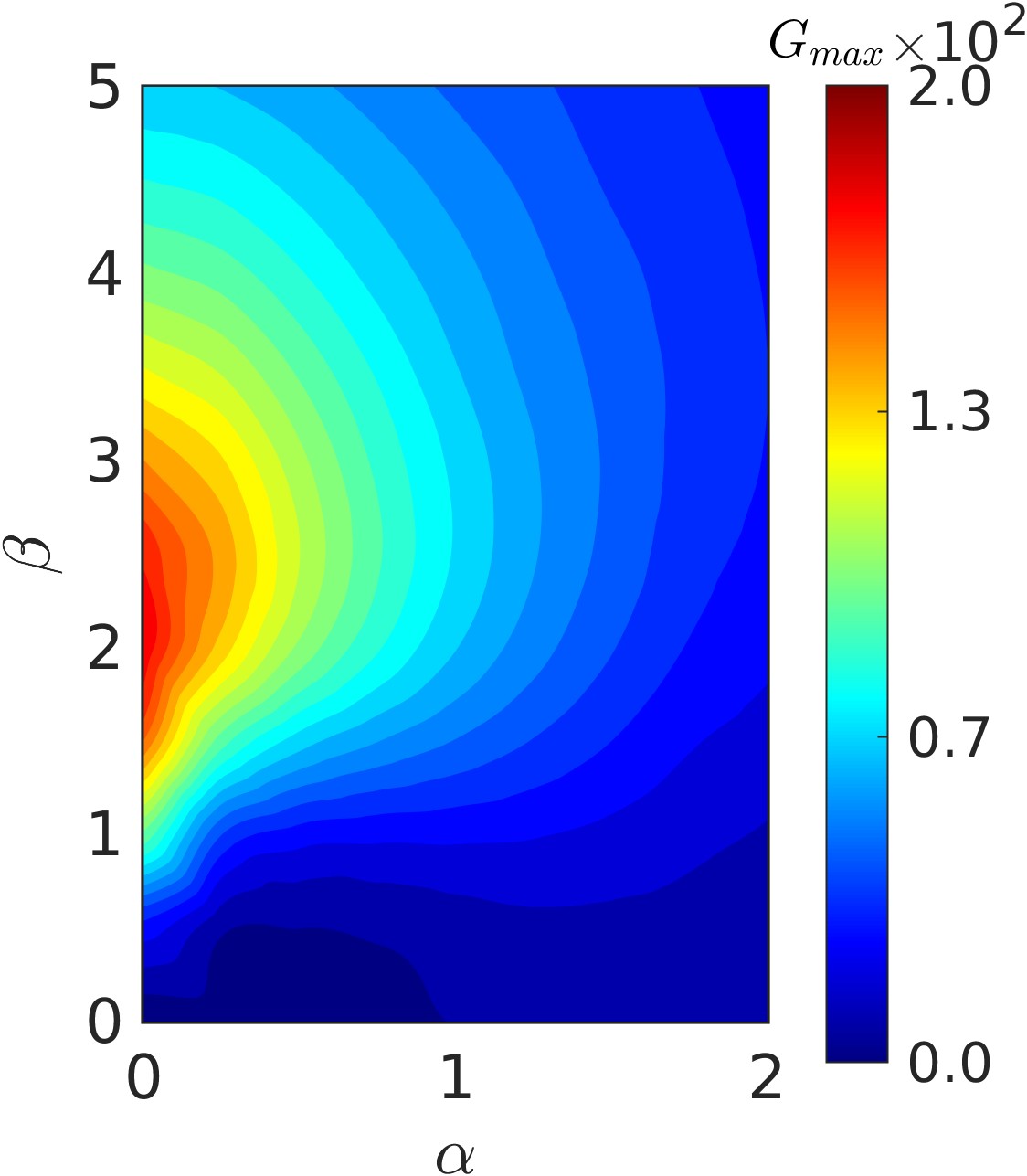}} \vspace{2mm} \\ 
	\caption{Transient growth maps at $Re = 1000$ and $Br = 0.07$ for $CO_2$ at isothermal conditions with (a) $T = 290\thinspace\textrm{K}$, (b) $T = 300\thinspace\textrm{K}$, and (c) $T = 310\thinspace\textrm{K}$.} 
 \label{fig:Transient_growth_CO2_validation}
\end{figure*}

%%%%%%%%%%%%%%%%%%%%
\bibliographystyle{jfm}
\bibliography{References}

\providecommand{\noopsort}[1]{}\providecommand{\singleletter}[1]{#1}%
\begin{thebibliography}{54}
\expandafter\ifx\csname natexlab\endcsname\relax\def\natexlab#1{#1}\fi
\def\au#1{#1} \def\ed#1{#1} \def\yr#1{#1}\def\at#1{#1}\def\jt#1{\textit{#1}}
  \def\bt#1{#1}\def\bvol#1{\textbf{#1}} \def\vol#1{#1} \def\pg#1{#1}
  \def\publ#1{#1}\def\arxiv#1{#1}\def\org#1{#1}\def\st#1{\textit{#1}}

\bibitem[Alves(2016)]{Alves2016-A}
{\sc \au{Alves, L.~S.}} \yr{2016} A classical linear stability analysis of
  normal mode instability of the compressible planar mixing-layer flow of a
  supercritical fluid.  \bt{In {\em 46th AIAA Fluid Dynamics Conference\/}},
  \pg{pp. 1--12}.  \publ{AIAA}.

\bibitem[Andreolli {\em et~al.\/}(2021)Andreolli, Quadrio \&
  Gatti]{Andreolli2021-A}
{\sc \au{Andreolli, A.}, \au{Quadrio, M.} \& \au{Gatti, D.}} \yr{2021}
  \at{Global energy budgets in turbulent couette and poiseuille flows}.  \jt{J.
  Fluid Mech.}  \bvol{924},  \pg{A25}.

\bibitem[Bell {\em et~al.\/}(2014)Bell, Wronski, Quoilin \& Lemort]{Bell2014-A}
{\sc \au{Bell, I.~H.}, \au{Wronski, J.}, \au{Quoilin, S.} \& \au{Lemort, V.}}
  \yr{2014}  \at{Pure and pseudo-pure fluid thermophysical property evaluation
  and the open-source thermophysical property library coolprop}.  \jt{Ind. Eng.
  Chem. Res.}  \bvol{53}~(6),  \pg{2498--2508}.

\bibitem[Bernades {\em et~al.\/}(2022)Bernades, Capuano \&
  Jofre]{Bernades2022c-A}
{\sc \au{Bernades, M.}, \au{Capuano, F.} \& \au{Jofre, L.}} \yr{2022}  \at{Flow
  physics characterization of microconfined high-pressure transcritical
  turbulence}.  \jt{Proceedings of the Summer Program 2022, Center for
  Turbulence Research, Stanford University}  \pg{pp. 215--224}.

\bibitem[Bernades {\em et~al.\/}(2023{\natexlab{{\em a\/}}})Bernades, Capuano
  \& Jofre]{Bernades2023-A}
{\sc \au{Bernades, M.}, \au{Capuano, F.} \& \au{Jofre, L.}}
  \yr{2023{\natexlab{{\em a\/}}}}  \at{Microconfined high-pressure
  transcritical fluid turbulence}.  \jt{Phys. Fluids}  \bvol{35},  \pg{015163}.

\bibitem[Bernades {\em et~al.\/}(2024)Bernades, Capuano \&
  Jofre]{Bernades2024c-A}
{\sc \au{Bernades, M.}, \au{Capuano, F.} \& \au{Jofre, L.}} \yr{2024} Linear
  stability exploration of transcritical non-isothermal poiseuille flows.
  \bt{In {\em 9th European Congress on Computational Methods in Applied
  Sciences and Engineering\/}},  \pg{pp. 1--12}.  \publ{ECCOMAS}.

\bibitem[Bernades \& Jofre(2022)]{Bernades2022-A}
{\sc \au{Bernades, M.} \& \au{Jofre, L.}} \yr{2022}  \at{Thermophysical
  analysis of microconfined turbulent flow regimes at supercritical fluid
  conditions in heat transfer applications}.  \jt{J. Heat Transfer}
  \bvol{144},  \pg{082501}.

\bibitem[Bernades {\em et~al.\/}(2023{\natexlab{{\em b\/}}})Bernades, Jofre \&
  Capuano]{Bernades2023c-A}
{\sc \au{Bernades, M.}, \au{Jofre, L.} \& \au{Capuano, F.}}
  \yr{2023{\natexlab{{\em b\/}}}} Non-dissipative large-eddy simulation of
  high-pressure transcritical turbulent flows: formulation and a priori
  analysis.  \bt{In {\em 14th International ERCOFTAC Symposium on Engineering
  Turbulence Modelling and Measurements\/}},  \pg{pp. 546--551}.
  \publ{ERCOFTAC}.

\bibitem[Boomkamp \& Miesen(1996)]{Boomkamp1996-A}
{\sc \au{Boomkamp, P.~A.~M.} \& \au{Miesen, R.~H.~M.}} \yr{1996}
  \at{Classification of instabilities in parallel two-phase flow}.  \jt{Int. J.
  Multiph. Fl.}  \bvol{22},  \pg{67--88}.

\bibitem[Burcat \& Ruscic(2005)]{Burcat2005-TR}
{\sc \au{Burcat, A.} \& \au{Ruscic, B.}} \yr{2005}  \bt{Third millennium ideal
  gas and condensed phase thermochemical database for combustion with updates
  from active thermochemical tables}. {\em Tech. Rep.\/}.  \org{Argonne
  National Laboratory}.

\bibitem[Busse(1969)]{Busse1969-A}
{\sc \au{Busse, F.~H.}} \yr{1969}  \at{Bounds on the transport of mass and
  momentum by turbulent flow between parallel platesw}.  \jt{Z. Angew. Math.
  Phys.}  \bvol{20}~(1),  \pg{1--14}.

\bibitem[Cabrit(2020)]{Nastro2020-B}
{\sc \au{Cabrit, O.}} \yr{2020} {\em Non-modal stability of variable-density
  round jets\/}.  \publ{Toulouse (France): Universit\'e de Toulouse}.

\bibitem[Chu(1965)]{Chu1965-A}
{\sc \au{Chu, B.~T.}} \yr{1965}  \at{On the energy transfer to small
  disturbances in fluid flow (part {I})}.  \jt{Acta Mechanica}  \bvol{1},
  \pg{215--234}.

\bibitem[Chung {\em et~al.\/}(1988)Chung, Ajlan, Lee \& Starling]{Chung1988-A}
{\sc \au{Chung, T.~H.}, \au{Ajlan, M.}, \au{Lee, L.~L.} \& \au{Starling,
  K.~E.}} \yr{1988}  \at{Generalized multiparameter correlation for nonpolar
  and polar fluid transport properties}.  \jt{Ind. Eng. Chem. Fund.}
  \bvol{27},  \pg{671--679}.

\bibitem[Chung {\em et~al.\/}(1984)Chung, Lee \& Starling]{Chung1984-A}
{\sc \au{Chung, T.~H.}, \au{Lee, L.~L.} \& \au{Starling, K.~E.}} \yr{1984}
  \at{Applications of kinetic gas theories and multiparameter correlation for
  prediction of dilute gas viscosity and thermal conductivity}.  \jt{Ind. Eng.
  Chem. Fund.}  \bvol{23},  \pg{8--13}.

\bibitem[Drazin \& Reid(1981)]{Drazin1981-A}
{\sc \au{Drazin, P.~G.} \& \au{Reid, W.~H.}} \yr{1981}  \at{Hydrodynamic
  stability}.  \jt{J. Fluid Mech.}  \bvol{124},  \pg{529--532}.

\bibitem[Firoozabadi(2016)]{Firoozabadi2016-B}
{\sc \au{Firoozabadi, A.}} \yr{2016} {\em {T}hermodynamics and {A}pplications
  in {H}ydrocarbon {E}nergy {P}roduction\/}, 1st edn.  \publ{McGraw-Hill
  Education, New York (USA)}.

\bibitem[Ghosh {\em et~al.\/}(2019)Ghosh, Loiseau, Breugem \&
  Brandt]{Ghosh2019-A}
{\sc \au{Ghosh, S.}, \au{Loiseau, J-C.}, \au{Breugem, W-P.} \& \au{Brandt, L.}}
  \yr{2019}  \at{Modal and non-modal linear stability of poiseuille flow
  through a channel with a porous substrate}.  \jt{Eur. J. Mech. B Fluids}
  \bvol{75},  \pg{29--43}.

\bibitem[Govindarajan \& Sahu(2014)]{Govindarajan2014-A}
{\sc \au{Govindarajan, R.} \& \au{Sahu, K.~C.}} \yr{2014}  \at{Instabilities in
  viscosity-stratified flow}.  \jt{Annu. Rev. Fluid Mech.}  \bvol{46}~(1),
  \pg{331--353}.

\bibitem[Hanifi {\em et~al.\/}(1996)Hanifi, Schmid \& Henningson]{Hanifi1996-A}
{\sc \au{Hanifi, A.}, \au{Schmid, P.~J.} \& \au{Henningson, D.~S.}} \yr{1996}
  \at{Transient growth in compressible boundary layer flow}.  \jt{Phys. Fluids}
   \bvol{8}~(3),  \pg{826--837}.

\bibitem[Jofre {\em et~al.\/}(2023)Jofre, Bernades \& Capuano]{Jofre2023-A}
{\sc \au{Jofre, L.}, \au{Bernades, M.} \& \au{Capuano, F.}} \yr{2023}
  \at{Dimensionality reduction of non-buoyant microconfined high-pressure
  transcritical fluid turbulence}.  \jt{Int. J. Heat Fluid Flow}  \bvol{102},
  \pg{109169}.

\bibitem[Jofre {\em et~al.\/}(2020)Jofre, del Rosario \&
  Iaccarino]{Jofre2020b-A}
{\sc \au{Jofre, L.}, \au{del Rosario, Z.~R.} \& \au{Iaccarino, G.}} \yr{2020}
  \at{Data-driven dimensional analysis of heat transfer in irradiated
  particle-laden turbulent flow}.  \jt{Int. J. Multiph. Fl.}  \bvol{125},
  \pg{103198}.

\bibitem[Jofre \& Urzay(2020)]{Jofre2020-A}
{\sc \au{Jofre, L.} \& \au{Urzay, J.}} \yr{2020}  \at{A characteristic length
  scale for density gradients in supercritical monocomponent flows near
  pseudoboiling}.  \jt{Annual Research Briefs, Center for Turbulence Research,
  Stanford University}  \pg{pp. 277--282}.

\bibitem[Jofre \& Urzay(2021)]{Jofre2021-A}
{\sc \au{Jofre, L.} \& \au{Urzay, J.}} \yr{2021}  \at{Transcritical
  diffuse-interface hydrodynamics of propellants in high-pressure combustors of
  chemical propulsion systems}.  \jt{Prog. Energy Combust. Sci.}  \bvol{82},
  \pg{100877}.

\bibitem[Joseph \& Carmi(1969)]{Joseph1969-A}
{\sc \au{Joseph, DD} \& \au{Carmi, S}} \yr{1969}  \at{Stability of poiseuille
  flow in pipes, annuli, and channels}.  \jt{Q Appl Math.}  \bvol{26}~(4),
  \pg{575--599}.

\bibitem[Linstrom \& Mallard(2021)]{NIST-M}
{\sc \au{Linstrom, P.~J.} \& \au{Mallard, W.~G.}} \yr{2021} Thermophysical
  properties of fluid systems, {NIST} {C}hemistry {W}ebbook ({SRD} 69).

\bibitem[Mack(1976)]{Mack1976-A}
{\sc \au{Mack, L.~M.}} \yr{1976}  \at{A numerical study of the temporal
  eigenvalue spectrum of the blasius boundary layer}.  \jt{J. Fluid Mech.}
  \bvol{73}~(3),  \pg{497--520}.

\bibitem[Malik {\em et~al.\/}(2006)Malik, Alam \& Dey]{Malik2006-A}
{\sc \au{Malik, M.}, \au{Alam, M.} \& \au{Dey, J.}} \yr{2006}  \at{Nonmodal
  energy growth and optimal perturbations in compressible plane couette flow}.
  \jt{Phys. Fluids}  \bvol{18},  \pg{034103}.

\bibitem[Malik {\em et~al.\/}(2008)Malik, Dey \& Alam]{Malik2008-A}
{\sc \au{Malik, M.}, \au{Dey, J.} \& \au{Alam, M.}} \yr{2008}  \at{Linear
  stability, transient energy growth, and the role of viscosity stratification
  in compressible plane couette flow}.  \jt{Phys. Rev. E}  \bvol{77}~(3),
  \pg{036322}.

\bibitem[Massaro {\em et~al.\/}(2023)Massaro, Martinelli, Schmid \&
  Quadrio]{Massaro2023-A}
{\sc \au{Massaro, D.}, \au{Martinelli, F.}, \au{Schmid, P.} \& \au{Quadrio,
  M.}} \yr{2023}  \at{Linear stability of poiseuille flow over a steady
  spanwise stokes layer}.  \jt{Phys. Rev. Fluids}  \bvol{8},  \pg{103902}.

\bibitem[Moeleker(1998)]{Moeleker1998-B}
{\sc \au{Moeleker, P.~J.~J.}} \yr{1998} {\em Linear temporal stability
  analysis\/}.  \publ{Delf (The Netherlands): Delft University Press}.

\bibitem[Orr(1907)]{Orr1907-B}
{\sc \au{Orr, William~M'F.}} \yr{1907} The stability or instability of the
  steady motions of a perfect liquid and of a viscous liquid. part ii: A
  viscous liquid.  \bt{In {\em Proceedings of the Royal Irish Academy. Section
  A: Mathematical and Physical Sciences\/}}, ,  \vol{vol.~27},  \pg{pp.
  69--138}.  \publ{Royal Irish Academy}.

\bibitem[Orzag(1971)]{Orzag1971-A}
{\sc \au{Orzag, S.~A.}} \yr{1971}  \at{Accurate solution of the
  orr–sommerfeld stability equation}.  \jt{J. Fluid Mech.}  \bvol{50},
  \pg{689--703}.

\bibitem[Poling {\em et~al.\/}(2001)Poling, Prausnitz \&
  O’Connell]{Poling2001-B}
{\sc \au{Poling, B.~E.}, \au{Prausnitz, J.~M.} \& \au{O’Connell, J.~P.}}
  \yr{2001} {\em Properties of Gases and Liquids\/}, 5th edn.  \publ{McGraw
  Hill, New York (USA)}.

\bibitem[Potter \& Graber(1972)]{Potter1972-A}
{\sc \au{Potter, M.~C.} \& \au{Graber, E.}} \yr{1972}  \at{Estability of plane
  poiseuille flow with heat transfer}.  \jt{Phys. Fluids}  \bvol{15}~(3),
  \pg{387--391}.

\bibitem[Reddy \& Henningson(1993)]{Reddy1993-A}
{\sc \au{Reddy, S.~C.} \& \au{Henningson, D.~S.}} \yr{1993}  \at{Energy growth
  in viscous channel flows}.  \jt{J. Fluid Mech.}  \bvol{252},  \pg{209--238}.

\bibitem[Ren {\em et~al.\/}(2019{\natexlab{{\em a\/}}})Ren, Fu \&
  Pecnik]{Ren2019b-A}
{\sc \au{Ren, J.}, \au{Fu, S.} \& \au{Pecnik, R.}} \yr{2019{\natexlab{{\em
  a\/}}}}  \at{Linear instability of poiseuille flows with highly non-ideal
  fluids}.  \jt{J. Fluid Mech.}  \bvol{859},  \pg{89--125}.

\bibitem[Ren {\em et~al.\/}(2019{\natexlab{{\em b\/}}})Ren, Marxen \&
  Pecnik]{Ren2019-A}
{\sc \au{Ren, J.}, \au{Marxen, O.} \& \au{Pecnik, R.}} \yr{2019{\natexlab{{\em
  b\/}}}}  \at{Boundary-layer stability of supercritical fluids in the vicinity
  of the widom line}.  \jt{J. Fluid Mech.}  \bvol{871},  \pg{831--864}.

\bibitem[Reynolds \& Colonna(2019)]{Reynolds2019-B}
{\sc \au{Reynolds, W.~C.} \& \au{Colonna, P.}} \yr{2019} {\em {T}hermodynamics:
  {F}undamentals and {E}ngineering {A}pplications\/}, 1st edn.  \publ{Cambridge
  University Press, Cambridge (UK)}.

\bibitem[Sahu \& Matar(2010)]{Sahu2010-A}
{\sc \au{Sahu, K.~C.} \& \au{Matar, O.~K.}} \yr{2010}  \at{Stability of plane
  channel flow with viscous heating}.  \jt{J. Fluids Eng.}  \bvol{132},
  \pg{011202--1}.

\bibitem[Saikia {\em et~al.\/}(2017)Saikia, Ramachandran, Sinha \&
  Govindarajan]{Saikia2017-A}
{\sc \au{Saikia, B.}, \au{Ramachandran, A.}, \au{Sinha, K.} \&
  \au{Govindarajan, R.}} \yr{2017}  \at{Effects of viscosity and conductivity
  stratification on the linear stability and transient growth within
  compressible couette flow}.  \jt{Phys. Fluids}  \bvol{29}~(2),  \pg{024105}.

\bibitem[Samanta(2020)]{Samanta2020-A}
{\sc \au{Samanta, A.}} \yr{2020}  \at{Linear stability of a plane
  couette–poiseuille ﬂow overlying a porous layer}.  \jt{Int. J. Multiph.
  Fl.}  \bvol{123},  \pg{103160}.

\bibitem[Sarkar(1995)]{Sarkar1995-A}
{\sc \au{Sarkar, S.}} \yr{1995}  \at{The stabilizing effect of compressibility
  in turbulent shear flow}.  \jt{J. Fluid Mech.}  \bvol{282},  \pg{163--186}.

\bibitem[Schmid(2007)]{Schmid2007-A}
{\sc \au{Schmid, P.~J.}} \yr{2007}  \at{Nonmodal stability theory}.  \jt{Annu.
  Rev. Fluid Mech.}  \bvol{39},  \pg{129--62}.

\bibitem[Schmid \& Brandt(2014)]{Schmid2014-A}
{\sc \au{Schmid, P.~J.} \& \au{Brandt, L.}} \yr{2014}  \at{Analysis of fluid
  systems: stability, receptivity, sensitivity}.  \jt{Appl. Mech. Rev.}
  \bvol{66},  \pg{024803}.

\bibitem[Sommerfeld(1908)]{Sommerfeld1908-A}
{\sc \au{Sommerfeld, A.}} \yr{1908} Ein beitrag zur hydrodynamischen erklarung
  der turbulenten flussigkeisbewegung.  \bt{In {\em Atti Congr. Int. Math. 4th
  Rome\/}},  \pg{pp. 116--124}.  \publ{Academia dei Lincei}.

\bibitem[Srivastava {\em et~al.\/}(2017)Srivastava, Dalal, Sahu \&
  Biswas]{Srivastava2017-A}
{\sc \au{Srivastava, H.}, \au{Dalal, A.}, \au{Sahu, K.~C.} \& \au{Biswas, G.}}
  \yr{2017}  \at{Temporal linear stability analysis of an entry flow in a
  channel with viscous heating}.  \jt{Int. J. Heat Mass Transf.}  \bvol{109},
  \pg{922--929}.

\bibitem[Thomas(1953)]{Thomas1953-A}
{\sc \au{Thomas, L.~H.}} \yr{1953}  \at{The stability of plane poiseuille
  flow}.  \jt{Phys. Rev.}  \bvol{91}~(4),  \pg{780--783}.

\bibitem[Thummar {\em et~al.\/}(2024)Thummar, Bhoraniya \&
  Narayanan]{Thumar2024-A}
{\sc \au{Thummar, M.}, \au{Bhoraniya, R.} \& \au{Narayanan, V.}} \yr{2024}
  \at{Transient energy growth analysis of flat-plate boundary layer with an
  oblique and non-uniform wall suction and injection}.  \jt{Int. J. Heat Fluid
  Flow}  \bvol{106},  \pg{109275}.

\bibitem[Trefethen(2000)]{Trefethen2000-B}
{\sc \au{Trefethen, Ll.~.N.}} \yr{2000} {\em Spectral Methods in MATLAB\/}.
  \publ{Society for Industrial and Applied Mathematics}.

\bibitem[Trefethen {\em et~al.\/}(1993)Trefethen, Trefethen, Reddy \&
  Driscoll]{Trefethen1993-A}
{\sc \au{Trefethen, L.~N.}, \au{Trefethen, A.~E.}, \au{Reddy, S.~C.} \&
  \au{Driscoll, T.~A.}} \yr{1993}  \at{Hydrodynamic stability without
  eigenvalues}.  \jt{Science}  \bvol{261}~(5121),  \pg{578--584}.

\bibitem[Wall(1996)]{Wall1996-A}
{\sc \au{Wall, W.}} \yr{1996}  \at{The linear stability of channel flow of
  fluid with temperature-dependent viscosity}.  \jt{J. Fluid Mech.}
  \bvol{323},  \pg{107--132}.

\bibitem[Xie {\em et~al.\/}(2017)Xie, Karimi \& Girimaji]{Xie2017-A}
{\sc \au{Xie, Z.}, \au{Karimi, M.} \& \au{Girimaji, S.~S.}} \yr{2017}
  \at{Small perturbation evolution in compressible poiseuille flow:
  pressure–velocity interactions and obliqueness effects}.  \jt{J. Fluid
  Mech.}  \bvol{814},  \pg{249--276}.

\bibitem[Yoo(2013)]{Yoo2013-A}
{\sc \au{Yoo, J.~Y.}} \yr{2013}  \at{The turbulent flows of supercritical
  fluids with heat transfer}.  \jt{Annu. Rev. Fluid Mech.}  \bvol{45},
  \pg{495--525}.

\end{thebibliography}

\end{document}